%% file: THz_PhDThesis.tex
\documentclass[11pt,a4paper,titlepage,
chapterprefix,headsepline,parskip,pdftex,twoside,
pointlessnumbers, bibtotoc,listof=totoc]{scrbook}
\input{header}
\begin{document}

\pagenumbering{roman}
\includepdf{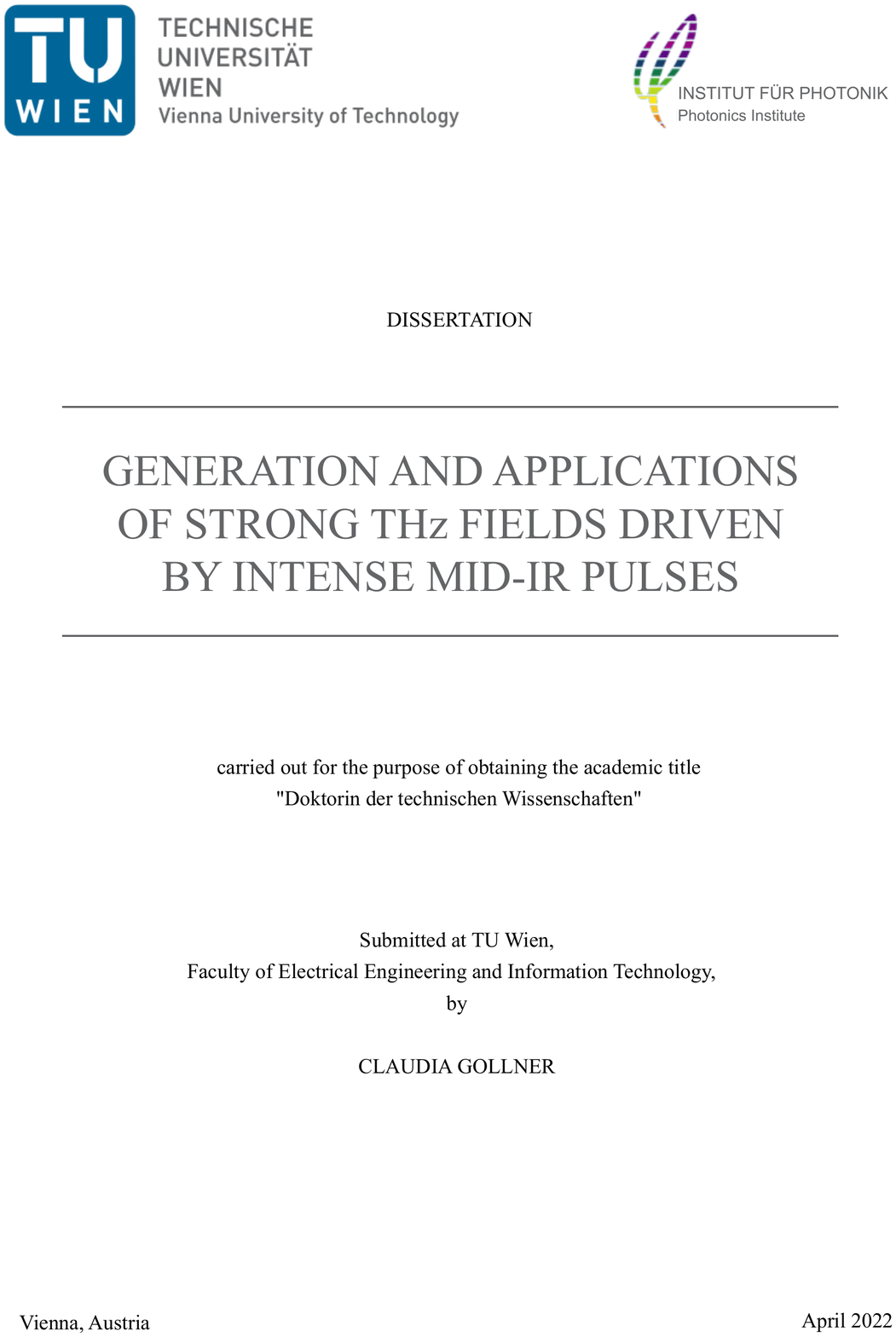}
\newpage\ \thispagestyle{empty}
\includepdf{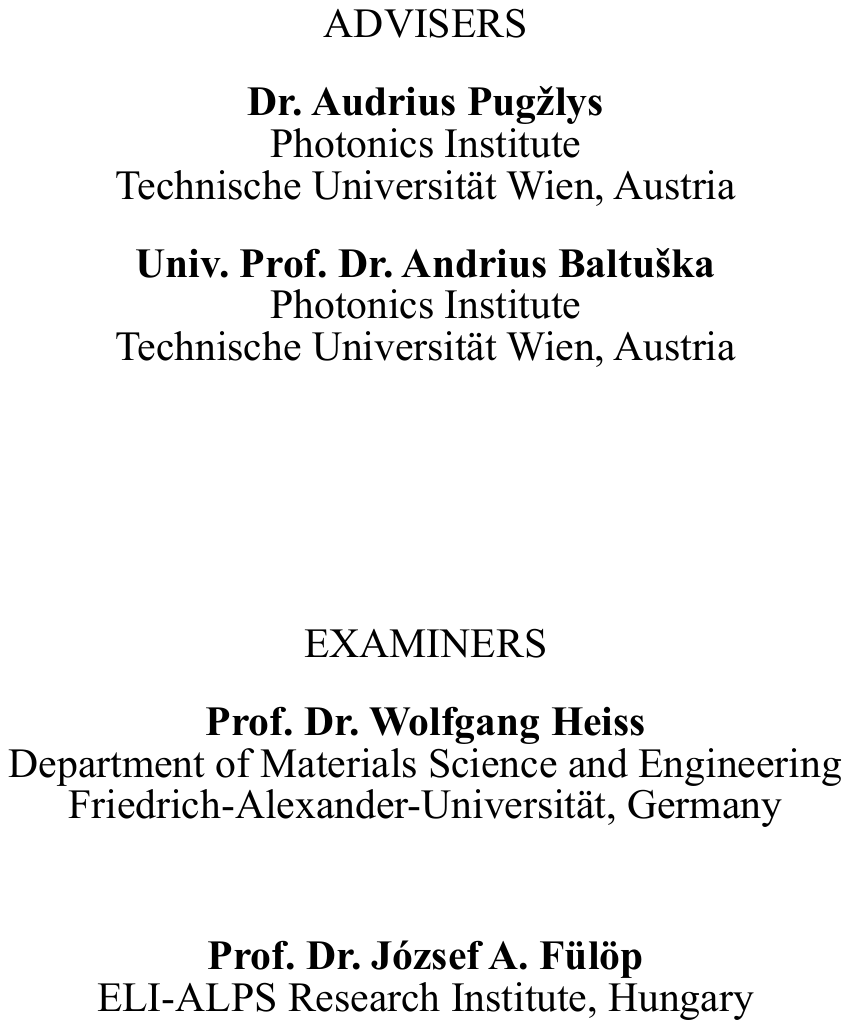}
\newpage\ \thispagestyle{empty}

\frontmatter
\input{Struktur/precontent}

\input{Struktur/acronyms}
\input{Struktur/verzeichnisse}

\mainmatter
\input{Chapters/Introduction}

\input{Chapters/THzCharacterization}

\input{Chapters/ExperimentalSetup}
\input{Chapters/OR}

\input{Chapters/QDs}
\input{Chapters/TwoColor}

\input{Chapters/XPM}
\input{Chapters/Summary}

\backmatter

\input{Struktur/acknow}


\nociteA{*}
\bibliographystyleA{unsrt} 
\bibliographyA{myPubsBib}

\nociteB{*}
\bibliographystyleB{unsrt} 
\bibliographyB{myProceedingsBib}

\nociteC{*}
\bibliographystyleC{unsrt} 
\bibliographyC{addPubs}

\bibliography{references} 
\bibliographystyle{unsrt}
\end{document}

%% file: header.tex
\usepackage{mathtools}
\usepackage{float}
\usepackage{emptypage}
\usepackage{geometry}
\usepackage[font={sf,footnotesize}, labelfont={color=myBlue,bf,sf}]{caption}
\usepackage{times}

\makeatletter 
\renewcommand{\paragraph}{\@startsection
   {paragraph} 
   {4} 
   {5mm} 
   {-\baselineskip} 
   {0.1\baselineskip} 
   {\normalsize\sfdefault}} 
\makeatother 

\makeatletter 
\renewcommand{\subparagraph}{\@startsection
   {subparagraph} 
   {5} 
   {5mm} 
   {-\baselineskip} 
   {0.1\baselineskip} 
   {\normalsize\sfdefault}} 
\makeatother 

\usepackage{braket}
\usepackage{setspace}
\usepackage{acronym}
\usepackage[pdftex]{graphicx}
\usepackage{tocbasic} 

\usepackage[dvipsnames]{xcolor}
\usepackage{pgf, float}
\usepackage[colorlinks=true,
    linkcolor=black,
    citecolor=myBlue,
    pagecolor=black,
    urlcolor=blue,
    breaklinks=true,
    bookmarksnumbered=true,
    bookmarks=false, 
    hypertexnames=true,
    pdfpagemode=UseOutlines,
    pdfview=FitH,
    plainpages=false,
    pdfpagelabels,
    bookmarks=true,
    linktocpage=true]{hyperref}

\addtokomafont{disposition}{\color{myBlue}}
\addtokomafont{chapterentry}{\normalcolor}

\usepackage[labeled,resetlabels]{multibib} 

\newcites{A,B,C}
{List of related Journal Articles,%
List of related Conference Proceedings,%
Additional Journal Articles and Proceedings}%

\usepackage{listings}
\newcounter{cfootnotecounter}

\setkomafont{chapter}{\Huge}

\RedeclareSectionCommand[
  beforeskip=0pt,
  afterindent=false,
  innerskip=0pt 
  ]{chapter}
  
\makeatletter
\renewcommand{\chapterlinesformat}[3]{%
  \ifstr{#2}{}{\centering}{}%
  \@hangfrom{#2}{#3}%
}
\makeatother

\usepackage[automark]{scrlayer-scrpage}

\clearscrheadings \clearscrplain \clearscrheadfoot
\ohead{\pagemark}
\ihead{\headmark}
\newcommand\chaptercolor{myDGrey}
\addtokomafont{pageheadfoot}{\color{\chaptercolor}}
\addtokomafont{pagenumber}{\color{\chaptercolor}}

\usepackage{array}

\usepackage[english]{babel}
\usepackage[utf8x]{inputenc}

\usepackage[T1]{fontenc}

\usepackage{url}

\usepackage{pdfpages}

\usepackage{subfig, float}													

\usepackage{cite} 
\usepackage{notoccite}
\usepackage[detect-family]{siunitx} 

\usepackage{subfiles}
\usepackage{upgreek}
\usepackage{enumitem}
\usepackage{nicefrac, amsmath,amsfonts,amsthm, amssymb} 

\usepackage{blindtext}	
\usepackage{lscape}
\usepackage{pdflscape}

\usepackage{booktabs}

\input{constants}

\input{styling}
\usepackage{ifthen}

\newcommand{\ee}{\mathrm{e}}
\newcommand{\hh}{\mathrm{h}}
\newcommand{\eh}{{\ee,\hh}}
\newcommand{\Op}[1]{\hat{\mathrm{#1}}}
\newcommand{\CdSe}{\mathrm{CdSe} }
\newcommand{\CdS}{\mathrm{CdS} }


%% file: constants.tex
\usepackage{nicefrac, amsmath}

\newcommand{\rv}{\mathbf{r}}

%% file: styling.tex
\definecolor{color01RED}{HTML}{8B1A0E}
\definecolor{color02GREEN}{HTML}{5e9c36}
\definecolor{color03BLUE}{HTML}{0a367e}
\definecolor{color04YELLOW}{HTML}{FBBB13}

\definecolor{color21GREEN01}{HTML}{006D2C}
\definecolor{color22GREEN02}{HTML}{31A354}
\definecolor{color23GREEN03}{HTML}{74C476}
\definecolor{color24GREEN04}{HTML}{BAE4B3}
\definecolor{color25GREEN05}{HTML}{EDF8E9}

\definecolor{color31BLUE01}{HTML}{810F7C}
\definecolor{color32BLUE02}{HTML}{8856A7}
\definecolor{color33BLUE03}{HTML}{8C96C6}
\definecolor{color34BLUE04}{HTML}{B3CDE3}
\definecolor{color35BLUE05}{HTML}{EDF8FB}

\definecolor{myBlue}{RGB}{45, 96, 181} 
\definecolor{myRed}{RGB}{176, 24, 2} 
\definecolor{myBGrey}{RGB}{153, 153, 152} 
\definecolor{myDGrey}{RGB}{77, 77, 77} 

%% file: Struktur/precontent.tex
\addchap{Dedication}
\begin{center}
\vspace{35mm}
$\dots$ {\large \textit{to my father}} $\dots$
\vfill
\end{center}
\newpage

\addchap{Abstract}
Although the THz spectral range experiences a tremendous grow of interest for at least two decades, it is still one of the least explored  but most exciting scientific fields to study light-matter interaction.
Due to the small photon energy of THz radiation, it can propagate through non-conductive materials and resonantly interact with low-energy excitations such as crystal lattice vibrations, molecular rotations or collective spin excitations, leading to a plethora of applications.
It can be utilized for biological tissue discrimination, 
pharmaceutical quality control, astronomy or high-speed communication, as well as for fundamental research to investigate carrier transport processes in solids, coherent control of quantum states or molecular alignment, to name only a few.
The THz pulse is thereby primarily used as a linear probe while an optical pulse excites the material that is examined.\\
In contrast, intense strong-field THz transients would allow on-demand control of the properties of matter and engineer new dynamic states in a wider range of materials.
Nonetheless, current table-top THz sources remain rather weak, with the most promising being optical rectification in nonlinear crystals and two-color plasma filaments pumped by near-infrared sources.
While the former is mainly restricted by multi-photon absorption of the short wavelength driving pulse, causing crystal damage, the latter suffers from pump pulse scattering in dense plasma and limited laser field-asymmetry. 
These limitations can be overcome with intense long wavelength driving pulses.
However,  until recently, such high-power mid-infrared sources were simply not available.\\
In this work, we exploit the recently developed mid-IR optical parametric chirped pulse amplifier (OPCPA) system to drive efficient THz generation in the organic crystal DAST (4-N,N-dimethylamino-4’-N’-methylstilbazolium tosylate) by optical rectification and in two-color plasma filaments, wherein we boost the optical- to THz conversion efficiency by almost an order of magnitude, compared to previous works with near-IR drivers.
The intense THz source is further applied to heterostructure quantum dots without any field enhancing structures to investigate the pure quantum confined Stark effect.
The possibility to manipulate optical properties of nano-scale semiconductors by direct THz radiation demonstrates the feasibility for an all-optical electro-absorption modulator with data rates in the range of Tbit/s.
\newpage

%% file: Struktur/acronyms.tex
\addchap{Acronyms}

\begin{acronym} \itemsep=-10pt
\acro{ase}[ASE]{amplified spontaneous emission}
\acro{Be}[Be]{Berryllium}
\acro{BBO}[BBO]{Beta Barium Borate}
\acro{bs}[BS]{beam splitter}
\acro{CdS}[CdS]{Cadmium sulfide}
\acro{CdSe}[CdSe]{Cadmium selenide}
\acro{cpa}[CPA]{chirped pulse amplifier}
\acro{CdTe}[CdTe]{Cadmium telluride}
\acro{cep}[CEP]{carrier envelope phase}
\acro{Cu}[Cu]{copper}
\acro{DAST}[DAST]{4-N,N-dimethylamino-4'-N'-methyl-stilbazolium tosylate}
\acro{dfg}[DFG]{difference-frequency generation}
\acro{dl}[DL]{delay line}
\acro{DSTMS}[DSTMS]{4-N,N-dimethylamino-4’-N’-methyl-stilbazolium 2,4,6-trimethylbenzenesulfonate}
\acro{ea}[EA]{electro-absorption}
\acro{eo}[EO]{electro-optic}
\acro{eoc}[EOC]{electro-optic crystal}
\acro{eos}[EOS]{electro-optic sampling}
\acro{facet}[FACET]{Facility for Advanced Acceleration Experimental Tests}
\acro{ft}[FT]{Fourier transformation}
\acro{fp}[FP]{flip mirror}
\acro{frog}[FROG]{frequency resolved optical gating}
\acro{fwhm}[FWHM]{full width at  half maximum}
\acro{fwm}[FWM]{four-wave mixing}
\acro{GaAs}[GaAs]{Gallium arsenide}
\acro{GaSe}[GaSe]{Gallium selenide}
\acro{GaP}[GaP]{Gallium phosphate}
\acro{gt}[GT]{Glan-Taylor prism}
\acro{gvd}[GVD]{group velocity dispersion}
\acro{hdpe}[HDPE]{high-density polyethylene}
\acro{hwp}[HWP]{half waveplate}
\acro{ir}[IR]{infrared}
\acro{KGW}[KGW]{Potassium Gadolinium Tungstate}
\acro{KTA}[KTA]{Potassium Titanyle Arsenate}
\acro{KTP}[KTP]{Potassium titanyl phosphate}
\acro{lcls}[LCLS]{Linac Coherent Light Source}
\acro{LN}[LN]{Lithium niobate}
\acro{lpf}[LPF]{long pass filter}
\acro{mir}[MIR]{mid-infrared}
\acro{mpa}[MPA]{multi-photon absorption}
\acro{nc}[NC]{nano-crystal}
\acro{nd}[ND]{neutral density filter}
\acro{Nd}[Nd]{Neodymium}
\acro{nep}[NEP]{noise equivalent power}
\acro{nl}[NL]{nonlinear}
\acro{nopa}[NOPA]{non-collinearly pumped optical parametric amplifier}
\acro{opa}[OPA]{optical parametric amplifier}
\acro{opcpa}[OPCPA]{optical parametric chirped pulse amplifier}
\acro{or}[OR]{optical rectification}
\acro{pc}[PC]{Pockels cell}
\acro{pca}[PCA]{photoconductive antenna}
\acro{ped}[PED]{pyro-electric detector}
\acro{pd}[PD]{photo diode}
\acro{pl}[PL]{photoluminescence}
\acro{qcse}[QCSE]{quantum confined Stark effect}
\acro{qd}[QD]{quantum dot}
\acro{qwp}[QWP]{quater waveplate}
\acro{ra}[RA]{regenerative amplifier}
\acro{sh}[SH]{second hamronic}
\acro{shg}[SHG]{second harmonic generation}
\acro{slac}[SLAC]{Stanford Linear Accelerator Center}
\acro{spm}[SPM]{self-phase modulation}
\acro{tem}[TEM]{transmission electron microscope}
\acro{tds}[TDS]{time-domain-spectroscopy}
\acro{THz}[THz]{terahertz}
\acro{tpc}[TPC]{transient photo-current}
\acro{to}[TO]{transverse optical}
\acro{wl}[WL]{white-light}
\acro{wp}[WP]{Wollaston prism}
\acro{xpm}[XPM]{cross-phase modulation}
\acro{YAG}[YAG]{yttrium aluminum garnet}
\acro{Yb}[Yb]{Ytterbium}
\acro{ZnSe}[ZnSe]{Zinc selenide}
\acro{ZnTe}[ZnTe]{Zinc telluride}
\end{acronym}

%% file: Struktur/verzeichnisse.tex

\listoffigures

\listoftables

\setcounter{tocdepth}{2}
\tableofcontents




%% file: Chapters/Introduction.tex
\chapter{Introduction}

Terahertz (THz) radiation is all around us, it is emitted as part of black-body radiation from anything with a temperature greater than a few Kelvin (0.1 THz $\simeq$ 4.8 K ), manifested as thermal emission.
The spectral content of THz radiation is sandwiched between microwaves and \ac{ir} waves (\SI{0.1}{\THz}-\SI{30}{\THz}), the so called \textit{THz gap}, wherein the technology for artificial generation and detection is still not sufficiently developed.
Despite great scientific interest since at least the 1920s \cite{Nichols:1923}, the THz frequency range remains one of the least exploited. 
At first, it was mainly utilized by astronomers due to the fact that about half the energy generated in the Universe since the Big Bang is concentrated in the THz spectral band \cite{Astro:2006}.
Back then, the spectral range was referred to as sub-millimeter wave or far-infrared radiation \cite{ RodamapDhillon:2017}.
Although the term \textit{terahertz} already occurred in 1974 \cite{Flemming:1974}, 
only in the mid-1990s this research area experienced a tremendous boost due to the pioneer work of Nuss and others in the field of THz \ac{tds} \cite{Hu:1995,Nuss:1996}.
Today, the THz spectrum covers many exciting areas, from fundamental science to 'real world' applications.

Because THz photon energies ($\sim$\SI{4}{\meV} at \SI{1}{\THz}) are far below typical electronic interband resonances, \textit{i.e.}  they are smaller than the band gap energy of non-conductive materials, THz radiation can propagate through matter.
For example, many materials like plastics, wood, paper, and clothing  
are transparent to THz radiation.
In addition, unlike X-rays, THz radiation is non-ionising and therefore doesn't cause any harmful effects to the human tissue. 
Moreover, many materials have unique resonant features (so called fingerprints) at THz frequencies that may lead to the identification and quantification of materials. 
Accordingly, there is a great potential in biomedical research, such as for blood cell detection, cancer cell characterization, bacterial identification and biological tissue discrimination \cite{MedsApp:2019, BioYang:2016}.
THz radiation can be furthermore applied for industrial quality control, food inspection,  homeland security, identification of hazardous chemicals, surveillance and security imaging, high-speed  broadband communication and many more \cite{THzSpectroscopyMasayoshi:2007, ExtremeTHz:2017,Honey:2013, DrugsPawar:2013, SalmonHindle:2018, THzHandbook:2015, THzComSuen:2016}. 

THz waves are also of great interest for purely scientific applications.
They can resonantly interact with low-energy excitations such as crystal lattice vibrations, molecular rotations 
or collective spin excitations \cite{Katayama:2012, SpinChoi:2020}. 
In contrast to optical photons, wherein the extra energy causes an unwanted temperature increase due to uncontrolled phonon excitations or hot electron distributions, THz radiation allows one to specifically target low energy excitations of interest. 
Note that, historically, the low-frequency phonon modes are referred to as the \ac{ir} modes, whereas the actual frequency of the mode is often in the THz region \cite{SalenNL_THz:2019}.
Within the last few decades, terahertz spectroscopy \cite{THzSpectroscopyMasayoshi:2007, THzSpectroscopyJepsen:2011,THzSpectroscopyBaxter:2011} 
has become a versatile tool in fundamental research to investigate structural dynamics as well as equilibrium properties of the four states of matter \cite{ApplicationKolner:2008, THzSpectroscopyBread:2002}. 
\begin{figure}[htb]
    \centering
    \includegraphics[width=\linewidth]{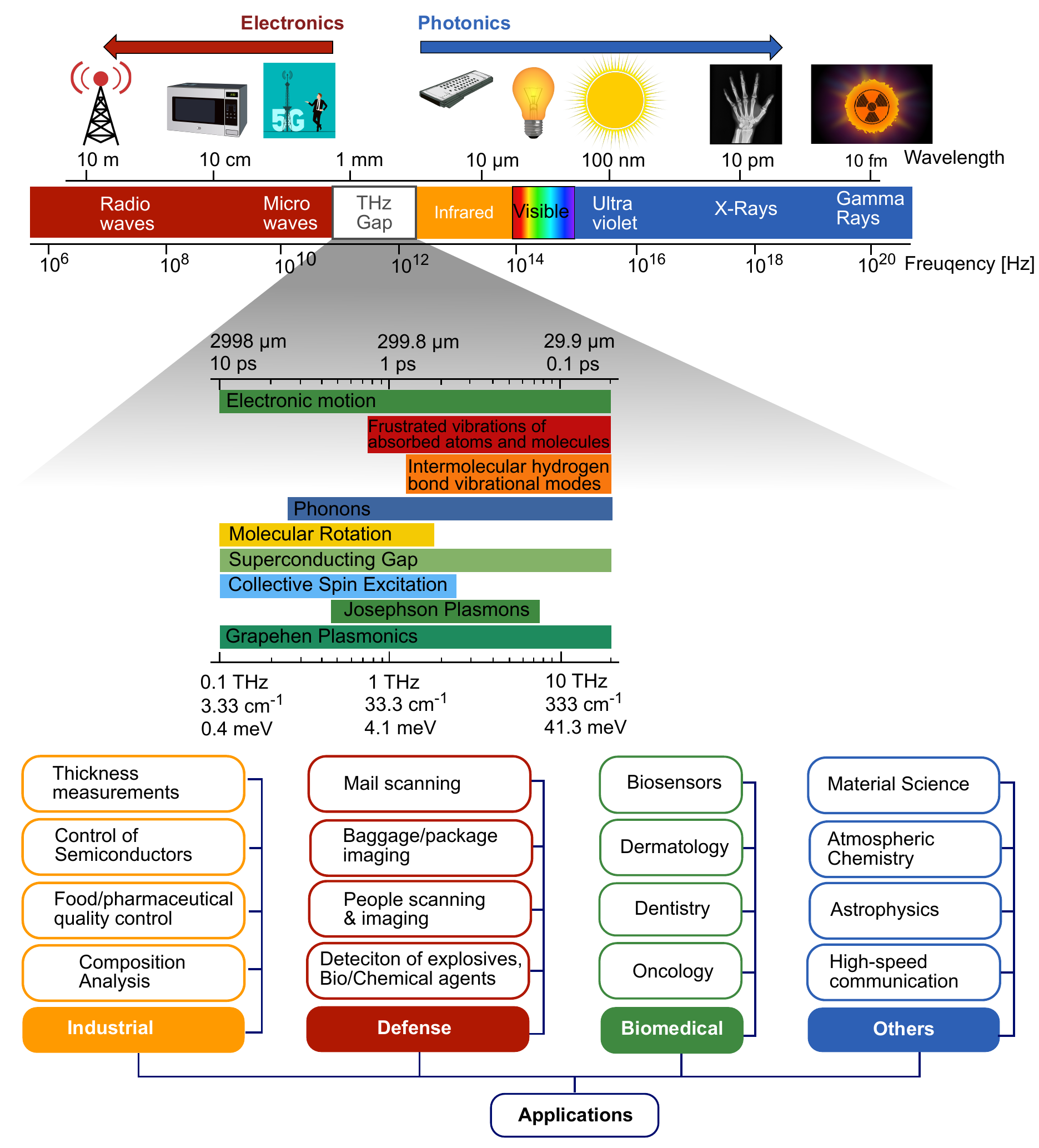}
    \caption[THz spectral range and applications]{Electromagnetic spectrum with a magnified view on fundamental excitations in solids and molecular systems in the THz range, adapted from ref.\cite{SalenNL_THz:2019}. The flowchart below illustrates a plethora of THz applications.
    }
    \label{fig:applications}
\end{figure}
Particular fields of interest range from carrier transport processes in solids \cite{Razzari:2009, Razzari:2010}, THz-driven nuclear dynamics in solids and molecules  \cite{Ham:2017}, over coherent control of quantum states \cite{ApplicationGreenland:2010}, to direct acoustic phonon excitation \cite{Phonons:2010} or molecular alignment \cite{Fleischer:2011}, to name only a few. 
Figure \ref{fig:applications} illustrates the spectral THz region and a magnified range for possible excitations in condensed matter and molecular gasses, taken from ref\cite{SalenNL_THz:2019}.

Despite the rapid development of THz science during the last two decades, the majority of available THz sources remains rather weak and the THz research area is mainly stuck in the realm of linear optics, where the THz pulse is primarily used as a linear probe while an optical pulse excites the material that is examined.
In contrast, strong-field intense THz pulses would allow on-demand control of the properties of matter and can induce an ultrafast electric- or magnetic-field switching, operating on a sub-pico second time scale, which is much faster than what can be achieved through conventional electronics \cite{IntenseHafez:2016}.
Thus, intense THz pulses with an electric field strength exceeding the intrinsic field
of matter have large potential to engineer new dynamic states in a wide range of materials and provide an opportunity to control matter by coherent lattice excitation, or to trigger transient phase transitions \cite{SalenNL_THz:2019}.
Recently, tremendous progress has been achieved in THz source development with \ac{or} of femto second (fs) pulses in \acp{eoc} and two-color plasma filaments, exceeding electric field strengths in the range of MV/cm
(a detailed overview on state-of-the-art THz sources is given in the following section). 
Consequently, groundwork has been laid by polarization switching of ferroelectric materials \cite{Grishunin:2017}, THz induced superconductivity \cite{Superconductivity:2012}, ultrafast control of magnetic domains \cite{Baierl:2016}, or table-top electron acceleration \cite{Eacc_Nanni:2015,Eacc_Zhang:2018}. 
Another new frontier of non-linear THz spectroscopy, where the THz field drives a \ac{nl} response of matter, is \ac{ea} switching in semiconductor materials, providing an opportunity to boost optical communication systems to Tbit/s data rates.

In this work, we report on THz sources with extreme electric field strengths and record optical- to THz conversion efficiencies, generated with (i) \ac{mir} two-color laser filaments, as reported in ref \citeA{TasosA:2020}, and (ii) by \ac{or} of high-energy mid-\ac{ir} pulses centered at \SI{3.9}{\micro \m} and \SI{1.95}{\micro \m} in the organic crystal DAST \citeA{GollnerAPL:2021}. 
In the case of two-color plasma filaments, the resulting THz field strength exceeds 100 MV/cm, with THz pulse energies of 0.185 mJ and conversion efficiency of 2.36\%, outpacing previous results, obtained with conventional drivers operating in the visible and near-IR spectral range, by more than an order of magnitude. 
For THz generation by OR in DAST, the long-wavelength driving pulses almost double the optical- to THz conversion efficiency - approaching 6\% - as compared to conventional pump sources at the telecommunication wavelength of \SI{1.5}{\micro \m}. 
The observed high sensitivity of  THz generation to the parameters of the mid-\ac{ir} driving pulses motivates an in-depth study of the underlying interplay of \ac{nl} wavelength- and intensity-dependent effects. 
With the obtained strong fields, it is possible to demonstrate a direct all optical encoding of a free-space, ultrafast THz signal onto an optical signal probing the absorption of CdSe/CdS core/shell \acp{qd}, based on the \ac{qcse}.
Without any field-enhancing structures, an extinction contrast of more than 6 dB and transmission changes in the visible of more than 15 \% are achieved, with the latter setting a new record for solution-processed \ac{ea} materials at room temperature. 
This method proves the feasibility for high speed modulators with transition rates in the Tbit/s range and paves the way for additional terahertz science featuring \ac{nl} light–matter interactions and applications.
\begin{figure}[htb]
    \centering
    \includegraphics[width=\linewidth]{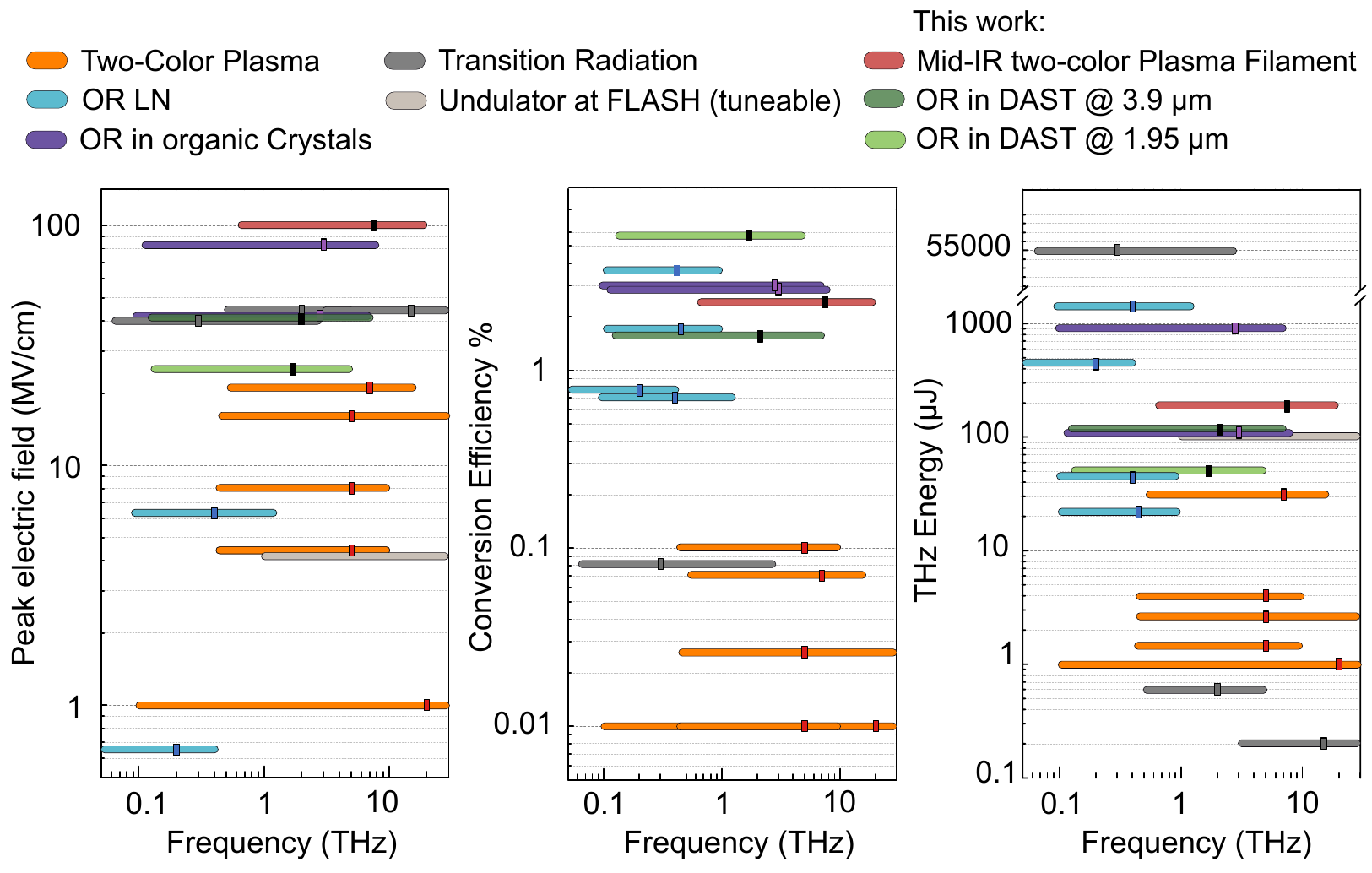}
    \caption[State-of-the-art THz sources]{Overview on state-of-the-art THz sources with respect to the electric field strength, optical- to THz conversion efficiency and generated THz energy, considering a spectral range from \SI{0.1}{\THz}-\SI{30}{\THz}. 
    Elongated bars represent the entire spectral content, small rectangles indicate the peak frequency.
    A color code is shown on top. 
    Data points are taken from the following references: two-color plasma filament \cite{PlasmaWidth:2013, SOTAplasma:2017, PlasmaKuk:2016, OhKim:2014, LongWLplasma:2013}, OR in \acf{LN} \cite{ORefficiencyHuang:2013, LiNbZhang:2021, Fulop:2014}, OR in organic crystals \cite{DSTMSVicario09:2014, BulletShalaby:2015}, transition radiation \cite{THz55mJ:2019} at the National Accelerator Laboratory (SLAC) \cite{SLAC:2013}, THz undulator beamline at FLASH \cite{FLASH:2013,FLASH:2008} and mid-\ac{ir} pump sources for two-color plasma filaments \citeA{TasosA:2020} and OR in DAST \citeA{GollnerAPL:2021}.
    The latter two techniques are presented in this work, which exhibit superior THz electric field strengths and conversion efficiencies.  
    }
    \label{fig:StateOfTheArt}
\end{figure}

\section{State of the Art THz Sources}

Despite remarkable advances in THz technology within the last decade, there is still an immediate need to develop intense THz sources and \ac{nl} THz spectroscopy techniques, providing the possibility to reveal a new category of NL phenomena and explore \ac{nl} effects in various materials.
Up until recently, intense THz radiation with MV/cm electric field strengths could only be achieved in large scale facilities, utilizing synchrotron radiation and free-electron lasers \cite{FLASH:2008, FLASH:2016}, or coherent transition radiation \cite{SLAC:2013, FACET:2011, THz55mJ:2019}.
Nowadays, strong electric fields become accessible with table-top  THz sources such as \ac{or} in organic  \cite{BulletShalaby:2015} as well as inorganic crystals \cite{Review:2021}, and two-color plasma filaments \cite{SOTAplasma:2017}.
In order to understand, compare and further develop THz sources, we briefly introduce the main state of the art techniques for generation of intense THz pulses.
However, it is important to note that a suitable THz source  is not solely determined by the pulse energy or electric field strength, but further  depends crucially on the spectral range, pulse duration and repetition rate of the radiated THz pulse.
Thus, depending on the experiment and requirements of subsequent applications, the THz source has to be chosen accordingly.
A summary of latest achievements in THz source development is provided in Fig.\ref{fig:StateOfTheArt}.  
Similar representations can be found in ref \cite{SalenNL_THz:2019, StrongFieldFulop:2020}.

\subsection{THz Radiation from accelerated Electron Bunches}

\subsubsection{Synchrotron Radiation}
\begin{figure}[htb]
    \centering
    \includegraphics[width=\linewidth]{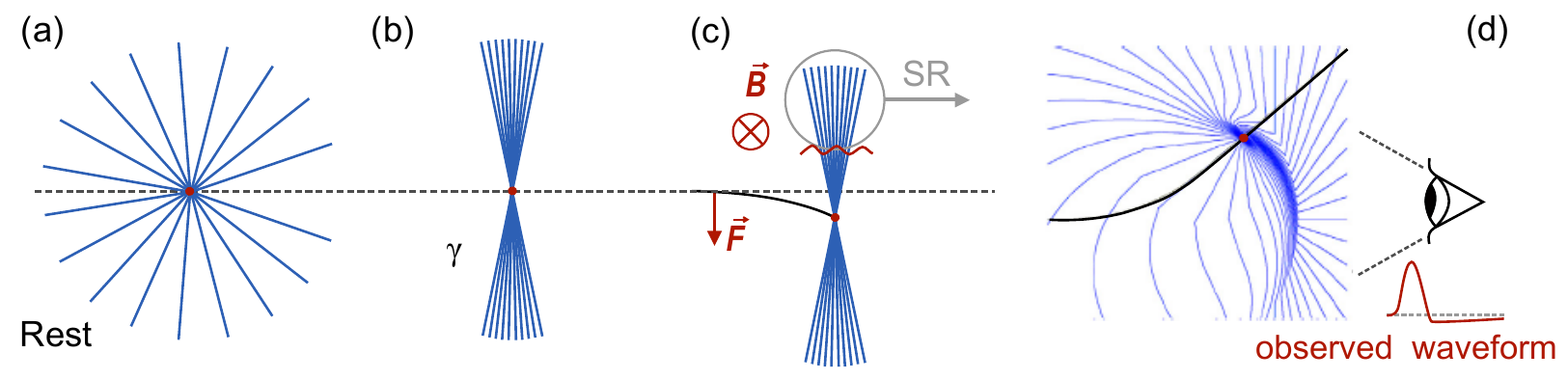}
    \caption[Basic principle of synchrotron radiation]{Electric field distribution for (a) a charged particle at rest, (b) relativistic charged particle moving with a Lorentz factor $\gamma$, (c) deflected relativistic charged particle emitting synchrotron radiation (SR) and (d) resulting electric field lines (blue) from an electron (red dot) moving along a curved path (black) in a bending magnet. The generated waveform shown next to the observer eye is a quasi-half-cycle. Taken from ref  \cite{SalenNL_THz:2019}. 
    }
    \label{fig:Synchrotron}
\end{figure}

Synchrotron radiation is defined as electro-magnetic waves emitted by relativistic charged particles in an external magnetic field.
The origin of such a synchrotron radiation lies in classical electrodynamics, wherein an accelerated charged particle, driven by the Lorentz force of the magnetic field, produces changing electric and magnetic fields.  
This interplay between induced electric and magnetic fields leads to propagating electromagnetic waves. 
Schematics of subsequent electric fields are shown in Fig.\ref{fig:Synchrotron}.
The Coulomb field is spherically symmetric in the instantaneous rest frame of the particle.
If the charged particle propagates in z-direction with velocity $v_z$ and Lorentz factor $\gamma=1/\sqrt{1-\beta^2}$, $\beta=v_z/c$  with $c$ as the speed of light, the electric field gets contracted along the propagation axis, whereas the transverse electric field is multiplied by $\gamma$ \cite{SalenNL_THz:2019}.
When the particle experiences a transverse magnetic field $\mathbf{B}$ and gets deflected, the resulting electric field distortion causes emission of synchrotron radiation.
Fig.\ref{fig:Synchrotron}(d) depicts the electric field lines of a moving electron in a bending magnet and the generated waveform experienced as a quasi-half-cycle THz pulse by the observer.
Temporal coherence is further achieved if the particle bunch length is much smaller than the radiation wavelength, such that the whole bunch radiates like a single particle, which is attainable for THz frequencies \cite{SR:1954, SR:1991}.
For the electron source, three main types are implemented in particle accelerators, namely field emission, thermionic emission and photoemission, with the latter being the state-of-the-art source. 
By exploiting the photoelectric effect, a short wavelength pump pulse is used to release electrons from a photo cathode which are further accelerated to gain relativistic velocities. 
One of the most powerful THz sources based on synchrotron radiation was demonstrated by 
the Jefferson Laboratory (US) with an energy-recovered linear accelerator (LINAC), achieving almost 20 W of average power at a repetition rate of up to 75 MHz (\SI{0.27}{\mu J} pulse energy) \cite{SR:2002}. They used 500-fs electron bunches in order to generate THz pulses with a central frequency at 0.6 THz with a \ac{fwhm} of about 0.5 THz.
Field strength measurements were not available but an order-of-magnitude estimate is about 0.1 MV/cm \cite{SalenNL_THz:2019}.

\subsubsection{Coherent Undulator Radiation and Free-Electron Lasers}
\begin{figure}[htb]
    \centering
    \includegraphics[width=\linewidth]{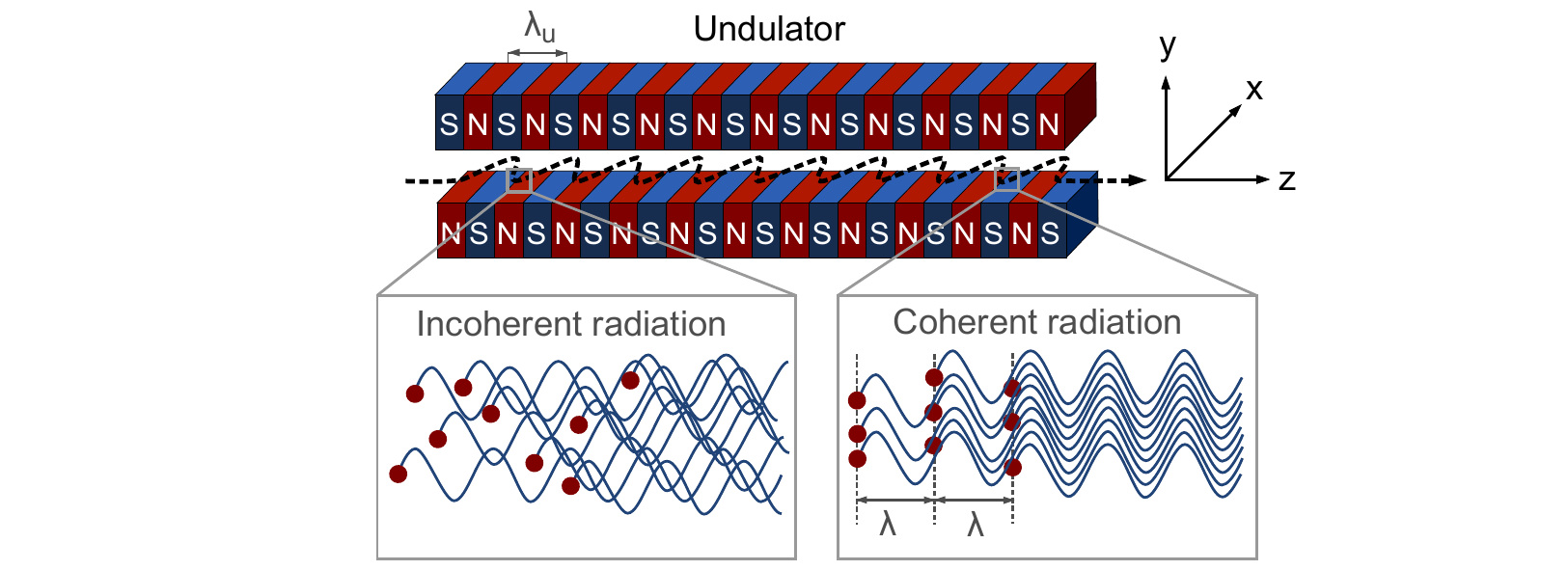}
    \caption[Working principle of a free-electron laser]{Working principle of a free-electron laser, adapted from from\cite{Mak:2019}. The undulator consists of an array of dipole magnets with alternating polarity, causing a spatially periodic magnetic field in y-direction. An electron bunch propagates along z, gets deflected repeatedly and hence oscillates in x-direction. Initially, the oscillating electrons radiate spontaneously with a random relative phase. Subsequently, the electrons are self-arranged into sub-wavelength long micro bunches which radiate coherently.   }. 
    \label{fig:FEL}
\end{figure}

Coherent undulator radiation and free-electron lasers can be understood as an extended synchrotron radiation, wherein an electron bunch is usually steered through a periodic magnetic field which causes periodic deflection of the bunch trajectory and thus increases the pulse energy, proportional to the number of oscillations performed by the bunch.  
Schematics of the undulator and amplification process are shown in Fig.\ref{fig:FEL}, with the static magnetic field $B_0$ parallel to the y axis, and periodicity along z.
When an electron propagates along z, oscillations in x direction cause the electron to radiate at the resonant wavelength \cite{Mak:2019}
\begin{equation}
\lambda= \frac{\lambda_u}{2\gamma^2}\left(1+\frac{K^2}{2}\right),
\end{equation}
with $\lambda_u$ as the undulator period (typically in the order of a few centimetres) and $K$ the dimensionless undulator parameter (in the order of unity) given by
\begin{equation}
K=\frac{e\lambda_uB_0}{2\pi m_e c},
\end{equation}
where $e$ is the absolute value of the electron charge and $m_e$ is the electron mass.
Initially, the resulting radiation is largely incoherent because the electrons are randomly distributed within the bunch, and the electro-magnetic waves emitted by individual electrons have random relative phases. This radiation is usually called spontaneous emission.
However, as the electrons continue to propagate along z, they can constructively interfere with the initial spontaneous emission, such that the radiation is significantly amplified. 
Because photons travel faster than the electrons, light emitted by the tail of the electron bunch propagates through the bunch and interacts with the electrons ahead of the tail.
Since electrons that emit radiation loose some energy and move slower, while electrons that absorb radiation move faster, the electron bunch becomes self organized into disk-shaped micro-bunches, which are separated by the resonant undulator wavelength and behave like single electrons. 
As in a conventional laser, the resultant radiation is largely coherent and the radiation power grows exponentially during the interaction. 
Gain saturation occurs when the micro-bunches are fully developed.
Thus, in analogy to conventional lasers, where the gain medium is a gas, liquid, crystal or semiconductor, for a free-electron laser, the gain medium is a beam of free electrons travelling through an undulator which experience self-amplified spontaneous emission.
Several user facilities based on THz free-electron lasers are operational worldwide, such as FLARE at Radboud University, CLIO at University of Paris or TELBE at Helmholtz-Zentrum Dresden Rossendorf (HZDR).
The latter report on a coherent THz source with a pulse energy of \SI{1.3}{\micro J} at a center frequency of 1 THz and an electric field strength of 100 kV/cm at a repetition rate of 1 MHz.
Moreover, the center frequency is tunable from 0.15 THz to 2.5 THz by carefully setting the strength of the magnetic field, undulator period and by adjusting the current in the electro-magnetic undulator \cite{Green:2016}.
In Hamburg, the THz undulator beamline FLASH utilizes a sequential design which allows soft x-ray and THz radiation being produced from the same electron bunch.
They achieve pulse energies reaching \SI{100}{\micro J} and peak electric fields up to 1 MV/cm.
The source is tunable from 1 THz to 30 THz with a spectral bandwidth (defined as $\Delta\lambda/\lambda$) of 10\% at a repetition rate of 10 Hz \cite{FLASH:2013,FLASH:2008, FLASH:2019}.

Hence, a big advantage of such scientific resources is the frequency tuneability of the generated THz emission via the electron bunch duration and undulator configuration, as well as their potential high repetition rate, which is beneficial for \ac{tds}. 
However, the greatest disadvantage is the large facility itself, with electron accelerators reaching over several km, being cost-intensive and hard to maintain.
Moreover, the narrow bandwidth and low THz energies at high repetition rates are a limiting factor for the electric field strength.

\subsubsection{Coherent Transition Radiation}
\begin{figure}[htb]
    \centering
    \includegraphics[width=\linewidth]{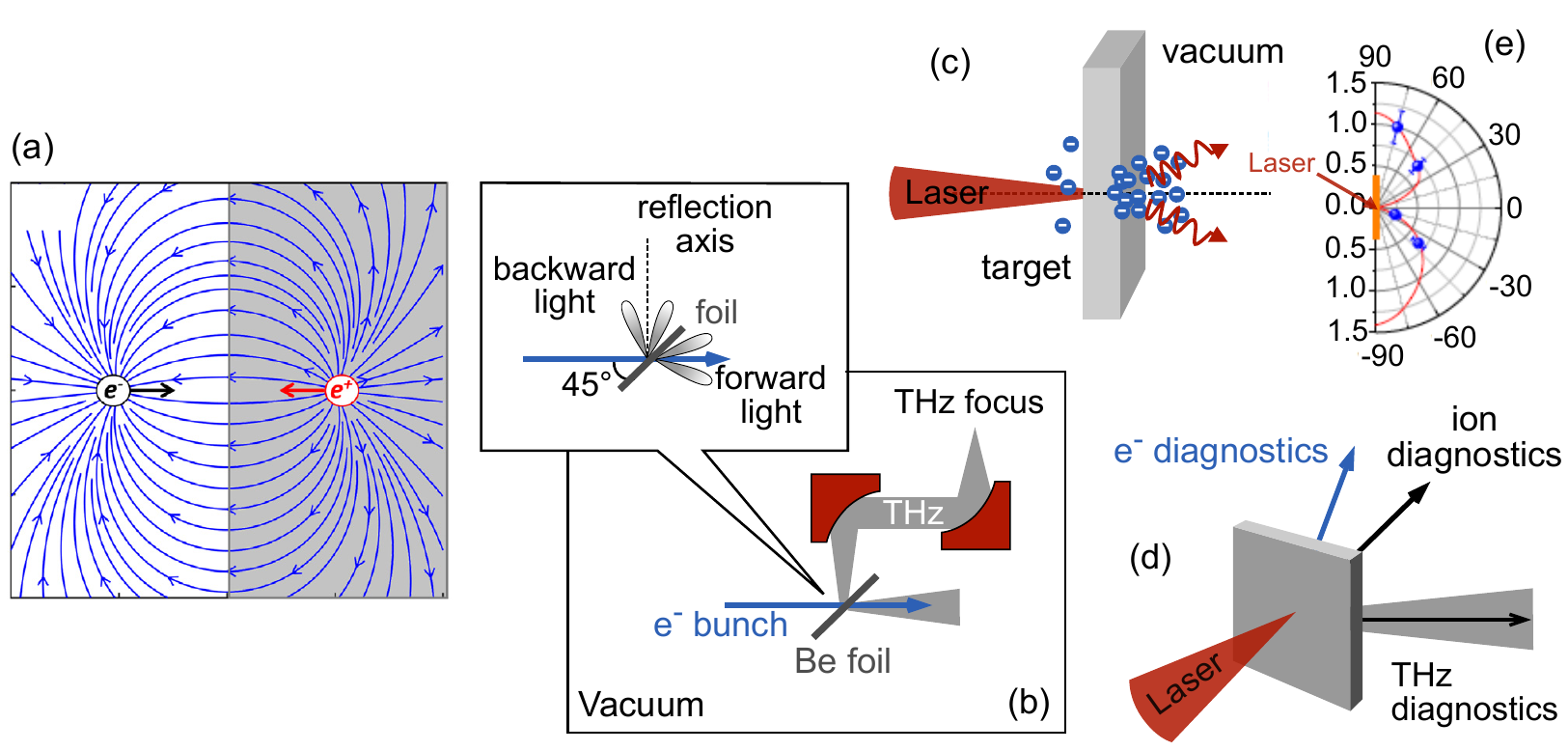}
    \caption[Coherent transition radiation]{(a) Electric field lines of a uniformly moving (from left to right) charge and its image in the metal, taken from \cite{SalenNL_THz:2019}. (b) Schematic setup for THz generation and detection by coherent transition radiation from relativistic electrons impinging on a thin metal foil. The inset shows the emission pattern, adapted from \cite{SalenNL_THz:2019}. (c) Schematics of THz emission from the rear surface of a thin metal foil, irradiated with a high-intense laser pulse. 
The THz dynamics depend on the hot electrons which are generated at the front surface during the interaction of a laser pulse with the foil  \cite{Herzer:2018}. 
(d) Experimental setup as described in ref \cite{THz55mJ:2019} and (e) resulting angular dependent THz emission by coherent transition radiation from the rear surface.
    }
    \label{fig:CTR}
\end{figure}

Another THz source utilizing accelerated plasma in large scale facilities is coherent transition radiation.
A time dependent change in the electric field and hence emission of an electro-magnetic wave is based on radiation from charged particles passing through discontinuities.
The simplest and widely used configuration for the production of transition radiation is that of a relativistic charge crossing an interface between vacuum and a metal.
Once again, as we know from classic electrodynamics, a negatively charged particle in vacuum induces a positive image charge on the metal surface, as shown in Fig.\ref{fig:CTR}(a).
The configuration of the moving charge and its image can be thought of as an electric dipole with a moment $p=2evt\mathrm{H}(z-vt)$, where $\mathrm{H}$ is the Heaviside unit-step function \cite{SalenNL_THz:2019}.
When the electron impinges on the surface, its field becomes screened and the dipole suddenly disappears. 
This abrupt change results in a non-zero value of the second time derivative of the dipole momentum, causing a spherical wave with $\mathbf{E \propto \ddot{p}}$.
In other words, in the far field, the electric field is a Coulomb field of a relativistic moving charge as shown in Fig.\ref{fig:Synchrotron}(b).
A sudden change in the charge velocity requires the electric field lines to reorganize, resulting in the emission of an electro-magnetic wave. 
The spectrum of the emitted field is dependent on the particle velocity and electron bunch configuration \cite{CTR:2003}.
For wavelengths longer than the bunch length, the emitted fields of all the electrons in the bunch add coherently, and the total radiated power scales with the square of the number of electrons \cite{SLAC:2013}.
The \ac{slac}  has two electron accelerators, the \ac{lcls} and \ac{facet}, both are perfectly suitable for generating intense broadband THz pulses via coherent transition radiation.
With an electric field strength of 44 MV/cm and corresponding magnetic field around 14 T, \ac{lcls} sets the record for accelerator-based THz sources\cite{SLAC:2013}.
A simplified schematic of the experimental setup is shown in Fig.\ref{fig:CTR}(b), wherein an electron bunch with a duration of 70 fs impinges on a \SI{10}{\micro m} thick \ac{Be} foil at \SI{45}{^{\circ}} with a repetition rate of 10 Hz.
Electrons approaching the foil produce fields of a collapsing dipole, and an expanding dipole as they leave, generating an impulse of dipole radiation on both sides \cite{CTRbook:1972}.
Half of this THz field is radiated in the forward direction as the beam exits the foil.
The THz energy is measured with a pyro electric detector accounting for $\sim$\SI{200}{\micro J}, and the THz spectrum is recorded with a Michelson interferometer spanning from 3-30 THz. Note that the total THz energy on the optical table is significantly larger but is hard to collect due to the angular distribution. 

However, the highest THz energy ever reported is also based on coherent transition radiation but utilizes an ultra-intense, picosecond (ps) laser pulse to irradiate a \SI{100}{\micro m} thick \ac{Cu} foil (see Fig.\ref{fig:CTR}(c)-(d)) \cite{THz55mJ:2019}.
The experiment was carried out at the Rutherford Appleton Laboratory using the Vulcan laser, with a central wavelength of 1053 nm and pulse duration $\sim$\SI{1.5}{ps}.
In 2005, the Vulcan laser was the highest-intensity focused laser in the world, operating at a repetition rate of 1 shot every 20 min \cite{Vulcan:1999,Vulcan:2006,Vulcan:2019}.
When the thin metal foil was irradiated with a laser energy of $\sim$\SI{60}{J}, intense THz radiation was observed at the rear face of the foil and measured to be $\sim$\SI{2.3}{mJ} at an angle of \SI{75}{^{\circ}} with respect to the rear target normal.
With additional measurements on the angular distribution and energy spectra of the freed electrons, it is possible to calculate the radiation spectrum and angular distribution from the coherent transition radiation theory.
Consequently, they evaluate the total THz pulse energy by extrapolating experimental measurements with model calculations and determine a total energy of the THz pulse emitted from the target rear to be $\sim$\SI{50}{mJ} ($\pm$20\%) within the frequency range up to \SI{3}{THz}.
Nevertheless, given that the THz radiation is emitted in a rather large divergence angle, it would be necessary to use collection optics with large acceptance angles to deliver more available THz energy for practical applications. 
Moreover, the applied extraordinary pump energy and low repetition rate makes this source rather impractical.

\subsection{Table-top THz Sources}

The main principle of a laser-based generation of THz radiation is to produce an electron current density $\mathbf{J_\mathrm{THz}}$, oscillating at the same frequency as the  electro-magnetic wave to be generated. The electric field is then proportional to the first time derivative of the current density
\begin{equation}
\mathbf{E_\mathrm{THz}} \propto \frac{\partial \mathbf{J_\mathrm{THz}}}{\partial t }.
\end{equation}

\subsubsection{Photoconductive Antennas}
\begin{figure}[htb]
    \centering
    \includegraphics[width=\linewidth]{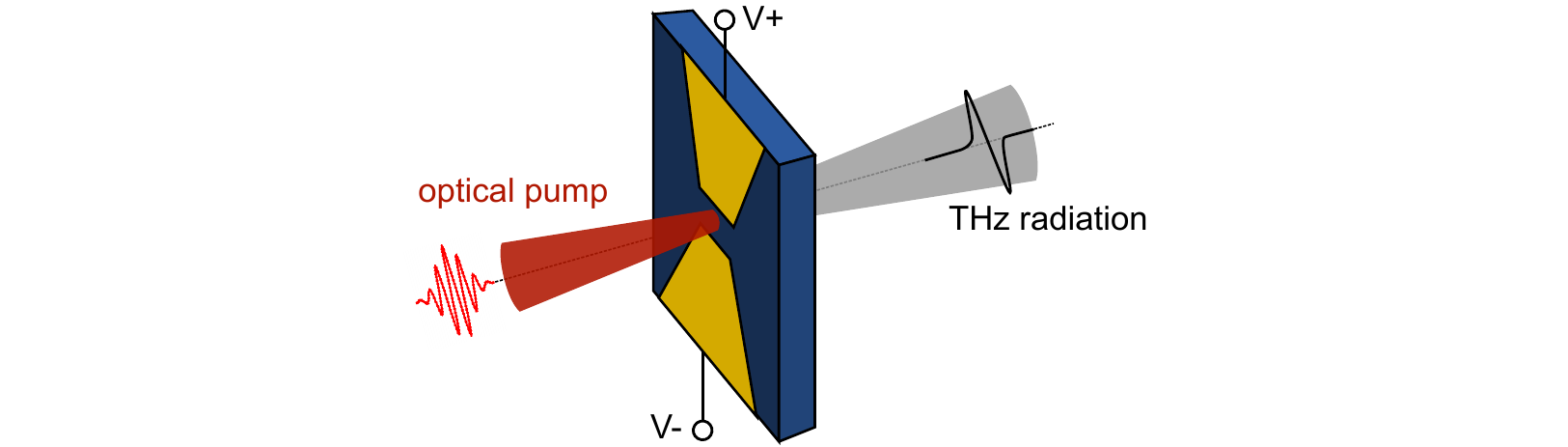}
    \caption[Basic principle of a photo conductive antenna]{Illustrative example of pulsed THz generation in a \ac{pca}. A short pump laser generates photocarriers in a thin semiconductor, which are accelerated by the applied bias voltage. The time dependent current density induces an electro-magnet field oscillating at THz frequencies.
    }
    \label{fig:PCA}
\end{figure}

Although \acp{pca} are not suitable for strong field THz applications, since the generated electric field doesn't reach \SI{500}{kV/cm}, it is worth mentioning their technology since photoconductive emitters have become a very popular tool for measurements where high THz fields are not essential.
The layout of a \ac{pca} and the concept of THz generation from a \ac{pca} is illustrated in Fig.\ref{fig:PCA}.
The device consists of a DC biased metal dipole, embedded on a semiconductor thin film.
When a femtosecond pulse with a pulse duration in the order of \SI{1}{ps} and photon energy higher than the bang gap of the semiconductor material is incident on the antenna gap, it generates electron-hole pairs.
The generated photocarriers are accelerated in the DC bias field, producing a transient photocurrent, which drives the dipole antenna and ultimately re-emits as a THz frequency pulse \cite{pca:2017}.
The transient response of the PCA depends on the optical pulse and the recombination time  of electron-hole pairs.
If the semiconductor has a long carrier lifetime, the generated photocarriers will continue to contribute to the photocurrent after the optical pulse is fully absorbed, which temporally broadens the output pulse and reduces the overall THz frequency bandwidth.
In addition to the limited THz bandwidth, the optical- to THz conversion efficiency easily saturates for high pump intensities due to photon bleaching and other \ac{nl} effects.
Nevertheless, Ropagnol \textit{et. al.} \cite{Ropagnol:2016, Ropagnol:2017} reported on THz fields of \SI{331}{kV/cm} from ZnSe based photoconductive large area emitters, with THz pulse energies of \SI{8.3}{\micro J} and a peak intensity at \SI{0.12}{THz} with a spectrum extending up to about \SI{2}{THz}.  
The optical pump is provided by frequency doubling the output of a Ti:Sapphire amplifier laser operating at a repetition rate of only \SI{10}{ Hz}.
A THz field in the same order of magnitude of $\geq$\SI{230}{kV/cm} was achieved by Singh and co-workers \cite{Singh:21} with a \ac{GaAs} large-area photoconductive emitter pumped with a Ti:Sapphire amplifier laser system at 800 nm wavelength and \SI{1}{kHz} repetition rate.
Although the THz pulse energy is significantly smaller ($\sim$\SI{70}{nJ}), the high electric field is explained by the broad THz spectrum extending up to \SI{4}{THz}. 

\subsubsection{Optical Rectification}
\begin{figure}[htb]
    \centering
    \includegraphics[width=\linewidth]{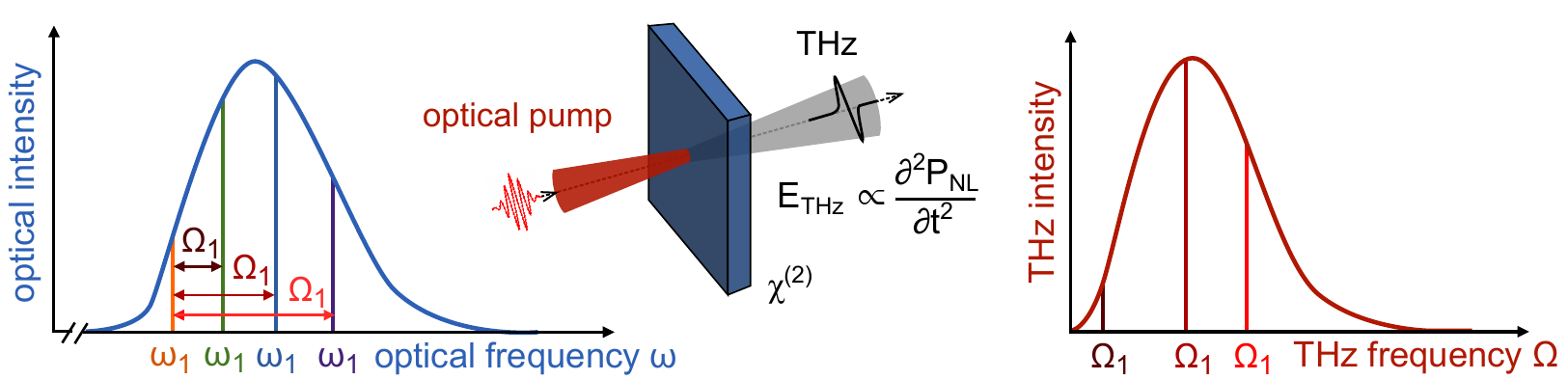}
    \caption[Basic principle of optical rectification]{Schematics of intrapulse \ac{dfg} between the frequency components of an ultrashort pulse incident on a second-order \ac{nl} optical medium, describing pulsed THz generation by \ac{or}. Because the spectral bandwidth and pulse duration are inversely proportional, the Fourier-transform-limited bandwidth of the generated THz pulse is restricted by the pulse duration of the driving laser. The electric field of the THz pulse is proportional to the second time derivative of the induced \ac{nl} polarization. Adapted from ref \cite{StrongFieldFulop:2020}. 
    }
    \label{fig:OR_DFG}
\end{figure}

Alongside of THz generation in two-color plasma filaments, one of the most promising table-top sources of pulsed THz radiation is based on  \ac{or} of ultrashort laser pulses in \ac{eo} crystals. 
When an electro-magnetic wave propagates through a medium, it causes the bound electrons to oscillate. 
Such induced Hertzian dipoles will emit electro-magnetic radiation with the same frequency as the propagating wave, featuring typical spherical phase fronts and dipolar directional characteristics in the far field.
The time-dependent current density, arising from the displacement of the charge density, can be written as $\mathbf{J}=\partial \mathbf{P}/\partial t$, with $\mathbf{P}$ as the induced dipole moment per unit volume.
However, for short laser pulses with high intensity, the oscillating electrons will not only follow the incident electro-magnetic field linearly, but will additionally exhibit a \ac{nl} response, resulting in radiation of additional frequency components with an electric field $\mathbf{E_{\mathrm{THz}}}\propto\partial^2 \mathbf{P_{\mathrm{NL}}}/\partial t^2$.
In the simplest case, this induced polarization scales with the square of the driving optical field, wherein the optical response is restricted to materials and structures lacking inversion symmetry \cite{Boyd:2003}.
Note that, the spherical waves emitted from all source locations of $\partial \mathbf{J_{\mathrm{NL}}}/\partial t$ have to superimpose constructively in a particular direction.  This requirement describes the necessity of the generated  and optical pulse to co-propagate with the same or at least similar propagation velocities, and is called phase matching condition.
In the particular case of THz generation by \ac{or} of an ultra short pulse in a second-order \ac{nl} crystal, the \ac{nl} response can be understood as intrapulse \ac{dfg} between all frequencies within the bandwidth of a fs laser pulse, as schematically illustrated in Fig.\ref{fig:OR_DFG}.
Each pair of optical frequency components generates a THz spectral component at their difference frequency.
The \ac{nl} polarization induced by the pump pulse can be calculated as \cite{StrongFieldFulop:2020, Boyd:2003}
\begin{equation}
P_{\mathrm{NL}}\left(\Omega\right)=\varepsilon_0\int_0^{\infty}\chi^{(2)} E(\omega+\Omega)E^*(\omega)\mathrm{d}\omega,
\label{equ:P_NL}
\end{equation}
where $\varepsilon_0$ is the vacuum permittivity, $E(\omega)$ is the Fourier-component of the pump pulse at optical angular frequency $\omega$, $\Omega$ is the difference frequency, and $\chi^{(2)}$  is the effective \ac{nl} susceptibility, depending on $\omega$ and $\Omega$ \cite{Boyd:2003}.
The complex conjugate $E^*(\omega)$ in Eq.\eqref{equ:P_NL} makes the resulting polarization independent of the carrier-envelope phase of the driving optical field, resulting in a carrier-envelope phase stable THz pulse \cite{Baltuska:1992}.
Further details and underlying physical mechanisms of \ac{or}, especially in organic \ac{nl} crystals and long wavelength driving pulses, are discussed in sections \ref{sec:OR} and \ref{sec:OCrystals}.  

In standard inorganic crystals such as \ac{LN}, \ac{or} allows one to generate high THz energies of \SI{1.4}{mJ} \cite{LiNbZhang:2021} with large optical- to THz conversion efficiencies (up to 3.7\%) \cite{ORefficiencyHuang:2013}.
However, due to the large mismatch of the optical group velocity and THz phase velocity, the spectral bandwidth is mainly restricted to below \SI{1}{THz}.
Recently, considerable progress in broadband THz generation has been achieved by OR in highly NL organic crystals, such as DAST (4-N,N-dimethylamino-4’-N’-methylstilbazolium tosylate) and its derivative DSTMS (4-N,N-dimethylamino-4’-N’-methyl-stilbazolium 2,4,6 trimethylbenzenesulfonate) driven by near-\ac{ir} pump sources \cite{DASTVicario:2015,DASTHauri:2011, Monoszlai:2013}, with an optical- to THz conversion efficiency reaching 3\% and  THz energy of \SI{0.9}{mJ} \cite{DSTMSVicario09:2014}. 
Due to favorable phase matching conditions and low linear absorption for pump wavelengths of \SI{1.25}{\micro\m}-\SI{1.5}{\micro\m}, it is possible to generate broadband THz radiation of up to \SI{8}{THz} and field strength of \SI{83}{MV/cm} \cite{BulletShalaby:2015}.
However, because the absorption edge of DAST is situated at around \SI{700}{nm}, the THz conversion efficiency is deteriorated by multi-photon absorption in combination with subsequent free-carrier absorption of the THz radiation. 
Moreover, the applicable pump fluence is fundamentally limited by the damage threshold of the crystal. 
Thus, the extraordinarily high THz energy of \SI{0.9}{mJ}, as reported in ref \cite{DSTMSVicario09:2014}, could only be achieved with a large-size partitioned crystal, wherein multiple parts of uncoated DSTMS crystals are fixed on a host substrate in a mosaic-like structure, resulting in a crystal surface area of of \SI{4}{cm^2}. 
Consequently, in order to overcome the need of large crystal surfaces to suppress NL absorption and to increase the damage threshold, the crystal can be pumped with longer wavelength driving sources.

\subsubsection{Two-Color Plasma Filaments}

\begin{figure}[htb]
    \centering
    \includegraphics[width=\linewidth]{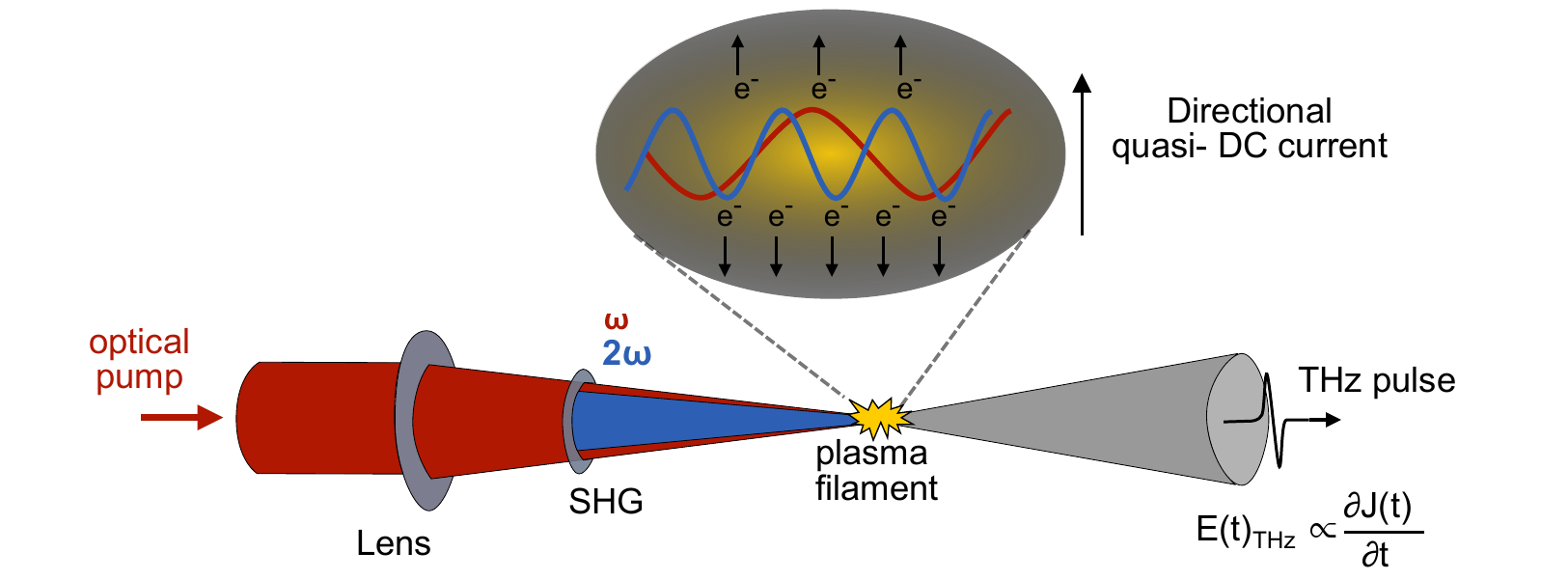}
    \caption[Schematic for THz generation by two-color plasma filaments]{Schematic for THz generation by two-color plasma filaments, reproduced from \cite{KimProceeding:2020}
    A high power fs laser and its second harmonic are focused to produce laser induced gas plasma.
The photo-ionized electrons create a current surge, which radiates a directional electro-magnetic field with an amplitude proportional to the time derivative of the current density. 
    }
    \label{fig:plasma_scheme}
\end{figure}

A schematic setup for THz generation  in two-color plasma filaments is shown in Fig.\ref{fig:plasma_scheme}, wherein a high-power fs laser and its \ac{sh} are focused to produce laser induced gas plasma and a subsequent photocurrent. 
The \ac{sh} is thereby needed to create a symmetry-breaking field in order to generate a non-vanishing net current of the photoionized electrons, exhibiting a quasi DC current \cite{Kim:2009}.
The THz emission is then  again proportional to the time derivative of the current density. 
As it is the case for \ac{or}, a sizeable THz field amplitude can only built up if the spherical waves emitted from local current densities interfere constructively.
Thus, the THz transient and optical pump pulse need to propagate with the same  velocity through the filament, which benefits long wavelength pump pulses in air plasma.
In contrast to \ac{or} in \ac{nl} solids, material damage is absent for the case of THz generation by laser induced plasma filaments.  
Because plasma is generated in a gaseous medium, such a medium is self-healing when a high energy laser is focused into a small volume. 
Moreover, the spectral bandwidth of the generated THz pulse is inversely proportional to the time scale of the tunneling ionisation rate, rather than the pulse duration of the pump source.
It is therefore possible to generate broadband THz pulses with a bandwidth exceeding \SI{50}{THz} \cite{PlasmaWidth:2013, PlasmaWidthKim:2008}. 
And hence, despite the small THz energies, ranging from a few \SI{}{\micro\J} to \SI{30}{\micro\J} \cite{PlasmaKuk:2016} up until recently, it is possible to generate large electric fields of more than \SI{20}{MV/cm}. 
Nonetheless, when driven by conventional \SI{0.8}{\micro\m} Ti:Saphire lasers, THz generation from laser filaments is constrained to conversion efficiencies of $\sim$  0.01\% \cite{PlasmaOhKim:2013}. 
With longer wavelength drivers, the THz generation efficiency scales up and an efficiency of 0.1\% can be achieved with \SI{1.8}{\micro\m} pump pulses \cite{LongWLplasma:2013}. 
The advantageous outcome is mainly attributed to a higher temporal asymmetry of the applied  laser field, smaller mismatch between the group and phase velocities of the optical and generated THz pulses, respectively, and stronger ponderomotive force. 
More details on the theoretical background for the THz generation process will be provided in sections \ref{sec:TPCM} and \ref{sec:LWD}.
Further experimental realization of wavelength-scaling  THz generation efficiency in laser filaments was so far hindered by a lack of high-energy ultrashort mid-\ac{ir} pulse sources.
 
In this work, we show that recently-developed mid-\ac{ir} \ac{opcpa} technology helps to overcome major challenges in efficient generation of broadband THz radiation, in the case of OR in the organic crystal DAST as well as by two-color filamentation in air.
The attained conversion efficiencies and electric field strengths set new records for table-top THz sources.

\section{Thesis Outline}

Prior to the presentation of our THz generation processes, in \textbf{Chapter \ref{ch:Characterization}} we describe basic detection techniques employed for the characterization of  the generated THz radiation.  
Here, the temporal profile of the THz pulse is measured with \ac{eos}  which provides information on the amplitude and phase of the THz transient. 
The THz energy is validated with a calibrated pyro-electric detector and a knife edge method is used to examine the radius of the THz beam at the focal plane.

In \textbf{Chapter \ref{ch:ExperimnetalSetup}} the high-power mid-IR \ac{opcpa} system operating at the central wavelength of \SI{3.9}{\micro m} is introduced, as originally presented in ref \cite{Andriukaitis:2011} and further extended as described in \cite{Valentina:2018} and \citeC{Shumakova:18}. 
The system generates pulses with  energies of more than \SI{30}{mJ} and pulse duration in the order of \SI{100}{fs} at the repetition rate of \SI{20}{Hz}.
We further provide details on the visible probe pulse, generated with a home-built \ac{nopa} and show typical \ac{eos} traces.

\textbf{Chapter \ref{ch:OR}} is devoted to THz generation by \ac{or} of mid-\ac{ir} pulses in the organic crystal DAST. 
Before presenting the main results, as published in ref \citeA{GollnerAPL:2021} and presented at \citeB{cleoE:21, Gollner:20, GollnerUF:20, Gollner:19, DASTirmmw:19, GollnerASSL:19},  underlying physical principles of \ac{or} in general, and \ac{or}  in organic crystals in particular, are expound.
We then present the experimental setup,  followed by a rigorous discussion on the experimental findings. 
We thereby elucidate the reason for the high THz conversion efficiency of almost 6\% and suggest a mechanism for the onset of saturation.

An application of the generated intense THz pulse is presented in \textbf{Chapter \ref{ch:QDs}}, wherein we manipulate optical properties of colloidal CdSe/Cds core/shell QDs. 
A  THz induced change in transmission of 15\% is observed at room temperature, without any field enhancing structures.
In order to understand the extraordinary performance, we first review  physical properties of confined-structures and basic principles of the  \ac{qcse}, followed by material characterization of the CdSe/CdS  QDs and an introduction to the experimental setup. 
A subsequent presentation of experimental results is supported by an intuitive and simple theoretical model, which admits further discussions on the energy band alignment of heterostructure QDs for the optimization of next-generation \ac{ea} modulators in optical communication systems.
The main results are published in \citeA{GollnerQD:2022}  and presented at \citeB{GollneCleoUSAr:21, GollnerCLEO:21, GollnerASSL:21}.

Efficient THz generation in two-color plasma filaments of mid-\ac{ir} pulses is presented and discussed in \textbf{Chapter \ref{ch:filaments}}.  
We first revise the transient photo current model and focus on the advantages of long-wavelength drivers. 
We then present the experimental setup and findings, revealing a record conversion efficiency, which is more than an order of magnitude higher than previously reported results, and extreme THz electric fields exceeding \SI{100}{MV/cm}.
The main results are published in \citeA{TasosA:2020} and presented at \citeB{GollnerASSL:19, InvitedIRMMW:2019, Koulouklidis:19, Koulouklidis:18}.

\textbf{Chapter \ref{ch:XPM}} is devoted to the demonstration of proof of principle experiments to confirm the extreme THz field generated by two-color filaments and to  illustrate the capability to modulate the optical properties of bulk semiconductors.
We report on \ac{xpm} of a visible probe and THz pump in \acs{ZnTe}, and THz induced \ac{pl} in \acs{ZnSe} and \acs{ZnTe}, as published in \citeA{KoulouklidisA:20} and presented at \citeB{THzInduced:19}.

Finally, all results are summarized in \textbf{Chapter \ref{ch:Sum}}.

%% file: Chapters/THzCharacterization.tex
\chapter{THz Pulse Characterization}
\label{ch:Characterization}

The properties of THz radiation can be measured with different techniques, depending
on the parameter of interest.
In general, detection of electromagnetic radiation can be broadly divided into two groups, namely coherent and incoherent detection systems. 

Coherent detection systems allow one to detect not only the amplitude, but also the phase of the THz field. 
The main principle of such a detection scheme is mixing of two signals, the THz signal to be detected and a reference signal.
As a consequence, the THz pulse is sampled in time domain and the power spectrum can be obtained by \ac{ft}.  
Prominent  detection techniques employ \acf{pca} and \acf{eos}.
While the former is impractical for intense THz sources due to possible undesired NL interactions and limited time resolution caused by semiconductor charge carrier lifetimes in the range of \SI{100}{}-\SI{300}{fs} \cite{TDStutorial:2018}, \ac{eos} is perfectly suited to measure sub-ps THz transients with a broad spectral bandwidth. 
The detectable bandwidth depends on the material parameters of the \acf{eoc}, like crystal thickness, dispersion and phonon resonances.
The time resolution is limited by the duration of the probe pulse, which is less than \SI{45}{fs} in this work.

Although not coherent by definition because the phase information is missing, another possible procedure to measure the THz spectrum is  based on a Michelson interferometer. 
An autocorrelation signal is thereby created by combining two replicas of the THz pulse  and the resulting interference signal is measured with a \ac{ped}.
Thus, this technique is not restricted by any material properties and the entire THz spectrum can be measured.
Alongside with \ac{eos}, it is also implemented for this work and further described below.   
  
In general, incoherent detection elements consist of a direct detection sensor, which essentially measures the energy released by the photon as the detector absorbs it. 
Such detectors allow only signal amplitude detection but lack phase information. 
Nonetheless, the main advantage of such a direct detection is the relative simplicity and large detection bandwidth.
Such detectors are often referred to as 'thermal', wherein a change in temperature is converted into a measurable voltage drop, change in current or resistance. 
Prominent examples are Golay cells, semiconductor bolometers or pyro-electric detectors.
However, due to the low THz photon energy of $\sim$ \SI{4}{meV}, as compared to room temperature thermal energy of \SI{26}{meV}, the largest challenge lies within the suppression of background noise, which can be described with the \ac{nep}.
The NEP can be drastically improved, of up to seven orders of magnitude,  by cooling the detector \cite{Sizov:2018}. 
Though, as a side product, the detectors become more bulky because of non-practical low-temperature equipment needed for operation. 
In this work, the detector of choice is an uncooled pyro-electric detector with a low NEP and a large dynamic range, sufficient to detect high energy THz pulses.
A detailed description of the detector and its calibration is elucidated in the following chapter.

Finally, to evaluate the THz electric field strength, it is necessary to measure the THz transient in time domain as described in section \ref{sec:Spectrum}, the THz energy as outlined in \ref{sec:Pyro} and beam profile as demonstrated in \ref{sec:BP}.
A conclusive explanation for the evaluation of the THz field strength is provided in the last section of this chapter.

\section{Time Domain Measurements of a THz Transient}
\label{sec:Spectrum}
\subsection{Electro-optic Sampling (EOS)}
\label{sec:EOS}
\subsubsection{Basic Principle}
\begin{figure}[h]
    \centering
    \includegraphics[width=\linewidth]{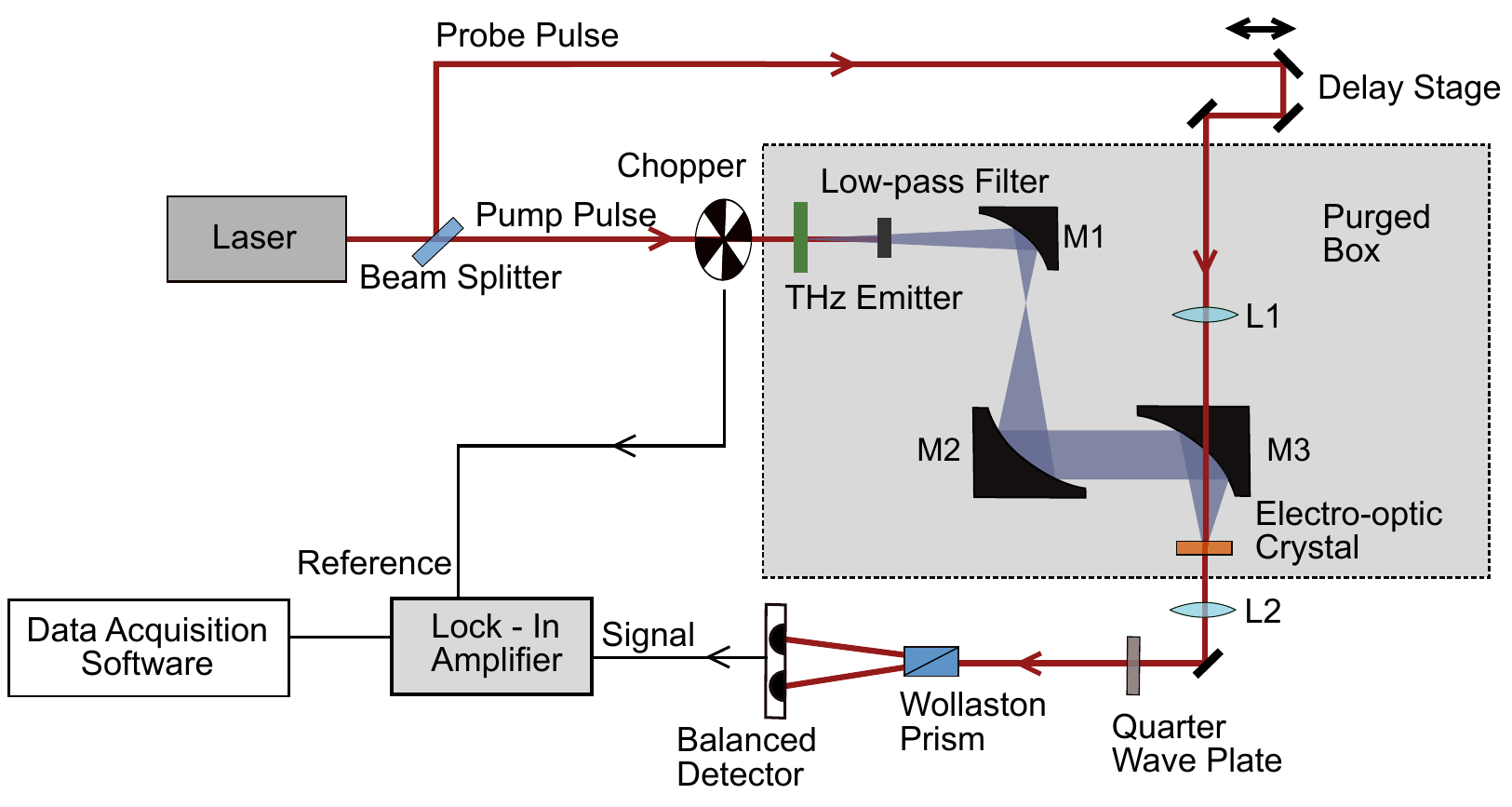}
    \caption[Experimental schematic for electro-optic sampling]{Experimental schematic for \ac{eos}. The main part of THz generation and detection is covered by a box filled with nitrogen to avoid THz absorption by water molecules in ambient air. 
     M1, M2, M3 - parabolic THz beam steering mirrors; L1, L2 - lenses to focus and collimate the probe pulse.
    }
    \label{fig:EOS_scheme}
\end{figure}

The main idea of \ac{eos} is illustrated in Fig.\ref{fig:EOS_scheme}.
The common source can be a fs - laser system with the output being split into two parts after passing a beam splitter. 
A small fractions is used as a probe pulse, which gates the \ac{eoc} used for detection.
The main portion is used to generate THz radiation, after being modulated by an optical chopper. 
Details on the THz source are presented in chapter \ref{ch:OR} and  \ref{ch:filaments}.
To separate the pump pulse and THz radiation, a low pass filter is placed into the beam path.
Several parabolic mirrors are used to steer the THz beam and focus it onto the \ac{eoc}.
The geometry of the focusing mirrors are chosen such that a large beam area results at the final mirror M3 to ensure tight focusing at the crystal position.
Electro-optic detection of a THz transient is possible when the THz and probe pulse coincide in a co-propagating geometry in the \ac{eoc}, which can be achieved with a hole in the center of the parabolic mirror M3.
The underlying mechanism of \ac{eos} is based on the Pockels effect, which introduces a change in refractive index with respect to an applied electric field.
As a consequence, the electric field of the THz transient induces birefringence, resulting in an anisotropic crystal structure and phase delay between the two orthogonal components of the linearly polarized probe pulse.
Thus, the polarization of the probe pulse becomes elliptical after propagation through the \ac{eoc} when the THz and probe pulse spatially and temporally overlap.
The magnitude of the change in polarization is thereby proportional to the amplitude and sign of the electric field of the THz pulse. 
The entire time profile of the THz transient can be traced by changing the time delay between the THz and the probe pulses  with an optical delay stage.
A Wollaston prism is used to split the two orthogonal polarization components of the probe pulse, which are then steered to a balanced detector consisting of two photo diodes.
The difference between the signal of the photo diodes $S(t)$ is used as a measure for the phase delay and can be described as
\begin{equation}
S(t)\propto I_{\mathrm{opt}}E_{\mathrm{THz}}(t),
\label{Equ:EOSprop}
\end{equation}
with $I_{\mathrm{opt}}$ being the intensity of the laser pulse and $E_{\mathrm{THz}}(t)$ the electric field of the THz pulse at time $t$, which corresponds to the time delay between the THz and probe pulse. 
Thus, because the time dependent amplitude of the elctric field $E_{\mathrm{THz}}(t)$ is measured, in contrast to other techniques which measure only the intensity $|E(t)|^2$, the phase information is captured.
Before the Wollaston prism, a \ac{qwp} is installed to equalize the probe intensity at the diodes when the THz pulse is blocked.
The balanced detection scheme is basically used to reduce noise originating from the background and intensity fluctuations of the probe pulse.
In order to further reduce the background noise, a Lock-In amplifier is used which selectively amplifies a signal at a given reference frequency. 
The input signal is thereby multiplied with a sine wave of that particular reference signal and sent through a low-pass filter. In such a way, any signal which arrives at the photo diodes with a different frequency other than the reference frequency, will be cut away and only the signal of interest is detected.
The reference frequency is usually provided by the optical chopper.
Subsequently, the Lock-In signal is processed with a data acquisition software.
In order to reduce humidity and avoid THz absorption by water molecules in ambient air, the THz generation and detection setup can be purged with nitrogen. 
The outcome of the measurement is crucially dependent on the used \ac{eoc}, which will be further discussed below.

\subsubsection{Pockels Effect in an electro-optic Crystal}
\label{sec:Pockels}
As mentioned above, the underlying physics of \ac{eos} is the linear electro-optic effect (Pockels effect), which describes an electric field induced change in polarization of a probe pulse in an \ac{eoc}.
Note that, the term polarization can be easily misunderstood. 
For photons, or more general, transverse waves, it defines the direction of the oscillating electric field. 
Whereas in the case of crystals and molecules, the phrase describes a charge distribution which is expressed as the density of permanent or induced electric dipole moments.
Thus, it can be readily inferred that the Pockels effect can only occur in non-centrosymmetric materials which lack inversion symmetry. 
Otherwise, in the case of centrosymmetric crystals, on average, the macroscopic dipole moments cancel each other and the net polarization is zero.
Another obvious requirement for an \ac{eoc} is the necessity of being sufficiently transparent at THz and optical frequencies.

The linear electro-optic effect originates from the same underlying principle as \ac{or}, which is a second order NL effect. 
The relation between the electric field vector $\mathbf{E}$ and the resulting polarisation density $\mathbf{P}$ is given as \cite{Wilke:2008}
\begin{equation}
\mathbf{P}=\hat{\chi}\left(E\right)\mathbf{E},
\label{equ:linPol}
\end{equation}
with $\hat{\chi}\left(E\right)$ as the electric susceptibility tensor.
The NL optical properties of the material are described by expanding $\hat{\chi}\left(E\right)$ in powers of the field $\mathbf{E}$:
\begin{equation}
P_i=\left(\chi^{(1)}_{ij}+\chi^{(2)}_{ijk}E_k+\chi^{(3)}_{ijkl}E_kE_l+\cdots\right)E_j,
\label{equ:NL_Pol}
\end{equation}
where $\chi^{(1)}$ is the linear susceptibility, while $\chi^{(2)}$ and $\chi^{(3)}$ are the NL susceptibilities of the second and third order, respectively. 
The second term in Eq.\eqref{equ:NL_Pol} describes the \textit{i}-th component of the NL polarization related to the components \textit{j} and \textit{k} of the electric fields $E_j$ and $E_k$ by the susceptibility tensor coefficients $\chi^{(2)}_{ijk}$.
The coefficients are all dependent on the frequency $\nu$ (or the angular frequency $\omega=2\pi\nu$ ) of the involved electric fields $\mathbf{E}$.
The electro-optic effect is most often described as a field-induced change of the refractive
index $n$ at $\omega$ as \cite{OC_THzPhotonics:2019}
\begin{equation}
\Delta\left(\frac{1}{n^2}\right)_{ij}=r_{ijk}E_k,
\end{equation}
with $r_{ijk}$ as the electro-optic tensor which is linked to the second-order susceptibility by
\begin{equation}
r_{ijk}=-\frac{2\chi^{2}_{ijk}(-\omega;\omega,0)}{n^2_{i}(\omega)n^2_j(\omega)},
\end{equation}
where $n_i(\omega)$ are the frequency-dependent refractive indices along the direction of propagation with respect to the corresponding dielectric axis $i$. 
In addition to a large EO coefficient, it is necessary to choose an appropriate crystal thickness and crystal orientation with respect to the linear polarization of the THz radiation. 

\begin{figure}[h]
    \centering
    \includegraphics[width=\linewidth]{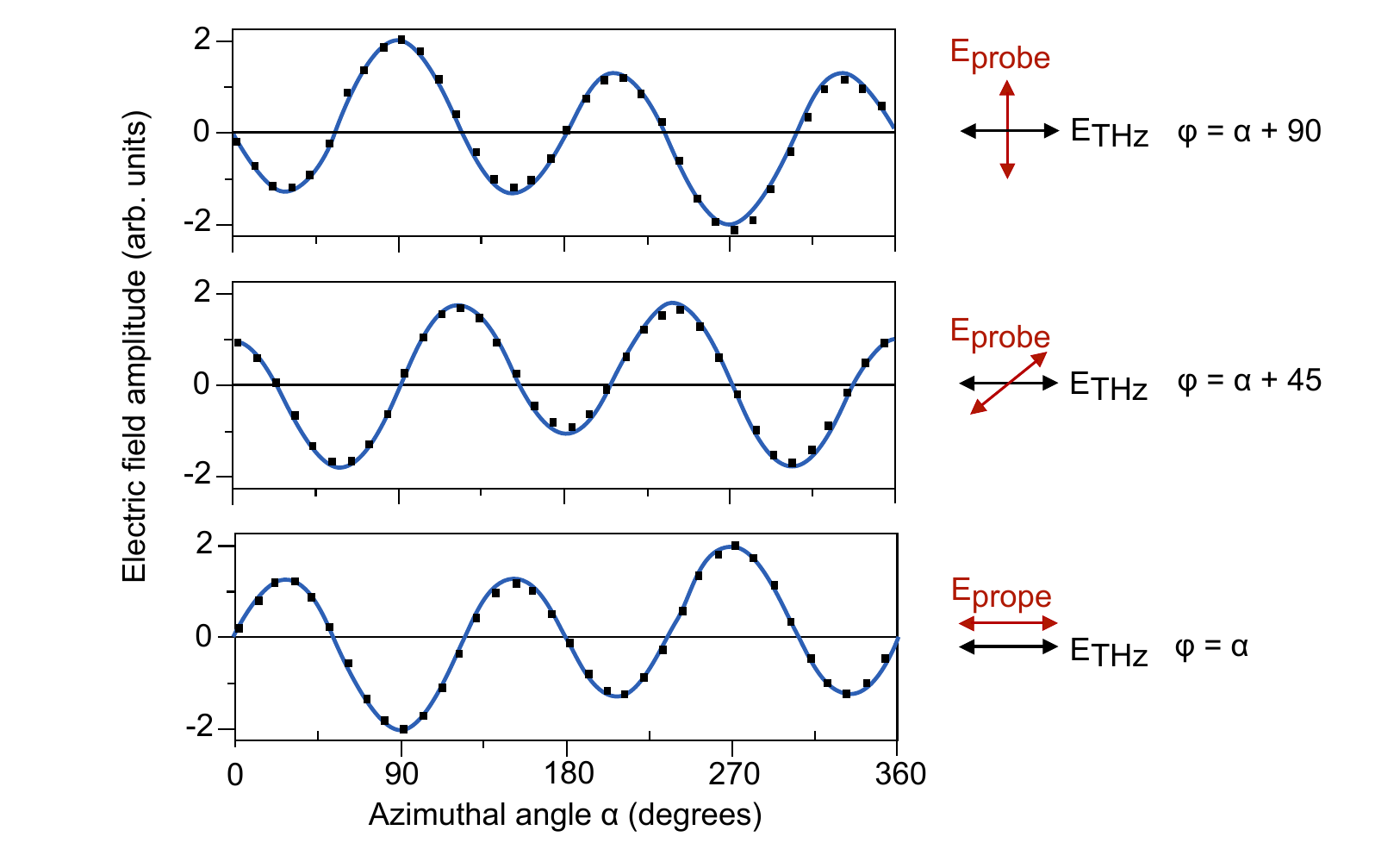}
    \caption[Dependence on the azimuthal angle for the electro-optic crystal ZnTe]{Measured dependence on the crystal’s azimuthal angle $\alpha$ of the detected THz electric field for three different fixed angles between the probe-beam polarization and the THz polarization, reproduced from ref \cite{PlankenOrientation:2001}.
    }
    \label{fig:EOC_orientation}
\end{figure}

In this work, a \SI{1}{mm} thick \ac{ZnTe} and \SI{50}{\micro m} thick \ac{GaP} crystals are used, both have a zinc blende crystal structure and are optically isotropic when no external field is applied ($n_i=n_0$).
Consequently, the electro-optic tensor is symmetric ($r_{ijk}=r_{jik}$) and the first two indices $i,j$ are conventionally replaced by a single index, for example $(2,3) \rightarrow 4$ and $r_{23k}=r_{32k} \rightarrow r_{4k}$.
The index $k$ indicates the polarization direction of the propagating electric wave.  
The electro-optic tensor $\mathbf{\hat{r}}$  of \ac{ZnTe} and \ac{GaP} can be therefore characterized by a single quantity $r_{41}$  and the refractive index $n_i(E)$ for the three principal axis of the crystal can be approximated to \cite{Casalbuoni:2008}:
\begin{subequations}\label{equ:n_E}
\begin{align}
n_1&=n_0+\frac{n_0^3r_{41}E}{2}\label{second}\\
n_2&=n_0-\frac{n_0^3r_{41}E}{2}\label{third}\\
n_3&=n_0.\label{fourth}
\end{align}
\end{subequations} 
For the approximation, $r_{41}E \ll 1/n_0^2$ is assumed and the ideal angle $\alpha = \pi/2$ between the THz electric field vector $\mathbf{E}$ with the crystal axis [-1,1,0] is considered, while the polarization of the probe pulse is parallel to the THz field.
This angle leads to the highest response in Eq.\eqref{eqn:PhaseShift} and can be found empirically during the experiment by simply rotating the \ac{eoc}.
However, as demonstrated by Planken \textit{et al.} \cite{PlankenOrientation:2001} and shown in Fig.\ref{fig:EOC_orientation}, large \ac{eos} signals can also be achieved if the THz electric field and probe polarization are perpendicular. 
The relative phase shift $\Gamma$ between the two orthogonal electric field components of the probe pulse (also called \textit{retardation parameter}), caused by the THz field induced birefringence, can be written as
\begin{equation}
\Gamma = \frac{2\pi(n_1-n_2)d}{\lambda_0}=\frac{2\pi n_0^3d}{\lambda_0}r_{41}E,\label{eqn:PhaseShift}
\end{equation}
with $\lambda_0$ as the wavelength of the probe pulse in vacuum and $d$ the thickness of the crystal.
The most widely employed \ac{eoc} for THz detection are \ac{ZnTe}, \ac{GaP} and \acs{GaSe} due to their high linear electro-optic coefficient \cite{Wilke:2008} (see table \ref{tab:EOC}).

From Eq.\eqref{eqn:PhaseShift} one could expect the highest EO signal for thick detection crystals.
However, the opposite might be the case.
The detectable THz bandwidth of an \ac{eoc} is determined by the coherence lengths and optical phonon resonances which absorb THz radiation.
Because the refractive indices at the frequencies of the optical probe and THz pulse are generally not identical, the two pulses travel with slightly different velocities through the crystal.
Thus, the temporal overlap and hence the detection efficiency decreases during propagation.
The distance over which a slight velocity mismatch of the THz phase velocity and optical group velocity can be tolerated is called the coherence length and is defined as \cite{CLength:1996} 
\begin{equation}
l_c=\frac{\pi c}{\omega_{\mathrm{THz}}|n_{\mathrm{gr}}(\omega_0)-n_{\mathrm{THz}}(\omega_{\mathrm{THz}})|},
\label{equ:coherneceL}
\end{equation}
with
\begin{equation}
n_{gr} = n_{\mathrm{opt}}(\lambda_0)-\lambda_0\left(\frac{\partial n_{\mathrm{opt}}}{\partial \lambda}\right)_{\lambda_0},
\end{equation}
where $c$ is the speed of light in vacuum, $\omega_0=2\pi c/\lambda_0$ is the center frequency of the optical probe pulse, $n_{\mathrm{THz}}$ and $n_{\mathrm{opt}}$ are the phase refractive indices at the THz and optical frequencies, respectively, and $n_{\mathrm{gr}}$ is the group refractive index of the probe pulse.
As a consequence, efficient THz detection (and generation) at a given frequency is only possible for a crystal thickness which is equal to or less than the coherence length.
For a larger crystal thickness, the detection signal eventually even decreases due to a phase slippage. 
Phenomenologically, this can be understood from the fact that the \ac{eos} process will be most efficient when the intensity profile of the optical pulse travels through the crystal at exactly the same speed as the electric field of the THz pulse. 
When these velocities differ, the probe pulse will average over several oscillations of the electric field, which leads to a strong reduction of the electro-optic signal\cite{BakkerEOS:1998}.
\begin{table}
\begin{center}
 \begin{tabular}{c l l c} 
\hline
\\
Material & EO coefficient & Lowest optical Phonon & GVM \\ 
 \	& (pm/V) & Resonance (THz) & (ps/mm) \\		
 \\
 \hline
 \hline
 \\
 ZnTe & $r_{41}$=\ 3.9 & 5.4 & 1.1 \\ 

 GaP & $r_{41}$=\ 0.97 & 10.96 & - \\

 GaSe & $r_{41}$=\ 14.4 & 7.1 & 0.1 \\
 \\
 \hline 
\end{tabular}
\caption[Material properties for common electro-optic crystals]{\label{tab:EOC}Material properties for common \ac{eoc}, taken from \cite{Wilke:2008}. Optical phonon lines are responsible for large THz absorption and abrupt velocity mismatch between the optical and THz pulse. GVM refers to the group velocity mismatch between optical and THz frequencies, away from phonon resonances.}
\end{center}
\end{table}
In other words, if the timing of the probe pulse is adjusted such that it coincides with the center of a positive half wave of the THz pulse, it might travels to the center of a negative half wave during propagation through the thick crystal, caused by group velocity mismatch. 
Thus, the positive phase retardation $\Gamma$ accumulated in the first half of the crystal will be largely cancelled by the negative phase retardation $\Gamma$ accumulated in the second half \cite{Casalbuoni:2008}.  
Nevertheless both, the detection strength and the detectable THz bandwidth depend on the crystal thickness, they exhibit a reciprocal relation.
This relation can be understood as follows: usually, an \ac{eoc} exhibits a certain dispersion in the optical and THz spectral range. 
Assuming that the phase mismatch $\Delta k$ is small for a particular narrow THz frequency range, a thick crystal can be used due to the correlation $l_c=2\pi/|\Delta k|$ and thus a large detection signal can be accumulated.
Hence, in the case of a thick crystal, only a narrow spectral bandwidth will satisfy the phase-matching condition (wherein the crystal thickness is less than the coherence length), leading to spectral focusing but a strong \ac{eos} signal.
Additional THz spectral components with a larger phase mismatch can only be measured with a reduced crystal thickness.
Otherwise, these spectral components will be cancelled out. 
As a consequence, the detectable signal will be smaller due to a shorter distance in which the THz and optical pulse propagate through the crystal.

Although ZnTe has a higher electro-optic coefficient and hence allows a very sensitive detection of small signals, GaP is mainly used for broad bandwidth detection because of beneficial phase matching conditions as well as the late appearance of phonon lines at large THz frequencies.
Despite the superior parameters of GaSe, the disadvantage lies in the softness of the material that makes it fragile during operation.
A detailed discussion on additional \ac{eoc} can be found in refs \cite{Wilke:2008, EOCs:1996}, including newly developed organic crystals.

\subsection{Michelson Interferometer}
\label{sec:Michelson}

\begin{figure}[htbp]
    \centering
    \includegraphics[width=\linewidth]{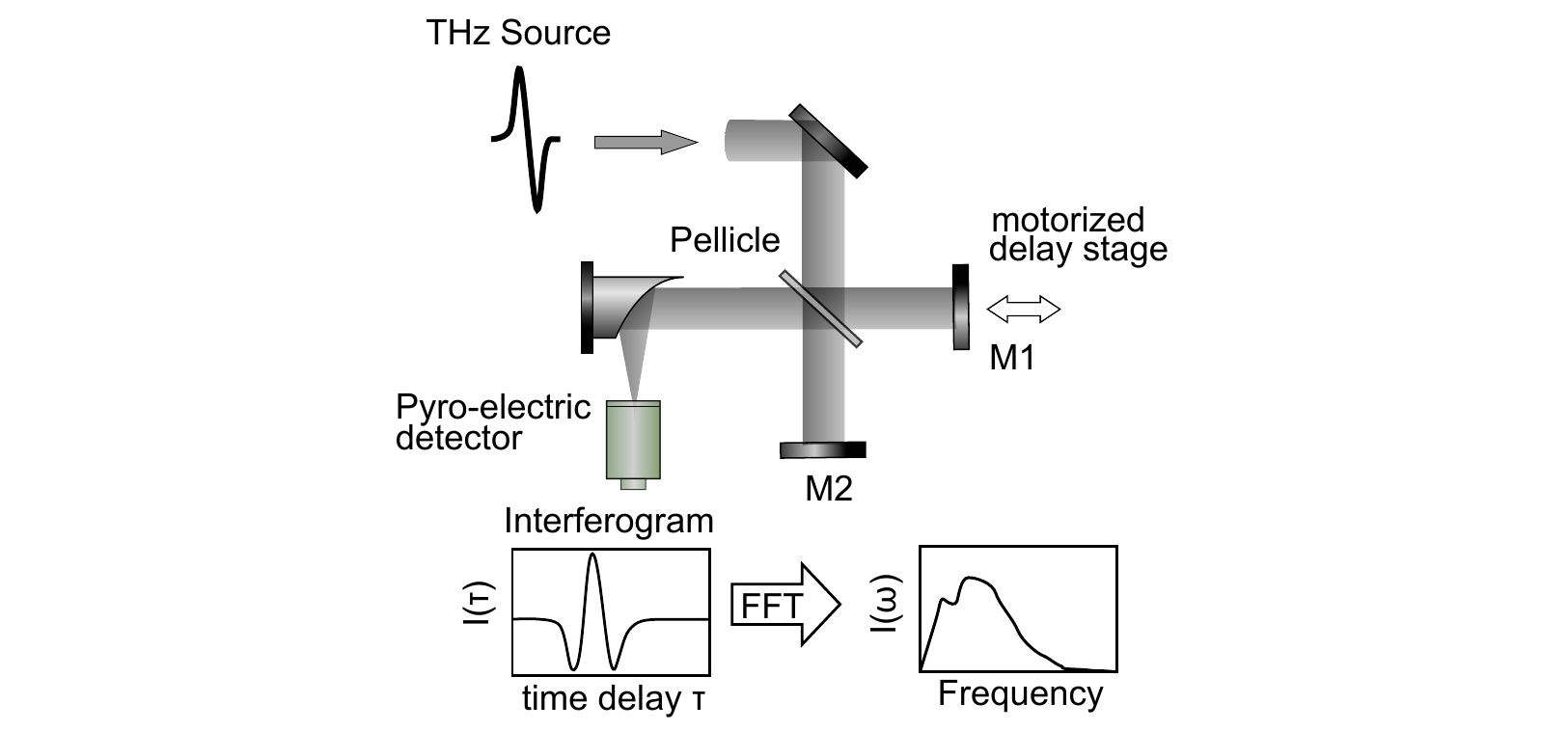}
    \caption[Schematics of a Michelson interferometer]{Schematics of a Michelson interferometer. M1,2 - plane gold mirrors, FFT - fast Fourier transformation.}
    \label{fig:Michelson}
\end{figure}

In contrast to \ac{eos}, a Michelson Interferometer is by definition not a coherent detection technique, because the THz signal is combined with a copy of itself instead of being mixed with a different reference signal, as shown in Fig.\ref{fig:Michelson}.
The THz radiation is thereby split with a beam splitter into two approximately equal parts and each replica propagates to a plane mirror.
The transmitted half of the original beam is reflected back by the stationary mirror \SI{}{M_2}, while the reflected field travels to a movable plane mirror \SI{}{M_1}, where it is reflected back to the beam splitter and to the detector.
A challenge in the mid-IR and THz spectral range is to find suitable optics. 
While gold mirrors are widely established to reflect THz radiation, beam splitters and other sophisticated optical elements might be harder to produce and cost intensive.
In this work, we use a pellicle beam splitter (Thorlabs, BP245B3) which is originally designed for the mid-IR but works surprisingly well in the THz range.
After the beam splitter, the collimated beam is focused at the THz detector (GenTec) with a parabolic gold mirror.

An interferogram is generated by measuring the signal at many discrete positions of the moving mirror \SI{}{M_1}. 
For a detector with a long response time (as compared to the pulse length, which is usually the case) the superposition of the field amplitudes from the two time delayed copies is recorded as
\begin{multline}
S(\tau) \propto \int_{-\infty}^{\infty}E(t)E(t)^*dt 
+ \int_{-\infty}^{\infty}E(t+\tau)E(t+\tau)^*dt \\
+\int_{-\infty}^{\infty}	
E(t)^*E(t+\tau)dt+\int_{-\infty}^{\infty}E(t)E(t+\tau)^*dt,
\label{Equ:Michelson}
\end{multline}
with $\tau$ as the time delay between the two copies and $E(t)^*$ the complex conjugate of $E(t)$.
The first two terms are proportional to a constant energy, which is not of interest here.
The last two terms represent the first order autocorrelation of $E(t)$. 
Because the resulting detector signal accounts for the intensity, rather than the field amplitude, the phase information is lost.
However, the intensity $I(\tau)$ as a function of the time delay $\tau$ between the recombined beam, is the Fourier transform of the spectral intensity of the light source $I(\tau)=\mathrm{FT}\{I(\omega)\}$.
Because of this mutual correlation between such 'transform pairs', a Michelson interferometer is also called Fourier transform spectrometer.
Nonetheless, the autocorrelation yields the spectrum, but not the pulse duration. 
Such a measurement provides the lower limit of the attainable pulse duration, the Fourier-Transform limit, but gives no inforamtion on the actual pulse shape and duration.
The pulse can, however, be arbitrary long (a light bulb yields a very broad spectrum and thus a very narrow interferogram, but it does not emit short pulses)\cite{Reimann:2007}.
Albeit the phase information is missing, it is simply an approach to measure the THz spectrum. Moreover, there are no limitations on the detectable spectral bandwidth.

\section{Pyro-Electric Detector Calibration and Filter Transmission}
\label{sec:Pyro}

The basic principle of a \ac{ped} is the ability of certain materials to generate a temporary voltage drop/rise when they are heated or cooled.
A change in temperature slightly modifies the crystal structure, changes the polarization and gives rise to a voltage across the crystal.
The sensitivity of a PED is usually increased by adding an absorbing layer (black layer) on top of the pyro-electric material and a heat sink on the opposite side, to boost the temperature difference between the electrodes.
However, the sensitivity is partially limited by the response time and signal to noise ratio, which are in part inversely related.
Moreover, because the detector needs a certain time to cool down again, the response time restrains the measurable repetition rate of the detected signal. 

The PED used in this work (GenTec, THz5I-BL-BNC) was calibrated by the company at a repetition rate of 5 Hz when the detector is irradiated with a continuous wave source at \SI{633}{nm} and chopped with an optical chopper at the desired repetition rate.
The resulting sensitivity reads \SI{205}{kV/W} for a peak-to-peak value measured with an oscilloscope.
In other words, a PED signal of \SI{1}{V} relates to an average irradiation power of \SI{4.9}{\micro\W}.
Nonetheless, because the mid-IR pump source and subsequent generated THz radiation operates at a repetition rate of \SI{20}{Hz} with sub -\SI{100}{fs} pulse durations, leading to a significantly smaller duty cycle as compared to the calibration pulses applied by the company,
the PED needs to be calibrated for the measurement conditions.
\begin{figure}[htbp]
    \centering
    \includegraphics[width=\linewidth]{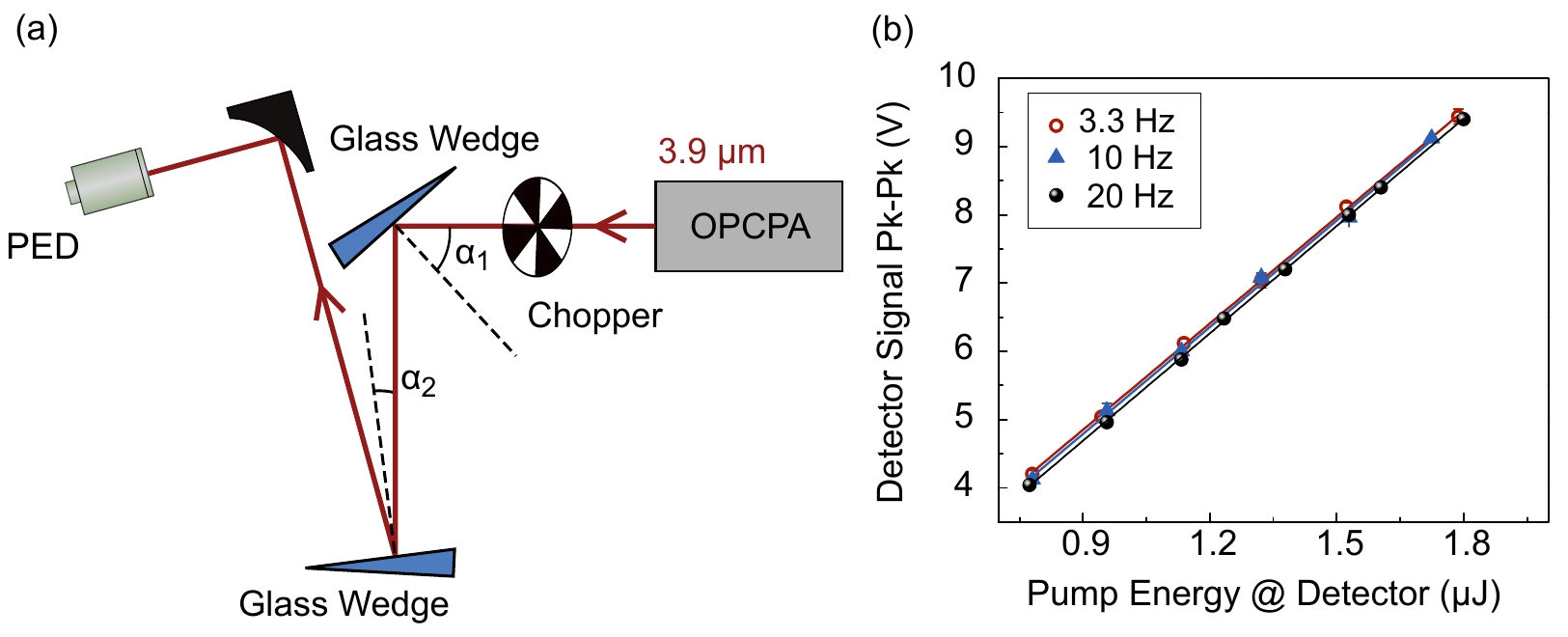}
    \caption[Experimental arrangement for the calibration of the pyro-electric detector]{(a) Experimental arrangement for the detector calibration. Reflections from two glass wedges are used to attenuate the mid-IR OPCPA output by more than three orders of magnitude. (b) PED signal and corresponding linear regression with respect to the attenuated pulse energy for different repetition rates (see legend). In the dynamic range form 3.3 - 20 Hz, no significant difference can be observed.}
    \label{fig:Calibaration}
\end{figure}

A schematic setup to obtain only a fraction of the \SI{3.9}{\micro \m} driving pulse used to calibrate the PED is shown in Fig.\ref{fig:Calibaration}(a).
The total pulse energy is measured at the OPCPA output and attenuated with two glass wedges. 
The known refraction index and reflection angles allow one to extract the reflection coefficients of the p-polarized driving pulse, resulting in an attenuation factor of more than three orders of magnitude.
The pump repetition rate is controlled with an optical chopper and the PED signal is recorded for several laser pulse energies. 
No significant difference in the detector response could be observed for 3.3, 10 and 20 Hz, respectively, as shown in Fig.\ref{fig:Calibaration}(b). 
A least square fit of the measured data allows one to determine an average coefficient $C_r(\lambda=\SI{3.9}{\micro \m})=\SI{0.1906}{\micro\J/V}$ which relates the PED signal in volt to the energy of the incident pulse in \SI{}{\micro\J}. 
Note that, according to the product specifications, the detector signal should be kept below 10 V in order to prevent saturation.
The manufacturer provides a  wavelength dependent response function, which describes the sensitivity of the detector in the range from 0.63 to \SI{395}{\micro\m}.
Thus, the provided correction factor is re-normalized at \SI{3.9}{\micro \m} and multiplied with the measured  coefficient $C_r$. 
\begin{figure}[htb]
\centering\includegraphics[width=\linewidth]{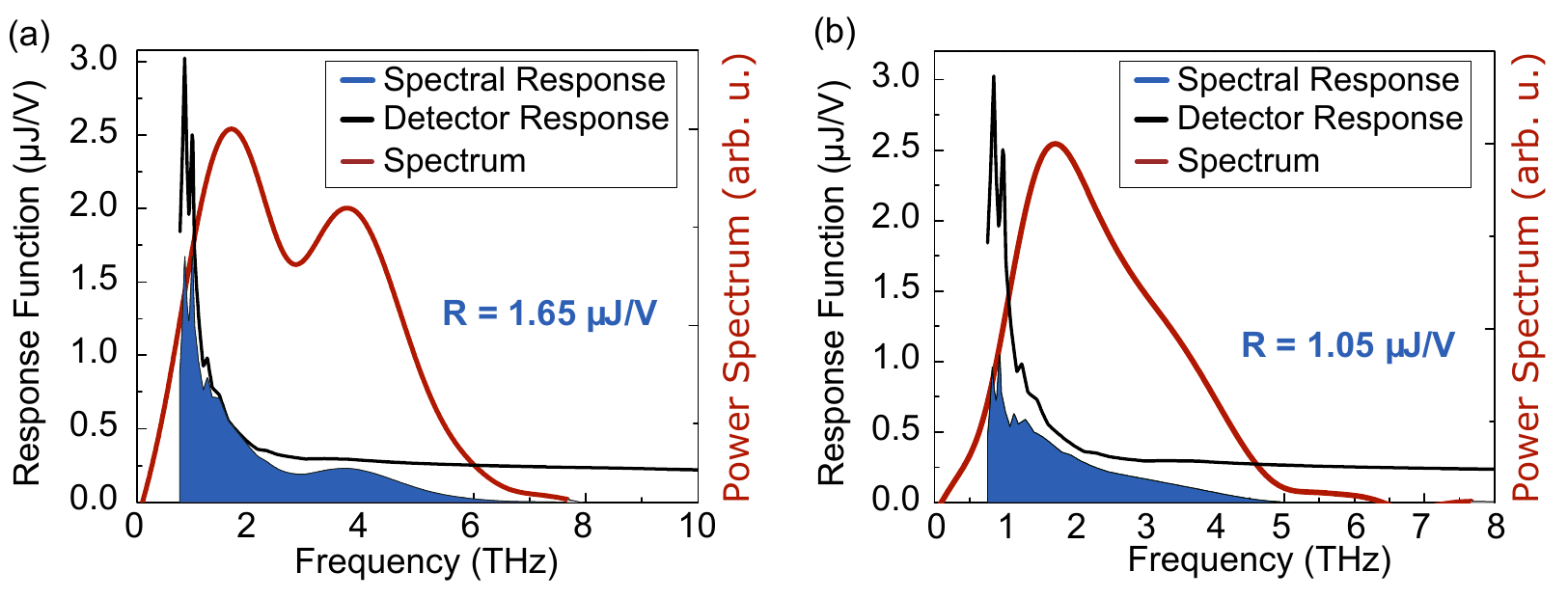}
\caption[Detector response function $F_D(f)$ and spectral response $S_f(f)$ of the pyro-electric detector]{Detector response function $F_D(f)$, spectral response $S_f(f)$ and
THz spectrum when the crystal is pumped with \SI{3.9}{\micro \m} (a) and 
\SI{1.95}{\micro \m} (b) pulses. The total response $R$ is given by the integral (blue
areas) of the spectral response. Note that, in both cases, $R$ is  
underestimated because the wavelength dependent detector sensitivity provided by the 
manufacturer is only given for frequencies above \SI{0.75}{THz}. Hence,
real THz pulse energies are slightly higher.}
\label{fig:Calibration}
\end{figure}
The resulting detector response function $F_D(f)$ in units of \SI{}{\micro\J/V}
with respect to the frequency is shown in Fig.\ref{fig:Calibration} 
(black lines).

To gain an absolute response factor $R$ in units of \SI{}{\micro\J/V} for a corresponding THz pulse, the normalized power spectrum $S_n(f)$ of the THz pulse needs to be multiplied with the response function $F_D(f)$ of the detector:
\begin{equation}
S_f(f) = S_n(f)\cdot F_D(f).
\label{eq:Sf}
\end{equation}
Two examples of recorded spectra are shown as red areas in Fig.\ref{fig:Calibration}, measured with \ac{eos}, when THz radiation is generated by OR in the organic crystal DAST, as described in section \ref{sec:SetupOR}. 
The resulting spectral response $S_f(f)$ is shown in blue and describes how much THz pulse energy is contained at each specific frequency. 
The total detector response is given by the integral over all these contributions:
\begin{equation}
R = \int_{0}^{\infty}S_f(f) df.
\label{eq:R}
\end{equation}
Figure \ref{fig:Calibration}(a) displays the detector response function, the
THz spectrum when the crystal is pumped with \SI{3.9}{\micro\m} pulses and the resulting spectral response function. 
The upper limit for the integral in Eq.\eqref{eq:R} is determined by the highest frequency of \SI{7.7}{THz}, which is detectable above  the noise level. 
Because the wavelength dependent sensitivity of the detector is only provided for frequencies above \SI{0.75}{THz}, the integral is performed with a lower limit of \SI{0.75}{THz}. 
Thus, the resulting total response of R=\SI{1.65}{\micro\J/V} is slightly underestimated and even higher  THz energy values can be expected. 
The THz spectrum for a pump pulse at \SI{1.95}{\micro\m} is shown in figure  \ref{fig:Calibration}(b), together with the detector response function $F_D(f)$ and resulting spectral response $S_f(f)$.
Because the spectral bandwidth is narrower as compared to the case of  \SI{3.9}{\micro\m} drivers and limited at \SI{5.5}{THz}, the spectral response is smaller and the total response accounts for R=\SI{1.05}{\micro\J/V}. 
\begin{figure}[htb]
\centering\includegraphics[width=\linewidth]{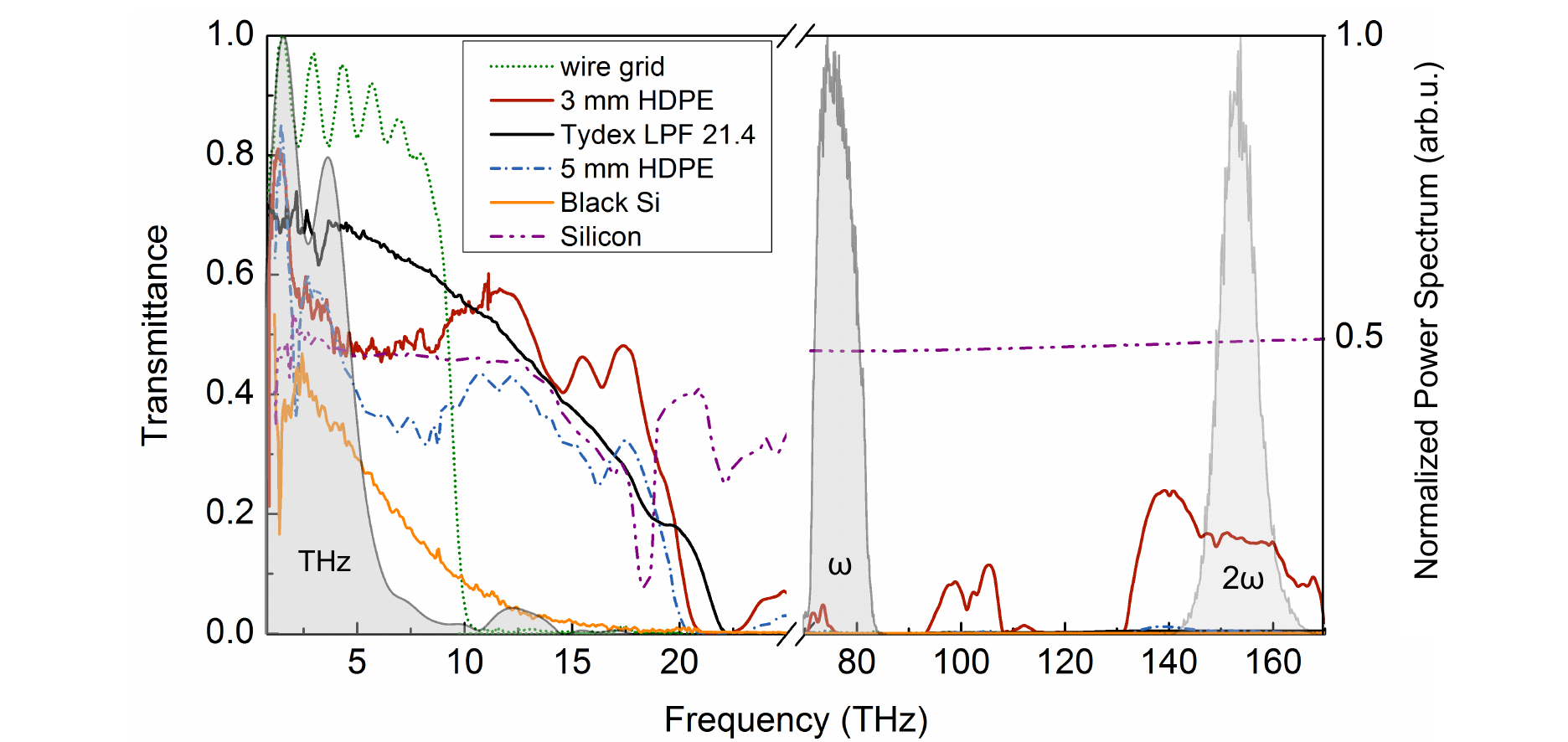}
\caption[Low pass filter transmittance]{Transmittance of a wire grid filter, HDPE sheets, Tydex, high resistivity float zone Silicon wafer, and a 0.5 mm thick low resistivity black Silicon wafer (left y-axis).
Gray shaded areas show the initial power spectra of the fundamental $\omega$ 
and second harmonic $2\omega$ pulses, as well as the power spectrum of a THz pulse generated by optical rectification.}
\label{fig:Filter}
\end{figure}

Note that, for the calibration measurement, no additional absorptive filters were used.
However, in order to separate the generated THz radiation from the driving pump pulses, and to prevent the PED from saturation, different sets of \acp{lpf} are used. 
Namely, in the case of THz generation by optical rectification in the organic crystal DAST, we use a wire grid filter with a cutoff frequency of 9 THz (Swiss Terahertz LLC), 3 mm thick \ac{hdpe} sheets (Goodfeller), and a long pass filter from Tydex (LPF 21.4).
For the work on THz generation by two-color plasma filaments, the set of long pass filters comprise of a 5 mm HDPE sheet, a 2 mm thick high resistivity float zone Silicon wafer, and a 0.5 mm thick low resistivity black Silicon wafer. 
The transmittance is depicted in Fig.\ref{fig:Filter} (left y-axis) together with a THz spectrum, generated by optical rectification, as well as a pump spectrum at \SI{3.9}{\micro\m} ($\omega=$\SI{76.9}{THz}) and at \SI{1.95}{\micro\m} ($2\omega$=\SI{153.9}{THz}) (grey areas). 
The transmittance for the optical range is several orders of magnitude lower compared to the THz range.
Specific values of the transmittance are summarized in table \ref{tab:filter}. 
\begin{table}[H]
\centering
\begin{tabular}{cccc}
\hline
\multicolumn{4}{c}{THz generation by optical rectification} \\
\hline
LPF & \SI{2}{THz} & $\omega$ & $2\omega$ \\
\hline
\hline
\\
wire grid & 0.83 & $1.74\times10^{-3}$ & $1.29\times10^{-3}$\\
3 mm HDPE & 0.7 & $3.52\times10^{-3}$ & 0.16\\
Tydex & 0.68 & $1.1\times10^{-5}$ & $4.25\times10^{-3}$\\
\\
\hline
\multicolumn{4}{c}{THz generation by plasma filaments} \\
\hline
LPF & \SI{7.5}{THz} & $\omega$ & $2\omega$ \\
\hline
\hline
\\
5 mm HDPE & 0.36 & $3.38\times10^{-5}$ &  $4.32\times10^{-3}$\\
Black Si & 0.17 & $5.42\times10^{-4}$ & $6.48\times10^{-4}$\\
Silicon & 0.47 & 0.473 & 0.49\\
\\
\hline
\end{tabular}
\caption[Transmittance of several low pass filters for the THz range, fundamental pulse
$\omega$ and its second harmonics $2\omega$]{Transmittance of several LPFs for the THz range, fundamental pulse
$\omega$ and its second harmonics $2\omega$. In the case of THz generation by optical rectification in DAST, the THz spectral range of interest peaks around \SI{2}{THz}, whereas for two-color plasma filaments the THz pulse is centred at \SI{7.5}{THz}.}
  \label{tab:filter}
\end{table}

\section{THz Beam Profile}
\label{sec:BP}

\begin{figure}[h]
    \centering
    \includegraphics[width=\linewidth]{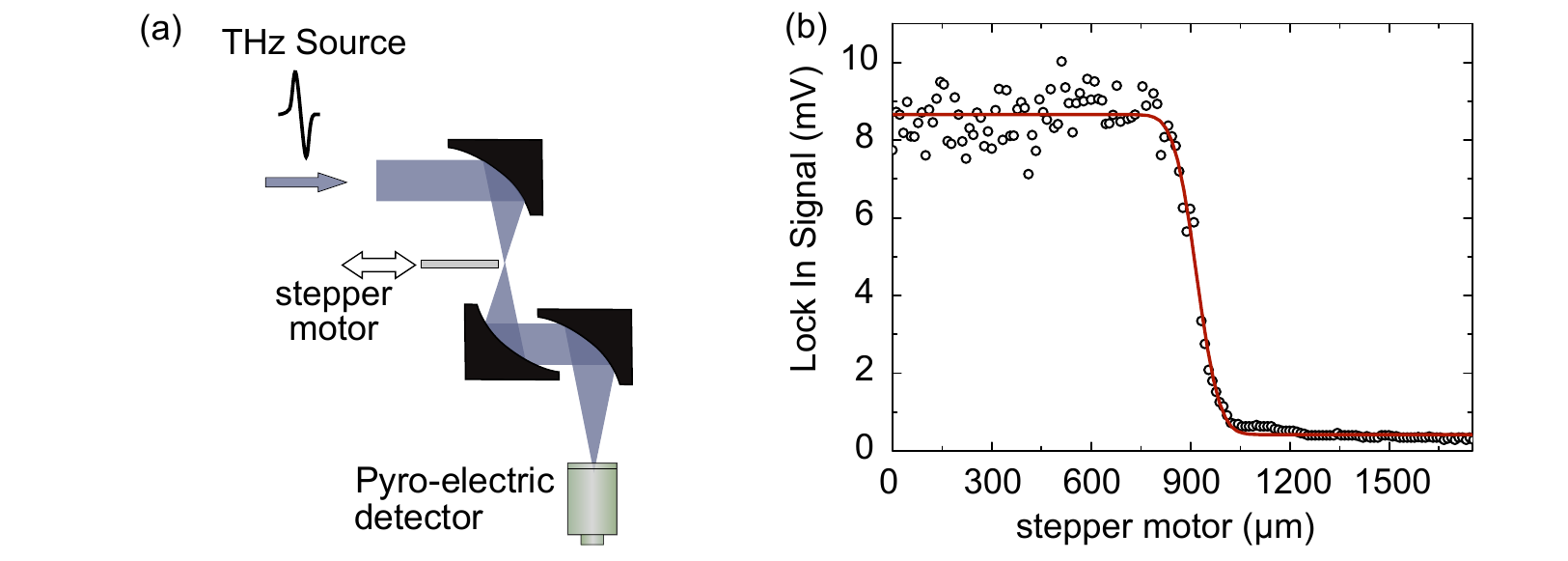}
    \caption[Illustration of the knife-edge method]{(a) Experimental sketch to measure the THz beam profile. (b) Recorded THz signal when a knife edge is moved across the focal plane (dots) and corresponding fit curve (red line).}
    \label{fig:KnifeEdge}
\end{figure}

To measure the spatial distribution of the THz pulse, the well established knife edge method is used.
In order to optimize and estimate the THz electric field strength at the position of the \ac{eoc} as shown in Fig.\ref{fig:EOS_scheme} and further sketched in Fig.\ref{fig:KnifeEdge}.(a), a sharp metal edge is moved across the THz beam in the focal plane of the parabolic mirror M3.
A short focal distance of \SI{5}{cm} is used to ensure tight focusing.
The THz beam is then re-collimated and focused on a detector with two additional parabolic gold mirrors.
The THz signal is recorded with a pyro-electric detector with respect to the position of the edge.
Assuming a Gaussian beam shape, the measured data can be fitted with the function \cite{KnifeEdge:2009}
\begin{equation}
    P=P_0+\frac{P_{\mathrm{max}}}{2}\left(1-\mathrm{erf}\left(\frac{\sqrt{2}\left(x-x_0\right)}{w}\right)\right),
\end{equation}
where $P_0$ is the background power, $P_{\mathrm{max}}$ is the maximal power, $x_0$ is the
position of the metal edge at half of the power, $\mathrm{erf}$ is the standard error function
and $w$ defines the beam radius at $1/e^2$ level. 
The example shown in Fig.\ref{fig:KnifeEdge}(b) represents a measured THz beam profile generated by OR in DAST, as described in section \ref{sec:ResultsOR}. The fitting curve reveals a beam radius of $w=\SI{107}{\micro\m}$. 

\section{Retrieval of the THz electric Field Amplitude}
\label{sec:Retrieval}
The THz electric field is estimated following the routine as described in supplementary information of ref \citeA{TasosA:2020}.
In general, the pulse energy $\epsilon$ can be written as the spatial and temporal integral of the pulse intensity
\begin{equation}
\epsilon=\iiint \limits_{-\infty}^{\infty} I(x,y,t)\mathrm{d}x\mathrm{d}y\mathrm{d}t.
\end{equation}
For a Gaussian beam propagating in z-direction, the intensity can be further expressed as
\begin{equation}
I(x,y,t)=\frac{1}{2}n_0 \varepsilon_0 c_0 \vert E(x,y,t)\vert^2,
\end{equation}
where $n_0$ is the refractive index at the pulse central frequency, $\varepsilon_0$ is the vacuum permittivity, $c_0$ is the speed of light in vacuum and $E(x,y,t)$ is  the spatio - temporal THz waveform on the detection plane. The electric field can be further expressed as
\begin{equation}
E(x,y,t)=E_{\mathrm{max}} E_t(t) E_A(x,y),
\end{equation}
where $E_{\mathrm{max}}$ is the maximum THz field amplitude, $E_t (t)$ is the normalized THz waveform as shown in Fig.\ref{fig:EOS_Sample}(a), 
and $E_A(x,y)^2$ is the normalized THz spatial profile on the detector plane \cite{OhKim:2014},
with
\begin{equation}
A=\iint \limits_{-\infty}^{\infty} E_A(x,y)^2 \mathrm{d}x\mathrm{d}y = \pi w^2
\end{equation}
as the beam spot area of the intensity profile at $1/e^2$ level.
The THz field strength is therefore determined from all three measurements of energy, spot size and temporal transient by
\begin{equation}
\epsilon = \frac{1}{2} \varepsilon c_0 A \int \limits_{-\infty}^{\infty}  E_t (t)^2 \mathrm{d}t E_{\mathrm{max}}^2,
\end{equation}
assuming $n_0=1$ in air.
Defining the integral of the squared normalized waveform as
\begin{equation}
G_t=\int \limits_{-\infty}^{\infty}  E_t (t)^2 \mathrm{d}t,
\end{equation}
The maximum THz field amplitude can be calculated with
\begin{equation}
E_{\mathrm{max}}=\sqrt{\frac{2\epsilon}{\varepsilon_0 c_0 G_t A}},
\label{equ:FielAmplidtude}
\end{equation}
wherein $\epsilon$ is measured with the PED, the beam spot $A$ is determined with knife edge and $G_t$ is resolved from the integrated temporal profile form \ac{eos}.

Note that, the evaluated field strength refers to the field at the focal position of the THz beam with a beam waist $w$.
Slight displacement of a sample of interest from the focal plane causes a drastic reduction of the peak electric field.
For instance, if the sample is misaligned by the Rayleigh length $z_{\mathrm{R}}=\pi w^2/ \lambda$, 
the field strength is already diminished by 30\%. 
A displacement of only 1.7 times the Rayleigh length results in half of the electric field strength at the sample position. 
In the case of a THz pulse with a long central wavelength of $\sim$ \SI{300}{\micro m} (\SI{1}{THz}) and beam waist in the order of \SI{100}{\micro m}, the Rayleigh length accounts for only \SI{105}{\micro m}.
Thus, the beam is highly divergent and the field strength reduces drastically if the sample is misplaced from the focal plane.
In addition, intensity fluctuations of the \SI{3.9}{\micro\metre} pump source cause a field strength error of approximately $\pm 5\%$.

%% file: Chapters/ExperimentalSetup.tex
\chapter{Experimental Setup}
\label{ch:ExperimnetalSetup}

Due to the lack of suitable broadband laser gain materials in the mid-IR spectral range, a common approach to generate energetic pulses with ultrashort pulse durations is optical parametric chirped pulse amplification.
It combines the technology of \acp{opa} and \acp{cpa} . 
Donna Strickland and Gérard Mourou received the Nobel Prize in 2018  for the invention of chirped pulse amplification \cite{STRICKLAND:1985}, a technique to create ultrashort, high-intensity laser pulses. 
The main idea of a \ac{cpa} is simple: because a laser pulse with intensities in the order of \SI{}{GW/cm^2} can cause serious damage to the gain medium through \ac{nl} processes such as self-focusing, the peak intensity needs to be diminished.
Reduction of the peak intensity is realised by temporal pulse broadening.
Prior to amplification, the pulse is stretched in time by splitting spectral components into different propagation path lengths, resulting in a chirped temporal profile. 
The pulse is then amplified and subsequently compressed in time to regain high peak powers.
In the case of an \acf{opcpa}, instead of material damage, another parameter is more important for sufficient pulse amplification by frequency down conversion in a \ac{nl} crystal:
powerful pump lasers exhibit pulse durations in the order of a few-tens to a few hundreds of picoseconds.
Thus, stretching of the seed allows efficient time overlap with the pump pulses.

This chapter is dedicated to a detailed description of our high-power mid-IR \ac{opcpa}  with a central wavelength of \SI{3.9}{\micro m}. 
The laser system was initially introduced in ref \cite{Andriukaitis:2011} and further upgraded as described in ref \citeC{Shumakova:18} and \cite{Valentina:2018}, resulting in a pulse energy of more than \SI{30}{mJ}, pulse duration of less than \SI{100}{fs} and peak power > \SI{300}{GW}.
In addition, we present the wavelength tunable \ac{nopa}, which is utilized to generate the visible probe pulse for \ac{eos} and used to sample THz induced changes in optical transmission in heterostructure \acp{qd} (see chapter \ref{ch:QDs}). 
We further present an overview layout of the entire mid-IR laser system, together with \ac{nopa} and synchronization of the different pump pulses.
Finally, experimental details on \ac{eos} are elucidated.

\section{Mid-IR Optical Parametric Chirped Pulse Amplifier}
\label{sec:OPCPA}

\begin{figure}[htb]
    \centering
    \includegraphics[width=\linewidth]{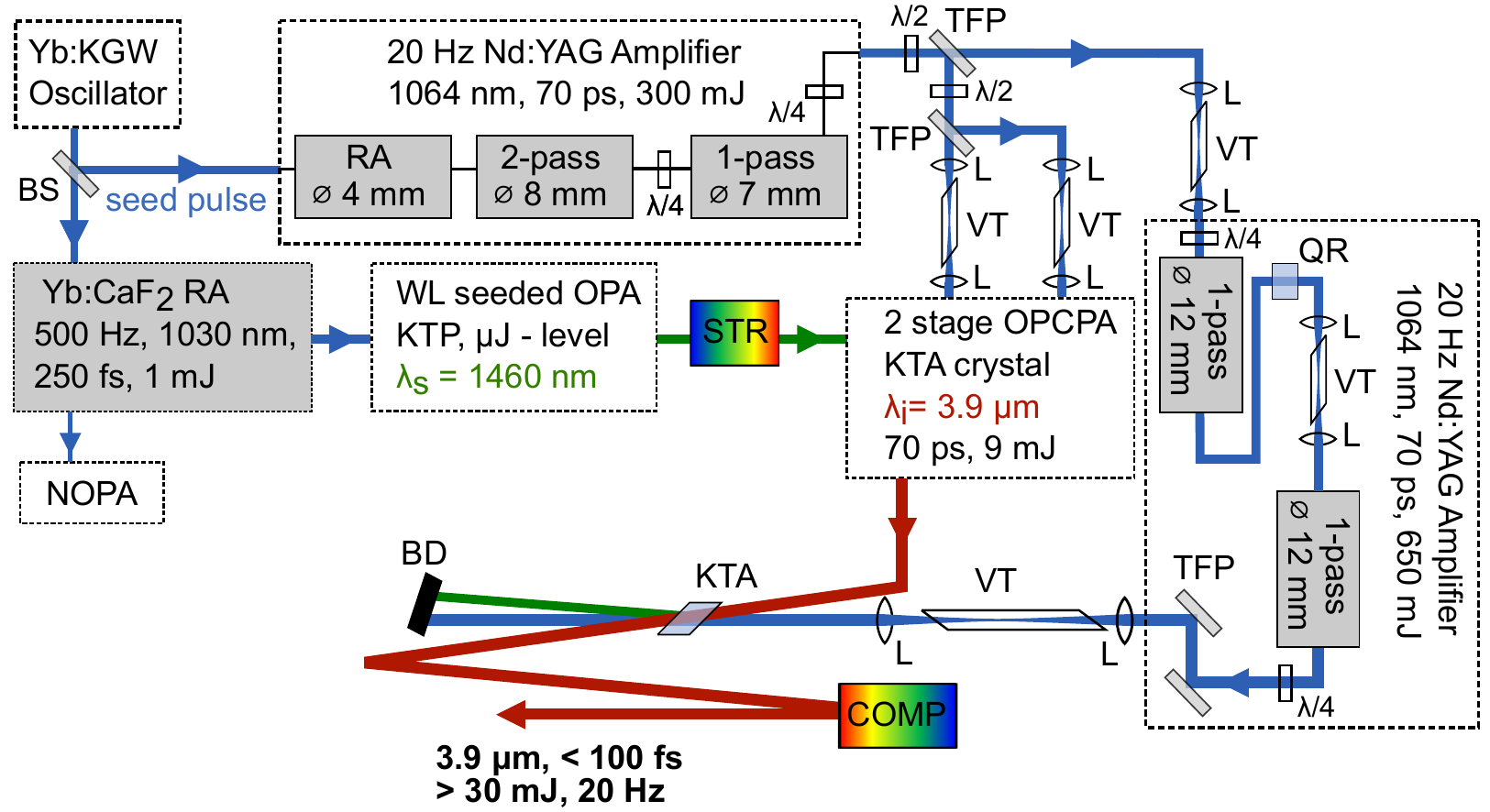}
    \caption[Layout of the mid-IR OPCPA]{Layout of the mid-IR \ac{opcpa}.TFP - thin film polariser, VT - vacuum tube, L - lens, STR - stretcher, COMP - compressor, BS - beam splitter, $\lambda$/4 - quarter waveplate, $\lambda$/2 - half waveplate}
    \label{fig:OPCPA}
\end{figure}

The layout of the laser system is depicted in Fig.\ref{fig:OPCPA}
The front-end of the mid-IR \ac{opcpa} is  a conventional Yb:KGW Kerr lens mode-locked oscillator (Light Conversion) with a repetition rate of \SI{76}{MHz}, providing optical synchronization and the seed pulses for two pump lasers, namely (i) a \SI{}{Yb:CaF_2} \ac{ra} based on a \ac{cpa} with a central wavelength of \SI{1030}{nm} and (ii) a Nd:YAG amplifier system with a central wavelength of \SI{1064}{nm}.
In order to separate those two seed beams, a low order half-wave plate and a  thin film polariser are  placed at the output of the oscillator and before the stretcher of the  \SI{}{Yb:CaF_2} \ac{ra}.

After the compressor, the pulse duration of the \SI{}{Yb:CaF_2} \ac{ra} is in the order of \SI{250}{fs} with an output energy of \SI{1}{mJ}, of which $\sim$\SI{100}{\micro J} will be later on used to drive the visible \ac{nopa}.
The major part of the output energy is used to drive three cascades of a \ac{wl} seeded \ac{opa} at a repetition rate of \SI{0.5}{kHz}.
The \ac{wl} is generated in a \SI{6}{mm} long \ac{YAG} crystal, providing  spectral brightness in the vicinity of \SI{1500}{nm}.
The \ac{wl} seed is amplified in three subsequent \ac{opa} stages based on \ac{KTP} crystals.
The final signal output from the OPA is centred at \SI{1460}{nm} with a pulse energy in the order of $\sim$ \SI{50}{\micro J}. 
It is then stretched in a negative dispersion stretcher based on a grism pair (consisting of  grating pairs and Brewster angled LAK16A prisms) to $\sim$ \SI{70}{ps} and used as an input signal for the first two stages of the following three stage mid-IR \ac{opcpa}.

The pump laser for the \ac{opcpa} is a \SI{20}{Hz} flash-lamp pumped Nd:YAG system  with two separate units, wherein one is used to pump the first two stages of the \ac{opcpa} and the second energetic entity pumps the final power amplification stage of the \ac{opcpa}.  
The first unit is based on a \ac{ra} with a \SI{65}{mm} long laser rod ($\varnothing$ \SI{4}{mm}), a double path amplifier ($\varnothing$ \SI{8}{mm}$\times$\SI{85}{mm} rod) and a single path booster amplifier ($\varnothing$ \SI{7}{mm}$\times$\SI{85}{mm} rod).
In order to limit \ac{nl} pulse spectrum broadening in the Nd:YAG \ac{ra}, a fused silica etalon is installed in the \ac{ra} cavity.
A quarter-waveplate is placed in front of the single path amplifier to change the polarization from linear to circular and hence reduce the peak intensity by a factor of 2 to minimize the risk of optical damage of the laser rod. 
The polarization of the \SI{300}{mJ} output pulse (\SI{1064}{nm}, \SI{70}{ps}) is again restored to linear after the laser rod.
With a variable \ac{bs}, consisting of a half-waveplate and a thin film polariser, approximately \SI{50}{mJ} are picked up to seed the second Nd:YAG pump unit, while the rest is used to pump the first two stages of the OPCPA.
In order to maintain the pump beam quality, the output surface of the laser rod is 4-f imaged (with certain de-magnification) onto the subsequent laser rods and \ac{nl} crystals. 
To prevent beam distortion from plasma, which is generated when the beam would be focused in air, several evacuated Brewster windowed cells are installed in the propagation path.\\
The second pump unit consists of two identical flash lamp pumped \SI{85}{mm} long Nd:YAG rods with a diameter of \SI{12}{mm}.
The peak intensity is again reduced by installing a quarter waveplate before the first rod in order to get circularly polarized light, and restored to linear polarization after the second rod.
Thermally induced depolarization in the first rod is compensated by a quartz rotator in between the rods, which swaps the tangential and radial polarization components such that depolarization accumulated in the first rod will be counteracted when propagating through the second rod.
Remaining depolarized light is filtered by thin film polarisers at the output of the amplifier with a pulse energy of \SI{650}{mJ}.

\begin{figure}[htb]
    \centering
    \includegraphics[width=\linewidth]{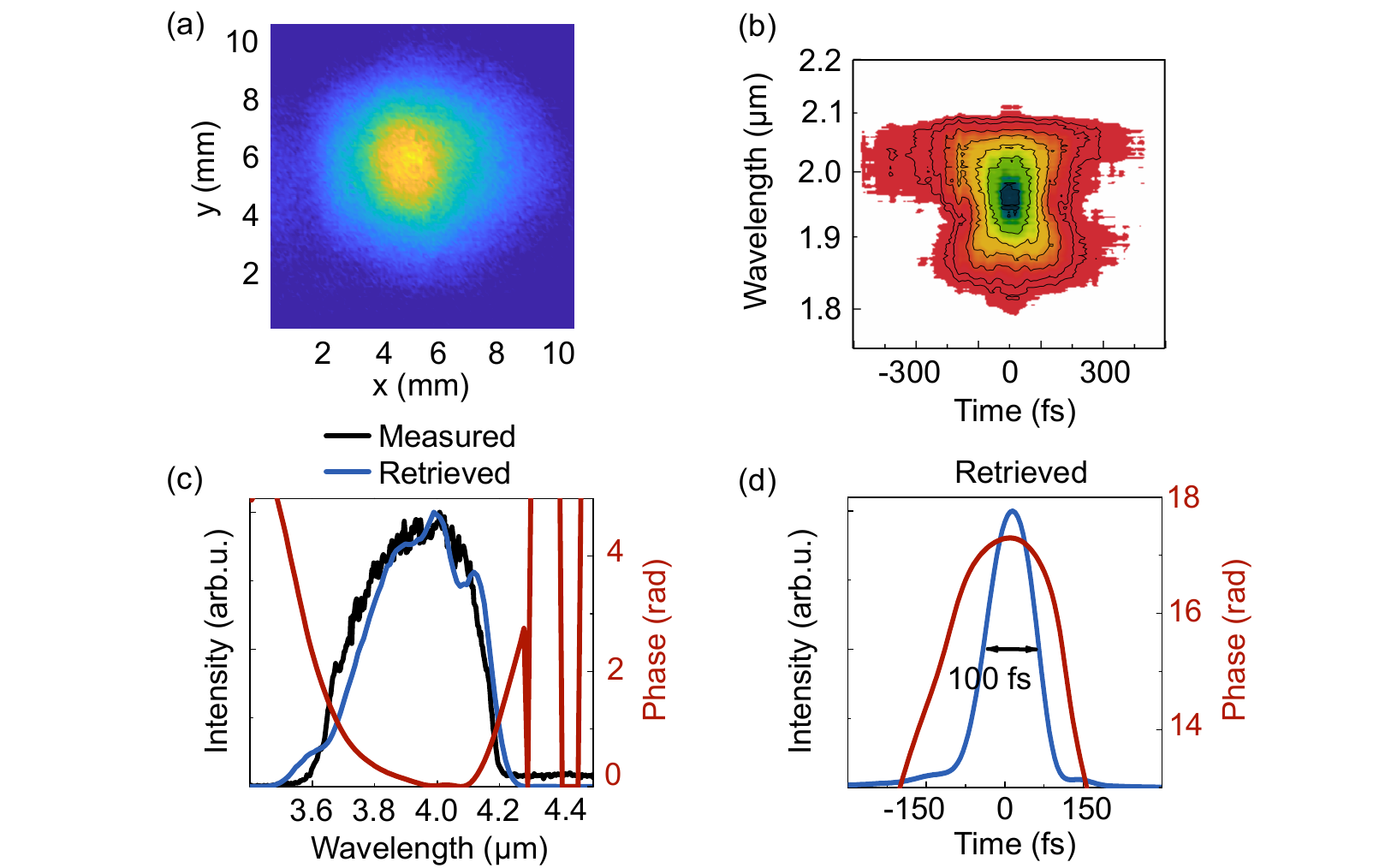}
    \caption[Characterization of the OPCPA output]{Characterization of the \ac{opcpa} output. (b) Beam profile measured with Spiricon Pyrocam III. (b) Measured \ac{frog} trace. (c) Retrieved spectrum from (b) (blue line) and measured with a spectrometer based on acousto-optic interactions (MOZZA - FASTLITE) (black line). (c) Retrieved pulse shape from (b).}
    \label{fig:OPCPA_character}
\end{figure}
The three stage \ac{opcpa} is based on \ac{KTA} crystals, which are transparent at around \SI{4}{\micro m}, and are suitable for amplification of  \SI{1460}{nm} signal and \SI{3.9}{\micro m} idler pulses while being pumped at \SI{1064}{nm}.
The first crystal is \SI{10}{mm} long  with an input surface of \SI{5}{mm}$\times$\SI{5}{mm}.
After parametric amplification, the idler at \SI{3.9}{\micro m} and pump at \SI{1064}{nm} are dumped while the signal pulse is steered to the second stage.
As multi-layer anti-reflection coatings are prone to optical damage, \ac{KTA} crystals of the the second and third stages feature uncoated Brewster cut faces that provide high transmission for the p-polarized pump and idler, and attenuate the s-polarized signal  by reflection losses of $\sim$25\% at each surface.
After the second \ac{opcpa} stage, the pump and signal pulses are discarded and the \SI{3.9}{\um} idler pulse (with an energy of \SI{9}{mJ}) is used to seed the final \ac{opcpa} stage consisting of a \SI{5}{mm} thick, Brewster angle cut type-II \ac{KTA} crystal.
Dichroic coatings in the mid-IR are notorious for optical breakdown and often only available as specialised costume design, which makes optical components cost intensive. 
Hence, in order to spatially separate the pump and signal pulses from the amplified idler, in the third stage the \SI{3.9}{\um} pulses are amplified non-collinearly, resulting in a pulse energy of > \SI{43}{mJ}.
In order to  avoid an appearance of angular chirp in the idler beam, in the second \ac{opcpa} stage parametric amplification takes place in a strictly collinear geometry. 
In the third, non-collinear OPCPA stage, which is seeded with angular chirp free \SI{3.9}{\um} pulses, the angular chirp  appears in the 1460 nm beam, generated during amplification.
Since the signal and idler pulses are phase conjugated, the compressor of the mid-IR pulses also consists of a  grating pair, which is arranged for negative dispersion. 
The  transmission efficiency of  the compressor is $\sim$70\%.
The generated mid-IR pulse is centred at \SI{3.9}{\micro m} with a spectrum extending from \SI{3.62}{\micro m} to \SI{4.19}{\micro m} at $1/e^2$ level, with a pulse duration of less than \SI{100}{fs} and pulse energies of more than \SI{30}{mJ}.
Output parameters of the mid-IR \ac{opcpa} are displayed in Fig. \ref{fig:OPCPA_character}

\section{Non-collinearly pumped Optical Parametric Amplifier}
\label{sec:NOPA}
\begin{figure}[htb]
    \centering
    \includegraphics[width=\linewidth]{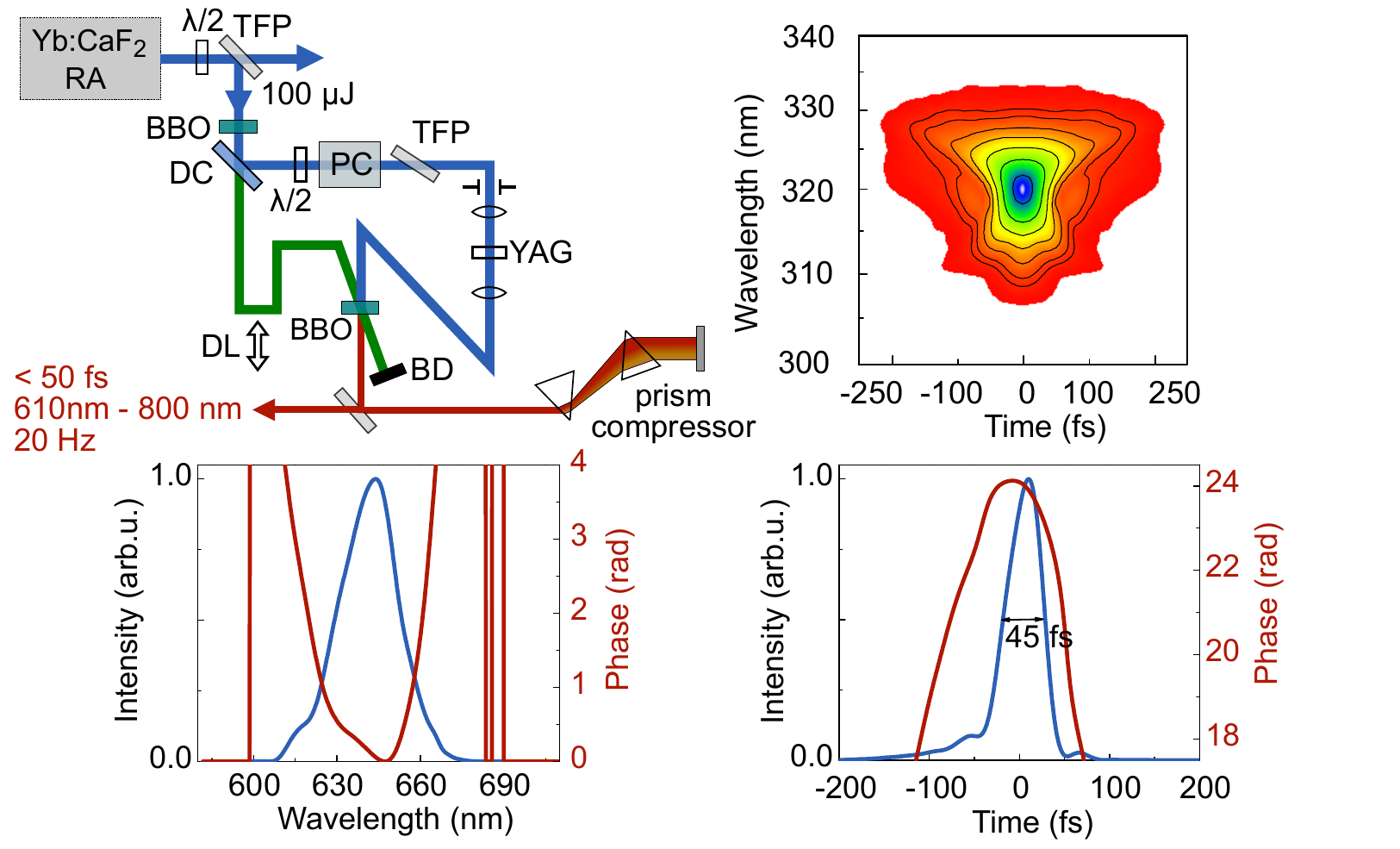}
    \caption[NOPA schematics and  SHG FROG characterization of the generated probe pulse]{\ac{nopa} schematics and  SHG FROG characterization of the generated probe pulse. TFP - thin film polariser, DC - dichroic mirror, PC -  Pockels cell, BD - beam dump, DL - delay line.}
    \label{fig:NOPA_FROG}
\end{figure}

A schematic setup of the visible \acf{nopa} as well as a characterization of the resulting probe pulse measured with \ac{shg} \ac{frog} are shown in Fig. \ref{fig:NOPA_FROG}.
A variable beam splitter consisting of a half waveplate and thin film polariser is installed at the output of the \SI{}{Yb:CaF_2} \ac{ra} operating at \SI{500}{Hz}.
The s-polarised reflected fraction of $\sim$\SI{100}{\micro J} is  frequency doubled in a \SI{2.5}{mm} thick \ac{BBO} crystal and further used to drive the \ac{wl} seeded single stage \ac{nopa} based on another \SI{2.5}{mm} thick type I \ac{BBO} crystal. 
\ac{wl} is generated with the near-IR pulse centred at \SI{1030}{nm} in a \SI{2}{mm} thick  \ac{YAG} crystal.
The synchronization between the mid-IR \ac{opcpa} system and the visible \ac{nopa}  is achieved by installing in the optical path of the \SI{1030}{nm} pulse, generating WL, an electro-optical chopper based on a half waveplate, Pockels cell and thin film polariser, which reduce the repetition rate of \ac{nopa} to \SI{20}{Hz}.
In order to broaden the spectral bandwidth and to achieve shorter probe pulses, a non-collinear pump arrangement is used \cite{Wilhelm:1997}. 
The NOPA generates broadband pulses, tuneable from \SI{610}{nm} to \SI{800}{nm}, which can be compressed to < 50 fs with  a pair of SF10 prisms. 
%
%

\section{Electro-Optic Sampling}
\label{sec:ProbePulse}
\begin{figure}[htb]
    \centering
    \includegraphics[width=\linewidth]{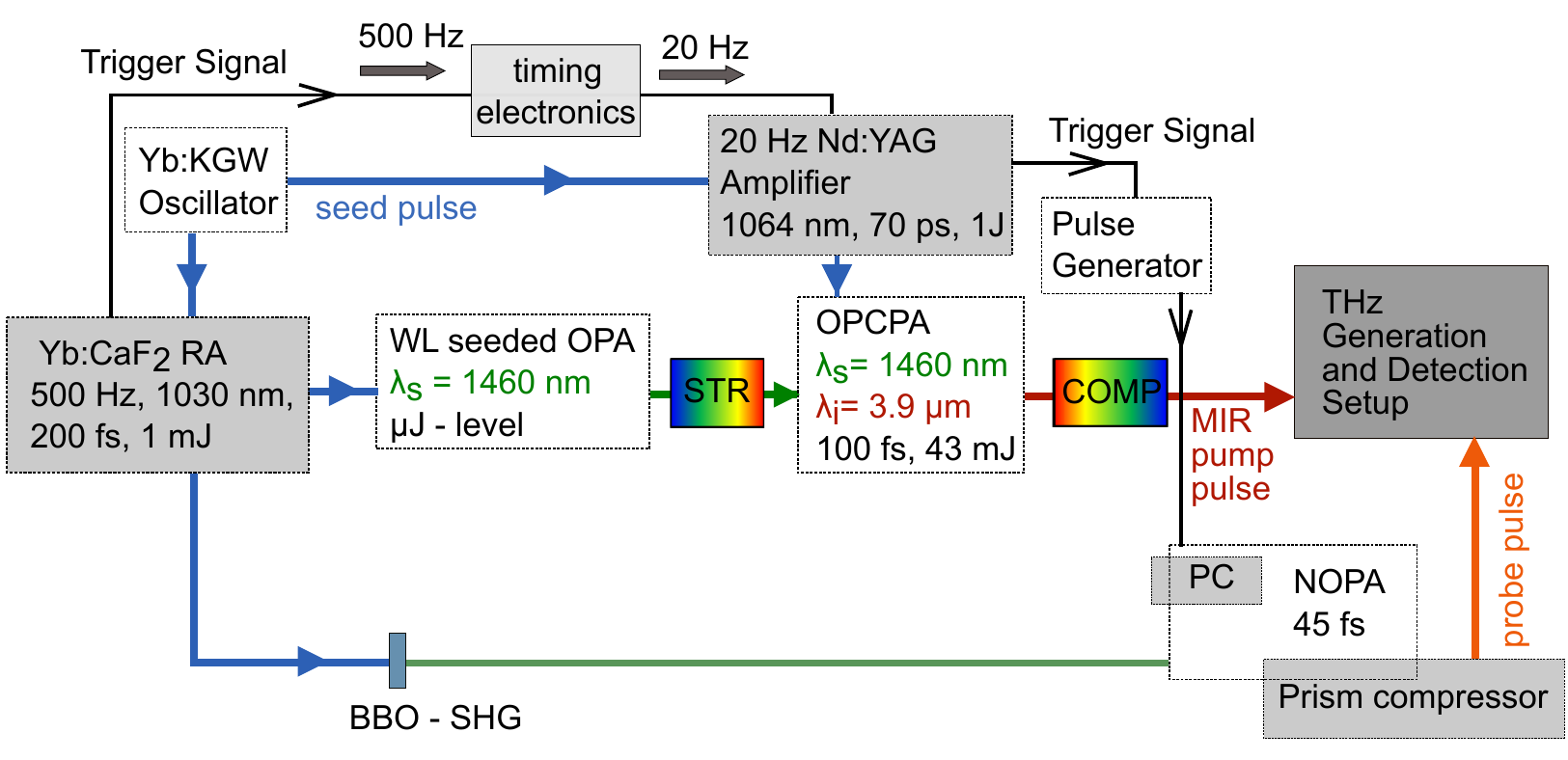}
    \caption[Schematics of the pump laser system and probe pulse generation, including synchronization between the driving pulses]{Schematics of the pump laser system and probe pulse generation, including synchronization between the driving pulses of the \ac{opa} operating at 500 Hz and \ac{opcpa} with a repetition rate of 20 Hz. STR - stretcher, COMP -compressor, PC - Pockels Cell}
    \label{fig:NOPA_EOS}
\end{figure}

In contrast to conventional THz generation and detection setups, wherein a Ti:Sapphire laser (centred around \SI{800}{nm}) is routinely utilized as a pump pulse for THz generation as well as probe pulse for EOS, here, the mutual source is provided by the \SI{}{Yb:CaF_2} \ac{ra} (see Fig.\ref{fig:NOPA_EOS}), which makes it necessary to introduce several meters of optical path length to temporally overlap the two pulses.
Synchronization between the components of the laser system is achieved as follows:
The oscillator triggers a digital delay generator which controls the Pockels cell of the \SI{}{Yb:CaF_2} \ac{ra}, operating at a repetition rate of \SI{500}{Hz}. 
This digital delay generator, providing a \SI{500}{Hz} trigger signal, is further used as an input signal for additional delay generators to synchronise the flash lamps and Pockels cell of the Nd:YAG pump system, as well as the Pockels cell of the \ac{nopa}, and to reduce the repetition rate  to \SI{20}{Hz}.

\begin{figure}[bht]
    \centering
    \includegraphics[width=\linewidth]{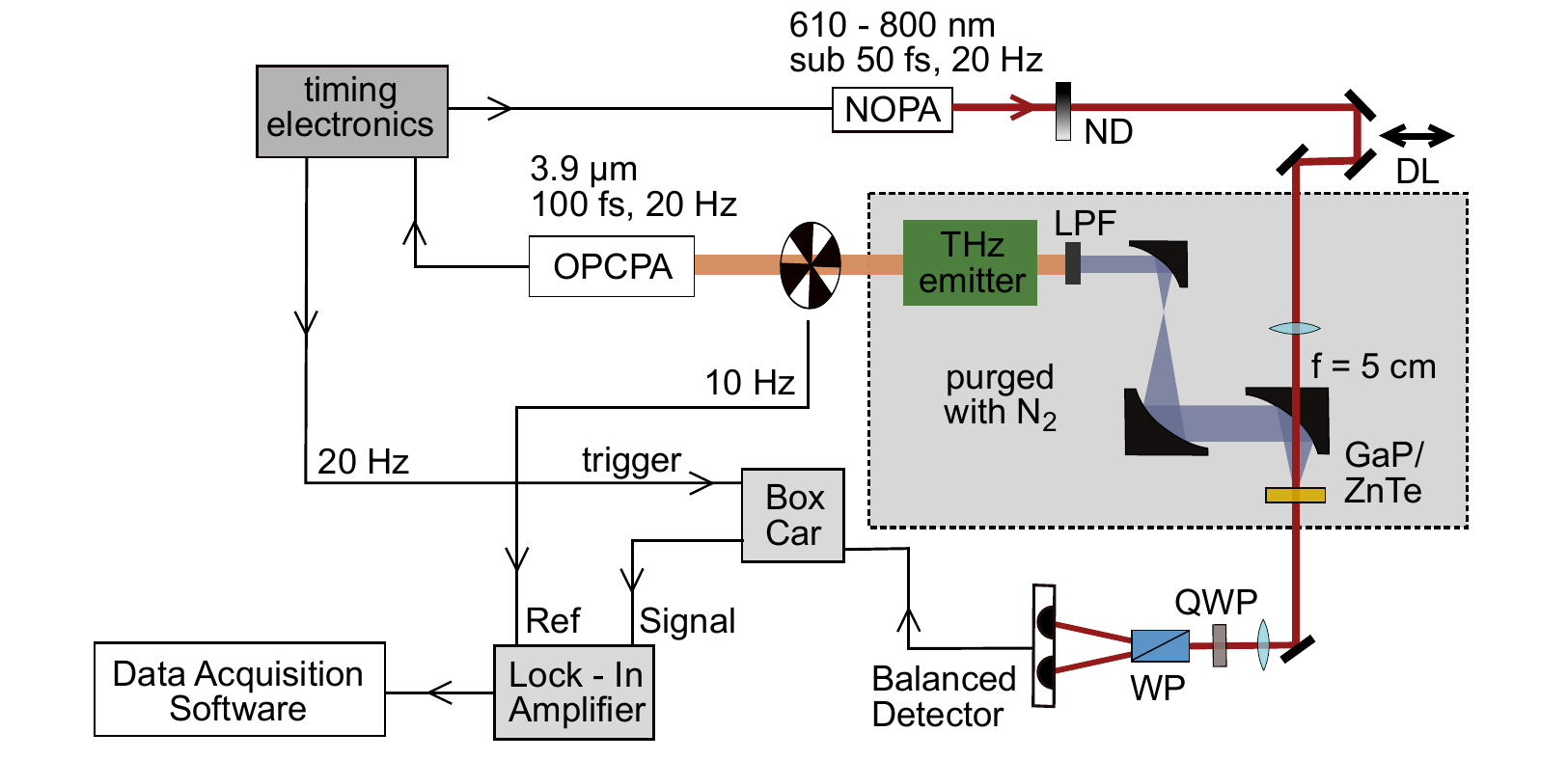}
    \caption[Details on the experimental setup for electro-optic sampling]{Experimental setup for \ac{eos}. ND - neutral density filter, LPF - low pass filter, DL - delay line, WP - Wollaston prism, QWP - quarter waveplate.}
    \label{fig:EOS_EOS_setup_Details}
\end{figure}

Details of the \ac{eos} setup are illustrated in Fig. \ref{fig:EOS_EOS_setup_Details}.
The THz emitter, indicated by a green rectangular, can be either THz generation by \ac{or} in DAST, as described in chapter \ref{ch:OR}, or THz generation by two-color plasma filaments (see chapter \ref{ch:filaments}).
The mid-IR pump pulse and generated THz radiation are separated with several \acp{lpf} (see Fig. \ref{fig:Filter} in section \ref{sec:Pyro} for more details on the filter transmission).
A \ac{nd} is used to attenuate the probe pulse intensity in order to avoid unwanted \ac{nl} effects.
To combine the THz pulse with the probe pulse onto the \ac{eoc}, a f = 5 cm parabolic mirror with a hole in the center is used. 
The THz induced birefringence in the \ac{eoc} is measured with a balanced detector (Thorlabs PDB210A) installed behind a \ac{wp} and \ac{qwp}, which are used to separate the orthogonal polarization components of the probe pulse,  and turn linear polarization into circular, respectively.
As the pulsed output of the balanced detector is short (in the order of \SI{0.5}{\micro s}), but is monitored for a long time period ($\sim$\SI{100}{ms}) determined by the frequency of the optical chopper (\SI{10}{Hz}), a boxcar module (SR250 Stanford Research Systems) was installed between the balanced detector and the Lock-In amplifier (SR830 by Stanford Research Systems), in order to reduce the background noise.

For initial alignment of the system we use a \SI{1}{mm} thick ZnTe crystal.
As phase matching in a broad spectral range is only possible in a thin \ac{eoc} crystal, the measured THz spectrum will be narrow.
However, such a spectrally narrow THz pulse can co-propagate with the probe pulse for a long time in the \ac{eoc}, which leads to a larger \ac{eos} signal. 
In order to detect the undistorted shape of the THz transient, a thin \ac{GaP} crystal of \SI{50}{\micro m} is placed into the focus of the parabolic mirror.
As described in section \ref{sec:EOS}, and further explained in detail in refs \cite{Casalbuoni:2008, Leitenstorfer:1999}, the \ac{eoc} response depends on the reflection losses at the crystal surface, absorption inside the crystal, mismatch between the THz phase velocity and group velocity of the optical gating pulse, and the NL electro-optic coefficient.
Because of that, during propagation through the crystal, the THz phase and amplitude will be distorted, which can be corrected in hindsight if the frequency dependent crystal response function is known.
In particular, a probe wavelength around \SI{800}{nm} is known to yield relatively undistorted THz signals in thin GaP crystals, especially for the spectral range  of $<$\SI{8}{THz} in the case of a \SI{50}{\micro m} thick crystal \cite{Casalbuoni:2008} and $>$\SI{20}{THz} for a crystal  thickness of \SI{13}{\micro m} \cite{Leitenstorfer:1999}.
The dip around $\sim$ \SI{8}{THz} of the detection bandwidth is mainly caused by \ac{to} phonon resonances. 
Nonetheless, higher frequencies are still measurable although the signal will be largely reduced.
In this work, probe pulses centred around \SI{620}{nm} to \SI{680}{nm} are mainly used for EOS.
Even tough it is possible to tune the NOPA to a central wavelength of \SI{800}{nm}, short probe wavelengths around \SI{650}{nm} are needed for tracing electro-absorption modulation, as discussed in chapter \ref{ch:QDs}.
However, it is determined that the detection error is marginal for a probe wavelength of \SI{650}{nm} in comparison to \SI{800}{nm} for THz pulses with spectral components of  $<$\SI{8}{THz}  \cite{Daniel:2021} .
\begin{figure}[htb]
    \centering
    \includegraphics[width=\linewidth]{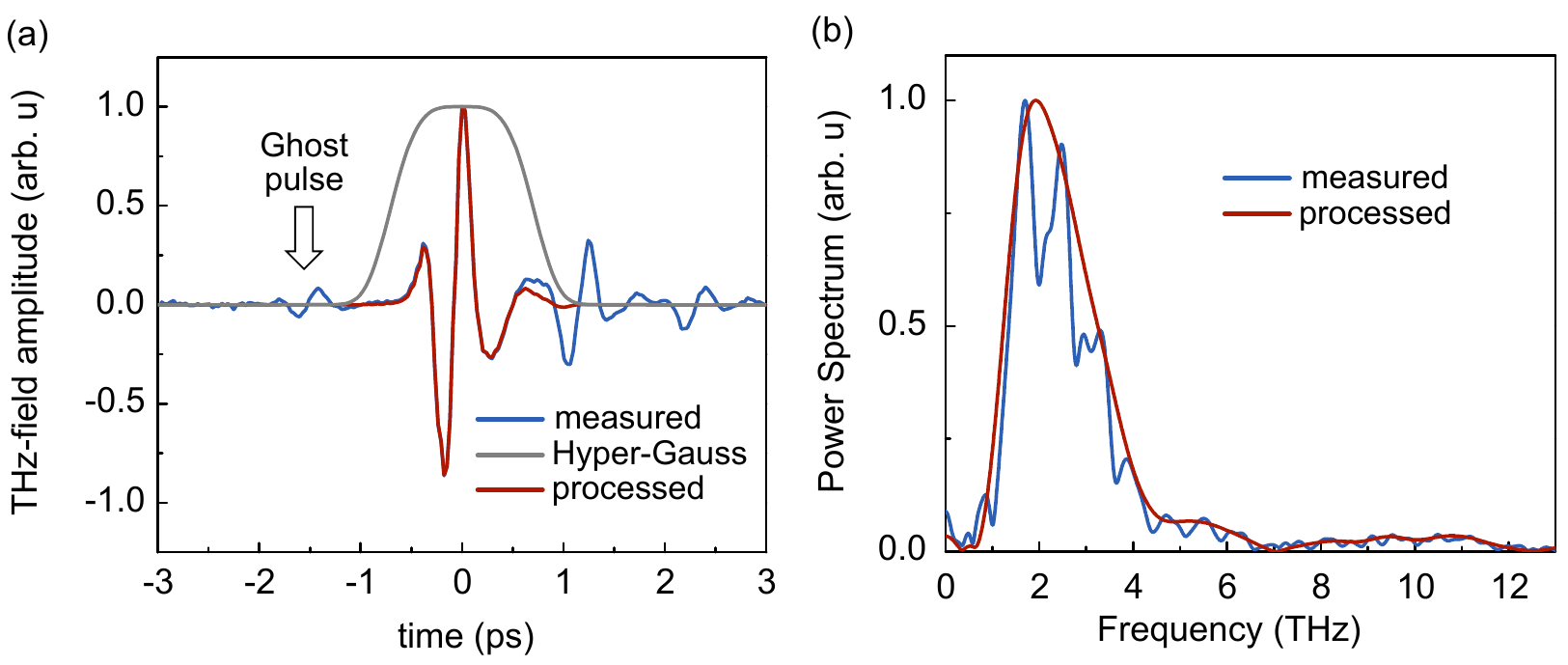}
    \caption[THz transient measured with electro-optic sampling and its Fourier transform]{(a) THz transient measured with EOS. The blue line represents the detected signal with respect to the time delay between the THz and probe pulse. The small signal before the main pulse depicts a ghost pulse, originating from the reflections at the crystal surface. The tail contains residual oscillations caused by the absorption in water vapour. The grey line indicates a Hyper-Gaussian filter function and the red curve shows the resulting processed signal. (b) FFT of the  measured signal  (blue), exhibiting etalon effects, and of the processed signal ( red).}
    \label{fig:EOS_Sample}
\end{figure}

An example of a sampled THz transient is shown in Fig.\ref{fig:EOS_Sample}, for a probe wavelength of \SI{620}{nm}, when the detection setup is purged with nitrogen to decrease the humidity down to 1.5 \%.
The THz pulse was generated by OR in the organic crystal DAST, as discussed in chapter \ref{ch:OR}. 
Because of the finite thickness of the detection crystal, multiple reflections of the THz  and probe pulse occur, which appear as 'ghost' pulses before and after the main signal.
When the entire detection window is used in time domain and Fourier transformed to the frequency domain, Fabry-Perot resonances are visible in the spectrum due to destructive interference with the additional pulses, as indicated by the blue line in Fig.\ref{fig:EOS_Sample}(b).
To avoid such features caused by etalon effects, a time filter can be applied by multiplying the time traces with a Hyper-Gaussian function (grey line), centred at the maximum of the THz-signal \cite{EOStutorial:2018}, which corresponds to  spectral filtering in frequency domain.
The processed signal in time domain is  Fourier transformed, resulting in a smooth THz spectrum exceeding \SI{6}{THz} (red line in Fig.\ref{fig:EOS_Sample}(b)).  
Such a pre-processing technique is also commonly used to suppress the influence of water vapor absorption in ambient air, which usually manifests in the time domain by an oscillating tail \cite{SchneiderOR:2006}.

%% file: Chapters/OR.tex
\chapter{THz generation by OR of mid-IR pulses in the organic crystal DAST}
\label{ch:OR}

Due to extraordinarily high nonlinearities, small dielectric constant and small
dispersion from low to optical frequencies, organic crystals are 
excellent materials for broadband THz generation \cite{Jazbinsek:2008,OC_THzPhotonics:2019}.
They provide high laser to THz conversion efficiencies at room temperature,
broad spectra and naturally collimated and aberration free THz beams \cite{CascadedVicario:2014}.
Because of beneficial phase matching conditions and high transparency in the near-IR, laser sources operating at the telecommunication wavelength of \SI{1550}{nm} are commonly used as drivers for efficient THz generation in DAST \cite{DASTVicario:2015,DASTHauri:2011}. 
Nonetheless, Vicario and co-workers \cite{DAST_TransVicario:2015} observed  a drop in transmission by more than a factor of two (from more than 60\% to less than 30\%) at a pump energy density of \SI{10}{mJ/cm^2} (corresponding to the intensity of \SI{0.15}{TW/cm^2}) when a \SI{200}{\micro\m} thick DAST crystal was pumped with \SI{65}{fs} pulses, centred at \SI{1.5}{\micro\m}.
Because the absorption edge of DAST is situated at around \SI{700}{nm} \cite{DASTCao:2016}, the crystal suffers from \ac{mpa}, which acts on the saturation of the conversion efficiency, as well as on the crystal damage threshold \cite{GaSeDamage1:2013,GaSeDamage2:2014} and restricts the applicable energy density of the driving pulses to about  \SI{20}{mJ/cm^2}.
Furthermore, \ac{mpa} of the driving pulse and subsequent free-carrier absorption of the THz radiation was identified to be responsible for the reduction of the THz generation efficiency via OR \cite{Zhen_YuCompetingNL:2008}.

In this work, we perform pioneering studies of THz generation in DAST, driven by intense mid-IR pulses centred at \SI{3.9}{\micro\m} and \SI{1.95}{\micro\m} in order to suppress MPA.
The following chapter is dedicated to elucidate the underlying principles of \ac{or} in general, and in particular for the case of the organic crystal DAST.
Subsequently, the experimental setup will be presented, followed by a rigorous discussion on the experimental findings.
We thereby validate that MPA in DAST can be suppressed to a large extent as compared to \SI{1.5}{\micro\m} drivers,  and crystal damage can be inhibited for significantly higher pump energy densities (up to almost five times), pushing the generation efficiency to significantly higher limits. An unprecedented high optical-to-THz generation efficiency approaching 6\% is observed in the case of a \SI{1.95}{\micro\m} pump pulse.

\section{Optical Rectification}
\label{sec:OR}
In 1961, \ac{shg} was first discovered by Franken \textit{et al.} \cite{Franken:1961}, followed by a 'gold rush' period in the mid 1960s for the finding of multiple nonlinear processes \cite{Geoffry:2014}. 
Amongst many others, \ac{or} was found as a side product of SHG in non-centrosymmetric materials.

\begin{figure}[htb]
    \centering
    \includegraphics[width=\linewidth]{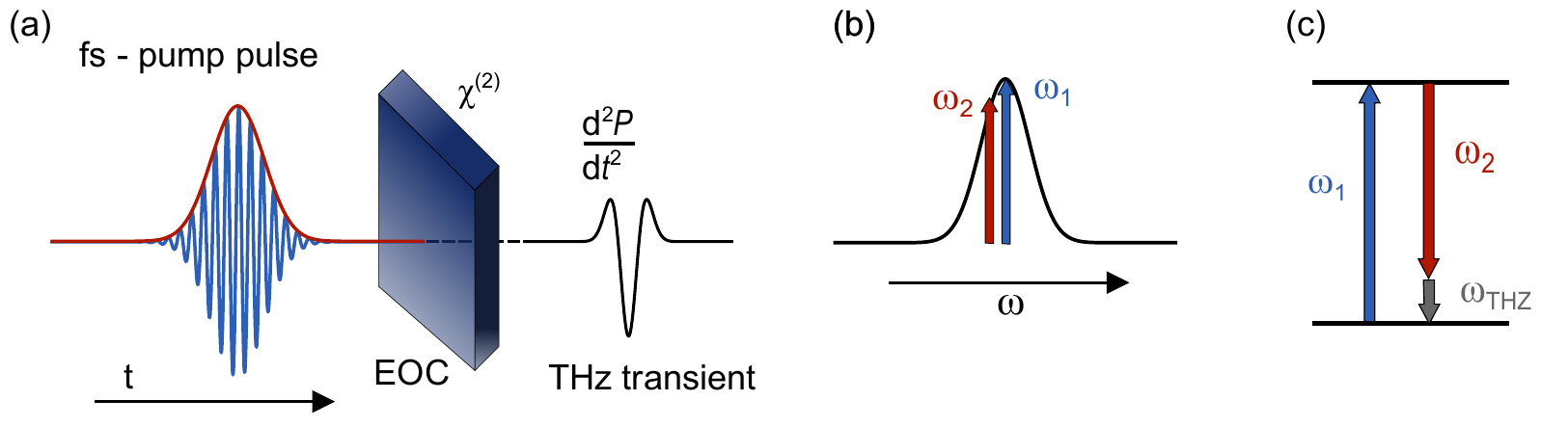}
    \caption[Illustration of optical rectification of a fs - pump pulse in a second order NL crystal]{Illustration of \ac{or} of a fs - pump pulse in a second order NL crystal. The right side indicates difference frequency mixing of all frequencies within the bandwidth $\Delta\omega$ of a fs laser pulse. 
    }
    \label{fig:ORscheme}
\end{figure}

As described in section \ref{sec:Pockels}, the response of molecules and atoms to an applied electromagnetic field can be expressed as the polarisation \textbf{P} of the medium, \textit{i.e.} as charge displacement or dipole moment per unit volume
\begin{equation}
P=\chi^{(1)}E+\chi^{(2)}E^2+ \chi^{(3)}E^3+\dots ,
\label{equ:NLpol}
\end{equation}
where \textit{E} is the applied electric field,  $\chi^{(1)}$ is the linear, $\chi^{(2)}$ is the second order and $\chi^{(3)}$ is the third order susceptibility. 
Note that the scalar notation for the relation of the polarisation and electric field is used here because of convenience. 
In general, $\textbf{P}$ and $\textbf{E}$ are represented by vector  fields, whereas the susceptibility $\chi^{(n)}$ is depicted by a multidimensional tensor, wherein $(n)$ denotes the order of the susceptibility.
As an example for the mathematical description of the dipole moment per unit volume, we
assume a simplified picture of an incoming plane wave $E(t)=E_0\cos\omega t$ with angular frequency $\omega$ and field amplitude $E_0$. 
We only take the first and second order susceptibility in Eq.\eqref{equ:NLpol} into account and rewrite the expression for the polarization (with $\cos^2 \theta=\left(1+\cos \theta \right)/2$)  as
\begin{equation}
P=\underbrace{\chi^{(1)}E_0\cos\omega t}_\text{linear term}+\underbrace{\frac{1}{2}\chi^{(2)}E_0^2(1+\cos 2\omega t)}_\text{nonlinear term}.
\label{equ:NLpol2}
\end{equation}
Thus, the oscillating dipoles do not only experience a linear response and re-emit with angular frequencies $\omega$ and $2 \omega$, but further experience a DC charge separation with respect to the intensity envelope $E_0^2$. 
As a consequence, a small voltage can be measured across a nonlinear crystal when irradiated by a continuous laser beam and is therefore also referred to as inverse process of the electro-optic effect \cite{Wilke:2008}. 
In turn, in the limit of very short laser pulses with a time dependent pulse envelope $E_0(t)$, a short-lived polarization is produced, acting as a source term according to Maxwell's equations:
\begin{equation}
\underbrace{\nabla^2E-\mu_0\epsilon_0 \frac{\partial^2E}{\partial t^2}}_\text{'final' outcoming wave}=\underbrace{\mu_0\frac{\partial^2P}{\partial t^2}}_\text{source term}.
\end{equation}
In other words, when an ultra-short laser pulse is incident on a non-centrosymmetric crystal, it causes a shift of the ions' average position and therefore induces a change in polarization, a quasi DC-polarization. 
A change in polarization is nothing else than a current. 
This current changes with respect to the pump pulse envelope. 
As a result, the pulse gets rectified end emits an electro-magnetic wave in the far field.   
Thus, if the pulse duration of the applied laser field is in the fs regime, a single cycle THz pulse can be generated, as indicated in Fig.\ref{fig:ORscheme}. 

Another possibility to illustrate THz generation by OR is intra-pulse difference frequency mixing between the frequency components of an ultra short pulse in a NL optical medium.
The spectral bandwidth associated with fs laser pulses is very large.
Hence, one can consider two optical fields oscillating at frequencies $E_1(t)=E_0\cos\omega_1 t$ and $E_2(t)=E_0\cos\omega_2 t$, leading to a NL second order term of
\begin{equation}
P_2^{NL}(t)=\chi^{(2)}E_1(t)E_2(t)=\chi^{(2)}\frac{E_0^2}{2}\left[\cos\left(\omega_1-\omega_2\right)t+\cos\left(\omega_1+\omega_2\right)t\right].
\label{equ:NLpol3}
\end{equation} 
The DC component consists of the term $P_2^{\omega_1-\omega_2}$ proportional to difference frequency generation, which describes the production of THz radiation by OR. 
The mechanism is illustrated with a so called 'photon picture' in Fig.\ref{fig:ORscheme}(c), where the frequency-mixing process is conventionally presented in a diagram with up-arrows indicating waves that are depleted, and down-arrows that are enhanced or generated \cite{Geoffry:2014}.
Note that, for $\omega_1 \rightarrow \omega_2$, Eq.\eqref{equ:NLpol3} results in the NL term of Eq.\eqref{equ:NLpol2}. And, in the same way as SHG is not relevant for THz generation, neither is sum frequency generation originating from the term second term $P_2^{\omega_1+\omega_2}$ in Eq.\eqref{equ:NLpol3}.
In fact, due to discrimination via phase matching, it is unlikely that all three processes take place simultaneously equally efficient in a dispersive medium.

Because THz radiation originates from a time varying polarization current which further arises from a change in polarization, the time profile of the radiated THz pulse from \ac{or} of the fs laser pulse is proportional to the second time derivative of the difference frequency term of the polarisation, as expressed in Eq.\eqref{equ:NLpol3}, whose time dependency is determined by the Gaussian time profile of the optical laser pulse \cite{Wilke:2008}. 
Accordingly, for a \SI{100}{fs} pulsed excitation, the non- centrosymmetric medium may emits electromagnetic radiation with a bandwidth
of $\leq$ 1/100 \SI{}{fs^{-1}}, that is, several THz. 

In analogy to efficient THz detection, as described in section \ref{sec:Pockels}, in order to achieve efficient THz generation, the phase matching condition needs to be fulfilled.
It is therefore required that the generated THz radiation propagates with the same velocity through the medium as the original femtosecond laser pulse to add up coherently over the length of the crystal. 
This can be accomplished with birefringent crystals, such as \ac{ZnTe}, \ac{CdTe}, \ac{GaAs}, \ac{GaSe}, or \ac{LN} to name only the most prominent inorganic semiconductors. 
Depending on the crystal structure, the angle of incidence of the fs pulse and its polarisation direction can be chosen such that sufficient phase matching is attained.
In the same manner as for THz detection based on the Pockels effect, perfect phase matching is never accomplished in a dispersive medium and the distance over which the slight velocity mismatch can be tolerated is defined as the coherence length as described in Eq.\eqref{equ:coherneceL}. 
Consequently, efficient THz generation only occurs in crystals that are thinner or equally thick as the coherence length. 
For a crystal thickness longer than the coherence length, the amplitude of the generated  spectral component decreases and eventually vanishes due to destructive interference with THz fields that are generated at different points in time.
Thus, the THz emission strength and THz emission bandwidth has a reciprocal relationship with respect to the crystal length. 
For a thin crystal with a crystal thickness smaller than the coherence length, the generated THz pulse can be broadband.
In turn, in a thick crystal, only THz pulses with a narrow spectral bandwidth will build up coherently, but they can co-propagate with the pump pulse for a longer time through the \ac{nl} medium and thus can accumulate a stronger THz field.

In general, for efficient generation and detection of THz radiation, the NL medium needs to be sufficiently transparent at the THz and optical frequency, to exhibit a high second-order susceptibility, a carefully chosen crystal thickness and proper crystal orientation with respect to the THz radiation.
Furthermore, because the pulse duration and spectral bandwidth experience a reciprocal dependence, the THz bandwidth is limited by the pump pulse duration. 
The generated spectrum further depends on the phase matching condition and THz absorption in the EO crystal. 
Whereas the THz energy and conversion efficiency is limited by the damage threshold of the EO crystal and competing higher order NL processes, such as self-phase modulation, Kerr lensing, NL absorption or cascaded effects \cite{Ravi:2014,MPA_OR_Hoffmann:2007, MPA_Polonyi:2016}.

\section{Nonlinear Optics in the organic Crystal DAST}
\label{sec:OCrystals}
\subsection{Microscopic and macroscopic Response}
DAST (4-N,N-dimethylamino-4'-N'-methyl-stilbazolium tosylate), which is the most well known and widely investigated organic \ac{eoc}, was  already first reported in 1989 \cite{Marder:1989} and is up to now recognized as one of the best state-of-the-art organic NL crystals \cite{OC_THzPhotonics:2019}.
In order to gain a deeper understanding of the NL properties of organic crystals, it is necessary to introduce the microscopic material properties, describing optical effects on the molecular level.
In analogy to the Taylor expansion of the susceptibility for the macroscopic polarization given in Eq.\eqref{equ:NL_Pol}, the molecular polarization $\mathbf{p}(t)$ is given by \cite{organicCrystalsBook:2015}
\begin{equation}
p_i(t)=\varepsilon_0\alpha_{ij}^{(1)}E_j(t)+\varepsilon_0 \beta_{ijk}^{(2)}E_j(t)E_k(t)+\cdots , 
\label{equ:microNL}
\end{equation}
where $\alpha^{(1)}$ is the linear polarizability tensor, and $\beta^{(2)}$ the second-order polarizability or first-order hyperpolarizability.
The relation between macroscopic optical properties with the microscopic polarization of the constituent molecules is commonly treated with the so called oriented-gas model \cite{MacroMicro:1982}, in which intermolecular contributions to the nonlinearity are neglected and only intramolecular interactions are taken into account. 
\begin{figure}[htb]
    \centering
    \includegraphics[width=\linewidth]{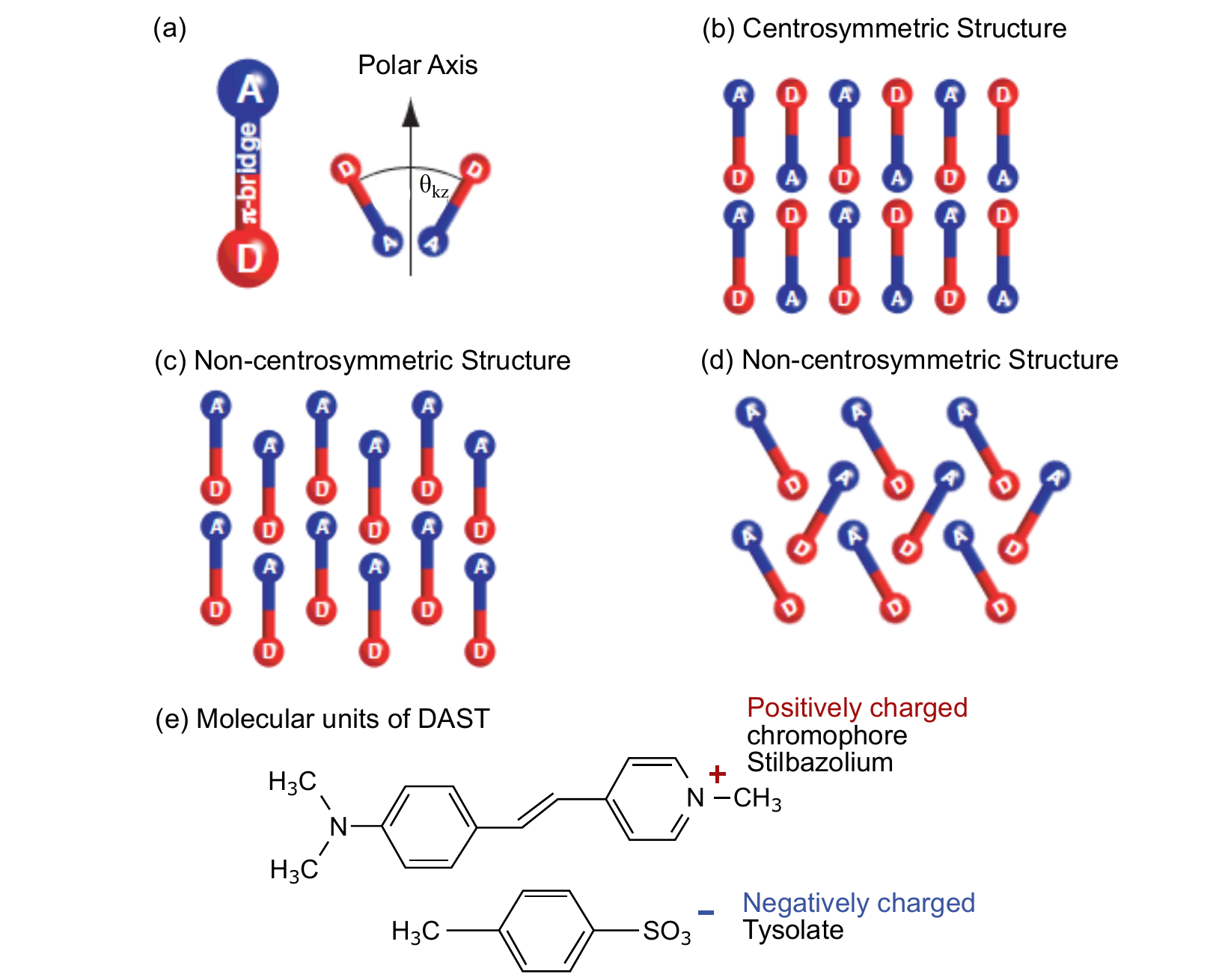}
    \caption[Schematics of organic crystals]{(a) Schematics of an induced dipole moment for a donor-acceptor structure with a $\pi$-conjugated bridge, and representation of the molecular-ordering angle $\theta_{kz}$, being the angle between the charge-transfer axis $z$ of the molecule and the polar axis $k$ of the crystal. (b)-(d) Examples of different molecular alignments in the case of (b) a centrosymmetric crystal structure with $\theta_{kz}=90^{\circ}$ and non-centrosymmetric structures for (c) perfectly parallel chromophores with $\theta_{kz}=0^{\circ}$ and (d) for $0 < \theta_{kz} < 90^{\circ}$, taken from ref \cite{OC_THzPhotonics:2019}. (e) Molecular units of DAST consisting of a positively charged stilbazolium cation to ensure a large molecular hyperpolarizabilty $\beta$, and a tysolate counter anion to promote non-centrosymmetric crystallization \cite{Jazbinsek:2008}. 
    }
    \label{fig:MOrientation}
\end{figure}
The assumption can be justified by the fact that intramolecular forces, such as covalent bonds, are much stronger than forces that are present between neighbouring molecules.
However, a correction factor $f_{\mathrm{local}}$ accounts for the electric field at the position of the molecule which is influenced by its neighbours and hence differs from the externally applied electric field.
For a NL optical process involving three different frequencies, the corresponding relation is given by \cite{organicCrystalsBook:2015, OC_THzPhotonics:2019, PhDMutter:2007} 
\begin{equation}
\chi_{kkk}^{(2)}(-\omega_3,\omega_2,\omega_1)=Nf_{\mathrm{local}}\langle \cos^3\theta_{kz} \rangle \beta_{zzz}(-\omega_3,\omega_2,\omega_1),
\end{equation}
where $N$ is the number of molecules per unit volume, $\theta_{kz}$ is the angle between the main charge transfer axis $z$ of the
molecule and the polar axis $k$ of the crystal, as illustrated in Fig.\ref{fig:MOrientation}(a), and $\langle \cos^3\theta_{kz} \rangle$ is the the orientational factor, the so-called acentric order parameter.
For molecules with strong nonlinearity along a single charge-transfer axis, it is reasonable to consider only one dominant tensor component $\beta_{zzz}$ along this axis. 
Therefore, to attain a large diagonal component of the macroscopic nonlinear susceptibility $\chi^{(2)}$, densely packed molecules with a high nonlinear molecular activity $\beta_{zzz}$ are required, which are parallel arranged so that the orientational factor is close to unity \cite{PhDMutter:2007}.

Molecules with large $\pi$-conjugated systems and strong donor - acceptor end groups are commonly used for organic crystals with high nonlinearities \cite{organicCrystalsBook:2015}.
When an electric field is applied to such a molecular structure with de-localized electrons, the charge distribution will move towards the acceptor, resulting in an asymmetric response upon an external oscillating field.
The extent of the linear redistribution of the charge density is denoted by the linear polarizability $\alpha$, and the the asymmetric
NL redistribution is expressed by the first-order hyperpolarizability $\beta$.
The molecular second-order polarizability can be maximized by the donor-acceptor strength, the planarity of the molecule and conjugation length.
A high second-order nonlinearity requires a high mobility of the $\pi$-conjugated electron system, which can only be achieved if the molecule is neither twisted nor bent.
Moreover, the molecular hyperpolarizability scales as $\beta \propto L_{\pi}^3$ with respect to the conjugation length of the $\pi$-electron system between the donor and acceptor group, leading to a higher nonlinearity for longer molecules\cite{ocBookBosshard:1995}.
However, molecules with a high second-order polarizability exhibit also a large ground state dipole moment which is prone to result in an antiparallel alignment of the molecules and therefore lead to centrosymmetric packing with vanishing macroscopic second order susceptibility $\chi^{(2)}=0$. Schematics of the induced dipole moment and possible molecular alignments are shown in Fig.\ref{fig:MOrientation}, taken from ref \cite{OC_THzPhotonics:2019}. 
To overcome antiparallel alignment of the contributing highly NL molecules, it is possible to use an ionic type of molecules, wherein the cation of the organic salt usually exhibits a high second-order activity, while the anion is used to ensure non-centrosymmetric crystal packing by strong Coulomb interaction.
In addition, the anion acts as a spacer in order to reduce the interaction between the NL active molecules \cite{PhDMutter:2007}.
The molecular units of the organic salt DAST are shown in Fig.\ref{fig:MOrientation}(e), consisting  of a positively charged chromophore stilbazolium and its negatively charged counter-anion tysolate. 
Stilbazolium assures a large hyperpolarizability $\beta$ due to the extended $\pi$-conjugated system, whereas the counter ion tosylate is used to promote a non-centrosymmetric crystallization.
As a result, the chromophores are packed with their main charge transfer axis oriented at about $\theta_{kz}=20^{\circ}$ with respect to the polar \textit{a}-axis of the crystal, leading to a high diagonal order parameter of $\cos^3\theta=0.83$, which is close to optimum parallel alignment \cite{Jazbinsek:2008}.
Thus, DAST crystals have high second-order NL  coefficients, being respectively ten times and twice as large as those of the inorganic standard crystal \ac{LN} \cite{Jazbinsek:2008} in the near-IR spectral range. 
Due to the high diagonal acentric order parameter, DAST crystals are strongly anisotropic with a refractive index difference $n_1-n_2 > 0.5$ for incident light polarized along the crystal's \textit{a} or \textit{b} axis, respectively \cite{refInd_Pan:1996}.
The refractive index is the key element for efficient phase matching between the optical pump and generated THz pulse.
For Ti:Sapphire wavelengths, the coherence length is larger for an emitter beam polarized parallel to the \textit{b}-axis of the crystal, compared to polarization parallel to the \textit{a}-axis.
Thus, although the nonlinearity is almost an order of magnitude larger for \textit{a}-axis emission, the \textit{b}-axis must be used to generate a  broad THz bandwidth \cite{Cunningham:2010}.
In contrast, excellent phase matching can be achieved for a pump beam polarized along the \textit{a}-axis with a central frequency in the vicinity of the spectral range of the telecommunication wavelength at \SI{1550}{nm}.

Aside from the higher NL coefficient, organic crystals provide superior phase matching conditions as compared to their inorganic counter parts.
The reason lies within the low material dispersion for the entire spectral range from long to short wavelengths.
In the case of DAST, the optical refractive index along the polar axis \textit{a} is about 2.1, and in the THz range it is about 2.3.
In comparison, for LN, the optical refractive index along the polar \textit{c} axis is about 2.2 and in the THz range it is about 5 \cite{OC_THzPhotonics:2019}.
The main difference between organic and inorganic materials considering the linear and NL polarizability response, as well as the dielectric constant, originates from the different crystal structure.
In the case of organic materials, the nonlinearity is mainly determined by the electronic polarizability of the molecular units. As discussed above, intermolecular interactions are of minor importance, which leads to a very small contribution of optical phonons.
Since the dynamics of ions is much slower compared to electrons, lattice vibrations contribute to the polarizability response only at small frequencies.
Thus, because organic crystals lack such an addition from lattice vibrations in the THz range,  the electro-optic effect is almost dispersion free and of electronic origin.
In contrast, inorganic crystals are based on a strong bonding between the lattice components (ions) and hence, lattice vibrations play a dominant role at low frequencies, since the ions act as additional polarizable elements. 
In turn, at high frequencies, the ions are too slow to follow, resulting in a drastic change of optical and NL paramters, which leads to a large dispersion difference between the THz and optical spectral range.

\subsection{Coherence Length and THz Generation Efficiency in the mid-IR}
\label{sec:cohLe}
In order to compare the optical- to THz conversion efficiency when THz generation is driven by pulses centred at the telecommunication wavelength of \SI{1.5}{\micro\m} and  in the mid-IR spectral range, the coherence length in DAST with respect to the driving wavelength and generated THz frequency is calculated.
Unfortunately, to the best of our knowledge, no reliable data concerning the refractive index, dispersion or absorption coefficient of DAST around \SI{3.9}{\micro\m} can be found in literature. 
Although Bosshard \textit{et al.} \cite{InfraredBosshard:2002} report on optical properties of DAST in the far- and mid-IR spectral range, they could only resolve resonances above \SI{1600}{cm^{-1}} (\SI{6.25}{\micro\m}) and even the C-H 
stretching modes around \SI{3000}{cm^{-1}} (\SI{3.33}{\micro\m}) were completely invisible by their infrared reflectometric measurements. 
Thus, we focus on a comparative study of \SI{1.5}{\micro\m} and \SI{1.95}{\micro\m} driving pulses, wherein we rely on the Sellmeier equation taken from ref\cite{Jazbinsek:2008,refInd_Pan:1996}. 
The refractive index of interest is thereby $n_1$ along the crystal a-axis, since the \SI{1.95}{\micro\m} pulses exhibit the same preferential crystal orientation for efficient THz generation as \SI{1.5}{\micro\m} pulses, which was experimentally verified in our lab.

\begin{figure}[htb]
\centering\includegraphics[width=\linewidth]{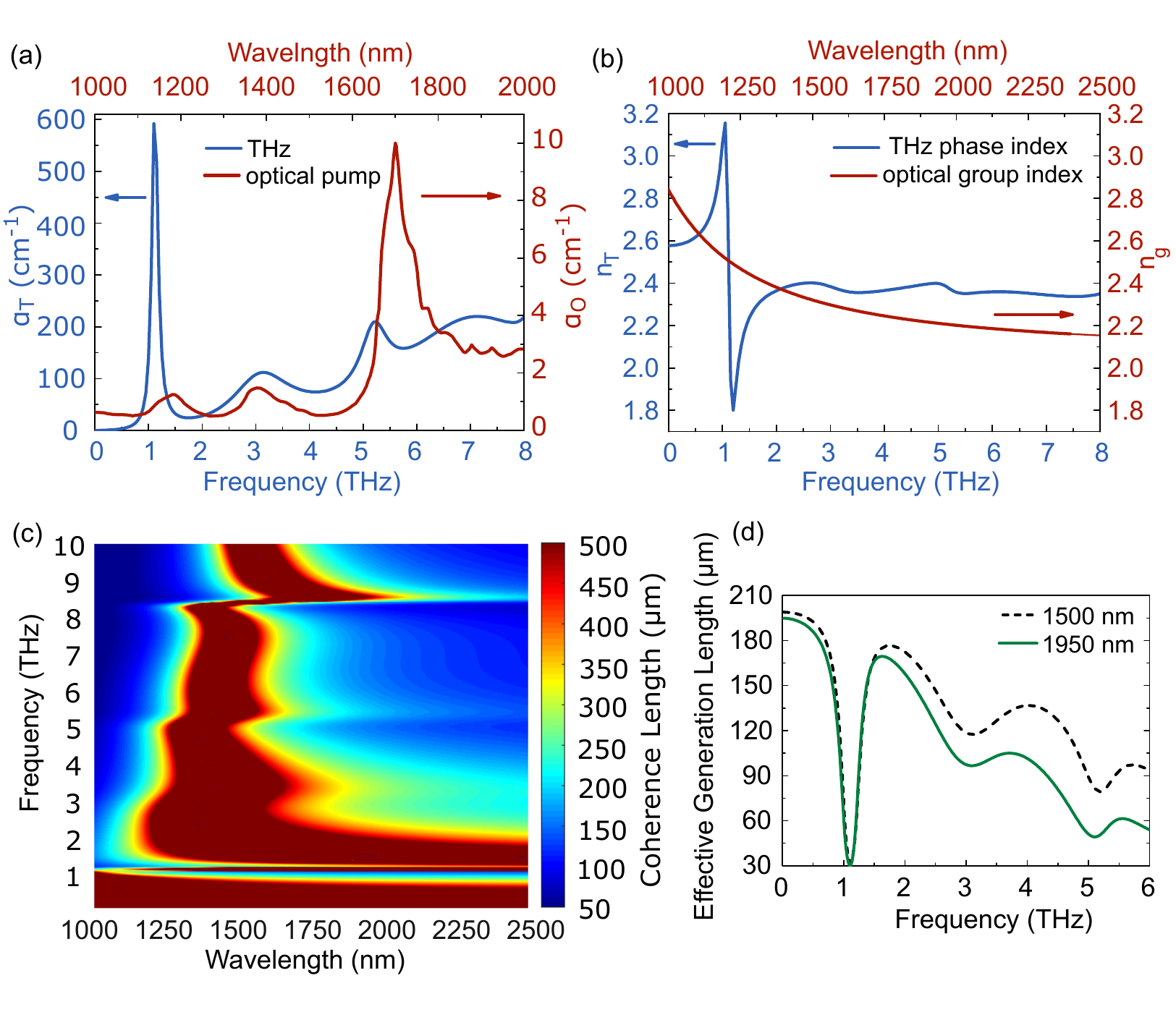}
\caption[Optical properties of DAST in the mid-IR and THz spectral range]{(a) Optical absorption (red) and THz absorption (blue) taken from ref \cite{Cunningham:2010} and \cite{Pan1996CrystalGA}, respectively.
(b) THz phase refractive index \cite{Cunningham:2010} (blue)
and optical group refractive index (red) calculated from the Sellmaier
euqation provided in ref\cite{Jazbinsek:2008}. (c) Coherence
length with respect to the pump wavelength and generated THz frequency.
(d) Effective generation length for a \SI{200}{\micro\m} thick DAST crystal
with respect to the generated THz spectrum for driving pulses centred at \SI{1.5}{\micro\m} (black - dashed) and \SI{1.95}{\micro\m} (green).}
\label{fig:opticalProperties}
\end{figure}

The conversion efficiency is defines as 
\begin{equation}
\eta=\frac{\mathrm{Fluence_{THz}}}{\mathrm{Fluence_{pump}}},    
\end{equation}
where the pump fluence is given by \cite{Vodopyanov:2006}
\begin{equation}
F_{\mathrm{p}}=\frac{c\varepsilon_0 n_o}{2A}\int_{-\infty}^{\infty}\left| E(t,0)\right|^2dt=\sqrt{\frac{\pi}{2A}}\frac{c\varepsilon_0 n_o}{2}E_0^2\tau=\sqrt{\frac{\pi}{2}}\frac{I_0\tau}{A},   
\end{equation}
with \textit{c} as the speed of light, $n_o$ the optical refractive index, $I_0$ the
peak intensity at the input of the crystal, and $\tau$ the pulse duration. In
the same way, the THz fluence is defined as
\begin{equation}
F_{\mathrm{THz}}=\frac{c\epsilon_0 n_{\mathrm{THz}}}{2A}\int_{-\infty}^{\infty} \left|E_{\mathrm{THz}}(t,L)\right|^2dt= \frac{c\epsilon_0 n_{\mathrm{THz}}}{2A}\int_{0}^{\infty}\left|E(\Omega,L)\right|^2d\Omega,   
\label{equ:THzFluence}
\end{equation}
with $n_{\mathrm{THz}}$ being the refractive index at THz frequencies and \textit{L} the crystal thickness.
As the same beam area $A$ for the optical pump and generated THz pulse is assumed at the crystal, it is neglected in the following calculations.
The interchangeability of time and frequency domain is supported by Parseval's theorem.
In the case of a plane-wave approximation, wherein the propagation of the pump pulse is not affected by any nonlinear effects such as multi-photon absorption (MPA), cascaded processes, pump depletion, and pump pulse broadening, the absolute value of the rectified field at THz frequencies can be written as \cite{Stillhart:2008,Brunner:2008}
\begin{equation}
\left|E(\Omega,L)\right|=\frac{\mu_0\chi^{(2)}\Omega I_0(\Omega)}{n_o(\lambda)\left[n_{\mathrm{THz}}(\Omega)+n_g(\lambda)\right]}L_{\mathrm{eff}}(\Omega ,\lambda,L). 
\label{equ:Efield}
\end{equation}
Here, $n_o$ and $n_g$ are the refractive and group indices at the given pump wavelength, $\chi^{(2)}$ is the effective NL susceptibility that is relevant for \ac{or}, and $I_0(\Omega)$ is the Fourier transform of the intensity envelope of the pump pulses at THz angular frequency  $\Omega$, which can be further separated into $I_0(\Omega)=I_0\cdot S(\Omega)$ with $I_0$ as the peak intensity and $S(\Omega)$ as the normalized power spectrum.
Because we aim for a comparison on the effect of the driving wavelength, we assume Gaussian pump pulses with the same \ac{fwhm} pulse duration of \SI{76}{fs} (measured with FROG for the \SI{1.95}{\micro\m} driving pulse, see section \ref{sec:SetupOR}).
The dependence of the THz field on the crystal length \textit{L} is given by the effective generation length $L_{\mathrm{eff}}$ described in refs\cite{SchneiderOR:2006,Stillhart:2008,Brunner:2008} 
\begin{multline}
L_{\mathrm{eff}}(\Omega,\lambda,L)=\\
\left(\frac{\exp\left(-2\alpha_0(\lambda)L\right)+\exp\left(-\alpha_{T}(\Omega)L\right)-2\exp\left(-\left[\alpha_0(\lambda)+\frac{\alpha_T(\Omega)}{2}L\right]\right)\cos\left(\frac{\pi L}{l_c(\Omega,\lambda)}\right)}{\left(\frac{\alpha_T(\Omega)}{2}-\alpha_0(\lambda)\right)^2+\left(\frac{\pi}{l_c(\Omega,\lambda)}\right)^2}\right)^{1/2},    
\end{multline}
including the absorption coefficient $\alpha_0(\lambda)$ of the optical pump, THz absorption coefficient $\alpha_T(\Omega)$ and coherence length $l_c$ as defined in Eq.\eqref{equ:coherneceL} and further discussed in section \ref{sec:OR}.
Figure \ref{fig:opticalProperties}(a) depicts the absorption coefficient of DAST in the near-IR range (red) and THz range (blue), taken from ref  \cite{Pan1996CrystalGA} and \cite{Cunningham:2010}, respectively. 
For a crystal thickness of less than \SI{200}{\micro\m}, which is the case in this work, the transmission of both driving pulses
 centred at \SI{1.5}{\micro\m} and \SI{1.95}{\micro\m} is more than 90\% and hence plays a minor role in the conversion efficiency.
The THz phase refractive index \cite{Cunningham:2010} $n_{\mathrm{THz}}$ (blue) and optical group refractive index $n_g$ (red) are shown in Fig.\ref{fig:opticalProperties}(b).
The latter is calculated from the Sellmeier equation provided in 
ref\cite{Jazbinsek:2008}. 
Due to the close proximity of the THz phase and optical group refractive index, the organic crystal DAST is such an excellent candidate for efficient THz generation by \ac{or}. 
Figure \ref{fig:opticalProperties}(c) depicts the coherence length with respect to the pump wavelength and generated THz spectrum. 
The colorbar is only shown up to \SI{500}{\micro\m} for a better visualization.  
The best phase matching conditions for the generation of a broad THz spectrum is supported by a pump wavelength of \SI{1.5}{\micro\m} with a coherence length of more than \SI{500}{\micro\m}, as already reported in literature \cite{DASTVicario:2015}.
However, the coherence length for \SI{1.95}{\micro\m} is still larger than \SI{300}{\micro\m} in the low THz frequency range and efficient THz generation can be expected of up to \SI{4}{THz} with a crystal thickness of \SI{200}{\micro\m}.
The effective generation length for a \SI{200}{\micro\m} thick DAST crystal is shown in Fig.\ref{fig:opticalProperties}(d) for a driving pulse centered at \SI{1.5}{\micro\m} (black-dashed) and \SI{1.95}{\micro\m} (green) with respect to the generated THz spectrum, taking the crystal absorption into  account.
We can summarize the THz fluence to
\begin{equation}
F_{\mathrm{THz}}=\frac{c\varepsilon_0 n_{\mathrm{THz}}\mu_0^2}{2}S_iI_0^2,
\end{equation}
where $S_i$ is calculated from the integral in Eq.\eqref{equ:THzFluence} for a spectral range from 0  to \SI{6}{THz} and pump wavelength $\lambda_i=\SI{1.5}{\micro\m}; \SI{1.95}{\micro\m}$: 
\begin{equation}
S_i=\int_{0}^{6\mathrm{THz}}\left[\frac{\chi^{(2)}\Omega S(\Omega)}{n_o(\lambda)\left[n_{\mathrm{THz}}(\Omega)+n_g(\lambda)\right]}\right]^2L_{\mathrm{eff}}(\Omega,\lambda_i,\SI{200}{\micro\m})^2d\Omega. 
\end{equation}
Finally, if we neglect pump wavelength independent constants and assume the same pump pulse duration, which is the case for the given example, the conversion efficiency can be written as 
\begin{equation}
\eta \propto S_i I_0.    
\end{equation}
We use $\chi^{(2)}_{111}=\SI{480}{pm/V}$  at $\lambda=\SI{1.5}{\micro\m}$\cite{DASTHauri:2011,Marder:1989} and \SI{420}{pm/V} at $\lambda=\SI{1.95}{\micro\m}$ \cite{organicCrystalsBook:2015} and find
\begin{equation}
\eta\left(\lambda=\SI{1.5}{\micro\m}\right)\propto 4.37\times 10^{-2}\ I_0 \quad \textrm{and} \quad \eta\left(\lambda=\SI{1.95}{\micro\m}\right)\propto 2.16\times 10^{-2}\ I_0.
\end{equation}
Hence, with respect to the coherence and effective generation length, the conversion efficiency at \SI{1.5}{\micro\m} is expected to be two times larger as compared to \SI{1.95}{\micro\m} driving pulses.

However, due to the suppression of \ac{mpa} for longer wavelength driving pulses, as demonstrated in ref\cite{Gollner:2021} and described in detail in section \ref{sec:ResultsOR}, it is possible to pump the crystal with peak intensities that are at least three times higher, before the conversion efficiency starts to saturate (\textit{i.e.}, intensities of \SI{0.4}{TW/cm^2} for \SI{1.95}{\micro\m} and \SI{0.95}{TW/cm^2} for \SI{3.9}{\micro\m} driving pulses, corresponding to pump fluences of more than \SI{30}{mJ/cm^2} and \SI{100}{mJ/cm^2}, respectively). 
Therefore, because the conversion efficiency scales linearly with the peak intensity until the onset of MPA \cite{MPA_OR_Hoffmann:2007}, the conversion efficiency is expected to be superior by a factor of 1.5 in the case \SI{1.95}{\micro\m} driving pulses.

Note that, for the given approximation, additional NL effects such as pump depletion, SHG, cascaded \ac{or} or pulse broadening are neglected.
Nonetheless, MPA does not only deteriorates THz generation due to a loss of pump energy, related crystal heating and subsequent decreased crystal damage threshold. 
In addition, because of MPA, free carriers are created in the crystal which strongly absorb THz radiation \cite{Vodopyanov:2006}.
Therefore, the possibility to apply a higher pump fluence with long wavelength driving pulses, in combination with a lack of THz absorption by freed carriers can result in a superior conversion efficiency compared
to near-IR driving pulses.

\subsection{Cascaded Optical Rectification}
\label{sec:cascaded}

As depicted in Fig.\ref{fig:ORscheme}, a red shifted pump photon is left behind due to the photon energy conservation law when a THz photon is generated via OR caused by frequency mixing of the spectral components from a fs pump pulse. Once the optical pulse becomes red-shifted, it can still contribute to another cycle of THz generation, leading to multiple shifts of the initial pump spectrum towards longer wavelengths. Hence, because a single optical photon of the pump pulse can contribute to the generation of multiple THz photons, the Manley-Row limit can be surpassed. 
\begin{figure}[htb]
\centering\includegraphics[width=\linewidth]{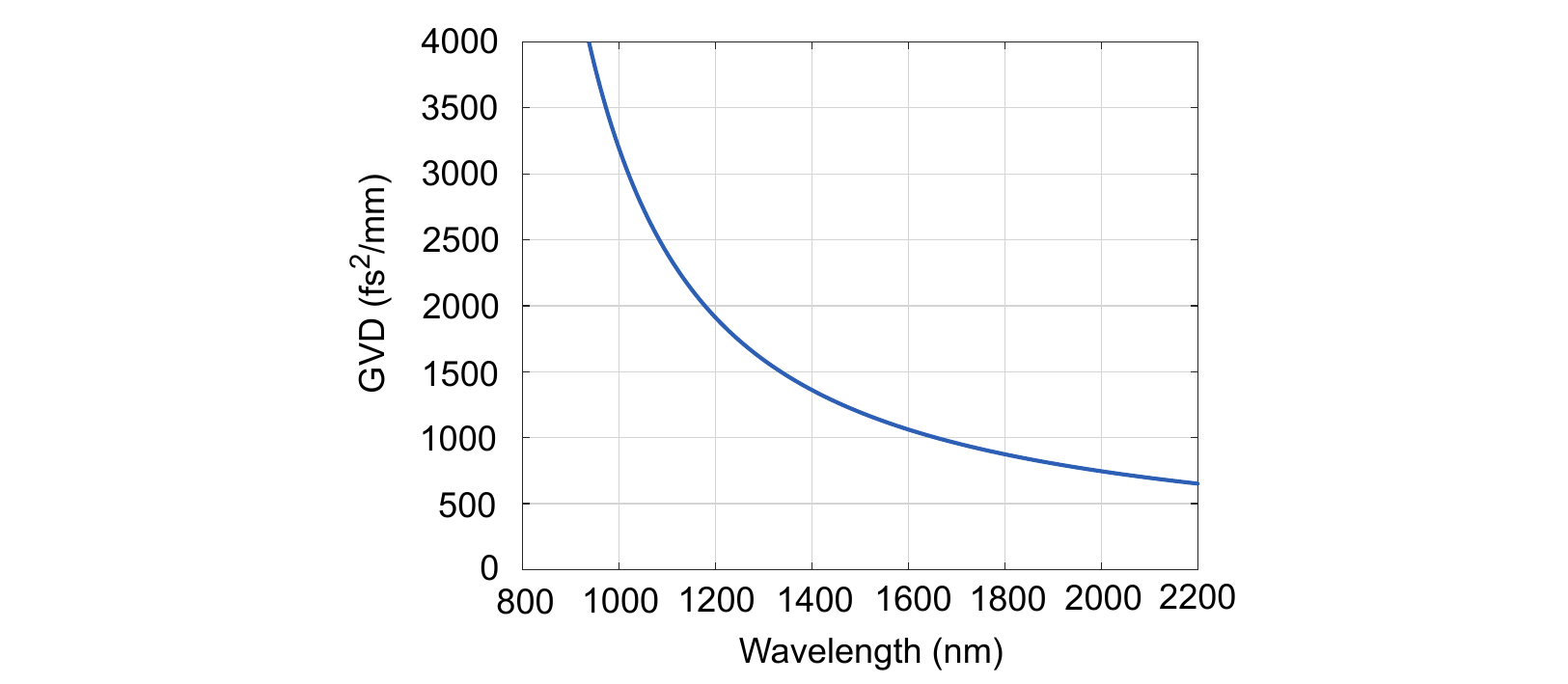}
\caption[Group velocity dispersion of DAST in the optical range]{Group velocity dispersion of DAST in the optical range, calculated from the Sellmeier equation provided in \cite{PucGVD:2020}. A smaller \ac{gvd} is observed for longer wavelength.}
\label{fig:DAST_GVD}
\end{figure}
The photon conversion efficiency, which is defined as the ratio between the number of THz and pump photons $N(h\nu_{\mathrm{THz}})/N(h\nu_{\mathrm{p}})$ can thus become substantially larger than 1.
This process of cascaded optical down-conversion continues to transfer optical energy to lower frequencies as long as the phase mismatch $\Delta k$ is small \cite{Vodopyanov:2006}.
Therefore, an efficient cascaded process and hence small phase mismatch can be maintained if the group velocity dispersion (GVD) of the optical pulse is small. 
Otherwise, the newly generated frequencies will travel with different group velocities, temporally broaden the pulse, and terminate the time overlap with the generated THz pulse. 
Thus, on the one hand, cascading effects can lead to optical-to-THz photon conversion levels of more than 100 \%.
On the other hand, the spectral broadening of the optical pump associated with cascading effects accentuates dispersive effects and strongly limits further increase in conversion efficiency \cite{Ravi:2014}.
Figure \ref{fig:DAST_GVD} shows the GVD in the optical spectral range, calculated from the Sellmeier equation taken from supplementary information of \cite{PucGVD:2020}. A lower GVD can be found at \SI{1.95}{\micro\m} (\SI{773}{fs^2/mm}) as compared to \SI{1.5}{\micro\m} (\SI{1191}{fs^2/mm}), which hints towards advantageous conditions for cascaded OR in the case of longer wavelength drivers.

\section{Experimental setup}
\label{sec:SetupOR}
\begin{figure}[htb]
\centering\includegraphics[width=\linewidth]{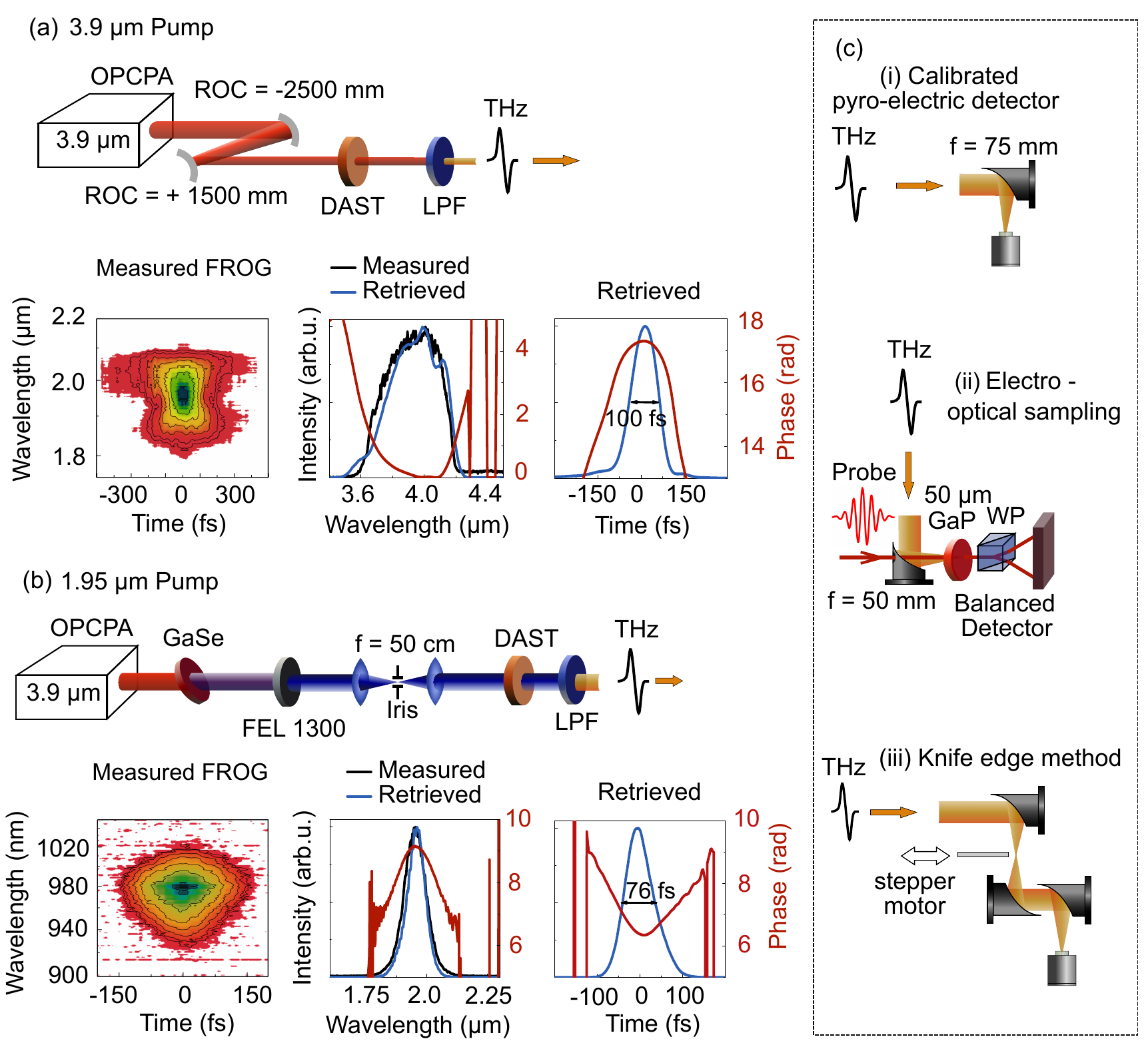}
\caption[Experimental setup for optical rectification and characterization of the mid-IR driving pulses]{Experimental setup and characterization of the driving pulses centred at (a) \SI{3.9}{\micro\m} (b) \SI{1.95}{\micro\m}. (c) Diagnostic setup for the THz-pulse characterization: (i) the THz energy is measured with a calibrated pyro-electric detector, (ii) the THz electric field is recorded  by EOS in a \SI{50}{\micro\m} thick GaP crystal, and (iii) the spatial distribution of the THz pulse is measured with the knife edge method; WP - Wollaston prism.}
\label{fig:setup}
\end{figure}

\begin{figure}[htb]
\centering\includegraphics[width=\linewidth]{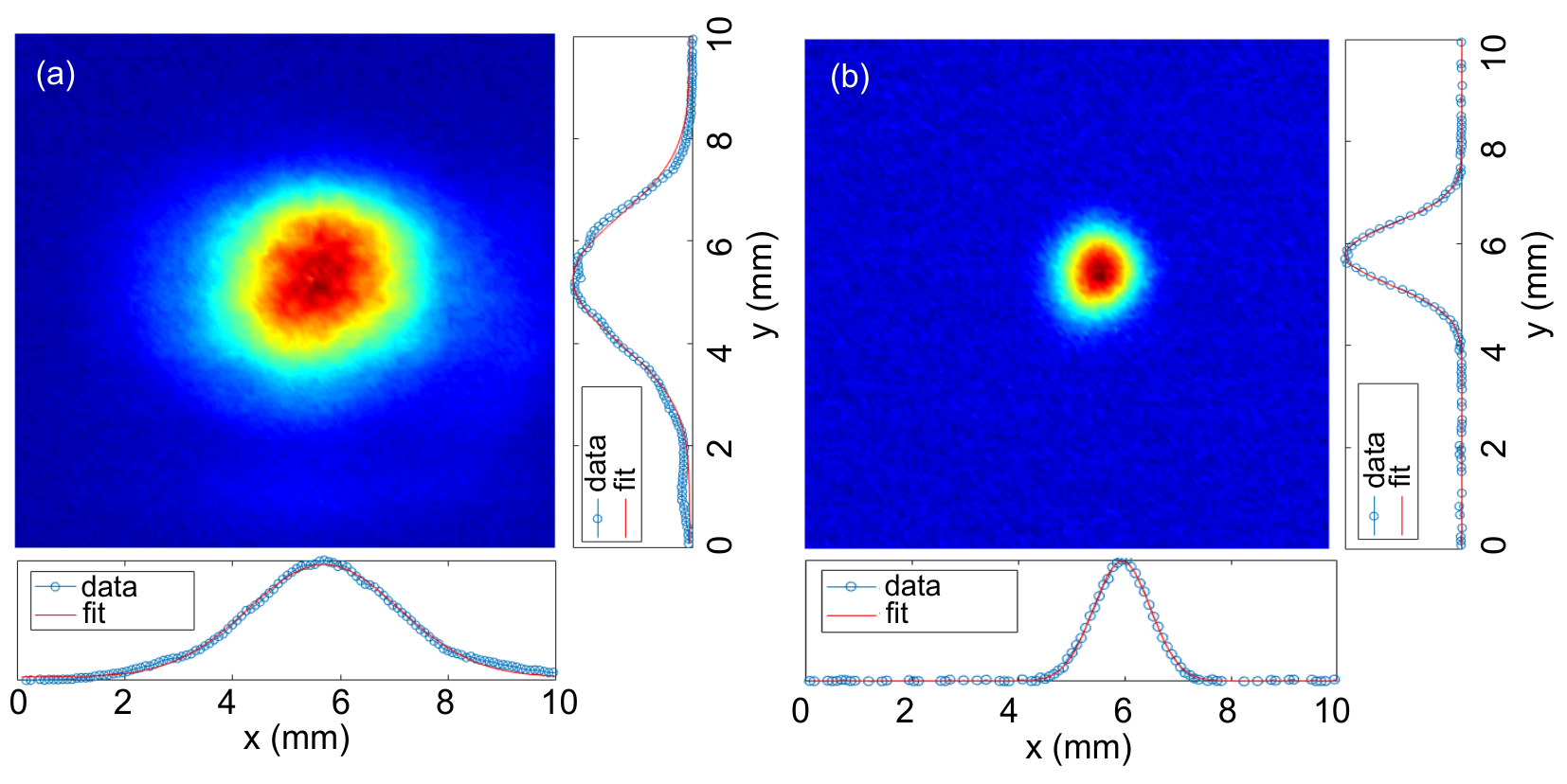}
\caption[Beam profile of the mid-IR driving pulses]{(a) Beam profile of the \SI{3.9}{\micro\m} driving pulse
with a diameter of \SI{3.2}{mm} at FWHM level, and (b) 
\SI{1.95}{\micro\m} pulse with a diameter of \SI{1.34}{mm}, measured with Spiricon Pyrocam III at the crystal position.}
\label{fig:BeamProfile}
\end{figure}

The experimental arrangement for THz generation by OR in a $\sim$ \SI{170}{\micro m} thick DAST crystal (Swiss Terahertz LLC), as well as  characterization of the driving pulses and THz transient are shown in Fig.\ref{fig:setup}.
The pump source is  a high power mid-IR \ac{opcpa} system, as presented in section \ref{sec:OPCPA}, 
operating at a repetition rate of \SI{20}{Hz} with pulse energies of more than \SI{30}{mJ} and a pulse duration of \SI{100}{fs} centred at \SI{3.9}{\micro\m}. 
The pulse duration as well as the chirp can be adjusted by changing the distance between the gratings of the pulse compressor.
Due to the limited clear aperture size of $\sim$\SI{6}{mm} of the DAST crystal, the beam size of the \SI{3.9}{\micro\m} driving pulse needs to be reduced with a telescope consisting of a pair of spherical mirrors, resulting in a diameter of \SI{3.2}{mm} at FWHM level. 
The beam profile of the mid-IR driving pulse is measured with a pyro-electric camera (Ophir Photonics, Spiricon Pyrocam III) and shown in Fig.\ref{fig:BeamProfile}(a). 
Although optimization of the \SI{3.9}{\micro\m} pump beam would be desirable, the implementation of spatial filtering is not straight forward due to the large peak intensity at the focal position, which would ionize air molecules and generate plasma.
To prevent beam disturbance originating from plasma scattering, it would be necessary to install a vacuum tube around the focal position, which was not available in the lab at this point.
Nonetheless, although beam inhomogeneities are a known limiting factor acting on the optical to THz conversion efficiency \cite{DASTVicario:2015}, the intensity distribution of the \SI{3.9}{\micro\m} beam is sufficiently smooth for efficient THz generation. 
In order to separate the THz radiation from the mid-IR driving  pulse and to prevent saturation of the pyro-electric detector (THZ5I-BL-BNC, GenTec), several low pass filters (LPF) are used, as described in section \ref{sec:Pyro}. 
After filtering, the THz pulse is steered to the diagnostic setup shown in Fig.\ref{fig:setup}(c).
As described in chapter \ref{ch:Characterization}, the generated THz pulses are characterized by measuring the energy with a calibrated pyro-electric detector, and by recording the THz electric field via \ac{eos} in a \SI{50}{\micro\m} thick GaP crystal.
The diameter of the THz beam is measured with a knife edge technique.

Because our \ac{opcpa} is not wavelength tuneable, another infrared wavelength was accessed by generating the \ac{sh} at \SI{1.95}{\micro\m} of the OPCPA output in a \SI{100}{\micro\m} thick z-cut  GaSe crystal (Fig.\ref{fig:setup}(b)). 
During the SHG process, the temporal pulse profile cleans up as pulses shorten to \SI{76}{fs}.
After SHG, fundamental pulses at \SI{3.9}{\micro\m} are blocked by a long pass filter (FEL 1300, Thorlabs). 
In order to improve the SH beam quality, which is deteriorated because of the photo-induced inhomogeneities of the GaSe crystal \cite{GaSeDamage1:2013}, spatial filtering of the  generated SH is performed.
Two \SI{}{CaF_2} lenses with a focal distance of f=\SI{50}{cm} are utilized in order to focus and re-collimate the beam. A pinhole is placed at the focal point of the objective lens to pass only the central maximum of the beam pattern (\textit{i.e} central part with a low spatial frequency) and to remove spatial noise from the system (defined as higher spatial frequencies).
The \ac{sh} beam size on the DAST crystal accounts for \SI{1.4}{mm^2} and is shown in  Fig.\ref{fig:BeamProfile}(b). 
Subsequent to the pulse separation of THz and driving pulse, characterisation of the THz transient is performed with the same diagnostic setup as described above.

\section{Results and Discussion}
\label{sec:ResultsOR}
\subsection{\label{sec:3.9um} THz Generation by 3900 nm driving Pulses}
\begin{figure}[htb]
\centering\includegraphics[width=\linewidth]{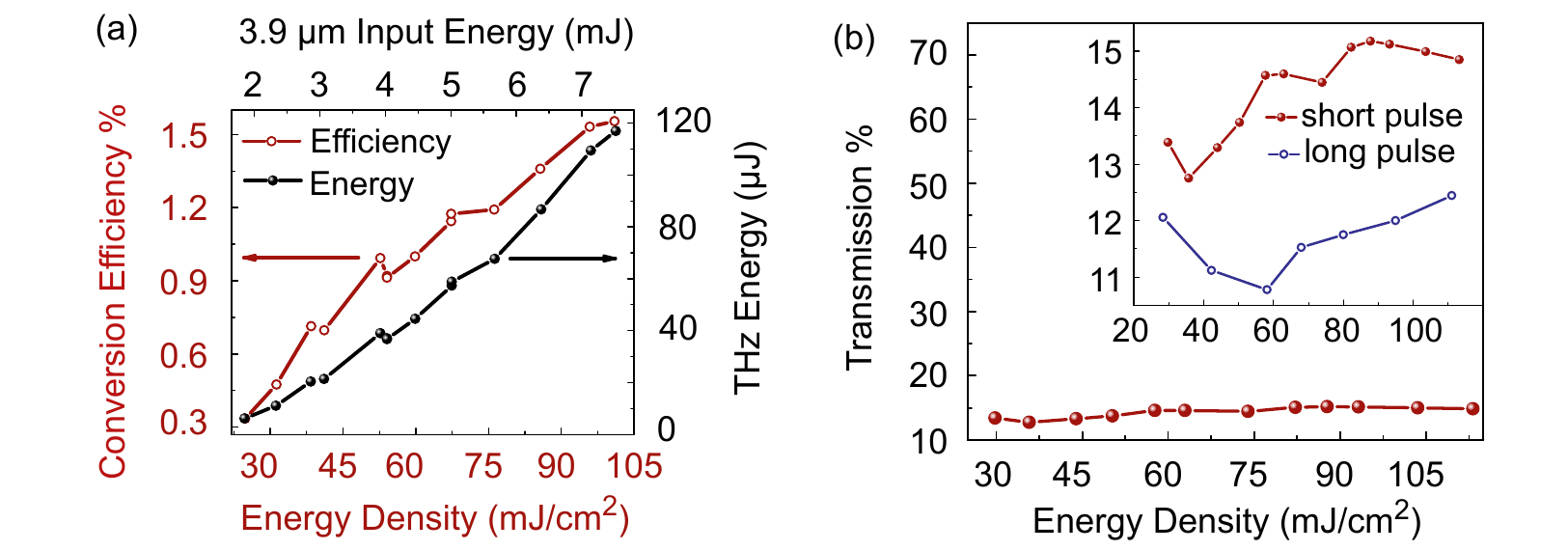}
\caption[Optical to THz Conversion efficiency and transmission of DAST for a \SI{3.9}{\micro\m} driving pulse]{(a) Optical to THz Conversion efficiency (red circles) 
and THz energy (black dots) in dependency
of the energy density and input energy, respectively, for \SI{3.9}{\micro\m}
driving pulses. 
(b) Transmission of the \SI{3.9}{\micro\m} pump pulse with respect to the pump fluence.The inset shows a magnified detail from 10.5-15.5\%. 
Because the transmission slightly increases for higher pump energy densities (red dots),
the experiment is repeated with long stretched to more than 3 ps pulses (blue circles) 
to confirm features of NL
photon bleaching.}
\label{fig:efficiency}
\end{figure}

We first examine THz generation with \SI{3.9}{\micro\m} driving pulses,  where the transmission of DAST drops to  $\sim$\SI{10}{\%} \cite{DAST_TransVicario:2015}. 
Fig.\ref{fig:efficiency}(a) depicts the mid-IR to THz conversion efficiency and generated THz pulse energy with respect to the applied pump fluence and pulse energy. 
Surprisingly, despite strong linear absorption, high conversion efficiencies of 1.5\% and THz energies of up to \SI{116}{\micro J} can be achieved. 
When the THz pulse energies as well as the conversion efficiencies are evaluated, we account for the transmission of the filters and \SI{10}{cm} of air, through which the generated THz pulses propagate before reaching  the pyro-electric detector.
If to take the absorption of the crystal into account, 
the efficiency is commensurate or even superior as compared to conventional drivers at 
the telecommunication wavelength. 
In addition to phase-matching of the driving and generated THz fields, 
a necessary condition for high conversion efficiencies is a high second order NL  susceptibility $\chi^{(2)}$ of the crystal. 
In the case when the virtual energy levels of the parametric process are close to a real energy level, the NL susceptibility is considered to be resonantly enhanced \cite{OC_THzPhotonics:2019}. 
Thus, one of the possible reasons for the high conversion efficiency is a resonant enhancement of the electro-optical coefficient in the vicinity of \SI{3.9}{\micro\m}.
Though, given that the complex permittivity of DAST at \SI{3.9}{\micro\m} is actually unknown, we can only hypothesize about advantageous phase matching conditions and enhanced electro-optical coefficient.

Nonetheless, the onset of saturation only appears at extraordinarily high pump fluences. 
Furthermore, despite the high linear absorption, optical damage was not observed for pump energy densities up to \SI{110}{mJ/cm^2}, which can be attributed to a suppression of MPA.
In order to verify this statement, we measure the transmission of DAST with respect to the pump fluence.
As it can be seen in Fig.\ref{fig:efficiency}(b), when the crystal is pumped with compressed \SI{100}{fs} pulses (red dots), the transmission, instead of decreasing, which is characteristic for MPA, even slightly increases at higher energy  densities. 
This can be a feature of NL photon-bleaching, wherein the leading edge of the pulse depletes the ground state of DAST.
When the experiments are repeated with stretched to a few ps -pulses (blue dots), the transmission appears to be slightly lower and virtually independent on the pump fluence.
Thus, we confirm that MPA of \SI{3.9}{\micro\m} pulses does not occur at almost an order of magnitude higher pump intensities as compared to the 
case of \SI{1.5}{\micro\m} pump pulses \cite{DAST_TransVicario:2015}.

Moreover, linear absorption could potentially lead to a substantial local heating inside the crystal, which might affect THz generation and act on the optical damage threshold of the crystal \cite{MostafaReRate:2015,RepRate:2013}.
In order to examine the effect of local heating, the experiments are repeated at different repetition rates by reducing the repetition rate of the pump pulses  with a mechanical chopper while keeping the pump fluence constant. 
Since no changes in the conversion efficiency, neither for the onset of saturation nor for the absolute values can be detected when reducing the repetition rate from \SI{20}{Hz} to \SI{3.3}{Hz}, heating due to linear absorption of DAST can be further ruled out as a possible limiting factor for THz  generation at 20 Hz repetition rate.

\begin{figure}[htb]
\centering
\includegraphics[width=\linewidth]{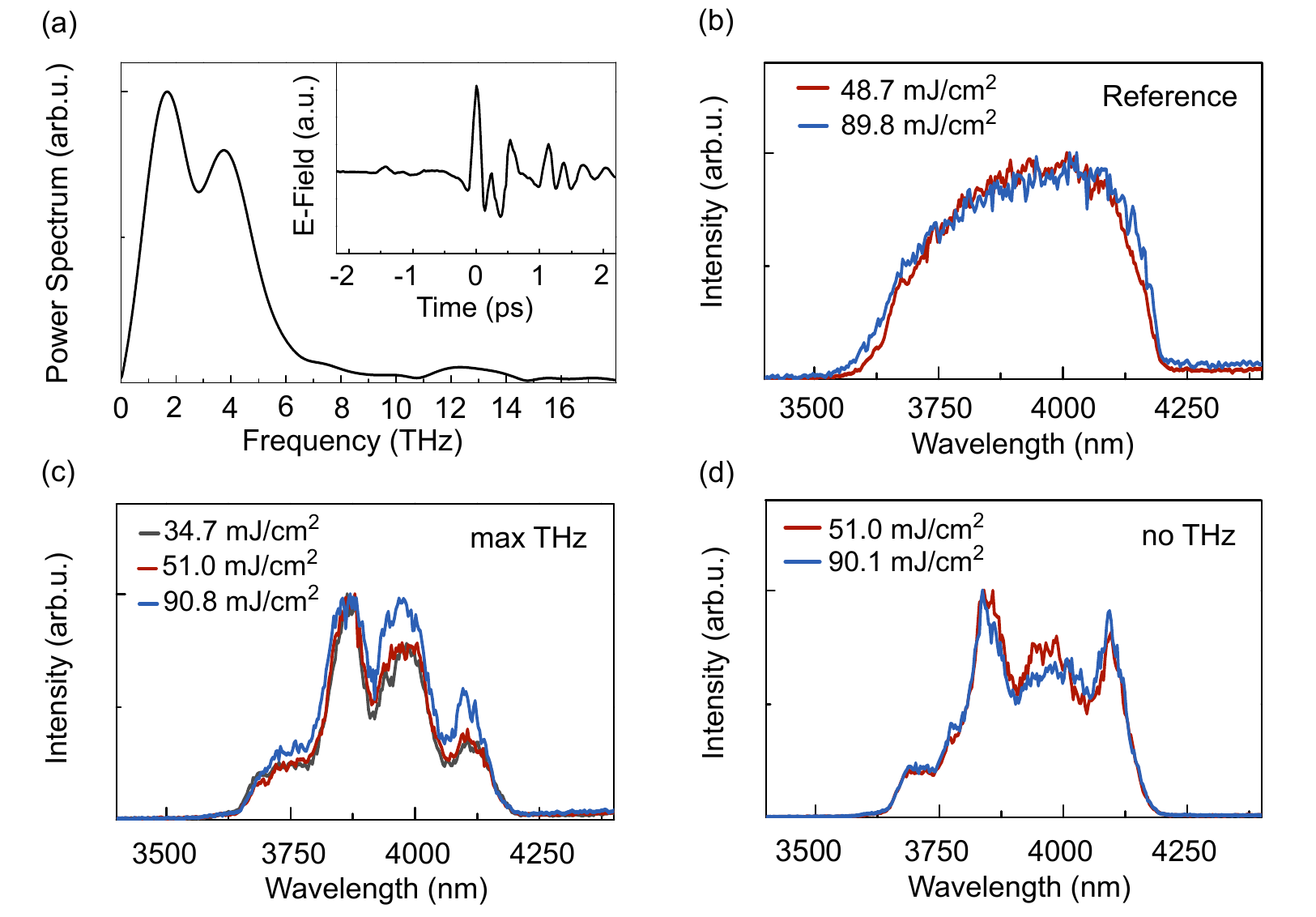}
\caption[THz spectrum generated by optical rectification of \SI{3.9}{\micro \m} driving pulses and spectrum of the driving pulse.]{(a) THz power spectrum measured with EOS in the case of \SI{3.9}{\micro \m} driving pulses. The inset shows the temporal THz transient. (b-d) Normalized spectra of the \SI{3.9}{\micro \m} driving pulse for different pump energy densities (see legend). (b) Reference pulse without a crystal, (c) when the crystal is aligned for max THz generation, and (d) when the crystal is rotated to inhibit THz generation. No spectral red shift can be observed}
\label{fig:4umSpectra}
\end{figure}
Figure \ref{fig:4umSpectra}.(a) shows the THz spectrum generated by \SI{3.9}{\micro\m} pump pulses, measured with EOS. 
The experimental setup is purged with \SI{}{N_2}, resulting in a reduced humidity of <10\%. 
The measurement reveals a multiple octave-spanning spectrum with a maximum at \SI{2.1}{THz}, FWHM bandwidth of \SI{4.2}{THz} and spectral content extending up to \SI{7.8}{THz}. 
In combination with a measured beam radius of $w=$\SI{107}{\micro\m} at $1/e^2$ level, the absolute electric field strengths can be obtained at the focal position of a f= 5cm parabolic mirror and accounts for \SI{40}{MV/cm}. 
The measured central frequency is similar to that of \SI{1.5}{\micro\m} driving pulses reported in \cite{DASTHauri:2011}, wherein a high optical- to- THz conversion efficiency of up to 2.2\% is achieved, but with only half the spectral bandwidth of \SI{2}{THz}. 
The high conversion efficiency in \cite{DASTHauri:2011} is ascribed to cascaded OR, wherein the pump pulse experiences a significant red shift, as described in section \ref{sec:cascaded}.
Although this process can initially result in higher optical- to THz conversion efficiencies, it can also terminate THz generation because the phase mismatch between the THz and newly generated pump photons enhances  \cite{Ravi:2014}.
In the case of \SI{3.9}{\micro \m} driving pulses, the photon conversion efficiency is evaluated to be 56\%, which reveals minor, if any contribution of cascaded effects for the THz generation process. 
An absence of cascaded OR is further confirmed by the lack of noticeable spectral broadening of the pump pulse. 
Spectra of the \SI{3.9}{\micro \m} driving pulses are measured with a spectrometer based on accousto-optic interaction (MOZZA-FASTLITE) before and after DAST. 
Figure \ref{fig:4umSpectra} presents the recorded spectra for different pump intensities in the cases when (b) DAST crystal is removed, referred to as reference, when (c) the crystal is aligned for maximum THz generation and (d) when the crystal is rotated to inhibit THz generation. 
The different spectral shape between (c) and (d) is attributed to the inhomogeneity of the crystal and preceding different crystal position when it is rotated by \SI{90}{^{\circ}}.
In all cases, the spectra do not show any energy density dependence, though signatures of linear absorption can be observed.

\begin{figure}[thb]
\centering
\includegraphics[width=\linewidth]{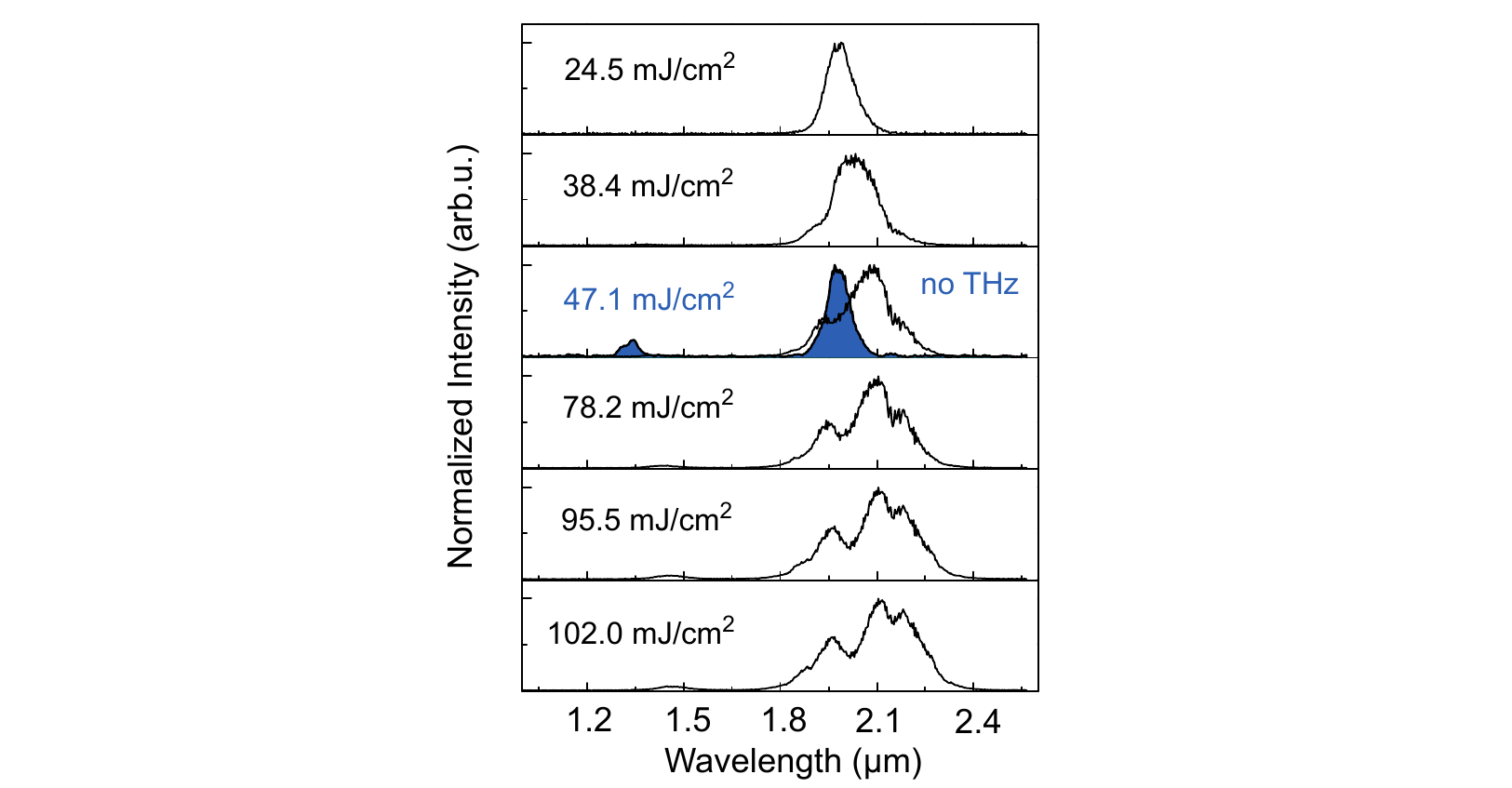}
\caption[Spectra of the driving pulse after propagation through DAST measured in the spectral range of 1100-\SI{2600}{nm}]{Spectra of the driving pulse after propagation through DAST measured in the spectral range of 1100-\SI{2600}{nm} (corresponding to SHG of the pump pulse) for different pump fluences when the crystal is aligned for maximum THz generation. When the crystal is rotated such that THz 
generation is inhibited, the corresponding spectrum (blue area) coincides with the spectrum that is measured at low pump fluences (top panel) when THz generation is less pronounced.}
\label{fig:SHGinDAST}
\end{figure}
Due to the the high second order nonlinearity of DAST, aside from THz generation by OR, also SHG takes place.
SH spectra in the spectral range of 1100-\SI{2600}{nm} at different pump fluences are shown in Fig.\ref{fig:SHGinDAST} when the crystal is aligned for maximum THz generation.
For higher pump energy densities, the spectrum of the generated SH continuously broadens to the red side. 
Modification of the SH spectrum takes place only when efficient THz 
generation is present. 
The spectrum shown by the blue area is measured when the crystal axis is rotated (misaligned) to inhibit THz generation. 
It resembles the one which was measured at low energy densities, when THz generation is rather inefficient.
All of this gives a strong indication that an efficient nonlinear interaction between the generated SH and THz pulses takes place in DAST. 
However, the spectral content in the vicinity of \SI{2}{\micro\m} only accounts for $\sim$\SI{1}{\%} of the total pump energy. 
Hence, neither THz generation by SH, nor pump depletion of the \SI{3.9}{\micro \m} driving pulse due to SHG significantly influence the efficiency of THz generation. 
Note that, even for pump energy densities of more than \SI{100}{mJ/cm^2}, saturation of the conversion efficiency cannot be distinctively observed.

\subsection{\label{sec:1.95um} THz Generation by 1950 nm driving Pulses}

\begin{figure}[htb]
\centering
\includegraphics[width=\linewidth]{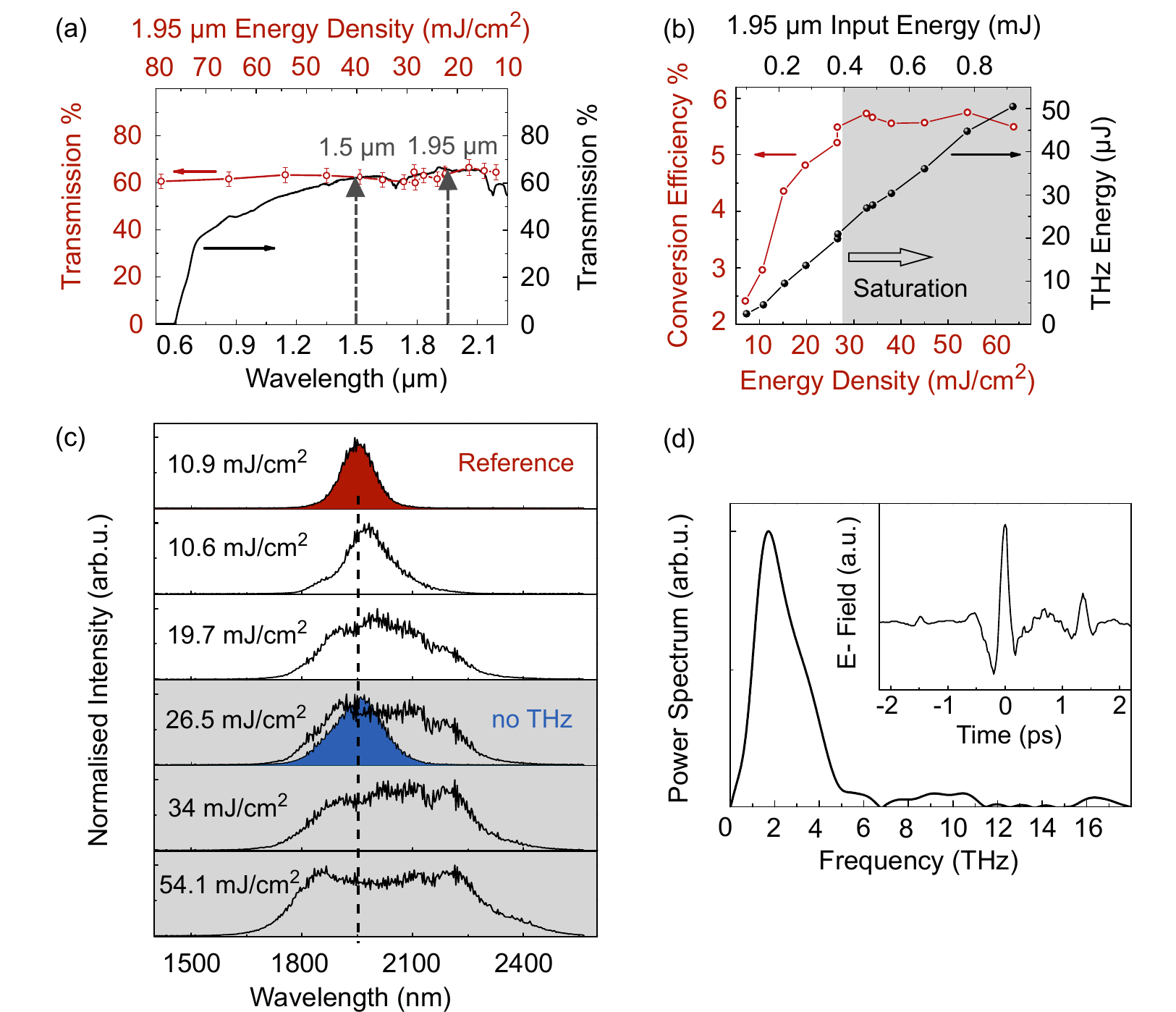}
\caption[DAST transmission spectrum, optical- to THz conversion efficiency and spectra of the \SI{1.95}{\micro m} driving pulse after propagation in DAST]{(a) DAST transmission spectrum (black line) and dependence of the transmission on the pump fluence  at \SI{1.95}{\micro\m} (red circles).
(b) Optical- to THz conversion efficiency (red circles) and THz pulse energy (black dots) with respect to the energy density for \SI{1.95}{\micro\m}driving pulses. 
(c) Pump spectra after propagation in DAST for different pump 
fluences when DAST is optimized for maximum THz generation. The red filled area in the top panel shows a reference pump spectrum without the crystal. 
The blue area in the middle panel shows the output spectrum when the crystal axis is rotated to inhibit THz generation.
(d) THz power spectrum measured with EOS. The inset depicts the temporal THz transient.
}
\label{fig:2umDAST}
\end{figure}

In order to investigate THz generation in DAST with long wavelength drivers but in a spectral range with less linear absorption, experiments with \SI{1.95}{\micro\m} pulses are performed.
Figure \ref{fig:2umDAST}(a) depicts the transmission spectrum (black line) measured with a UV-VIS-NIR spectrophotometer (Cary 5G), revealing a comparable transmission of $\sim$\SI{60}{\%} at \SI{1.5}{\micro\m} and \SI{1.95}{\micro\m}.
Fresnel losses at the input and output surfaces of the crystal are not taken into account.  
However, the measurement of the transmission at different pump energy densities reveal that, despite a relatively small difference in wavelength, MPA is completely suppressed in the case of \SI{1.95}{\micro\m} pulses when the crystal is pumped with energy densities of up to \SI{80}{mJ/cm^2}, corresponding to an intensity of $\sim$\SI{1}{TW/cm^2}.
Mind that in the case of a \SI{1.5}{\micro\m} driver, more than 50\% decrease in transmission due to MPA was observed at a ten times lower  intensity ($\sim$\SI{0.1}{TW/cm^2}) \cite{DAST_TransVicario:2015}. 
Figure \ref{fig:2umDAST}(b) illustrates the THz conversion efficiency with respect to the pump fluence, when the crystal is pumped with \SI{1.95}{\micro\m} pulses. 
At the lowest pump fluence of \SI{7}{mJ/cm^2}, an optical- to THz conversion efficiency of $\sim$\SI{2.4}{\%} is evaluated. 
This is in good agreement with previously published values for shorter wavelength drivers.
A conversion efficiency exceeding 2\% could be achieved when a DAST crystal  was pumped by a Cr:forsterite laser with a central wavelength of \SI{1.25}{\micro\m} and pump fluence around \SI{10}{mJ/cm^2} \cite{DASTVicario:2015}, while Hauri and co-workers \cite{DASTHauri:2011} demonstrated THz generation for \SI{1.2}{\micro\m}-\SI{1.5}{\micro\m} driving pulses in DAST with a wavelength independent efficiency of 2.2\%.
Similar results are reported for a derivative of DAST, namely DSTMS (4-N,N-dimethylamino-4’-N’-methyl-stilbazolium 2,4,6-trimethylbenzenesulfonate),   when the crystal is pumped with a \SI{1.25}{\micro\m} driver, wherein conversion efficiencies of more than 3\% at a 
pump fluence of $\sim$\SI{6.6}{mJ/cm^2} are achieved, followed by a reduction of the efficiency for higher pump energy densities \cite{DSTMSVicario09:2014}.
Novelli \textit{et al.} \cite{Novelli:2020} reported on wavelength independent THz generation conversion efficiencies for \SI{1.5}{\micro\m} and \SI{2}{\micro\m} when DSTMS is pumped at a fluence of \SI{6.4}{mJ/cm^2}.
Thus, the THz conversion efficiency in the case of our \SI{1.95}{\micro\m} driving source is not surprising for low pump fluences, because in addition to the commensurate transmission at
\SI{1.5}{\micro\m} and \SI{1.95}{\micro\m}, DAST 
exhibits NL optical coefficients in the same order of magnitude for both wavelength
($d_{11}(1542\ \mathrm{nm}) = 290,\ d_{11}(1907\ \mathrm{nm}) = 210$)
\cite{organicCrystalsBook:2015}. 
Moreover, as shown in section \ref{sec:cohLe}, the effective generation length in a \SI{200}{\micro\m} thick DAST crystal is comparable in the case of \SI{1.5}{\micro\m} and \SI{1.95}{\micro\m} drivers.
However, for a \SI{1.95}{\micro\m} pump source, the absence of MPA allows one to apply substantially larger pump fluences of \SI{30}{mJ/cm^2} (compared to  \SI{10}{mJ/cm^2} for near-IR drivers) before the onset of saturation. 
Thus, because the conversion efficiency scales linearly with respect to the pump fluence, exceptionally high THz conversion efficiencies of 5.7\% and generated THz energies of \SI{50}{\micro\J} are observed.

To further investigate the mechanism of this efficient THz generation, spectral transformations of pump pulses are monitored after propagation in DAST, when the crystal is aligned for maximum THz generation. 
The spectrum at the top panel of Fig.\ref{fig:2umDAST}(c) represents the reference without the DAST crystal in the beam path. 
When the crystal is inserted into the beam, the spectrum slightly broadens at low pump fluences, which is a feature of \ac{spm}.
With further increase of the pumpe fluence, spectral broadening dominated on the red side of the spectrum gets increasingly pronounced and stabilizes at a pump fluence of about \SI{30}{mJ/cm^2}, corresponding to the onset of saturation of the THz generation efficiency, as indicated by the shaded grey area (Fig.\ref{fig:2umDAST}(b)). 
In contrast, when THz generation is inhibited at an intermediate pump fluence (blue filled area in Fig.\ref{fig:2umDAST}(c)), which is realized by rotating the crystal axis
with respect to the polarization of the pump pulse, the spectrum exhibits a minor symmetrical spectral broadening, ascribed to SPM, and emulates the spectrum at low pump fluences. 
The significant red shift of the spectrum, related to efficient THz generation, strongly indicates cascaded OR. 
This observation is further confirmed by the spectral content of the THz pulse, measured with EOS (Fig.\ref{fig:2umDAST}(d)). 
The Fourier transformation reveals a maximum intensity at \SI{1.7}{THz} with a spectral width of \SI{2.6}{THz} at FWHM, and spectral components extending up to \SI{5}{THz}. 
Thus, the photon conversion efficiency, defined as the ratio of the number of THz to optical photons, is evaluated to be more than 500\%. 
Aside from the high pump fluence applied to the EO crystal, a lower group velocity dispersion at \SI{1.95}{\micro\m} can be an additional reason for the advanced photon conversion efficiency compared to 235\% \cite{DASTHauri:2011} at the telecommunication wavelength.
Nonetheless, besides the positive contribution to the conversion efficiency, cascaded OR results in a strong spectral broadening which enhances phase mismatch, and can be identified as a possible limiting factor for further efficient THz generation at higher pump fluences
\cite{Ravi:2014}. 

\section{Conclusion}

In conclusion, we demonstrated that suppression of \ac{mpa} allows to  shift the onset of saturation of the optical-to THz conversion efficiency for DAST to higher pump fluences by nearly an order of magnitude when the crystal
is pumped at longer wavelength driving sources as compared to the telecommunication wavelength around \SI{1.5}{\micro m}. 

In the case of \SI{3.9}{\micro m} driving pulses a conversion efficiency of 1.5\%, similar to the case of conventional driving sources operating in the transparency region of DAST, was measured. 
If to take into account that more than 80\% of the input energy is absorbed during propagation through the crystal, the effective conversion efficiency is even higher. 
As one of the possible reasons for the high efficiency and generated THz pulse energies we name a resonantly enhanced electro-optical coefficient, in addition to the high pump fluence. 
No indication of thermal effects, cascaded processes or parasitic SHG, influencing THz generation, is observed.
Moreover, although exceptionally high pump fluences of more than \SI{100}{mJ/cm^2} are applied to the crystal, the conversion efficiency shows no signs of saturation.  

In the case of \SI{1.95}{\micro m} driving pulses, record conversion efficiencies approaching 6\% are determined.
This beneficial outcome can be mainly attributed to the capability of applying higher pump intensities as compared to \SI{1.5}{\micro m} sources, and to efficient cascaded processes caused by the low GVD. 
Cascaded OR is identified as both, a mechanism through which a high conversion efficiency is achieved, and a limiting factor for the THz generation efficiency, when the newly generated spectral components promote phase mismatch between the pump and THz pulses. 
Because of the ongoing development of intense long wavelength laser sources, the gained insights on THz generation by OR of mid-IR drivers are of great interest in general, and in particular in light of the recently developed Cr:ZnSe laser systems generating multi-mJ femtosecond pulses in the spectral range of 2-\SI{2.5}{\micro m} \cite{2umLaser:2018}.

%% file: Chapters/QDs.tex
\chapter{THz induced electro-absorption switching in CdSe/CdS core/shell QDs}
\label{ch:QDs}

Rapidly developing modern optical communication systems demand small-scale electro-optic devices with large and fast changes in optical properties. 
Such nano-scale devices can be used, for example, as optical interconnects for data-storage or on-chip data links \cite{Kirchain:2008}.
Within the last few decades, \ac{ea} modulators based on the \ac{qcse} in quantum-well structures came to prominence for high-speed optical networks \cite{OmodulatorReed:2010, OmodulatorKuo:2005}.
The optical properties are thereby changed by an external electric field applied along the axis of confinement.
With such an external field, it is possible to manipulate the optical absorption of the confined material and hence alter the intensity profile of a transmitted signal pulse.
The modulation speed and signal contrast in such modulators is fundamentally limited by the time scale necessary to change the electric field and the optimization of the RF electrodes, respectively \cite{RF_Stepanenko:2019}.
The performance of the modulators can be quantified by their bandwidth, data rate, and extinction ratio.
In contrast to optics, wherein bandwidth refers to a precisely defined spectral content, in communication and computing the word bandwidth and data rate are somehow interchangeable and describe not only the rate at which data is transferred but also the rate at which it is processed.
Data rate is thereby expressed in bits per second and requires a short time period between the ON--OFF signal in order to be large.
As a consequence of Fourier transformation, such a short time period and hence fast data rate results in a large frequency bandwidth.
The extinction ratio describes the contrast between the ON--OFF signal
and is determined by the ratio between the energy (power) used to transmit a logic level '1' and energy used to transmit a logic level '0'.
State of the art \ac{ea} modulators based on the \ac{qcse} mainly consist of epitaxially grown multiple quantum well structures with a maximum bandwidth of \SI{23}{GHz}\cite{EOM_QW_Chaisakul:2012}, data rates of \SI{40}{Gb/s} \cite{EOM_QW_Fukano:2007}, and extinction ratio of \SI{10}{dB}.

While the modulation speed is restricted by the switching time of the electric field, the extent to which an ac dielectric field can manipulate quantum states by increasing the electric field strength, is limited by the dielectric breakdown of the material.
Consequently, THz wave-forms can be utilised as a driving field in order to overcome these physical limits and to implement systems with data rates in the order of Tbit/s \cite{THzCommunitication:2010} whilst avoiding dielectric breakdown owing to the low duty cycle.

Moreover, because the \ac{qcse} acts on the discrete energy levels of the nano-device, largest modulation of the absorption behaviour can be achieved for structures with a strong confinement, resulting in large excition binding energies, which further leads to sharp absorption features.
Therefore, low-dimensional materials like \acp{qd} are promising candidates for \ac{ea} modulators.

Pioneering work was performed by Hoffmann and co-worker \cite{GaAs_Hoffmann:2010}, wherein they report on sub-ps optical absorption switching in an InGaAs/GaAs QD-based saturable absorber mirror, induced by an incident THz waveform.
A few years later, Pein \textit{et al.} \cite{Pein:2017,Pein:2019} demonstrated THz-field induced \ac{qcse} in colloidal  CdSe/CdS  QDs by depositing the nano material on a micro-slit array consisting of parallel gold lines, in order to amplify the THz field.
The field enhancing structure was needed to apply an electric field strength in the MV/cm range, necessary to alter the QD band gap and to change the optical transmission.
Consequently, the resulting THz field is inhomogeneous and additional processes, such as electrode driven charge injection, can not be entirely decoupled from the QCSE, which further requires a complex theory to describe the underlying physical mechanisms.

Because field strengths in the range of MV/cm are accessible in our lab, as described in chapter \ref{ch:OR}, we report for the first time on a direct all optical encoding of a free space THz signal onto an optical probe signal in CdSe/CdS core/shell QDs without any field enhancing structures.
In contrast to previous works, the colloidal QDs are deposited onto a glass substrate by a simple drop-casting method.  
The possibility to manipulate optical properties of nano-scale semiconductors by direct THz radiation significantly simplifies the experimental setup, device design, and theoretical description,  and opens new perspectives in studying the THz induced \ac{qcse} in a large variety of samples.
A simple theoretical model is employed, which matches the experimental data remarkably well and rigorously discusses the influence of the energy band structure of the QDs on the modulation efficiency. 

The achieved absolute change in transmission of the probe pulse exceeds astonishing 15\% when THz radiation is incident on the QD sample, outperforming previously reported modulators based on solution processed materials with an applied ac field \cite{Pero_Walters:2018}. 
Furthermore, we investigate the interplay between the spectral bandwidth  and pulse duration of the probe pulse with respect to the modulation contrast and demonstrate that the signal amplitude can be controlled by the shape of the applied THz field, resulting in an extinction ratio of more than \SI{6}{dB}, which is on a par with state-of-the-art epitaxial quantum well EA modulators. 
The attained results prove the feasibility of high speed optical communication systems with \SI{}{Tbit/s} data rates and propose a route to optimize \ac{ea} modulation by band gap engineering of colloidal \acp{qd}.

This chapter is organised as followed: we first introduce some physical principles of colloidal \acp{qd}, 
with a subsequent discussion on the \ac{qcse} in general, and in particular with a view on the developed theoretical model for CdSe/CdS core/shell QDs.
The synthesis of the colloidal \acp{qd} is elucidated in the next subsection, followed by a material characterization of the resulting \acp{qd}.
Finally, we present the experimental setup and discuss the obtained results in detail.

\section{Excitons and Quantum Confinement in Semiconductors}
\label{sec:QDs}
\begin{figure}[htb]
\centering\includegraphics[width=\linewidth]{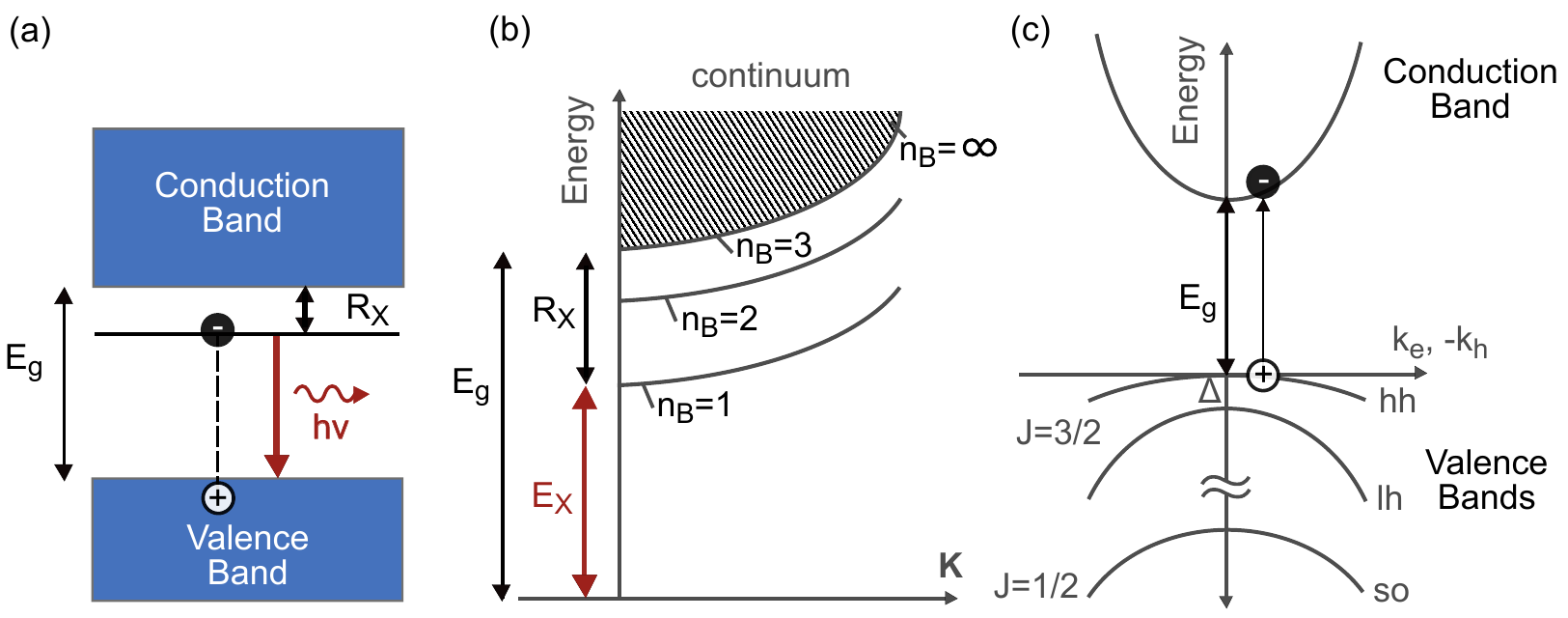}
\caption[Schematics of the exciton picture for a bulk semiconductor]{Schematics of the exciton picture for a bulk semiconductor, overly simplified (a) and a more detailed illustration (b) including energy levels of the exciton binding energy $R_{\mathrm{X}}$. For large principle quantum numbers $n_{\mathrm{B}}$ the difference between the energy levels becomes smaller, devolving into the continuum  conduction band. $E_{\mathrm{g}}$ denotes the energy gap of the semiconductor bulk material, $\mathbf{K}=\mathbf{k}_{\mathrm{e}}+\mathbf{k}_{\mathrm{h}}$ presents the exciton wave vector, and $E_{\mathrm{X}}$ describes the exciton energy which can be released as a photon when the exciton dissociates. In general, an exciton can be created in a direct gap semiconductor by absorption of a photon $h\nu=E_{\mathrm{g}}-R_{\mathrm{X}}$ (or $h\nu=E_{\mathrm{g}}-R_{\mathrm{X}}\pm h\Omega$ for indirect semiconductors including phonons). Higher photon energies $h\nu\geqslant E_{\mathrm{g}}$ are not necessary, though possible in order to generate an excition via band-to-band excitation and subsequent relaxation of carriers into the exciton state.(c) Pair excitation in the scheme of valence and conduction band for a wurtzite CdSe crystal structure around $\mathbf{K}=0$, indicating the splitting of the valence band caused by spin-orbit-coupling of the crystal atoms. The three subbands are referred to as heavy-hole (hh, J=3/2), light-hole (lh, J=3/2) and split-off-hole (so, J=1/2). The small splitting between the heavy- and light-hole band of $\Delta =\SI{25}{meV}$ in bulk CdSe is often neglected in QD calculations\cite{Klimov:2010}.}
\label{fig:SC_BandStructure}
\end{figure}

The most important parameters defining the optical properties of semiconductor materials is the band gap, corresponding to the difference energy between valance and conduction band, as well as the exciton binding energy, determining the transition energy of an electron into an excited state. 
Although the band gap can slightly change for different temperatures, it is basically determined by the chemical bond between the constituent atoms of the crystal lattice.
When an electron persisting in the valence band absorbs a photon, it is lifted into an excited state and leaves behind a hole in the valence band. 
Initially, such excited electrons can have larger energies than the band gap.
Due to thermalization, they will loose energies and end up near  the bottom of the conduction band, leading to the formation of excitons.
In this sense, an optical excitation is a two-particle transition, in which the electron and hole experience a Coulomb interaction describing the exciton binding energy. 
Note that an exciton is a bound state of an electron and a hole, resulting in a transition energy which is less than the band gap energy by the exciton binding energy, as indicated in Fig.\ref{fig:SC_BandStructure}.
This binding energy depends on the dielectric constant $\varepsilon$ of the material and accounts for the screening effect caused by the surrounding electrons. 
In other words, in vacuum, a negatively charged electron and positively charged hole would attract each other, resulting in a recombination.
In contrast, because a hole in the valence band is surrounded by other electrons, the excited electron experiences a repulsive force (screening), leading to a stabilizing energy balance.
Thus, the strength of the coupling or binding energy determines the stability of the exciton. 
With an effective mass approximation, wherein the movement of electron and holes caused by the periodic crystal potential is accounted by the effective masses $m_e^*$ and $m_h^*$, 
the Coulomb interaction between electron and hole leads to a hydrogen-like problem with a Coulomb potential term $-\mathrm{e}^2/(4\pi\varepsilon_0\varepsilon|\mathbf{r}_e-\mathbf{r}_h|)$, with $\mathbf{r}_e$ and $\mathbf{r}_h$ as the spatial coordinates of electron and hole, respectively. 
For simple parabolic bands and direct-gap semiconductor, the dispersion relation of excitons can be written as \cite{SC_Optics:2012}
\begin{equation}
E_{\mathrm{ex}}\left(n_B,\mathbf{K}\right)=E_{\mathrm{g}}-R_{\mathrm{X}}\frac{1}{n_B^2}+\frac{\hbar^2\mathbf{K}^2}{2M},
\label{equ:ExctionDispersion}
\end{equation}
with $E_{\mathrm{g}}$ as the band gap energy, $R_{\mathrm{X}}$ the exciton Rydberg energy (the binding energy of the electron and the hole) and $n_B=1,2,3\dots$ as the exciton principle quantum number.
The third term in Eq.\eqref{equ:ExctionDispersion} describes the kinetic energy of the exciton, with $M=m_e^*+m_h^*$ as the translational mass and $\mathbf{K}=\mathbf{k}_e+\mathbf{k}_h$ the wave vector of the exciton.
A schematic of valence and conduction band in the exciton picture is shown in Fig.\ref{fig:SC_BandStructure}, reproduced from ref \cite{SC_Optics:2012}.
The exciton Rydberg energy $R_X$ can be expressed with respect to the Rydberg energy of a hydrogen atom $R_y^*=\SI{13.6}{eV}$ as
\begin{equation}
R_{\mathrm{X}}=R_y^*\frac{\mu}{m_0}\frac{1}{\varepsilon^2},
\label{equ:Rydberg}
\end{equation} 
with $\mu=m_e^*m_h^*/(m_e^*+m_h^*)$ as the reduced exciton mass and $m_0$ the mass of a free electron.
In analogy, the Bohr radius $a_\mathrm{B}^\mathrm{H}$, defined as the most probable distance of proton and electron in a hydrogen atom in its ground state, can be used to describe the exciton Bohr radius
\begin{equation}
a_\mathrm{B}^{\mathrm{ex}}=\varepsilon\frac{m_0}{\mu}a_\mathrm{B}^\mathrm{H}.
\label{equ:BohrR}
\end{equation}
Since the dielectric function and the Bohr radius of an exciton are sufficiently larger than in a hydrogen atom, it follows from Eq.\eqref{equ:BohrR} and \eqref{equ:Rydberg} that the exciton Rydberg energy is much smaller.
Using material parameters of typical semiconductors, one finds \cite{SC_Optics:2012}
\begin{equation}
\SI{1}{meV} \leq R_{\mathrm{X}} \leq \SI{200}{meV}\ll E_{\mathrm{g}} \quad \textrm{and} \quad \SI{50}{nm}\gtrsim a_\mathrm{B}^{\mathrm{ex}} \geq \SI{1}{nm} > a_{\mathrm{lattice}}.
\end{equation}
Thus, the large dielectric constant results in a reduced Coulomb interaction and hence a spatial distribution of the electron and hole which is larger than the lattice unit cell.
Such excitons experience the crystal as an effective medium, justify the effective mass approximation in a self-consistent way and are called Wannier-Mott excitons.

\begin{figure}[hbt]
\centering\includegraphics[width=\linewidth]{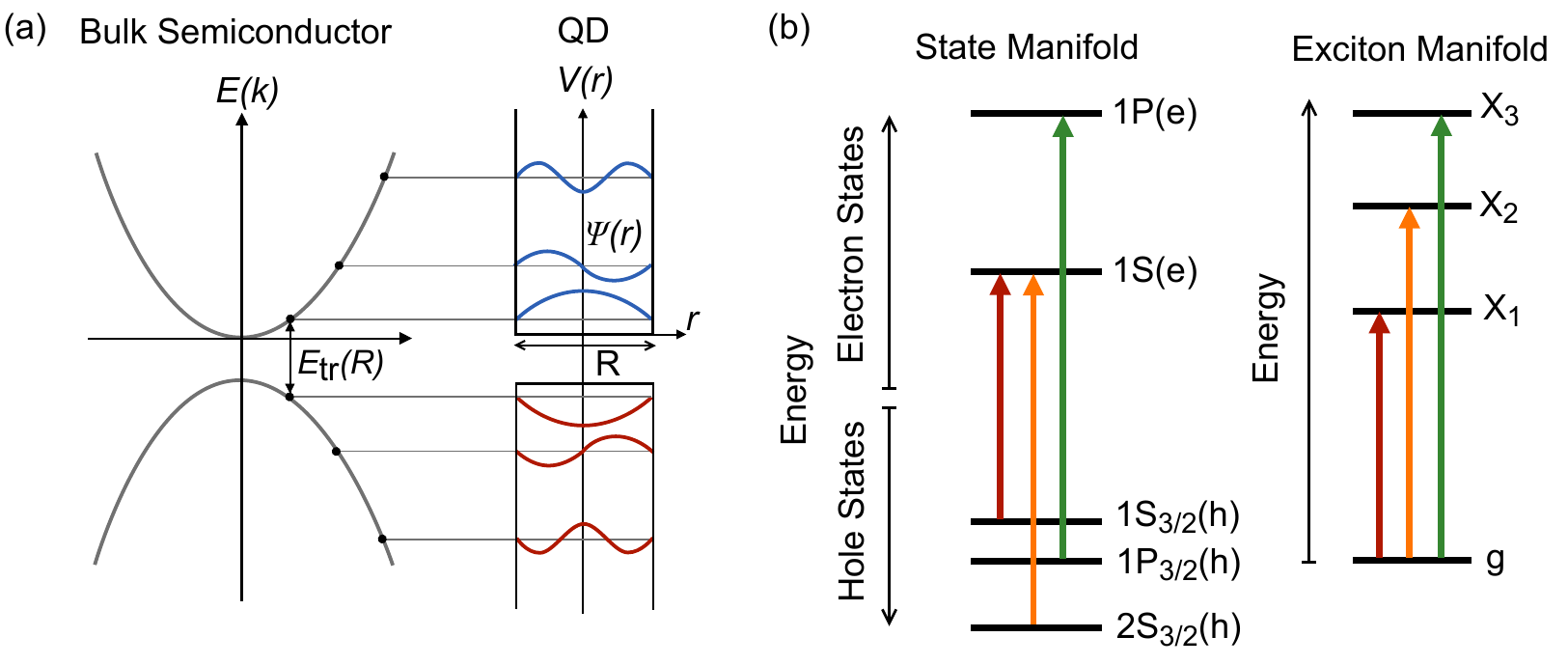}
\caption[Schematics of the energy quantization from continuous parabolic bands in bulk semiconductors to discrete levels in QDs]{(a) Schematics of the energy quantization from continuous parabolic bands in bulk semiconductors to discrete levels in \acp{qd}, reproduced from ref \cite{schmidNanotechnology:2008}. The transition energy $E_{\mathrm{tr}}(R)$ of the first excited state is larger than the band gap energy of the bulk material and depends on the radius $R$ of the QD. (b) Electronic energy levels of CdSe QDs, reproduced from \cite{TA_Zhang:2016}, representing the three lowest-energy transitions, denoted as \SI{}{X_1}, \SI{}{X_2} and \SI{}{X_3}, corresponding to the transitions $1\mathrm{S}_{3/2}(\mathrm{h}) \rightarrow 1\mathrm{S}(\mathrm{e})$,  $2\mathrm{S}_{3/2}(\mathrm{h}) \rightarrow 1\mathrm{S}(\mathrm{e})$ and $1\mathrm{P}_{3/2}(\mathrm{h}) \rightarrow 1\mathrm{P}(\mathrm{e})$, respectively \cite{Caram:2014}. The electronic transitions are labelled by a principal quantum number $n$ and angular momentum state $l$, originating from the confinement energy and referring to the envelope function, as well as a total angular momentum term $J$, indicating the splitting of the valence band caused by spin-orbit-coupling of the crystal lattice atoms.}
\label{fig:Quantization}
\end{figure}

Another important consequence of the Bohr radius is the approximate dimension for the onset of confinement effects at which physical properties of a material are noticeable affected.
In particular, if the size of the semiconductor structure becomes comparable to or smaller than the natural length scale of the electron-hole pair, the carriers experience the so called \textit{quantum size effect}, which leads to an atomic-like optical behaviour in \acp{nc} as the bulk bands become quantized \cite{Klimov:2010}.
As a first approximation, the energy levels in such a regime can be described using the particle in a box model, wherein the carriers are confined by the boundaries of the material.
The simplest form of the particle in a box model considers an infinite square potential. For confinement in all three dimensions, the discrete energies are found by
\begin{equation}
E_{n_x n_y n_z}=\frac{\hbar^2\pi^2}{2m_0}\left[\left(\frac{n_x}{L_x}\right)^2+\left(\frac{n_y}{L_y}\right)^2+\left(\frac{n_z}{L_z}\right)^2\right],
\label{equ:cubic}
\end{equation}
with $L_{x,y,z}$ as the dimension of the NC. 
Thus, the kinetic energy term for bulk semiconductors in Eq.\eqref{equ:ExctionDispersion} becomes quantized and each parabolic energy band of valence and conduction band will be split into a ladder of hole and electron levels, respectively, as sketched in Fig.\ref{fig:Quantization}.
Note that, the ground state is twofold degenerate due to spin, while the first excited state is sixfold degenerate ($(n_x,n_y,n_z)=1,1,2$ and permutation), which resembles the twofold degenerate S and sixfold degenerate P state of an atom.
Solving the Schr\"{o}dinger equation of the spherical problem of a QD, the  energy of the confined  particle  can be attained  \cite{Klimov:2010, Fluegge:1974}
\begin{equation}
E_{nl}=\frac{\hbar^2}{2 m_0}\left(\frac{X_{nl}}{R}\right)^2,
\label{equ:sphere}
\end{equation}
with $X_{nl}$ as the $l^{th}$ zero of the $n^{th}$ spherical Bessel function and $R$ the radius of the quantum dot.
Due to the discrete atomic-like energy levels (labelled as 1\textit{S}, 1\textit{P}, 1\textit{D}, 2\textit{S}, ect.), QDs are often referred to as \textit{artificial atoms}.
From Eq.\eqref{equ:sphere} it can be seen that the confinement energy scales as $1/R^2$, and is largest for small QDs, as is the distance between the discrete energy levels.

However, the particle in Eq.\eqref{equ:sphere} with free electron rest mass $m_0$ is confined to an empty sphere, while the \acp{nc} consist of semiconductor atoms in a periodic crystal lattice.
Nonetheless, the NC problem can be again reduced to the particle-in-a-sphere if to consider the effective mass approximation, as mentioned above, wherein periodic Bloch functions are used to describe the electronic wave functions of the bulk crystal structure.
Graphically, the effective mass accounts for the curvature of the conduction and valence bands at $\mathbf{K}=0$. Physically, the effective mass attempts to incorporate the complicated periodic potential felt by the carrier in the lattice \cite{Klimov:2010}.
Utilizing additional approximations, namely the \textit{envelope function approximation} \cite{bastard:1988}, and assuming a weak \textit{k} dependence of the Bloch functions, the total single particle wavefunction can be written as linear combination of a periodic function $u_n$ and single particle envelope function $\Phi(\mathrm{\mathbf{r}_e},\mathrm{\mathbf{r}_h})$.
Ultimately, the NC problem is reduced to determining this envelope function which is addressed by the particle-in-a-sphere model, wherein the free particle mass $m_0$ in Eq.\eqref{equ:sphere} is replaced be the effective or reduced mass of the particle.

Yet, the Coulomb attraction between the electron and hole is not incorporated in the treatment above. 
Though, how it is included depends on the confinement regime, defined as \cite{Klimov:2010}:

\textit{Strong confinement} ($R<a_e, a_h < a_{\mathrm{B}}^{\mathrm{ex}}$):  the NC radius $R$ is much smaller than the Bohr radius of the electron $a_e$, hole $a_h$ and exciton $a_{\mathrm{B}}^{\mathrm{ex}}$ (defined as the probability density of electron, hole and exciton, respectively).\\
\textit{Weak confinement} ($a_e,a_h < R < a_{\mathrm{B}}^{\mathrm{ex}}$): $R$ is larger than the electron and hole Bohr radius, but smaller than the exciton Bohr radius. Thus, only the center of mass motion is confined.\\
\textit{Intermediate confinement} ($a_h < R < a_e, a_{\mathrm{B}}^{\mathrm{ex}}$): the QD radius is in between $a_h$ and $a_e$. Hence, one particle (e.g. electron) is strongly confined while the other (e.g. hole) is not.

The Bohr radius of electron $a_e$ and hole $a_h$ are thereby defined as the width of their probability distribution (in bulk semiconductors), determined by their wave functions.
Depending on the size of the QDs, CdSe QDs can be in either the strong confinement or the intermediate confinement regime, since $a_{\mathrm{B}}^{\mathrm{ex}}=\SI{6}{nm}$ \cite{Klimov:2010}.

In general, the transition energy comprises of three terms, namely the band gap energy of the semiconductor bulk material $E_{\mathrm{g}}$, confinement energy $E_{nl}$, and a Coulomb term $E_{\mathrm{c}}$
\begin{equation}
E_{\mathrm{tr}}=E_{g}+E_{nl}+E_{\mathrm{c}}.
\label{equ:transition}
\end{equation}
In the strong confinement regime, the quadratic confinement term ($1/R^2$) dominates compared to the Coulomb interaction ($1/R$), and thus, the electron and hole can be treated independently, wherein each is described as a particle-in-a-sphere.
The Coulomb term than may be added as a first-order energy correction and the energies can be written as
\begin{equation}
E(n_h,l_h,n_e,l_e)=E_{\mathrm{g}}+\frac{\hbar^2}{2R^2}\left[\frac{X_{n_el_e}^2}{m_e^*}+\frac{X_{n_hl_h}^2}{m_h^*}\right]+E_{\mathrm{c}},
\label{equ:NLtransitoin}
\end{equation}
with $E_{\mathrm{c}}=-1.786e^2/\varepsilon R$ as the first-order Coulomb correction for the electron in the $1S_e$ level.
The lowest pair state can than be written as $1S_h1S_e$ with $n_h=n_e=1$ and $l_h=l_e=0$, ($n$ and $l$ refer to the principle and angular quantum number, respectively) and the transition energy from ground to first excited state  simplifies to
\begin{equation}
E_{11}=E_{\mathrm{g}}+\frac{\hbar^2\pi^2}{2\mu R^2}-1.786\frac{e^2}{\epsilon R}.
\label{equ:groundState}
\end{equation}
Such a solution for the energy of the first excited state in semiconductor QDs in the strong confinement regime was first calculated by Brus in the early 1980s \cite{Brus:1983,Brus:1984} by means of a variational method and effective mass approximation.
The first semiempirical calculation, using a tight-binding approach, was then published in 1989 by Lippens and Lannoo \cite{Lippens:1989} and further refinements of the theoretical description where achieved by other groups in the 90s, including the linear combination of atomic orbitals \cite{Delure:1993}, the semiempirical pseudopotential calculation \cite{Wang:1996} and \textit{kp} method \cite{Wang:1998}.
Despite their differences, all theories result in the same statement: the transition energy of the quantum dot increases with decreasing size of the QD.
Moreover, albeit many effects such as crystal anisotropy, spin-orbit coupling of crystal lattice atoms or mixture of hole states from  the three valence subbands   have to be considered in a more sophisticated manner, as a first approximation, Eq.\eqref{equ:groundState} reveals two important insights, namely (i) the confinement energy is always a positive term and thus the energy of the lowest possible state is always higher as compared to the bulk situation (ii) the Coulomb interaction between an electron and hole pair lowers the energy (see Eq.\eqref{equ:groundState}), but because of the $1/R^2$ dependence, the quantum confinement effect becomes the predominant energy term for very small quantum dot sizes.     
Hence, the size dependent energy gap of QDs is an effective tool for material design with well-controlled optical properties.
Note that, due to the discrete energy levels, the absorption spectrum would exhibit delta function features.
However, because of inhomogeneous broadening of the energy lines, QD size distribution, as well as thermal effects, the absorption spectrum can be best described with multi-Gaussian components and a polynomial background describing the remaining unresolved absorption continuum \cite{TA_Zhang:2016, Norris:1996}. 

Spectral broadening is also relevant for the emission spectrum of \acp{qd}.
Usually a redshift between the first excitonic absorption peak and the \ac{pl} is observed, a so called Stokes shift. 
Important processes for this red shift are charge transfer from smaller \ac{nc} to larger ones before re-emission of the photon in densely packed NC films, as well as ultra-fast relaxation via shallow trap states before the photon emission takes place.
Moreover, as the dimension of the QD decreases, the non radiative Auger effect becomes more pronounced, wherein an electron-hole pair recombines non-radiatively by transferring the recombination energy to a third carrier. 
This third carrier can be an electron, for example, which is ejected from the \ac{qd} \cite{Auger:2015}.
In addition, smaller \acp{qd} result in an increased surface to volume ratio, leading to additional surface trap states and enhanced non radiative decay.
In order to overcome such a reduced photoluminescence quantum yield, as well as to increase the chemical and photochemical stability, it is common practice to passivate the \ac{nc} surface with an inorganic shell and appropriate ligands.

\begin{figure}[htb]
\centering\includegraphics[width=\linewidth]{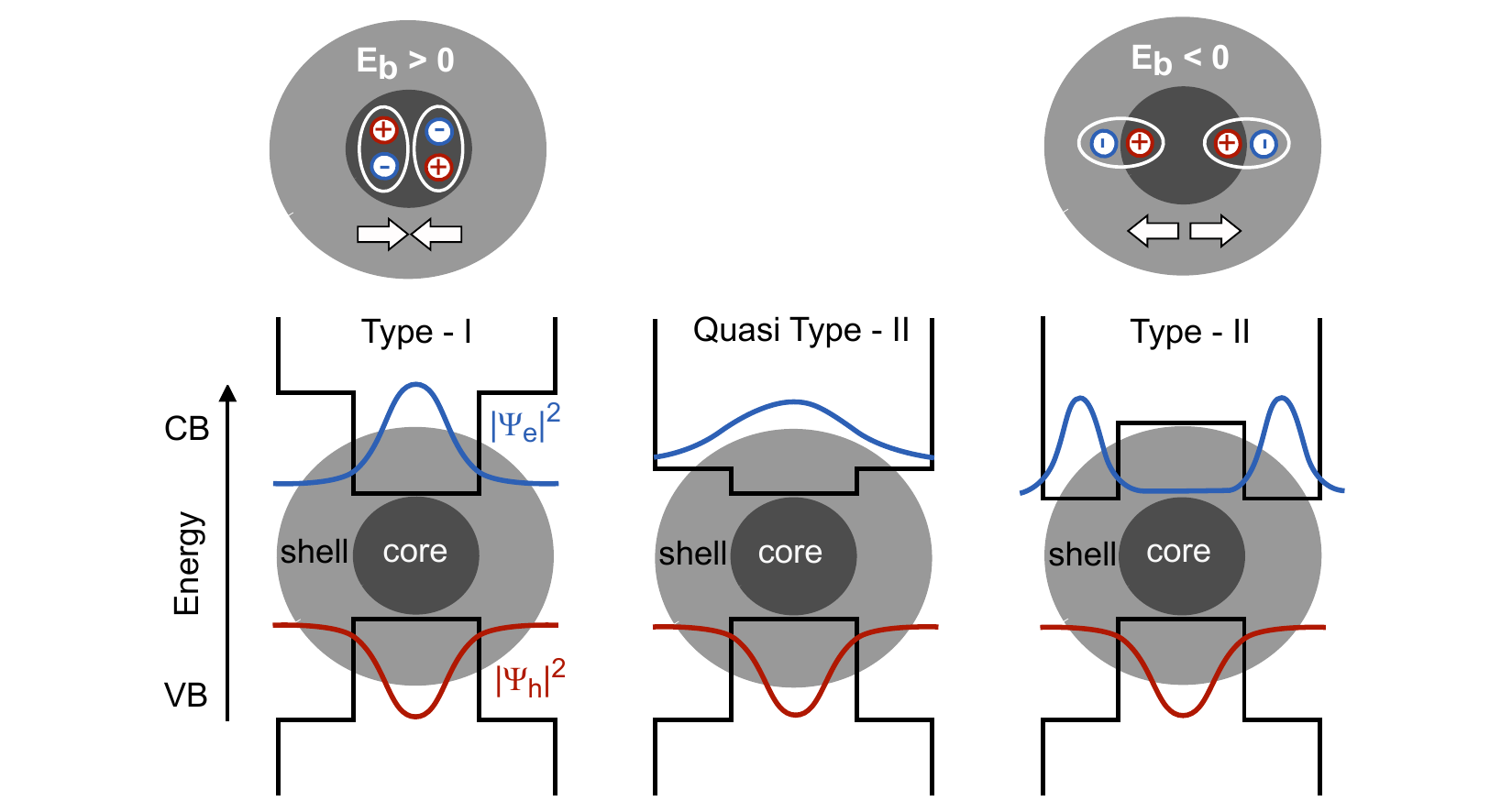}
\caption[Schematics of three different types of energy band alignments for  heterostructure QDs]{Schematics of three different types of energy band alignments for  heterostructure \acp{qd}, and its corresponding probability density for holes $|\Psi_\mathrm{h}|^2$ (red) in the valence band (VB) and electrons $|\Psi_\mathrm{e}|^2$ (blue) in the conduction band (CB), respectively. The picture above illustrates exciton--exciton interaction, depending on the energy band structure. Type-I QDs cause attractive interaction with a resulting positive biexciton binding energy $E_\mathrm{b}>0$, while type-II leads to exciton--exciton repulsion.}
\label{fig:QD_BandStructure}
\end{figure}

Such a heterostructure semiconductor further allows to engineer the energy band alignment of the core and shell material by changing the composition, introducing stress or strain via lattice mismatch or varying the layer thickness.
The resulting step-like band structures, caused by the offset between the band gap energy of core and shell material, are classified as \textit{type-I}, \textit{quasi type-II} and \textit{type-II} band alignment \cite{Mastria:2016, KlimovTypeII:2007}, as schematically shown in Fig.\ref{fig:QD_BandStructure}. 
In the case of a type-I energy band configurations, the electron and hole are confined in the same region of the QD, yielding a large overlap integral of electron and hole wavefunction. 
In turn, the band edges of the core and shell material are staggered in type-II, wherein both the low-energy electron and hole states are
in different semiconductors, resulting in the segregation of electron and hole wavefunctions between the core and the shell material.
A quasi type-II structure exhibits an intermediate band alignment with a finite offset for only one carrier, causing de-localization,   and a zero offset for the opposite one.
In the case of CdSe/CdS QDs, quasi type-II strongly reduces the confinement of the electron and allows it to penetrate the shell, while the hole is always localized in the CdSe core with a positive band offset of \SI{0.31}{eV}-\SI{0.48}{eV} at the core-shell interface.
In contrast, the reported band offset of the electron varies from \SI{-0.25}{eV} to \SI{0.27}{eV} in dependence on the morphology, crystal structure and dimension of the core and shell material \cite{BandOffsetSteiner:2008}.
Thus, it has been shown that CdSe/CdS QDs can  be tuned from type-I over quasi type-II to type-II by varying the core radius and shell thickness \cite{CihanBandAlignment2:2013,ASE_Dong:2013}.

Due to quantum confinement, resulting in atomic-like discrete energy levels, the ground state 1S(h) can only be by occupied by two electrons.
Thus, population inversion can be achieved if at least two electron-hole pairs are created simultaneously which further interact with each other and generate a biexciton.
Hence, the most probable mechanism for optical gain is given by biexciton emission.
When a biexciton is annihilated, it decomposes into a photon and a single exciton $E_{\mathrm{XX}} \rightarrow \hbar\omega+E_\mathrm{X}$, wherein the energy of the photon $\hbar\omega\!=\!E_\mathrm{X}- E_\mathrm{b}$ differs from the exciton energy $E_\mathrm{X}$ by the biexciton binding energy $E_\mathrm{b}$.
Therefore, after the onset of \ac{ase}, a wavelength shift of the biexciton emission with respect to the exciton emission can be observed in the PL spectrum, which can be ascribed to the biexciton binding energy $E_\mathrm{b}\!=\!2E_\mathrm{X}-E_\mathrm{XX}$, with $E_\mathrm{XX}$ as the biexciton energy \cite{Annual_Klimov:2007}.
The amount and sign of $E_\mathrm{b}$ allows one  to infer the band alignment of the core/shell structure.
If the electrons and holes are both confined in one region (type-I), the excitons experience an attractive interaction, as indicated in Fig.\ref{fig:QD_BandStructure}, resulting in a positive biexciton binding energy $E_\mathrm{b} > 0$.
In turn, if the electrons are separated in the core and shell material (type-II), the generated excitons exhibit a repulsive interaction and hence a negative binding energy $E_\mathrm{b} < 0$.
In the case of a border-line quasi type-II band structure, where the electrons are de-localized within the entire QD while the hole states are confined in the core material, neither exciton--exciton repulsion nor attraction is supported. The binding energy will be close to zero and the biexciton emission appears close to the center of exciton emission \cite{CihanBandAlignment2:2013}.
 
\section{Quantum confined Stark effect}
\label{sec:QCSE}
The \ac{qcse} was first reported by Miller \textit{et al.} in 1984 \cite{Miller:1984} and further discussed in 1985 \cite{Miller:1985}, describing the electric-field dependence of optical absorption (electro-absorption) near the optical band edge in quantum well structures.
Similar to the Franz-Keldysh effect in bulk material, an applied electric field tilts the energy bands.
When an incident photon with an energy just below the absorption edge raises an electron to higher energies, the electron can then tunnel to allowed empty states in the conduction band due to the tilt of the band. 
Hence, photon absorption with energies smaller than the band edge is possible and the absorption spectrum shifts to smaller energies (\textit{i.e} longer wavelengths). 
However, such an effect in a 3D bulk material is fundamentally limited by the classical (static) ionization field $E_i=E_B/8ea$, with $E_B$ as the zero-field exciton binding energy and $a$ as the Bohr radius \cite{IonizationField:1972}.
For fields of the order of a few times this ionization field, the exciton resonance is severely broadened because field ionization drastically reduces the exciton lifetime until the peak becomes unresolvable.
In other words, \ac{ea} is deteriorated by exciton field ionization.

\begin{figure}[ht]
\centering\includegraphics[width=\linewidth]{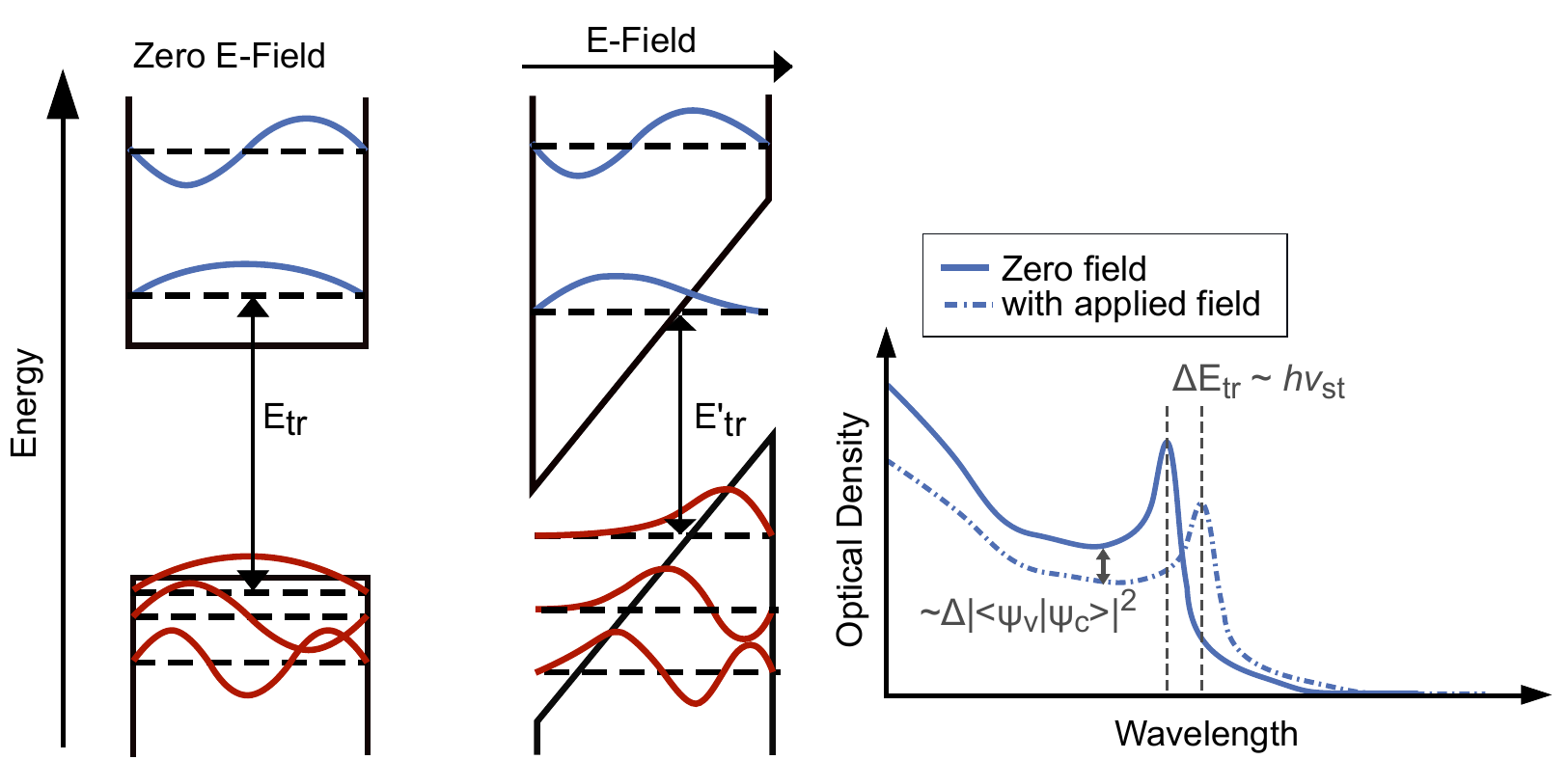}
\caption[Schematics of the QCSE in a quantum-well structure]{Schematics of the QCSE in a quantum-well structure. Due to an applied electric field the energy bands are tilted, the transition energy is reduced and the overlap integral of electron and whole wavefunction is diminished, leading to a red-shifted decreased optical density spectrum. }
\label{fig:QCSE}
\end{figure}

In turn, the QCSE, describing a quantum-confined Franz-Keldysh effect including electron-hole interaction \cite{Miller:1986}, inhibits exciton field ionization and thus allows to apply much stronger electric fields.
Moreover, because the absorption spectrum of confined structures exhibit distinct absorption features due to discrete energy levels, such steep changes in the absorption spectrum will result in a substantial modulation of the absorption properties, even for small shifts of the absorption edge.
The optical properties of such quantum structures are thereby changed by an external electric field applied along the axis of confinement. 
The field skews the potential well and tilts both the valence and conduction bands.
As a consequence, the electron and hole wavefunctions are pulled in opposite directions, occupying lower energy states and typically leading to a modification of the absorption spectrum in the vicinity of the band-edge.
Simultaneously, because the electron and hole move to opposite sides of the quantum structure, while the well barriers prevent field ionization, the overlap integral between the wavefunctions, as well as the oscillator strength of the contributing transitions are reduced.  
Schematics of the QCSE in a quantum well structure are shown in Fig.\ref{fig:QCSE}, wherein $E_{\mathrm{tr}}'$ depicts the altered transition energy leading to a Stark shift $h\nu_{\mathrm{st}}$ in the absorption spectrum, and the squared overlap integral of electron and hole wavefunction$|\langle\Psi_{\mathrm{h}}|\Psi_{\mathrm{e}}\rangle|^2$ acting as the main driving force for a reduced absorption.
Modulation of the absorption spectrum induced by an external electric field can be further caused by transition broadening originating from energy level splitting of originally degenerate states, the appearance of optical transitions that have been initially forbidden due the symmetry of the wavefunctions or reduction of the transition lifetime due to exciton ionization. 
Changes of the absorption spectrum are also accompanied by a change of the exciton binding energy due to a reduction of the Coulomb interaction as the electron and hole become spatially separated \cite{Miller:1989}.
However, the Stark shift is usually manifested in a net decrease in  transition energy and thus a red-shift of its optical absorption peak.
The magnitude of the QCSE energy shift of the electron and hole levels increase monotonically with the well width in quantum well structures \cite{Matsuura:1986,Hiroshima:1986}.
Though, an upper limit on the scalability of the QCSE is given by the exciton Bohr radius \cite{Miller:1985}. 
Increasing the size of nano-devices diminishes the quantum confinement and reduces the exciton binding energy, such that the exciton eventually field-ionizes and the bulk electric field response can be expected.

In order to understand the QCSE in heterostructure QDs, a simple theoretical model was employed.
The numerical simulations describe the energy states and wave functions for  CdSe/CdS core/shell QDs for different types of energy band structures, with the main focus on a quasi-type II band alignment emerging from a CdSe core with a diameter of \SI{4.1}{nm} and CdS shell with a thickness of \SI{2.4}{nm}.
The electron and hole wavefunctions are first calculated for an unperturbed case and when an external field is present.
Consequently, the change in optical density can be determined with respect to an applied electric field, wherein the  key elements are provided by the change in transition dipole moment as well as the Stark shift.

Recently, Nandan and Mehata \cite{nandan2019wavefunction} calculated the Eigen-energies and overlap integrals within a mean field Hartree approach for a three-dimensional  CdSe/CdS core/shell quantum dot without an external field,  and demonstrated the feasibility to tune the energy band structure from type-I to quasi type-II band alignment by increasing the shell thickness.
Here, the absorption spectrum is calculated in a simplified manner.
Due to the periodic crystal structure of CdSe and CdS, the full exciton wave-function is represented via the electron $u_{\mathrm{e}}$ and hole $u_{\mathrm{h}}$ Bloch wave-functions
\begin{equation}
    \Psi(\mathbf{r}_{\mathrm{e}},\mathbf{r}_{\mathrm{h}}) = \Phi(\mathbf{r}_{\mathrm{e}},\mathbf{r}_{\mathrm{h}}) u_{\mathrm{e}}(\mathbf{r}_{\mathrm{e}})u_{\mathrm{h}}(\mathbf{r}_{\mathrm{h}}),
\end{equation} 
with $\Phi(\mathbf{r}_{\mathrm{e}},\mathbf{r}_{\mathrm{h}})$ as the slowly varying exciton envelope wave function  \cite{peter2010fundamentals}.
In order to obtain the envelope wave function, an effective Hamiltonian for electrons and holes is used\cite{velizhanin2016renormalization,peter2010fundamentals}
\begin{equation}
\left[ \Op{H}_\ee^0 + \Op{H}_\hh^0 + \Op{V}_\mathrm{col} +  \Op{V}_F \right] \Phi(\rv_\ee,\rv_\hh) = E\Phi(\rv_\ee,\rv_\hh),
\end{equation}
with the single particle Hamiltonian
\begin{equation}
\Op{H}_\eh^0 =   -\frac{\hbar^2}{2 }\nabla_\eh \frac1{m_\eh^*(\rv_\eh)} \nabla_\eh + \Op{V}_\eh(\rv_\eh)
\end{equation} 
consisting of a kinetic term with an effective mass $m_\eh^*(\rv_\eh)$ for electrons and holes, respectively,  and confinement potential $\Op{V}_\eh(\rv_\eh)$. 
The attractive Coulomb interaction between electron and hole is governed by \(\Op{V}_\mathrm{col}\) and the applied electric field is represented by \(\Op{V}_F\!=\!\pm e\,F\,\Op{z}\). 
This external electric field \(F\) gives a preferential spatial direction to the otherwise isotropic \acp{qd}. 
The dominant resonance  --- often referred to as X\(_1\) (as indicated in Fig.\ref{fig:Quantization}) --- is caused by the s-wave like valence band state of the hole and the s-wave like conduction band state of the electron,  both are spherically symmetric.
It is therefore sufficient to use a one-dimensional model in order to discuss the underlying physics. 
The full envelope wave-function is approximated by \(\Phi\!\approx\!\Phi_x\Phi_y\Phi_z\) which is then solved individually with the energies \(E\!=\!E_x+E_y+E_z\). 
Due to strong screening of the semiconductor materials, the Coulomb interaction between the electron and hole is sufficiently small, further supporting this simplification. 
Thus, the theoretical description is focused on the wave function in \(z\)-direction, others can be obtained equivalently.

\begin{figure}[htb]
    \centering
    \includegraphics[width=\linewidth]{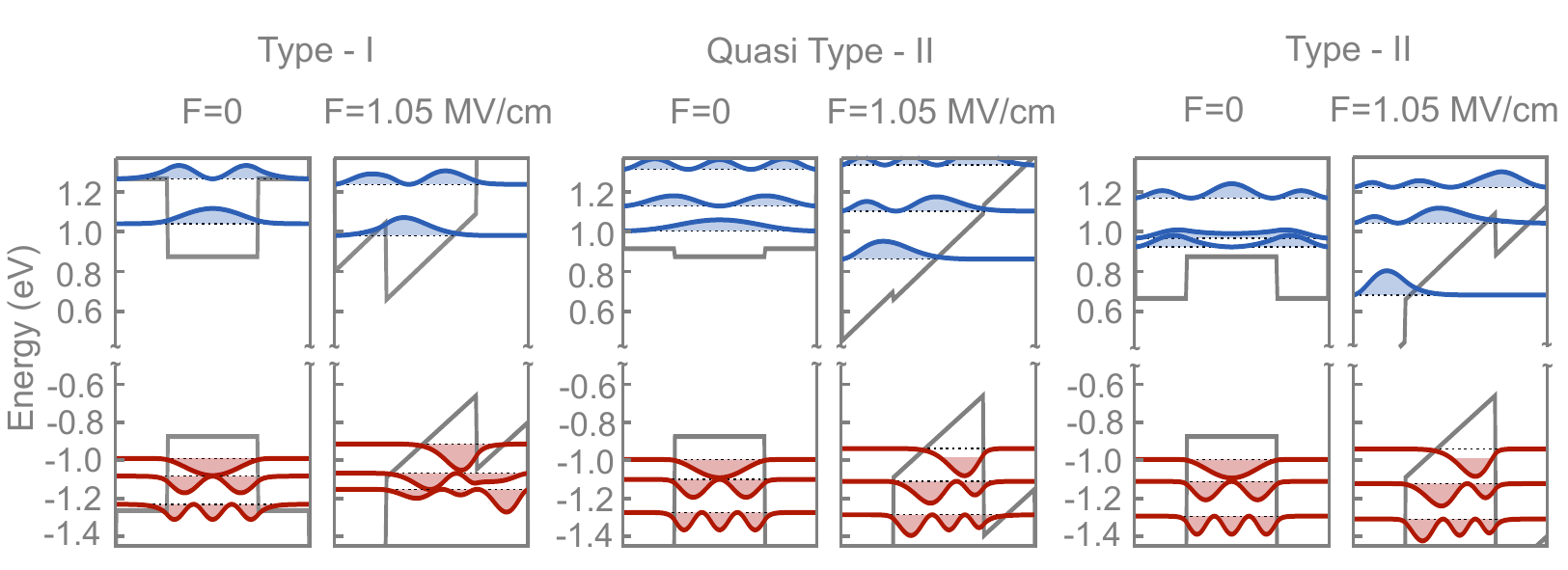}
    \caption[Energy band alignments for type-I, quasi type-II and type-II  CdSe/CdS core/shell QDs without and with an applied electric field \(F\)]{Energy band alignments for type-I, quasi type-II and type-II  CdSe/CdS core/shell \acp{qd} without and with an applied electric field \(F\). The resulting discrete energy states are displayed in the otherwise continues conduction and valence band, as well as the probability densities of electron (blue) and hole states (red).}
    \label{fig:ProbDense}
\end{figure}

In the strong-confinement limit, which is fulfilled in the present case, the envelope function 
\begin{equation}
    \Phi_z(\rv_\ee,\rv_\hh) \approx \varphi_\ee(\rv_\ee)\,\varphi_\hh(\rv_\hh)
\end{equation}
can be approximated by the product of the electron \(\varphi_\ee\) and the hole wave-function \(\varphi_\hh\),  respectively \cite{tyrrell2011effective}. 
Both, electron and hole, are confined by the external potential wells of the heterostructure
\begin{equation}
\Op{V}_\eh(z_\eh)=\begin{cases}
V^\mathrm{C}_\eh & z_\eh <R_1 \\ 
V^\mathrm{S}_\eh & R_1 < z_\eh < R_2 \\ 
\pm \infty & z_\eh > R_2
\end{cases}
\end{equation}
where \(V^\mathrm{C}_\eh\) is the core potential given by the CdSe band-gap and \(V^\mathrm{S}_\eh\) is the potential of the shell arising from the band-gap of CdS and the band alignment controlled by strain between the different constituents \cite{li2004first}. 
The radius of the core is given by \(R_1\) and the full quantum dot radius is \(R_2\).  
Due to the weak interaction, a mean field like ansatz is chosen wherein the electron is moving in an effective field produced by the hole and vice versa.
This effective Hartree potential \(\phi_\eh^\mathrm{H}\) has to fulfil Poisson's equation \cite{csahin2012reordering}
\begin{equation}
\frac{\partial}{\partial z} \kappa(z) \frac{\partial}{\partial z} \phi^\mathrm{H}_\eh(z) = \pm e\, \rho_\eh(z)
\label{equ:Hatree_pot}
\end{equation}
with the electron/hole probability density \(\rho_\eh\!=\!\left|\varphi_\eh \right|^2\) and the relative dielectricity \(\kappa(z)\). 
Eq.\eqref{equ:Hatree_pot} is solved numerically with a finite element integrator. 
 With the Hartree potential, the Schr\"odinger equation separates into two coupled differential equations for the electron and hole. 
 In order to obtain the correct interaction, a self-consistent approach is chosen. 
 In the first step, the electron and hole states are calculated without interaction. 
 In the next step, the effective Hartree potentials are calculated with the obtained wave-functions used to get more precise states. 
 This procedure is repeated until convergence is reached. 
 The exciton energy spectrum is then approximated by the individual Eigen-energies \(E_\eh\) as 
\begin{equation}
    E = E_\ee - E_\hh \ .
    \label{equ:TransEnergy}
\end{equation}
Figure \ref{fig:ProbDense} shows the calculated energy levels and probability densities $\left|\varphi_\eh \right|^2$ for electron (blue) and hole (red) states for the unperturbed system and for an external field of F=\SI{1.05}{MV/cm} inside the QDs, for three types of energy band alignment.
The graphs illustrate the change in transition energy as well as the reduced overlap integral, wherein type-II exhibits the largest Stark shift.
Material constants are taken from ref \cite{nandan2019wavefunction} and are summarized in table \ref{tab:MaterialCons}. 
\begin{table}[htb]
\centering
\begin{tabular}{c|ccccc}
Material & \(m_\ee^*\) / \(m_0\) & \(m_\hh^*\) / \(m_0\) & \(\kappa / \epsilon_0\) & \(E_\mathrm{g}\) / eV & \\\hline
CdSe & 0.13 & 0.45 & 9.3 & 1.75 \\
CdS & 0.21 & 0.8 & 8.9 & 2.49
\end{tabular}
\caption[Material constants for bulk CdSe and CdS]{\label{tab:MaterialCons} Material constants for bulk CdSe and CdS, respectively, taken from \cite{nandan2019wavefunction}. \(m_\ee^*\) and \(m_\hh^*\) are the effective electron and hole mass, given as ratio of the electron rest mass \(m_0\). \(\kappa\) is the relative dielectric constant with respect to the  vacuum permittivity and \(E_\mathrm{g}\) is the bulk energy gap.}
\end{table}

Nonetheless, the full wave-function consists of the envelope function, as well as a Bloch-wave contribution reflecting the unit cell properties of the crystal lattice. 
The Bloch wave-function symmetry has a direct consequence for the possible optical transitions. 
The transition dipole moment, and thus its square value and the oscillator strength can be calculated via
\begin{equation}
    \bra{0}e
    \,\Op{p}\ket{\Psi} = e
    \bra{u_\ee} \Op{p} \ket{u_\hh} \int_0^L  \mathrm{d}z \varphi_\ee(z)\varphi_\hh(z) \ . \label{equ:matrix_kane}
\end{equation}
with the Fermi--Sea state \(\ket{0} = \delta(\rv_\ee - \rv_\hh)\). 
The overlap of the Bloch wave function can be evaluated using the Kane momentum parameter \cite{kane1958influence, peter2010fundamentals}
\begin{equation}
    \left| \bra{u_\ee} \Op{p} \ket{u_\hh} \right|^2 = m_0\,E_p/4 ,
\end{equation}
where  \(E_p^\CdS = 19.6\)eV, \(E_p^\CdSe=17.5\)eV and \(m_0\) is the free electron mass \cite{efros1996band}. 
These Kane matrix elements are determined from bulk material. 
Though, Velizhanin \cite{velizhanin2016renormalization} demonstrated an overestimation by a factor of two compared to confined systems. 
However, in this work, we are going to compare the difference in absorption spectra to a normalized measurement result, as will be shown in section \ref{sec:ResultsQds}. 
Thus, this discrepancy may not play a major role.
The transition dipole moment is directly proportional to the overlap of the envelope wave-function.  
Using Eq.\eqref{equ:matrix_kane}, we find that the oscillator strength depends strongly on the band alignment of the conduction band between core and shell wave-functions \cite{shabaev2012fine}. 
Finally, we can evaluate the unit-less  oscillator strength by
\begin{equation}
    f_\mathrm{vc} = \frac{E_p}{2\,E} \left| \int_0^L \mathrm{d} z \varphi_\ee(z)\varphi_\hh(z)\right|^2,
    \label{equ:OsciStrength}
\end{equation}
with $E$ as the exciton transition energy \cite{csahin2012reordering}. 
The transition dipole moment then reads 
\begin{equation}
\left| \mu_\mathrm{vc} \right|^2= \frac{e^2\hbar^2}{2mE} f_\mathrm{vc}.
\label{equ:DipoleM}
\end{equation}

\section{Quantum Dot Synthesis}
\label{sec:Synthesis}
The synthesis procedure of CdSe/CdS core/shell \acp{qd} the  is based on the method established  by Chen \textit{et al.} \cite{Chen:2013}, with minor modifications.
CdSe QDs with wurtzite (WZ) structure are used as core material. 
CdO (\SI{180}{mg}, \SI{1.404}{mmol}), octadecylphosphonic acid (ODPA) (\SI{840}{mg}, \SI{2.508}{mmol}), and trioctylphosphine oxide (TOPO) (\SI{9.0}{g}) are mixed in a \SI{25}{mL} three-neck flask, heated to \SI{150}{^{\circ}C}, degassed under vacuum for 1 hour and then heated to \SI{320}{^{\circ}C} under a constant nitrogen flow. 
Once a colorless clear solution is formed, \SI{1.5}{mL} of trioctylphosphine (TOP) is added.
The reaction mixture is then heated to \SI{360}{^{\circ}C} and a TOPSe/TOP solution (\SI{174}{mg} Se in \SI{1.1}{mL} TOP) is swiftly injected into the flask.
After 10 min of growth time, the reaction is terminated by fast cooling. 
The resulting CdSe QDs are washed 3 times with acetone/hexane, re-dispersed in hexane and left in the fridge overnight. 
Afterwards, the solution is filtered with a \SI{0.2}{\mu m} PTFE filter. 
The size and concentration of QDs is estimated from the optical absorption spectra, using a method reported in ref \cite{Jasieniak:2009}.
In order to grow the CdS shell, \SI{300}{nmol} of WZ-CdSe in hexane are mixed with \SI{3}{mL} 1-octadecene (ODE) and \SI{3}{mL} oleylamine (OAm), degassed under vacuum at room temperature for 60 min, and at \SI{110}{^{\circ}C} for 15 min. 
Finally the mixture is heated to \SI{310}{^{\circ}C} under nitrogen flow and maintained for 2 hours. 
During the heating process, when the temperature reaches \SI{240}{^{\circ}C}, the solutions of cadmium oleate in ODE and hexanethiol in ODE are carefully injected drop-wise with a syringe pump with a rate of \SI{3}{mL/hr}. 
A total amount of \SI{1.92}{mmol} \SI{}{Cd(Oleate)_2} in \SI{6.4}{mL} ODE and \SI{335}{\mu L} hexanethiol in \SI{6.1}{mL} ODE are added during 2 hours. 
Afterwards, \SI{3}{mL} oleic acid is injected and the solution is
further annealed at \SI{310}{^{\circ}C} for 1 hour. 
The resulting CdSe/CdS core/shell QDs are washed 3 times with acetone/hexane, redissolved in hexane and filtered through a \SI{0.2}{\mu m}  filter.
\begin{figure}[htbp]
    \centering
    \includegraphics[width=\linewidth]{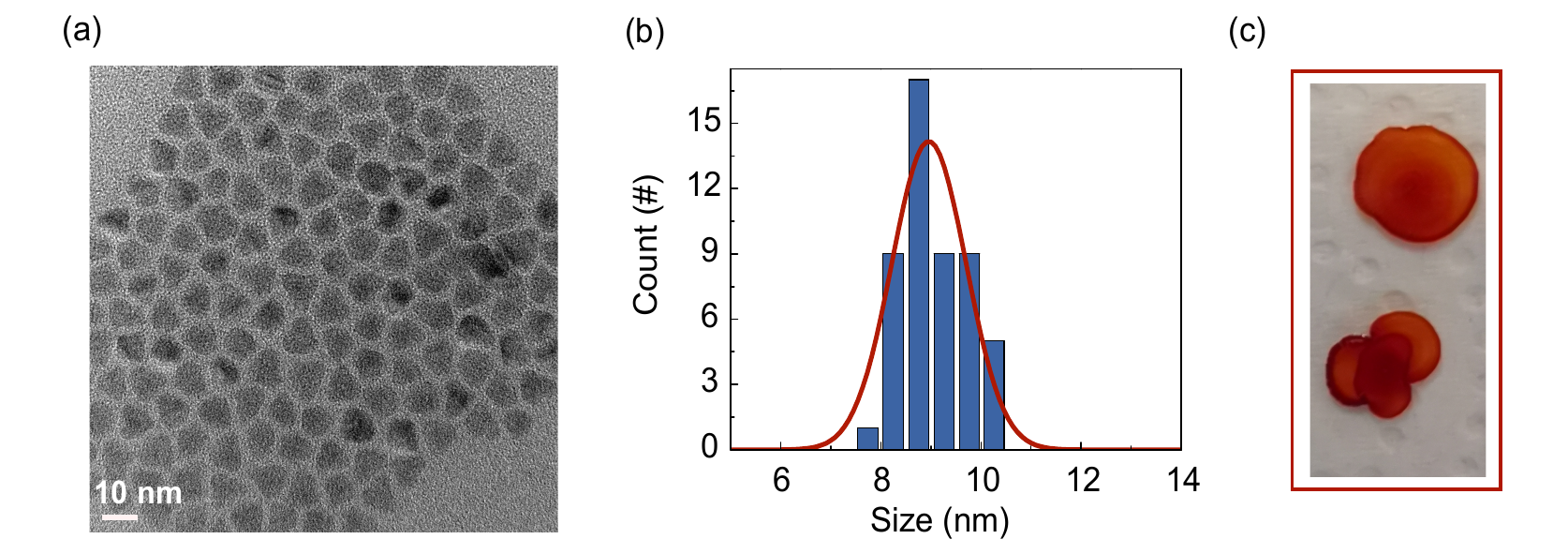}
    \caption[TEM image of CdSe/CdS core/shell QDs and colloidal QDs on a glass substrate deposited with drop-casting.]{(a) TEM image of CdSe/CdS core/shell QDs and (b) histogram of the corresponding size distribution revealing an average size of \SI{8.9}{nm} with a standard deviation of \SI{0.8}{nm} (c) Colloidal QDs on a glass substrate deposited with drop-casting.}
    \label{fig:synthesis}
\end{figure}

The overall procedure results in an average CdSe core diameter of \SI{4.1}{nm} and CdS shell thickness of \SI{2.4}{nm}.
A \ac{tem} image of the resulting colloidal QDs and a corresponding histogram of the size distribution are shown in Fig.\ref{fig:synthesis}(a) and (b), respectively, revealing  an average size of \SI{8.9}{nm} with a standard deviation of \SI{0.8}{nm}.
A measurement of the \ac{pl} quantum yield reveals 72\%.
The colloidal QDs are deposited on a glass substrate by simple drop casting, forming a thin film of limited homogeneity, as shown in Fig.\ref{fig:synthesis}(c).

\section{Optical Material Characterization}
\label{sec:QDsCharacterization}

The transmission spectrum of a thin film of colloidal CdSe/CdS QDs is measured with a UV-VIS-NIR spectrophotometer (Cary 5G).
In order to calculate the optical density, defined as the negative logarithm of the sample transmittance
\begin{equation}
	OD\left(\lambda\right)=-\log\left[\mathrm{T}(\lambda)\right],
	\label{equ:OD}
\end{equation}
the measured transmission is corrected for the reflection losses of the  substrate by normalizing the transmission to one at long wavelengths, for which it is known that neither the glass substrate nor the QD sample exhibits absorption.
Thus, it is possible to approximately  account for Fresnel losses at the interface between air/QD film and QD film/substrate.
Moreover, the normalization factor accounts for 0.54 and can be interpreted as the total reflection losses, which allows to estimate the dielectric constant of the QD film. We therefore assume a refractive index of $n_{\mathrm{G}}=1.5$ for the glass substrate, use $R=|(n_1-n_2)/(n_1+n_2)|^2$ for the reflection losses at the surface between different materials, and with the relation $\varepsilon=n^2$ evaluate $\varepsilon_{_\mathrm{QD}} \sim 10$ for the dielectric constant of the QD film. 
Note that, a dielectric constant of a factor of 10 is in agreement with the literature \cite{Pein:2019}, however, the exact value for the dielectric constant of the entire composite differs depending on the applied model~\cite{dielConstGrinolds:2015}, with the most prominent being the Maxwell–Garnett method~\cite{dielConstVahdani:2014, Pein:2019}, Bruggeman effective-medium theory~\cite{Garnet_Brugg:1981} or equivalent circuit describing a series of capacitors~\cite{Bozy:2013}.
Nonetheless, we find that, due to screening effects of the applied electric field, caused by the large dielectric constant of QDs, and spacing between the QDs induced by the surrounding inorganic ligands, in combination with Fresnel reflection losses, the electric field inside the QDs can be an order of magnitude weaker than the incident THz peak electric field. 
\begin{figure}[htb]
    \centering
    \includegraphics[width=\linewidth]{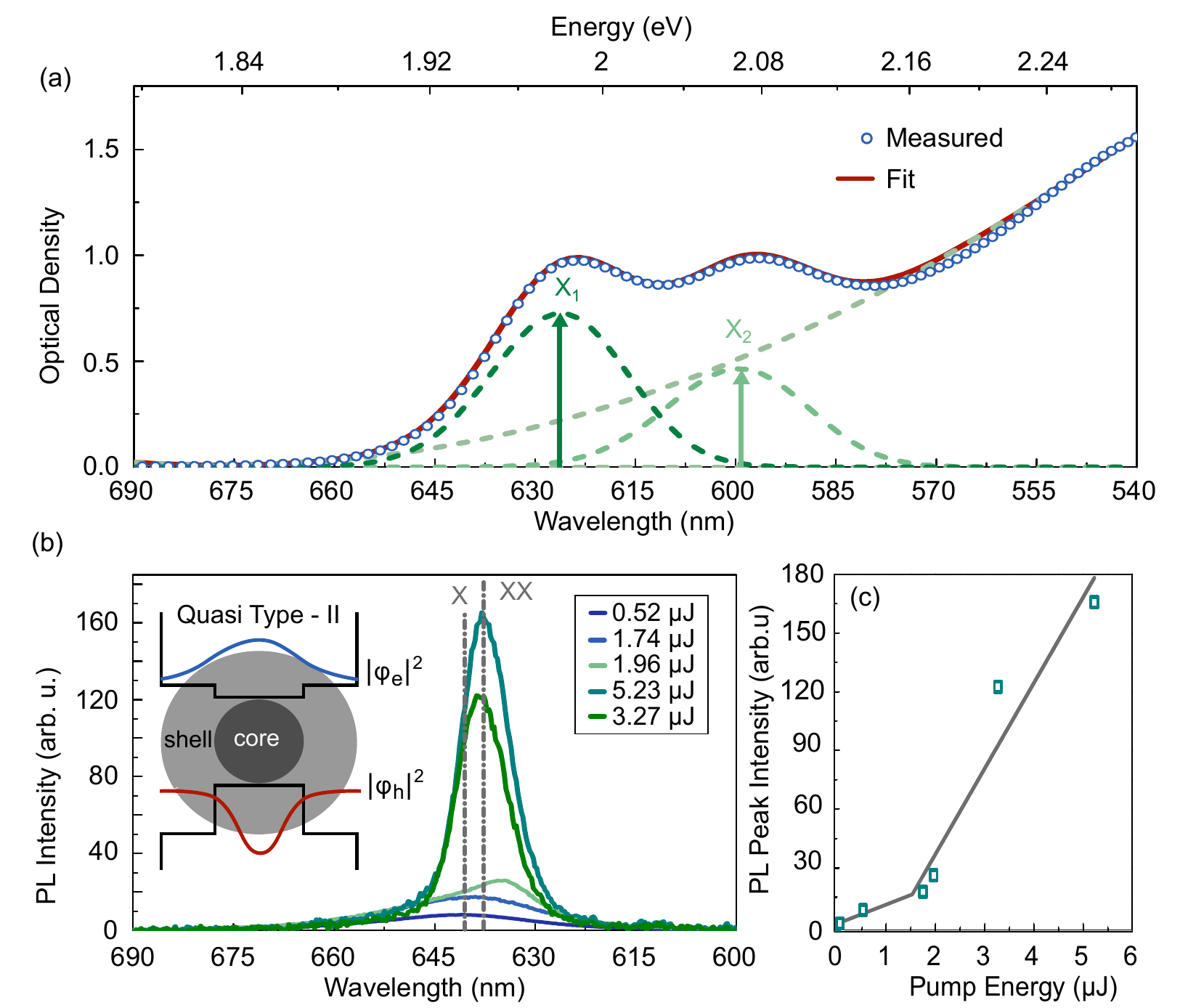}
    \caption[Optical density of the QD film, photoluminescence (PL) spectrum and peak intensity of the PL measurement with respect to the applied pump pulse energy]{(a) Measured optical density of the QD film (blue dots) and fit function (red line). The green dashed lines present the decomposed fit function including two discrete Gaussian energy transitions ($\mathrm{X_1}$ and $\mathrm{X_2}$). (b) Photoluminescence spectrum when the QD film is excited with \SI{515}{nm} pulses of different energies. The labels X and XX refer to exciton and biexciton emission, respectively. The difference in photon energy between X and XX allows to infer the energy band alignment, indicating a quasi type II band structure. (c) Peak intensity of the PL measurement with respect to the applied pump pulse energy. A typical kink confirms the presence ASE.}
\label{fig:OMC}
\end{figure}

The optical density, obtained from the measured transmission spectrum is shown in Fig.\ref{fig:OMC}(a) (blue dots), together with a fit (red line) consisting of two Gaussian components and a polynomial cubic background term describing the absorption continuum (green dashed lines).
The two lowest-energy transitions at the red side of the absorption spectrum are denoted as $\mathrm{X_1}$ and $\mathrm{X_2}$, as described in Fig.\ref{fig:Quantization}, wherein $\mathrm{X_1}$ is centred at \SI{626}{nm} and associated with the band edge.

In order to attain information on the energy band alignment of the heterostructure QDs, PL spectra are recorded for different pump pulse energies when the sample film is excited with \SI{250}{fs} pulses centred at \SI{515}{nm}.
At low excitation energies, the exciton emission is centred at \SI{640}{nm}, indicated by the dashed line labelled X in Fig.\ref{fig:OMC}(b).
At higher pump energies, a second peak appears at slightly higher photon energies,  originating from \ac{ase} arising from biexciton emission (XX) \cite{ASE1_Klimov:2000, ASE2_GiantRod:2018, CihanBandAlignment2:2013, Gollner:2015}. 
The onset of ASE is further confirmed by the superlinear dependence of the PL intensity on the pump energy \cite{ASE_Dong:2013} as shown in Fig.\ref{fig:OMC}(c).
The spectral shift of exciton and biexciton emission reveals the biexciton binding energy which further indicates the type of exciton--exciton interaction and hence unveils the energy band alignment.
The biexciton emission in Fig.\ref{fig:OMC}(b) at \SI{637}{nm} displays a small blue shift of \SI{-8.5}{meV} with respect to the exciton emission, illustrating  weak exciton--exciton interaction and thus a quasi type-II band structure \cite{CihanBandAlignment2:2013}.
 
\section{Experimental Setup}
\label{sec:setupQDs}
\begin{figure}[htb]
    \centering
    \includegraphics[width=\linewidth]{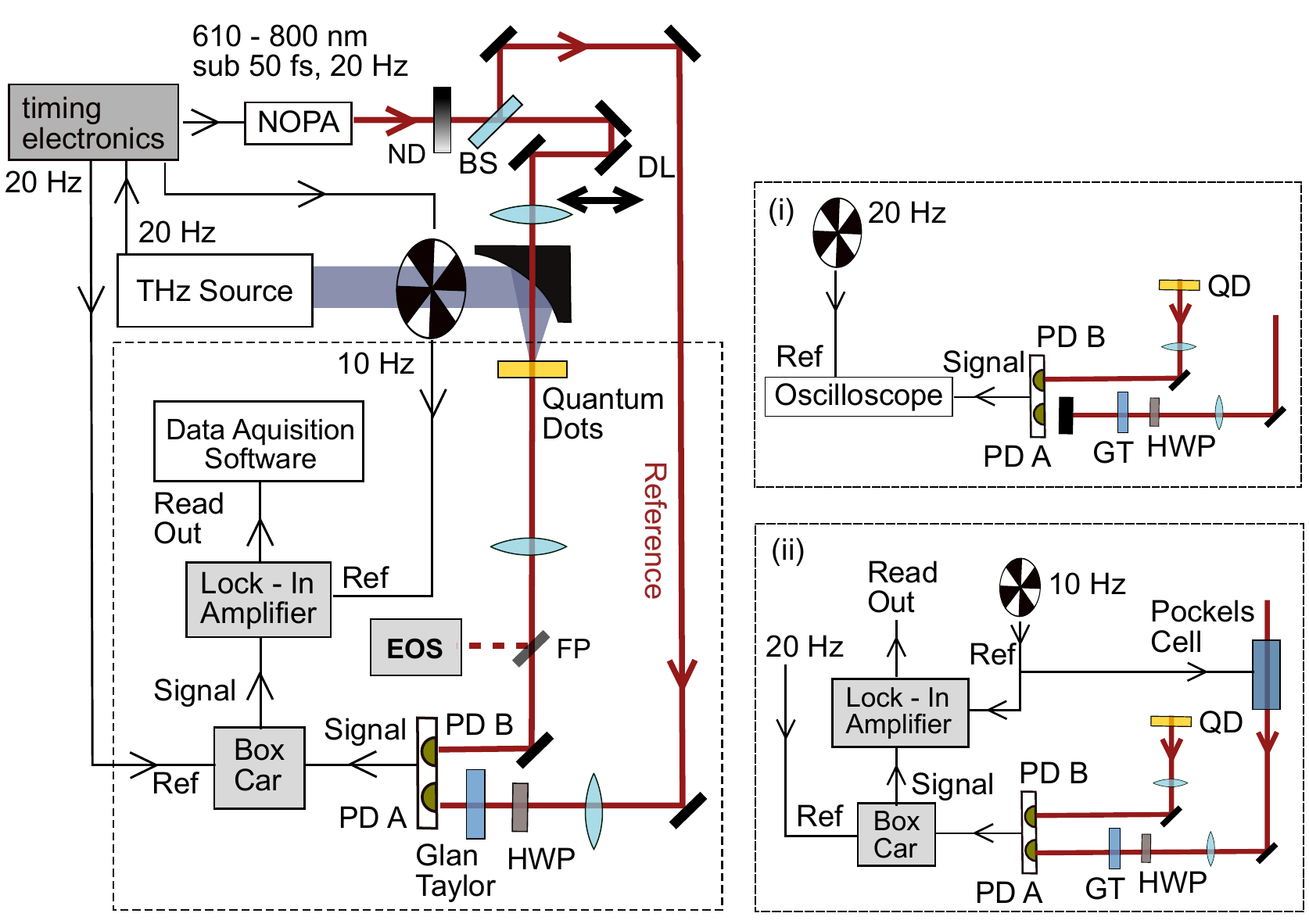}
    \caption[Experimental setup to measure a change in optical properties of colloidal CdSe/CdS QDs induced by a strong field THz transient]{Experimental setup to measure a change in optical properties of colloidal CdSe/CdS QDs induced by a strong field THz transient (left). Panels on the right sketch two methods to assign the acquired signal from the Lock-In amplifier to a change in the sample transmission. BS - beam splitter, DL - delay stage, GT - Glan Taylor prism, ND - neutral density filter, HWP - half waveplate, PD - photo diode, FP - flip mirror, EOS - electro-optic sampling configuration.}
\label{fig:QD_Setup}
\end{figure}

The experimental setup for an observation of THz induced changes in optical absorption of a thin QD film is based on a balanced detection scheme, as shown in Fig.\ref{fig:QD_Setup}.
The THz radiation is generated by optical rectification in DAST of a mid-IR pulse centred at \SI{3.9}{\mu m}, as described in section \ref{sec:ResultsOR} and ref \citeA{GollnerAPL:2021}.
The probe pulse is generated with a white light seeded \ac{nopa}, as outlined in section \ref{sec:NOPA}.
The NOPA supports a tuneabel spectral range of 610 nm - 800 nm with pulse durations of less than \SI{50}{fs}.
The intensity of the visible output pulse is controlled with a variable \ac{nd} after which it is divided into a probe and reference arm.
The probe pulse is focused on the QD sample and spatially overlapped with the THz pulse using a parabolic mirror with a hole in the center and afterwards guided to a photo diode (PD B).
The time delay between THz and probe pulse is controlled with a high precision \ac{dl}. 
The reference beam is steered to another photo diode (PD A), in order to account for intensity fluctuations of the probe pulse and to reduce noise.
In the absence of a THz transient, the signal on the detector is balanced with a variable attenuator made of a \ac{gt} and a \ac{hwp}, installed before the detector PD A.
The THz source is chopped at \SI{10}{Hz} with an optical chopper which is synchronized with the probe pulse and triggers the lock-in amplifier, which registers a difference between the signals from PD A and PD B. 
A boxcar integrator is installed in-between the balanced detector and the Lock-In \cite{BoxCar:2020} to further reduce noise and to amplify the detected signal.
To monitor the THz transient, the setup can be easily re-arranged with a \ac{fp} in order to switch to an \ac{eos} configuration.
The entire setup is purged with nitrogen, resulting in $\sim$ 3\%  relative humidity, to reduce THz absorption by water molecules in ambient air.

In order to assign the acquired signal from the Lock-In amplifier to a certain change in the sample transmission, the following procedures are applied: 
(i) When the probe pulse temporally overlaps with a strong THz field, the change in transmission can be large enough to measure it with an oscilloscope. 
Thus, the reference arm at PD A is simply blocked and the pk-pk value on the oscilloscope is averaged when the THz pulse is blocked and then unblocked.
For this set of measurement, the trigger signal for the chopper and oscilloscope is set to \SI{20}{Hz}.
(ii) For the second procedure, we install a \ac{pc} in the reference arm before the half wave-plate and Glan-Taylor prism  with a known change in transmission between the PC-ON and PC-OFF signal of 24\%. 
The PC and Lock-In amplifier are both triggered at \SI{10}{Hz} and both arms (reference to photo diode PD A and probe to PD B) are open while the THz signal is blocked. 
The measured difference signal on the Lock-In can then be attributed to 24\% change in transmission and the peak amplitude of the THz induced change in transmission can be scaled accordingly.
The two methods exhibit a discrepancy and thus measurement error of $\sim$ 5\%.

\section{Results and Discussion}
\label{sec:ResultsQds}
\subsection{Modulation of the Probe Signal controlled by the THz Waveform}
\begin{figure}[h]
    \centering
    \includegraphics[width=\linewidth]{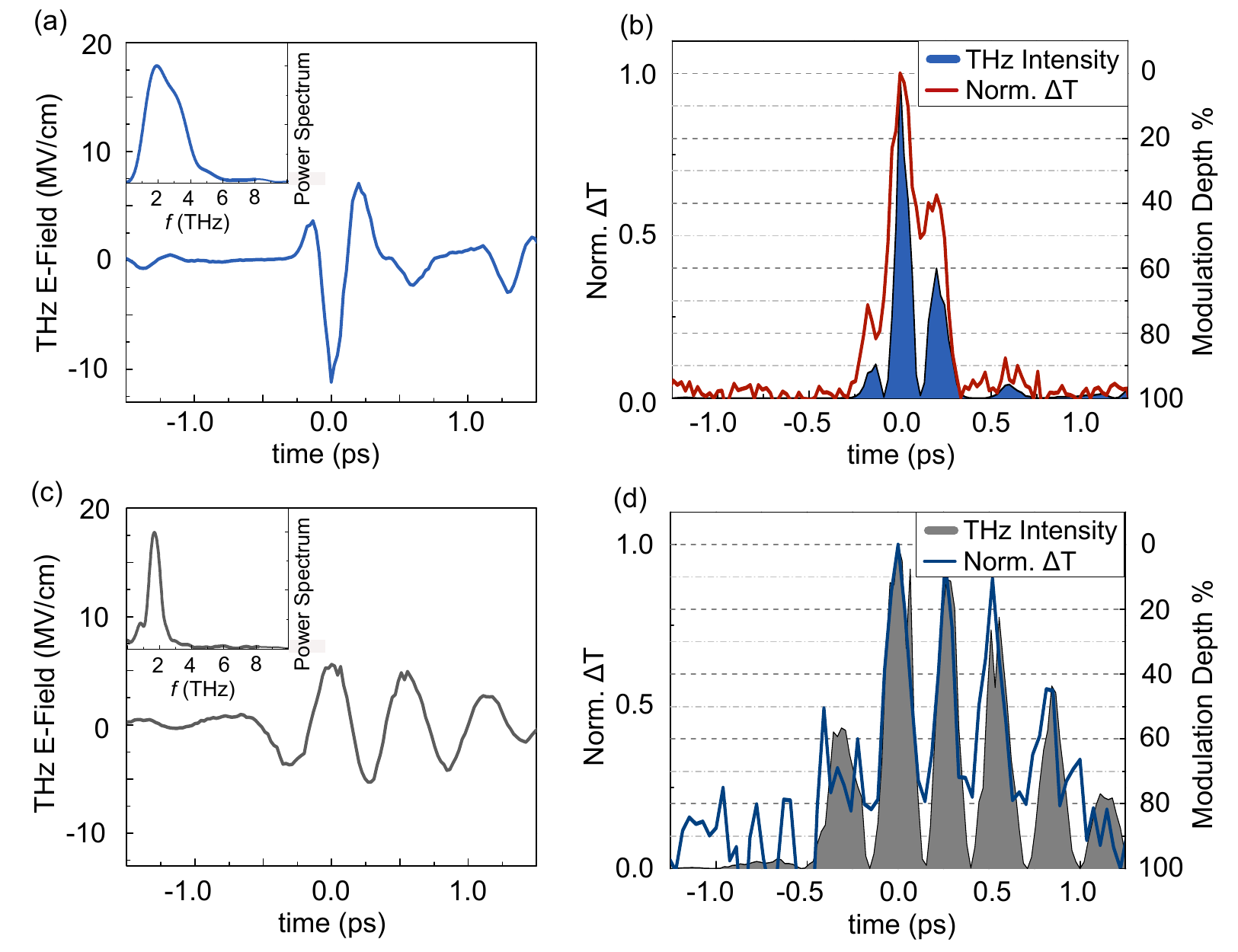}
    \caption[THz transient measured with electro-optic sampling and corresponding normalized change in transmission of the QD sample]{(a) THz transient measured with EOS and corresponding spectrum (inset) obtained by Fourier transformation of the measured time trace. (b) Normalized change in transmission of the QD sample at 622 nm (red line), and THz intensity (blue area) attained from the square of the THz waveform shown in (a). The right y-axis indicates the modulation depth. (c) THz waveform and spectrum (inset) when a long pass filter with a cut-off frequency of \SI{2}{THz} is used. (d) Corresponding THz intensity (grey area) and QD signal (blue line).}
\label{fig:WFshape}
\end{figure}

We first investigate a THz induced change in transmission of an optical probe pulse centred at \SI{622}{nm}. 
Because the photon energy is larger than the band gap of the QDs, an applied electric field will result in reduced absorption and hence negative change in optical density.
A THz field transient, measured with EOS is shown in Fig.\ref{fig:WFshape}(a). 
A replica of the main pulse at around $\pm$\SI{1.2}{ps} originates from surface reflections of the THz and probe pulse at the \SI{50}{\mu m} thick GaP crystal, used for \ac{eos}, as described in section \ref{sec:ProbePulse}.
The corresponding THz spectrum, obtained by  Fourier transformation of the main pulse, is shown in the inset. 
With the measured THz energy and beam radius of \SI{110}{\mu m} at the focal position of the parabolic mirror with a focal length of \SI{5}{cm}, the THz field strength is evaluated to exceed \SI{10}{MV/cm}.
Note that, the electric field felt by the \acp{qd} is substantially smaller than the applied electric field due to reflection losses and dielectric screening, as mentioned above.
If not stated otherwise, the declared field strengths in the following section refer to the incident THz electric field at the focal position.
The blue area in Fig.\ref{fig:WFshape}(b) represents the square of the THz field amplitude. 
The red line is the THz induced change in transmission of the QDs sample, which evidently follows the THz electric field oscillation, demonstrating the feasibility of ultrafast spectral manipulation on a sub-ps time scale  and consequently the possibility  of an all-optical EA modulator with data rates in the range of Tbit/s.
When the THz field changes its polarity and the THz intensity becomes zero, the QD signal does not follow entirely.
The maximum contrast between the ON--OFF signal is around 50\%, as indicated by the right y-axis.
In analogy to optical transmitters used for digital communication, wherein the extinction ratio is defined as the ratio between the energy (power) used to transmit a logic level ‘1’ and energy used to transmit a logic level ‘0’, a modulation contrast of 50\% corresponds to an extinction ratio of \SI{3}{dB}.
The small contrast is fundamentally limited by the duration and spectral width of the probe pulse. 
In principle, the modulation contrast could be improved by utilizing  shorter probe pulses. But, shorter pulses lead to a broader spectrum containing spectral components above and below the QD band edge, hereby probing \ac{ea} modulations of opposite sign and, hence, leading to a low modulation contrast.
In turn,  a narrow probe spectrum results in a longer pulse duration and therefore smears the time resolution and contrast.
Thus, the modulation speed, determined by the period of the THz field oscillation, can only be increased at the expense of the modulation contrast, and vice versa.
Because the THz spectral width and pulse duration are inversely proportional, one possibility to shape the THz field and elongate the pulse period is to narrow the THz spectrum by inserting a long pass filter into the THz path in order to suppress higher frequency components.
Here, we use a wire grid filter (Swiss Terahertz LLC) with a cut-off frequency at \SI{2}{THz}.
The resulting THz transient and intensity spectrum are shown in Fig.\ref{fig:WFshape}(c).
Due to the significant loss of spectral components and hence prolonged pulse duration, the maximum THz field drops down to $\sim$\SI{5.5}{MV/cm}.
The corresponding THz induced change in transmission of the QD sample and normalized THz intensity are shown in Fig.\ref{fig:WFshape}(d).
Although the signal-to-noise ratio of the measured QD signal slightly decreases, the contrast between the THz periods can be substantially increased, leading to an extinction ratio of \SI{6.8}{dB}, which is comparable with quantum well \ac{ea} modulators operating in the GHz range.

\subsection{THz Field Dependence}

\begin{figure}[htb]
    \centering
    \includegraphics[width=\linewidth]{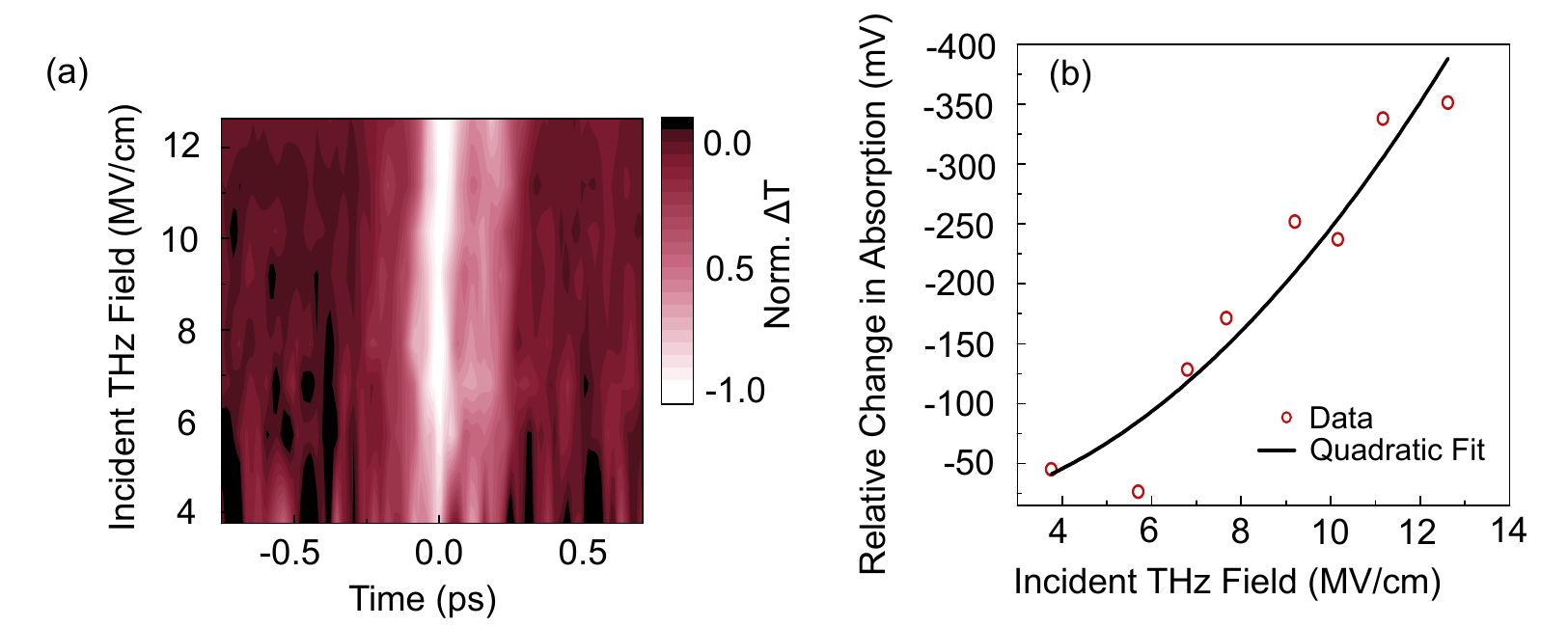}
    \caption[Contour plot of normalized changes in transmission with respect to the THz field strength and time delay between THz and probe pulse]{(a) Contour plot of normalized changes in transmission with respect to the THz field strength and time delay between THz and probe pulse. (b) Quadratic dependence of the QD signal with respect to the incident THz field.}
\label{fig:FieldDep}
\end{figure}

Pein and co-workers\cite{Pein:2017} reported on THz driven QD luminescence via charge separation between neighbouring QDs.
The THz induced voltage drop between QDs is thereby large enough such that electrons from the valence band can tunnel to the conduction band of the neighbouring QD and recombine via radiative decay. 
Such above threshold THz-induced luminescence indicate the existence of charged states which persist for longer than \SI{1}{ps} \cite{Pein:2019}. 
As a consequence, the absorption modulation could be temporally smeared by such longer-lived \ac{qd} states. 
To assure that the poor modulation contrast in the case of a high frequency THz field originates from the short THz period, rather than the strong electric field (which can potentially cause charged states), we repeat the experiment for different THz field strengths in the case of the THz transient shown in Fig.\ref{fig:WFshape}(a).
A color map of the normalized absorption modulations with respect to the incident electric field is shown in Fig.\ref{fig:FieldDep}(a).
Because the shape of the QD response does not change significantly with the applied field strength, it is evident that additional field dependent dynamics are absent and further confirms that the absorption spectrum modulation originates solely from  THz induced \ac{ea} effects.
Moreover, Fig.\ref{fig:FieldDep}(b) depicts the change in absorption, given by the Lock In signal in mV, when the maximum of the THz field temporally overlaps with the intensity envelope of the probe pule.
The quadratic dependence on the incident THz field is observed, as it is expected in the case of the QCSE, and indicates no noticeable influence of field ionization mechanisms\cite{CdS_StarkColvin:1994}.

When we assign the acquired signal from the Lock-In amplifier to a specific change in the sample transmission, as described in section \ref{sec:setupQDs}, we find 8.6\% change in transmission when the THz and probe pulse temporally overlap with an applied THz field of \SI{10.2}{MV/cm}.
Another QD response measured with a THz field strength of \SI{13.3}{MV/cm} reveals a change in transmission of astonishing 15.8\%, which follows the quadratic prediction fitted in Fig.\ref{fig:FieldDep}(b). 
To the best of our knowledge, this is the highest value ever reported for solution processed \ac{ea} materials at room temperature.

\subsection{Stark Spectroscopy}
\begin{figure}[htb]
    \centering
    \includegraphics[width=\linewidth]{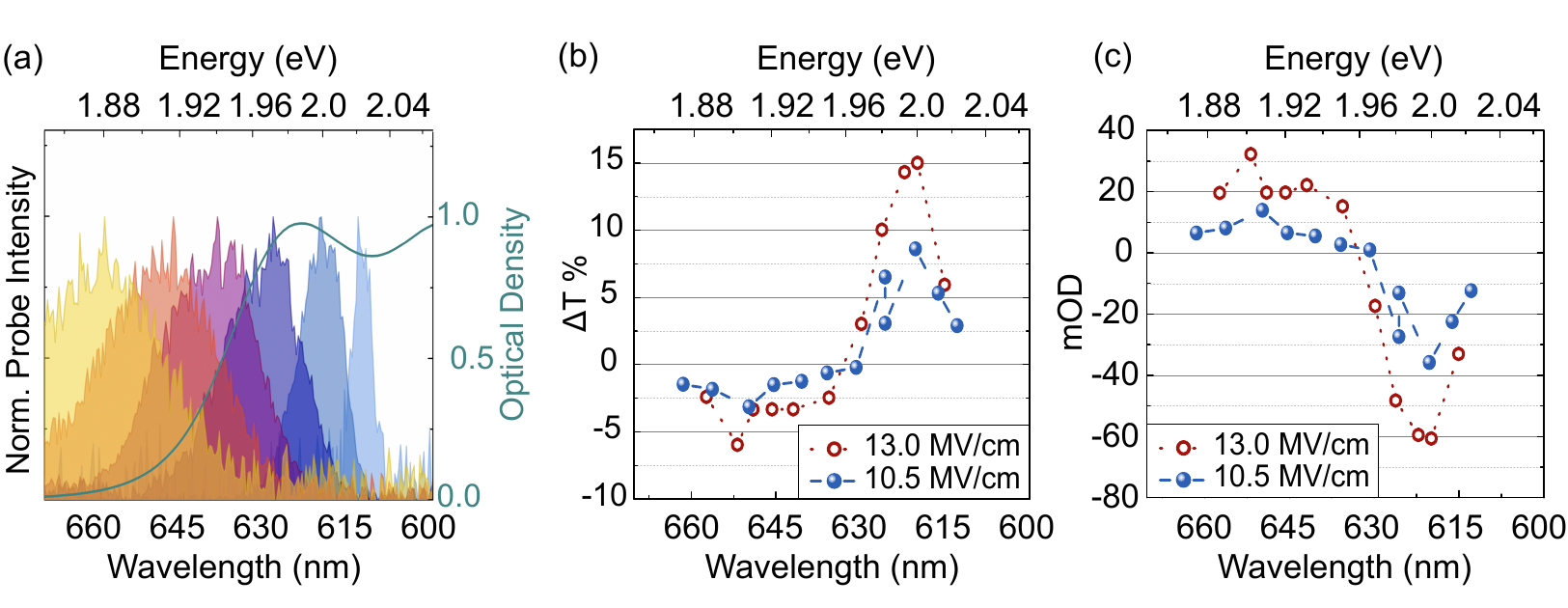}
    \caption[Optical density, normalized probe spectra and resulting change in optical density for an incident THz electric field]{(a) Optical density (green line) and normalized probe spectra (filled areas) with a central wavelength tuned across the band edge. (b) Resulting change in transmission for an incident electric field strength of \SI{13}{MV/cm} (red) and \SI{10.5}{MV/cm} (blue). (c) Estimated change in optical density.}
\label{fig:WavelengthDepExp}
\end{figure}

Stark spectroscopy, also known as \ac{ea} spectroscopy, studies the effect of an applied electric field on the absorption spectrum.
The formalism of Stark spectroscopy was originally applied to an ensemble of molecules, where field-induced alignment/orientation is possible. 
However, to date it is also successfully utilized on confined structures, such as layered hybrid perovskites\cite{Pero_Walters:2018} or CdSe \acp{nc} \cite{CdS_StarkColvin:1994}, where orientational effects are unlikely since the films are rigid. 
The principal requirement for the validity of the Stark formalism is that the electronic states of the system are discrete\cite{CdS_StarkColvin:1994}, as they are in QD structures.
Thus, in order to understand the origin of the THz induced modulation of the absorption spectrum, we qualitatively discuss the QCSE with respect to the probe spectrum when the central frequency is tuned across the band edge of the QD film, as illustrated in Fig.\ref{fig:WavelengthDepExp}(a).
While tuning the central wavelength of the probe pulse by adjusting the non-linear crystal and delay stage of the NOPA, the probe intensity is monitored with a photo diode and corrected with a neutral density filter if necessary, in order to assure the same intensity for each central wavelength. 
The QD response to the THz field is then sampled for each central wavelength with respect to the time delay between THz-pump and visible-probe pulse.
We then integrate the time trace in an interval of [-0.4;0.4] ps and plot it with respect to the central wavelength. 
Fig.\ref{fig:WavelengthDepExp}(b) and (c) depict two sets of measurements for an incident THz field of approximately \SI{10.5}{MV/cm} and \SI{13}{MV/cm}. 
The change in optical density is calculated as followed:
first, the integrated QD response is re-normalized to the maximum change in transmission at around \SI{622}{nm}, which is $\Delta \mathrm{T}\left(\mathrm{10.5\ MV/cm}\right)\simeq 8.6\%$ and $\Delta \mathrm{T}\left(\mathrm{13\ MV/cm}\right)\simeq15\%$ (see Fig.\ref{fig:WavelengthDepExp}(b)), as discussed in the previous section.
Hence, the measured QD response can be assigned to a certain change in transmittance for each central wavelength of the probe pulse, denoted as $\mathrm{\Delta T}(\lambda)$. 
Second, the change in optical density is defined as
\begin{equation}
\begin{split}
\Delta OD(\lambda) & = OD_1(\lambda)-OD_0(\lambda)=\log\left(\frac{\mathrm{T}_0(\lambda)}{\mathrm{T}_1(\lambda)}\right)\\
	&=\log\left(\frac{\mathrm{T}_0(\lambda)}{\mathrm{T}_0(\lambda)+\Delta \mathrm{T}(\lambda)\cdot\mathrm{T}_0(\lambda)}\right)=\log\left(\frac{1}{1+\Delta \mathrm{T}(\lambda)}\right),
\label{equ.DeltaOD}
\end{split}
\end{equation}
with $\mathrm{T}_0(\lambda)$ as the transmittance without an applied electric field and $\mathrm{T}_1(\lambda)$ when the THz field is present.
Consequently, the change in optical density can be calculated with the data provided in Fig.\ref{fig:WavelengthDepExp}(b) and is illustrated in Fig.\ref{fig:WavelengthDepExp}(c).

In order to identify the origin of the THz induced modulation of the absorption spectrum,  simulations on the underlying modifications of the electronic states in the case of various energy band structures are performed.
In analogy to Stark spectroscopy, the optical density can be described with respect to the optical frequency $\nu$ as
\begin{equation}
    O D(\nu)/\nu \sim \left| \mathbf{\mu}_{\mathrm{vc}} \right|^2 s(\nu) \ ,
    \label{equ:absorption}
\end{equation}
with the electric transition dipole moment $\mathbf{\mu}_{\mathrm{vc}}\!=\!e\bra{ \Psi_\mathrm{v}}\hat{\mathbf{\mathrm{r}}}\ket{\Psi_\mathrm{c}}$ from valence band state $|\Psi_\mathrm{v}\rangle$ to conduction band state $|\Psi_\mathrm{c}\rangle$, fundamental electric charge \(e\),  position state operator $\hat{\mathbf{\mathrm{r}}}$ and normalised line shape function $s(\nu)$ \cite{liptay1960bestimmung, liptay1976optical, Liptay:1969}.
The transition energy and dipole moment is calculated by applying a simple one-dimensional approach, as described in section \ref{sec:QCSE}.
In order to estimate the broadening of the line shape function due to size dispersion, the total quantum dot diameter of 8.9 \(\pm\)0.8 nm is used.
The ground-state energy of a quantum dot \cite{kayanuma1986wannier} without interaction can be approximated by 
\begin{equation}
    E_\mathrm{eff} \approx \frac{\pi^2\hbar^2}{2\mu L^2}
\end{equation}
with the reduced electron--hole mass \(\mu^{-1}=1/m_\ee^{*}+1/m_\hh^{*}\) and the effective diameter \(L\). The variation of the energy states thus reads
\begin{equation}
    \sigma \equiv \Delta E_\mathrm{eff} = 2\,E_\mathrm{eff}\frac{\Delta L}{L}  \approx 22.5\; \mathrm{meV}
\end{equation}
and is used in an approximated Gaussian  line-shape function 
\begin{equation}
    s(\nu) \sim  \mathrm{e}^{-\left(\frac{E-h\,\nu}{\sigma}\right)^2}, 
\end{equation}
with $E$ being the transition energy, calculated with Eq.\eqref{equ:TransEnergy}.
The good match with the experimentally obtained optical density (see Fig.\ref{fig:OMC}(a)) suggests that the FWHM of the line-shape function is dominated by the dispersion of the nano-particle size. 
Nonetheless, the calculations are very sensitive to the quantum dot parameters as well as the influence of the ligands and dielectric environment in general \cite{nazzal2009comparative, casas2017effect, dielEffectZeiri:2019}.
Thus, the band-gap of the calculation  differs by a few percent from the experimental data. 
For a better comparison between experiments and simulations, a small  energy shift is introduced. 
\begin{figure}[htb]
    \centering
    \includegraphics[width=\linewidth]{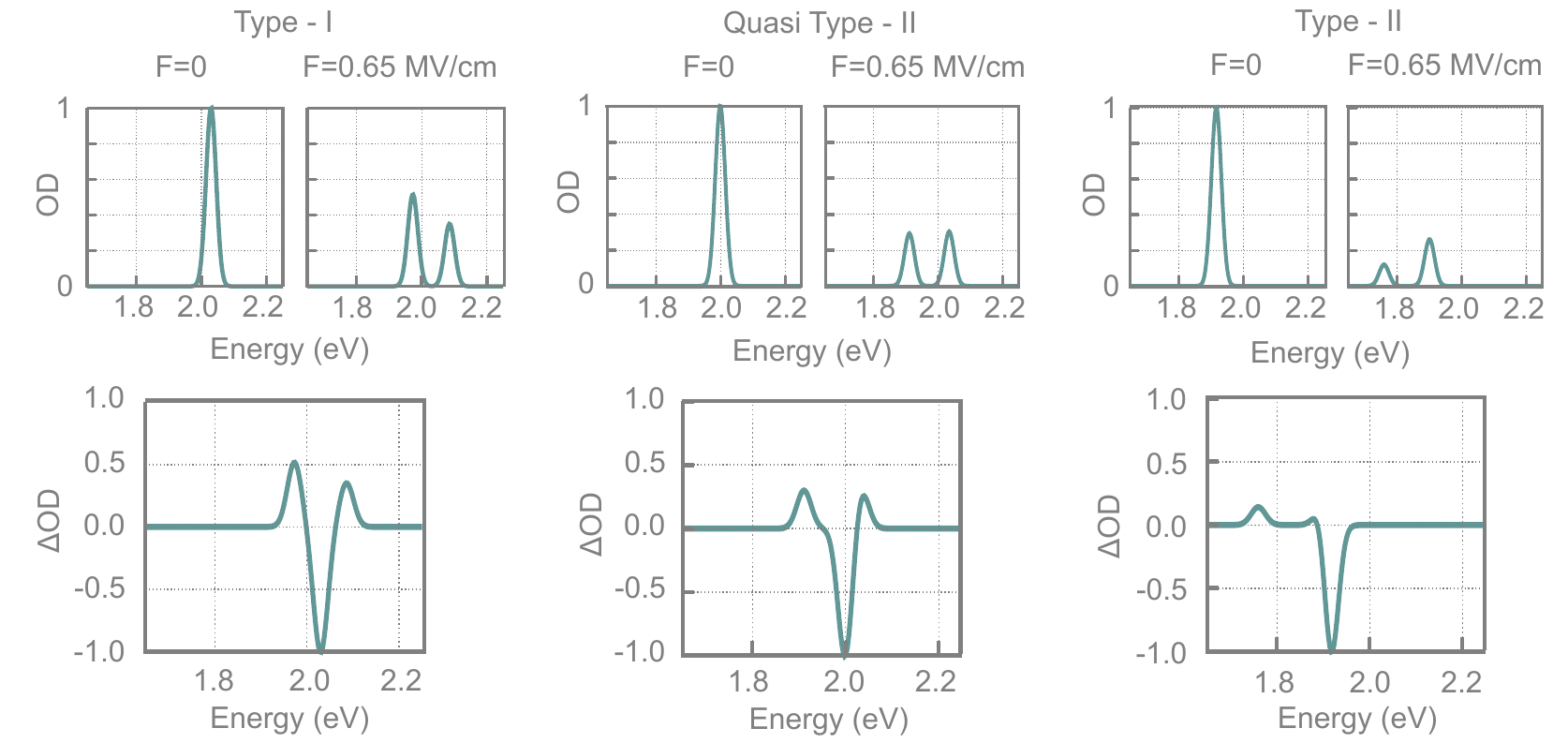}
    \caption[Simulated normalized optical densities]{Simulated normalized optical density (OD) spectra of the first transition  without an electric field (left panels) for (a) type-I, (b) quasi type-II and (c) type-II energy alignments. An applied electric field of  \(F=0.65\ \)MV/cm (right panel) results in a shift of the first transition to smaller energies with reduced amplitude and an appearance of an additional absorption band on the blue side of the spectrum. Bottom panels: normalized changes in optical density \(\Delta\)OD defined as the difference between optical densities with and without an external electric field, respectively (\(\mathrm{OD}^F-\mathrm{OD}^0\)).  }
\label{fig:ProbDens2}
\end{figure}
 
\begin{figure}[hbt]
    \centering
    \includegraphics[width=\linewidth]{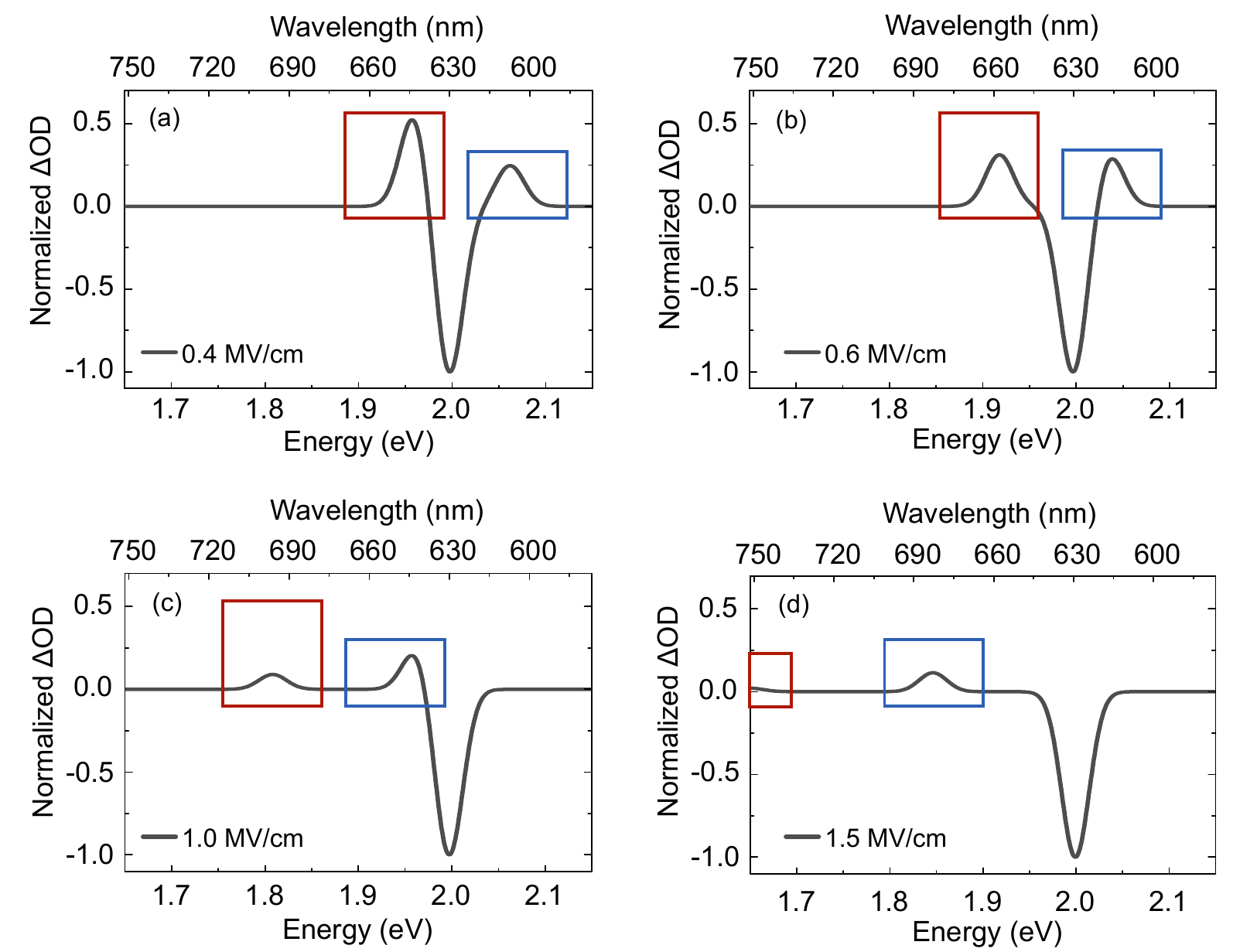}
    \caption[Evolution of the change in optical density with respect to the applied electric field strength]{Evolution of the change in optical density with respect to the applied electric field strength (see legends) for a quasi type-II energy band alignment. Red shifted first transition energy $1\mathrm{S}_{3/2}(\mathrm{h}) \rightarrow 1\mathrm{S}(\mathrm{e})$ are tagged by red rectangles, blue boxes indicate the higher lying transition $1\mathrm{P}_{3/2}(\mathrm{h}) \rightarrow 1\mathrm{S}(\mathrm{e})$. Both transitions move farther to the red side of the spectrum with increasing electric field
strength.}
\label{fig:OD_evolution}
\end{figure}

In the presence of an external electric field $F$ the absorption spectrum can be expressed as \cite{liptay1960bestimmung, liptay1976optical, Liptay:1969}
\begin{equation}
    O D^F(\nu)/\nu \sim \left| \mathbf{\mu}^F_{\mathrm{vc}} \right|^2 s^F(\nu) \ .
    \label{equ:A_with_E_gneral}
\end{equation}
In the case when the line shape function does not change significantly under the influence of the external field, it can be approximated by  $s^F\!(\nu) \!\approx\! s(\nu - \nu_\mathrm{st})$, with $h\nu_\mathrm{st}$ being the Stark energy shift~\cite{liptay1960bestimmung}, accounting for the energy shift of the electronic levels when an external electric field is applied.
The change in absorption amplitude is mainly caused by the altered transition dipole moment $\mu_{\mathrm{vc}}^F$ which is dominated by the  overlap integral of electron and hole wavefunctions 
$\braket{\varphi_\mathrm{e}|\varphi_\mathrm{h}}$ \cite{kane1958influence, peter2010fundamentals}, as evident by Eq.\eqref{equ:OsciStrength} and \eqref{equ:DipoleM}.
When Liptay first theoretically formulated Stark spectroscopy \cite{liptay1960bestimmung}, he used perturbation theory to account for the change in transition dipole moment as well as for the energy shift.
However, when the applied electric fields are in the MV/cm range, the assumption of small perturbations is not valid any more.
For the case of strong electric fields, Pein \textit{et al.} \cite{Pein:2019} used an elaborate full-atomic, semi-empirical tight binding model in order to describe the electron and hole states.
Despite the simplified one dimensional concept used in this work, the calculations reproduce the results of the Stark energy shifts reported by Pein and co-workers surprisingly well, emulate Liptay's model for small fields and provide a qualitative insight into the physical behaviour of the QDs.
In the simulation, focus is laid on the first transition labelled as $\mathrm{X_1}$ in Fig.\ref{fig:OMC}(a) and Fig. \ref{fig:Quantization}.
The higher-energy transitions are not included in the model based on their negligible contribution in the spectral range of the probe pulses, as well as the inverse dependence of the transition dipole moment on the energy gap.
Figure \ref{fig:ProbDens2} shows the calculated optical density (top panels) as well as the change in optical density for three types of energy band alignments of CdSe/CdS QDs when an electric field of \SI{0.65}{MV/cm} is applied (bottom panels).
In the case of a quasi type-II structure,  the change in optical density reveals two local maxima at \SI{1.9}{eV} and \SI{2.03}{eV}. 
The first peak represents the Stark shifted optical density  of the first transition $1\mathrm{S}_{3/2}(\mathrm{h}) \rightarrow 1\mathrm{S}(\mathrm{e})$, exhibiting a decreased transition energy and hence a red-shift of the absorption edge as compared to the unperturbed case.
The second maxima can be attributed to the transition $1\mathrm{P}_{3/2}(\mathrm{h}) \rightarrow 1\mathrm{S}(\mathrm{e})$, which is originally forbidden when no external field is applied due to the symmetry of the wavefunctions.
The applied electric field breaks the symmetry and increases the absorption for longer wavelengths.
The negative optical density changes around \SI{2}{eV} are ascribed to a reduction of the normalized overlap integral at the band gap with respect to the overlap integral of the unperturbed QD. 
An evolution of the change in optical density with respect to the electric field strength is illustrated in Fig.\ref{fig:OD_evolution} for a quasi type-II energy band alignment.
Red and blue boxes indicate the red shifted first excitation energy $1\mathrm{S}_{3/2}(\mathrm{h}) \rightarrow 1\mathrm{S}(\mathrm{e})$ and higher lying transition energy $1\mathrm{P}_{3/2}(\mathrm{h}) \rightarrow 1\mathrm{S}(\mathrm{e})$, respectively.
In the case of small electric fields (Fig.\ref{fig:OD_evolution}(a) and (b)), the initially forbidden transition contributes to a positive change in optical density on the blue side of the energy band gap.     
For stronger electric fields (> \SI{0.8}{MV/cm}), both transitions are shifted further to the red side, such that the positive addition to the optical density at the blue side of the spectrum vanishes.

\begin{figure}[thb]
    \centering
    \includegraphics[width=\linewidth]{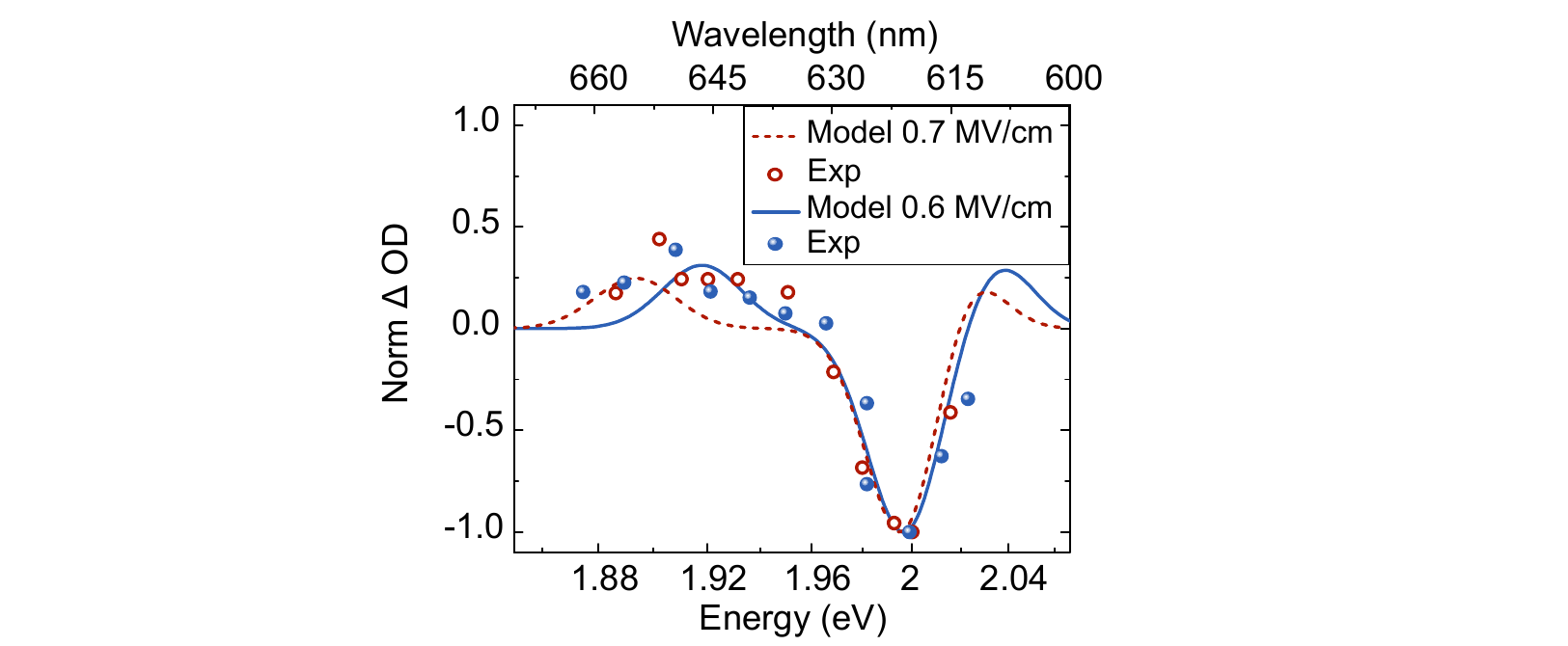}
    \caption[Measured  Stark spectra]{Measured  Stark spectra (dots) for two different field strengths experienced by the excitons, and simulated changes in optical density (lines) for \SI{0.7}{MV/cm} (red) and \SI{0.6}{MV/cm} (blue).}
\label{fig:WavelengthDepSimExp}
\end{figure}
Measured and simulated Stark spectra are depicted in Fig.\ref{fig:WavelengthDepSimExp} for the two different THz field strengths as shown in Fig.\ref{fig:WavelengthDepExp}(c).
In order to compare the results of the measurements and the calculations, the spectra are normalised.
The two sets of experimental data are fitted by varying the strength of the electric field experienced by the QDs.
The best match between experiment and theoretical model are found for an electric field strength between \SI{0.6}{MV/cm} (blue), and \SI{0.7}{MV/cm} (red), experienced by the QDs. 
Because of dielectric screening and reflection losses in combination with the uncertainty in placing the sample at the exact focus of the parabolic mirror, the electric field felt by the exciton of a QD can be more than an order of magnitude smaller than the incident THz electric field, which is in good agreement with the experimental and simulated electric field strengths.
Unfortunately, the positive change in optical density originating from the evolving transition $1\mathrm{P}_{3/2}(\mathrm{h}) \rightarrow 1\mathrm{S}(\mathrm{e})$ at the  blue spectral range, can not be experimentally verified due to the limited tunability of the probe pulse (610 nm - 800 nm).
A discrepancy between the calculated and the measured data at lower photon energies is attributed to the broad spectrum and finite pulse duration of the probe pulse, which drastically reduces the spectral sensitivity of the central wavelength, as well as intensity fluctuations. 
Although a full three-dimensional model would improve the accuracy of the calculated energy states, a sufficiently good overlap between the measured data and simulations validates the simple model and justifies further discussions on the energy band structure.

\begin{figure}[htb]
    \centering
    \includegraphics[width=\linewidth]{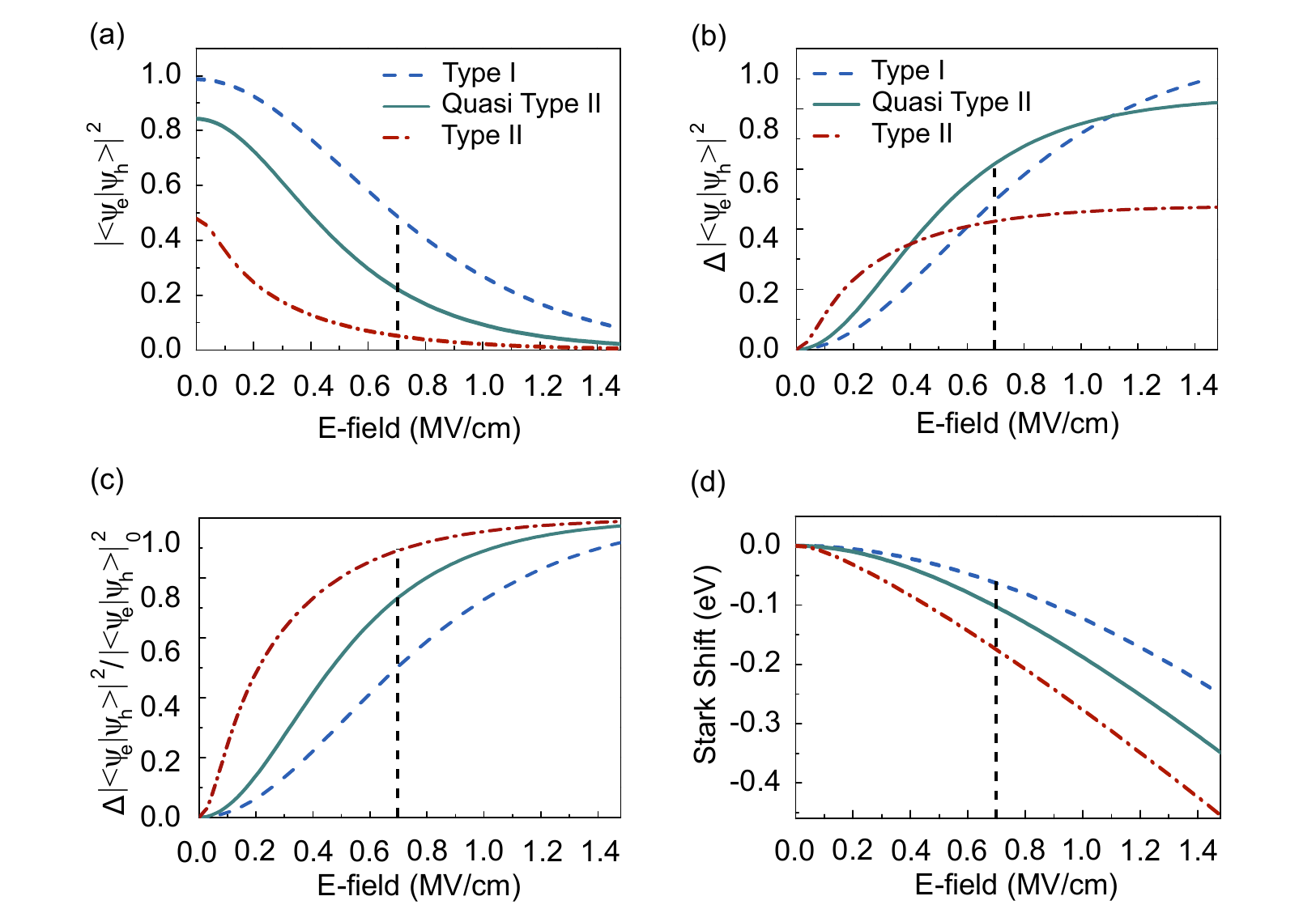}
    \caption[Simulations on the energy bands for heterostructure QDs with respect to an applied electric field]{(a) Calculated square of the overlap integral of electron and hole wavefunctions, (b) its absolute, and (c) relative change with respect to an applied electric field, for three different energy band alignments (see legend). (d) Simulated Stark shift as a function of the electric field strength. Dashed vertical lines indicate a field strength of \SI{0.7}{MV/cm}, which is relevant to the data presented in \ref{fig:WavelengthDepSimExp}.}
\label{fig:QD_Modle_Ext}
\end{figure}
Figure \ref{fig:QD_Modle_Ext} shows the square of the overlap integral, its absolute and relative change, as well as the Stark shift with respect to the applied electric field, simulated for three types of band alignments, revealing the origin of the measured extreme changes in optical absorption.
The vertical dashed lines indicate a field strength of \SI{0.7}{MV/cm}, which is relevant to the data presented in Fig.\ref{fig:WavelengthDepSimExp}.
As it follows from the simulations shown in Fig.\ref{fig:QD_Modle_Ext}(b), the strongest absolute change in the case of a strong electric field around \SI{0.7}{MV/cm} is expected for a quasi type-II band structure, when the relatively loosely confined de-localised electrons travel the furthest possible distance away from the holes, leading to a large spatial separation of both charge carriers. 
Our findings are further supported by Bozyigit \textit{et al.}\cite{Bozy:2013}, who investigated \ac{pl} quenching induced by an applied external field in colloidal CdSe/CdS QDs, with similar core and shell dimensions as in this work.
In contrast to previous works \cite{Quenching:2009,Quenching:2007}, and in line with our own observations, instead of field-induced charging of the QDs, they attribute the observed luminescence quenching to the spatial separation of electron and hole wavefunctions (causing the optical transition matrix element to decrease).
Nonetheless, Fig.\ref{fig:QD_Modle_Ext}(c) reveals that, although the relative changes are similar for type-II and quasi type-II in the case of strong electric fields exceeding \SI{1}{MV/cm}, a type-II energy band alignment reaches significantly larger values for smaller electric fields.
In addition, as shown in Fig.\ref{fig:QD_Modle_Ext}(d), largest changes in the transition energy for small applied electric fields can be also achieved with a type-II QDs.
Such a band structure can be attained with a very thick shell, which allows the electrons to delocalize freely over the entire QD volume, while the holes are confined in the core \cite{CihanBandAlignment2:2013}.
Thus, for an absorption spectrum with sharp and well pronounced features, a type-II structure can be beneficial since small fields can lead to large energy shifts, in addition to a reduced overlap integral, resulting in a substantial change in absorption.
However, although it was predicted by simulations \cite{Matsuura:1986,Hiroshima:1986} that the magnitude of the QCSE shifts of the electron and hole levels increase monotonically with the well width in quantum well structures, an upper limit on the scalability of the QCSE is given by the exciton Bohr radius \cite{Miller:1985}. 
Increasing the width of nano-devices diminishes the quantum confinement and reduces the exciton binding energy, such that the exciton eventually field-ionizes and the bulk electric field response would be expected.

\section{Conclusion}
\label{QDconclusion}

In summary, we report for the first time on a direct all-optical encoding of a free space THz transient onto a visible probe signal in heterostructure QDs without any field enhancing structures.
This simple approach allows to investigate the pure QCSE without any physical artefacts, such as electrode driven charge injection and demonstrates the capability of ultrafast optical interconnects with Tbit/s data rates.
A beneficial energy band alignment of the colloidal CdSe/CdS \acp{qd} under investigation lead to a change in transmission of outstanding 15\%, which is to the best of our knowledge, the highest value ever reported for solution processed materials. 
Moreover, we demonstrate the sensitivity of \ac{ea} modulation to the shape of the THz waveform and thereby achieve a modulation contrast of more than 6 dB, which is in line with state-of-the art quantum-well \ac{ea} modulators.
The absence of field enhancing structures further provides the possibility to employ a simple and intuitive theoretical model which matches the experimental data remarkably well, and allows us to explain the experimental findings. 
Based on the simulation results, we propose a route to further  improve the signal contrast and modulation depth, namely by utilizing QDs with thicker shells in order to exploit a type-II band alignment for next generation ultrafast optical switches.
The fact that CdSe/CdS QDs can be tuned between the type-I and type-II regime by varying the core radius and shell thickness \cite{ASE_Dong:2013, CihanBandAlignment2:2013}, makes it possible to precisely adjust the energy band structure.
The possibility to further tune the absorption spectrum of QDs from the visible to the telecommunication wavelength by an appropriate choice of composition, size and barrier width, along with their solution process-ability makes them excellent candidates for applications in high-speed optical communication systems.

%% file: Chapters/TwoColor.tex
\chapter{THz Generation by mid-IR two-color Plasma Filaments}
\label{ch:filaments}

For THz generation by two color plasma filaments, usually a high power femtosecond laser and its \ac{sh} are focused to produce laser induced gas plasma. The photo-ionized electrons create a current surge, which further radiates a directional EM field. 
It is thereby necessary to break the symmetry between the positive and negative amplitudes of the oscillating optical field in order to produce a non-vanishing net current.
In contrast to THz generation by optical rectification, wherein the generated spectral bandwidth is fundamentally limited by the pump pulse duration, and the produced THz energy is constrained by the damage threshold of the \ac{eoc} as well as competing higher order NL processes, THz radiation by two-color filaments allows one to deposit much higher pump intensities because gas plasma is self healing.
Moreover, because tunneling ionization is a highly NL process, the ionization time scale can be much shorter than the pulse duration of the pump source. Hence, the THz bandwidth can be much broader than the transform limited pump pulse bandwidth.
Thus, despite the small optical-to THz conversion efficiency of 0.01\% \cite{PlasmaKuk:2016} from a fundamental pump wavelength of \SI{0.8}{\mu m}, it is possible to generate strong THz electric fields exceeding \SI{20}{MV/cm}. 
The limit for the THz yield is given by plasma defocusing, plasma scattering and temporal walk-off between fundamental, SH and THz pulse \cite{SOTAplasma:2017}, as well as THz absorption in over-dense plasma channels \cite{Kim:2007, PlasmaWidthKim:2008}, when the frequency of the propagating THz wave equals the plasma frequency. 

Within the last decade, it has been shown that the conversion efficiency can be substantially increased if to use longer wavelength driving pulses.
The first breakthrough was achieved by Clerici and co-workers \cite{LongWLplasma:2013}, demonstrating an increase in efficiency by an order of magnitude to $\sim 0.1\%$ for \SI{1.8}{\mu m} driving pulses.
Due to a lack of  intense mid-\ac{ir} driving sources, the experimental investigation could not be pushed to longer wavelength and further publications are mainly based on  simulations.
For example, a similar one order of magnitude enhancement of the THz yield was theoretically predicted for laser pulses with a central wavelength of \SI{2}{\mu m} \cite{Nguyen:2017, Berge:2013}.
Recently, a theoretical analysis on two-color filamentation of mid-IR pulses centred at \SI{3.9}{\mu m} was performed by Fedorov \textit{et al.} \cite{Fedorov:2018} by taking \ac{nl} propagation processes into account. 
The study reveals extremely high achievable THz conversion efficiencies, which are more than two orders of magnitude higher than for \SI{0.8}{\mu m} driving pulses. 
The results got later numerically confirmed by Nguyen \textit{et al.} \cite{Nguyen:2018}.
The outstanding THz conversion efficiency in the case of mid-IR drivers is attributed to strong photocurrents due to larger ponderomotive forces, longer and wider plasma channels, negligible walk-off between the fundamental and second harmonic, and additional field symmetry breaking by generated high harmonics.

In this work, we provide for the first time the experimental evidence for such efficient THz pulse generation with a fundamental driving pulse centred at \SI{3.9}{\mu m}.
The high power mid-IR pump pulses are provided by a home-built OPCPA system, as described in section \ref{sec:OPCPA}.
We report on ultrashort sub-cycle THz pulses with sub-millijoule energy and THz conversion efficiency of \SI{2.36}{\%}, which exceeds by far any previously reported experimental values for plasma-based THz sources, and results in THz field strength amplitudes of more than \SI{100}{MV/cm}.

In the following chapter, we first discuss the transient photo-current model, which describes the underlying interplay between electric field ionization of molecular gasses and THz generation.
We thereby mainly refer to the pioneering achievements of Kim and co-workers, found in refs \cite{Kim:2007,Kim:2009, Kim:2012, PlasmaWidth:2013}.  
Subsequently, we focus on theoretical predictions for long-wavelength drivers.
And finally, we introduce the experimental setup and present our findings.

\section{Transient Photo-Current Model}
\label{sec:TPCM}
The main principle of THz pulse generation via two-color filaments relies on focusing a high power femtosecond laser and its second harmonic to produce laser-induced gas plasma.
Such a THz generation scheme was first observed by Cook and Hochstrasser in  2000 \cite{Cook:2000} and interpreted as \ac{fwm} process $(\omega+\omega-2\omega)$  in a gas medium. 
But soon it was realized that the third order nonlinearity $\chi^{(3)}$ of a gas medium is too small to explain the measured  THz field strength \cite{Kress:2004, Kim:2007}.
Moreover, the phase dependence of the THz yield does not correlate with the FWM-model, which predicts the highest optical to THz conversion efficiency for a relative phase of zero between the fundamental and second harmonic pulse.
Instead, it was experimentally observed that the highest THz yield occurs for a relative phase of $\pi/2$. 

\begin{figure}[htb]
    \centering
    \includegraphics[width=\linewidth]{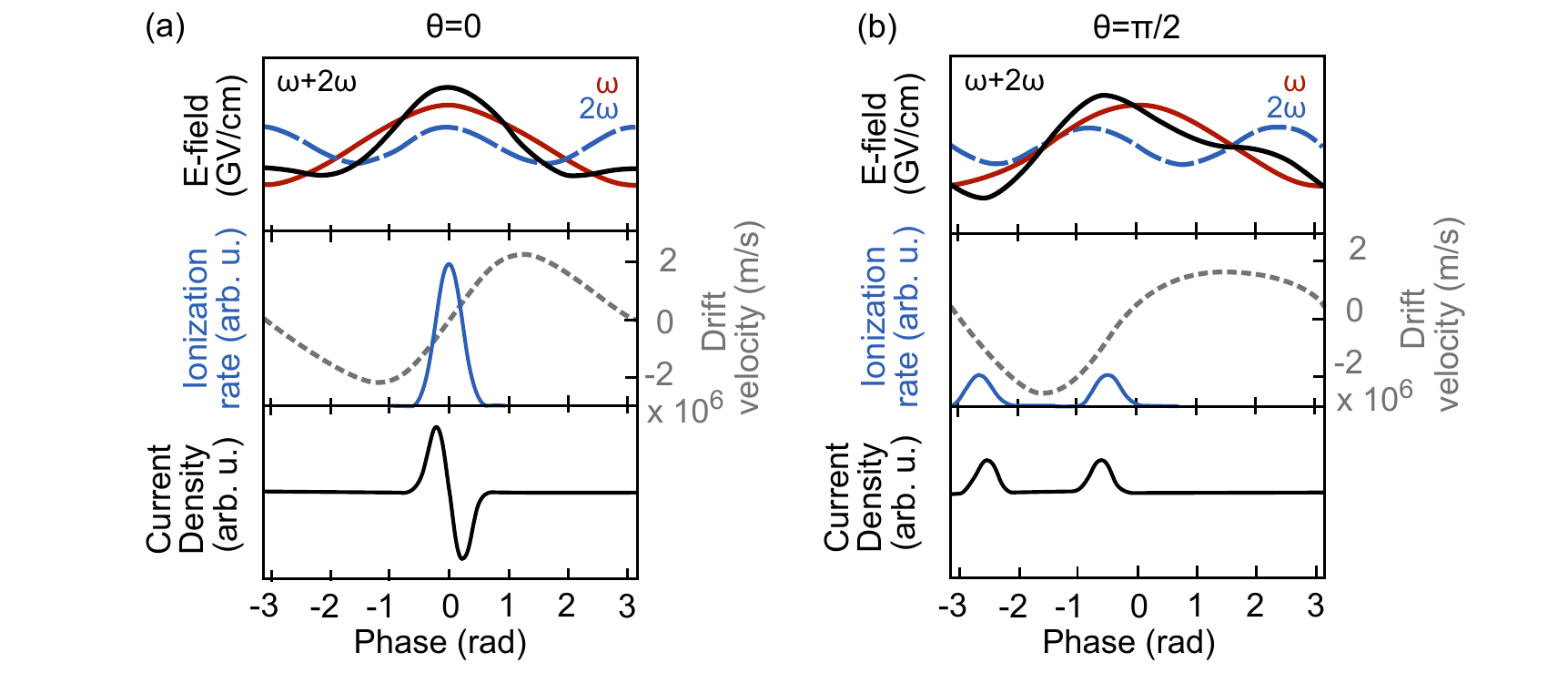}
    \caption[Fundamental, SH and combined laser field over one cycle of period for THz generation in two-color plasma filaments]{Fundamental (red solid line), SH (blue dashed line) and combined (black solid line) laser field over one cycle of period. Blue solid lines in the middle panels depict the ionization rate, grey dashed lines refer to the drift velocity. Bottom panels present the resulting current density for a relative phase of (a) $\theta=0$ and (b) $\theta=\pi/2$, reproduced from ref \cite{Kim:2009}.}
\label{fig:TPC}
\end{figure}

Another explanation of the phenomenon, which is now widely accepted, was first proposed by Kim \textit{et al.} \cite{Kim:2007}, and is based on the asymmetric transient photo current (also known as  plasma current model).
Due to photoionisation of gas molecules, a  time dependent current surge occurs, which further radiates a directional electromagnetic field proportional to the time derivative of the current density.
In order to generate a non vanishing net current density integrated over the entire pulse duration, it is necessary to apply an asymmetric electric field.
Otherwise, half of the freed electrons would drift up and the other half would drift down, resulting in a cancellation of the plasma current components.
In other words, THz fields which are generated at different times of the oscillating symmetric field amplitude of a  laser pulse are out of phase and experience destructive interference. 
In turn, THz radiation generated at different times by photo-ionized electrons from an asymmetric laser field can interfere constructively. Such an asymmetric laser field can be generated by combining the fundamental and its SH, expressed as \cite{PlasmaWidth:2013}
\begin{equation}
E_L\left(t\right)=E_{\omega}(t)\mathrm{cos}\left(\omega t\right) + E_{2\omega}(t)\mathrm{cos}\left(2\omega t+ \theta \right),
\end{equation}
with $E_{\omega}(t)$ and $E_{2\omega}(t)$ as the amplitude envelope of the fundamental and SH pulse, respectively, and $\theta$ being the relative phase between those fields.
Although $E_{\omega}(t)$ and $E_{2\omega}(t)$ are in general time dependent and often assumed to be Gaussian, for the following steps we define $E_{\omega}$ and $E_{2\omega}$ as time independent field amplitudes in order to simplify the illustration of the transient photo current model, but keep in mind that the generated net current density is considered for the entire time integral of the laser pulse duration.\\
This combined electric field creates electron plasma due to photoionization and further accelerates the freed electrons.
Newton's equation of motion in combination with the Lorentz force $F(t)$ for the electric field then leads to the time varying velocity \cite{Kim:2012}
\begin{equation}
F(t)=m_e a(t)=-e E_L(t) \quad \rightarrow \quad v(t)=-\frac{e}{m_e}\int_{t'}^t E_L(t)\mathrm{d}t,
\label{equ:velocity}
\end{equation}
with $t'$ as the time  when the electrons are released from the parent atoms (or molecules).
The electron charge and mass are given by $e$ and $m_e$, respectively, and $a(t)$ denotes the electron acceleration.
In contrast to high harmonic generation (HHG), where the accelerated electrons re-collide with the parent ions causing short - wavelength emission, 
the electrons drifting away from the ions without experiencing re-scattering account for THz generation \cite{PlasmaWidthKim:2008}.   
Assuming parallel linear polarized light of $\omega$ and $2\omega$, the freed electrons not only oscillate at the laser frequency, but also drift  in the transverse direction  at a velocity
\begin{equation}
v_d(t')=\frac{e}{m_e \omega}\left[E_{\omega}\sin \left(\omega t'\right) + E_{2\omega} \sin \left(
2\omega t' + \theta \right)/2 \right],
\label{equ:dirftVel}
\end{equation}
wherein zero initial velocity of the electrons at time $v(t')=0$ is assumed.
From Eq.\eqref{equ:dirftVel} we see that the drift velocity $v_d \propto \lambda$ is proportional to the wavelength of the driving pulse.
Quantitatively, the radiated THz field can be derived from the electron current density, given by \cite{Kim:2012}
\begin{equation}
J(t)=\int_{-\infty}^{t}dJ(t') \quad \mathrm{and} \quad dJ(t')=-edN_e(t')v_d(t'),
\label{equ:currentDens}
\end{equation}
with $dN_e(t')$ as the density of free elecrtons produced by the laser field in the interval between $t'$ and $t'+dt'$.
Each ionization event causes an ultrafast current surge and emits THz radiation in all directions. 
Because the maximum electron displacement during the interaction is much smaller than the generated THz wavelength, such a current surge can be considered as dipole radiation from a point source, with the far-field scaling as \cite{Kim:2009}
\begin{equation}
E_{\mathrm{THz}} \propto \frac{\mathrm{d}J(t)}{\mathrm{d}t}=e\frac{\mathrm{d}N_e(t)}{\mathrm{d}t}v_d(t),
\label{equ:E_THz}
\end{equation} 
where the change of electron density can be approximated by 
\begin{equation}
\mathrm{d}N_e(t)/\mathrm{d}t=w(t)\left[N_g-Ne(t)\right]\simeq N_gw(t),
\end{equation} 
with $N_g$ as the neutral gas density and $w(t)$ being the photoionization rate. 
When the Keldysh parameter fulfils $\gamma = \sqrt{U_i/(2 U_p)} < 1$ \cite{Keldish:1965}, with $U_i$ as the ionization potential of the gas constituent and $U_p$ as the laser ponderomotive potential energy, tunnel ionization becomes dominant.
The tunnel ionization rate for hydrogen-like atoms can be assumed, given by
\begin{equation}
 w(t)=4\omega_a\frac{E_a}{|E_L(t)|} \exp\left(-\frac{2}{3}\frac{E_a}{|E_L(t)|}\right),
\end{equation} 
where $\omega_a=4.134 \times 10^{16}\ \mathrm{s^{-1}}$  is the atomic frequency and $E_a=5.14 \times 10^9\ \mathrm{V/cm}$ is the atomic field seen by the ground-state electron in the hydrogen-like atom \cite{Landau:1977}.
Because the THz pulse energy as well as the THz pulse intensity, both commonly denoted as THz yield \cite{PlasmaWidthKim:2008, Nguyen:2019}, are proportional to the square of the THz field, it follows form Eq.\eqref{equ:E_THz} and \eqref{equ:dirftVel} that the THz yield scales with $\lambda^2$.
Thus, according to the photo current model, larger THz energies can be expected for longer wavelength driving pulses.\\
Fig.\ref{fig:TPC}(a) and (b) show a period of the combined laser field, drift velocity, ionization rate and resulting DC current for a phase difference between fundamental and SH pulse of $\theta=0$ and $\theta=\pi/2$, respectively.
In the case of zero phase difference, the drift velocity becomes an odd function and the product with the even ionization rate results in an asymmetric function.
Thus, the current summed over the laser cycle is zero.
In the case of $\theta=\pi/2$, the combination of fundamental and \ac{sh} pulse results in an asymmetric electric field, leading to a non-vanishing current density when integrated over the pulse period.
Consequently, the emerging sub-cycle current bursts add up over the entire pulse duration, forming a directional current surge.

\begin{figure}[htb]
    \centering
    \includegraphics[width=\linewidth]{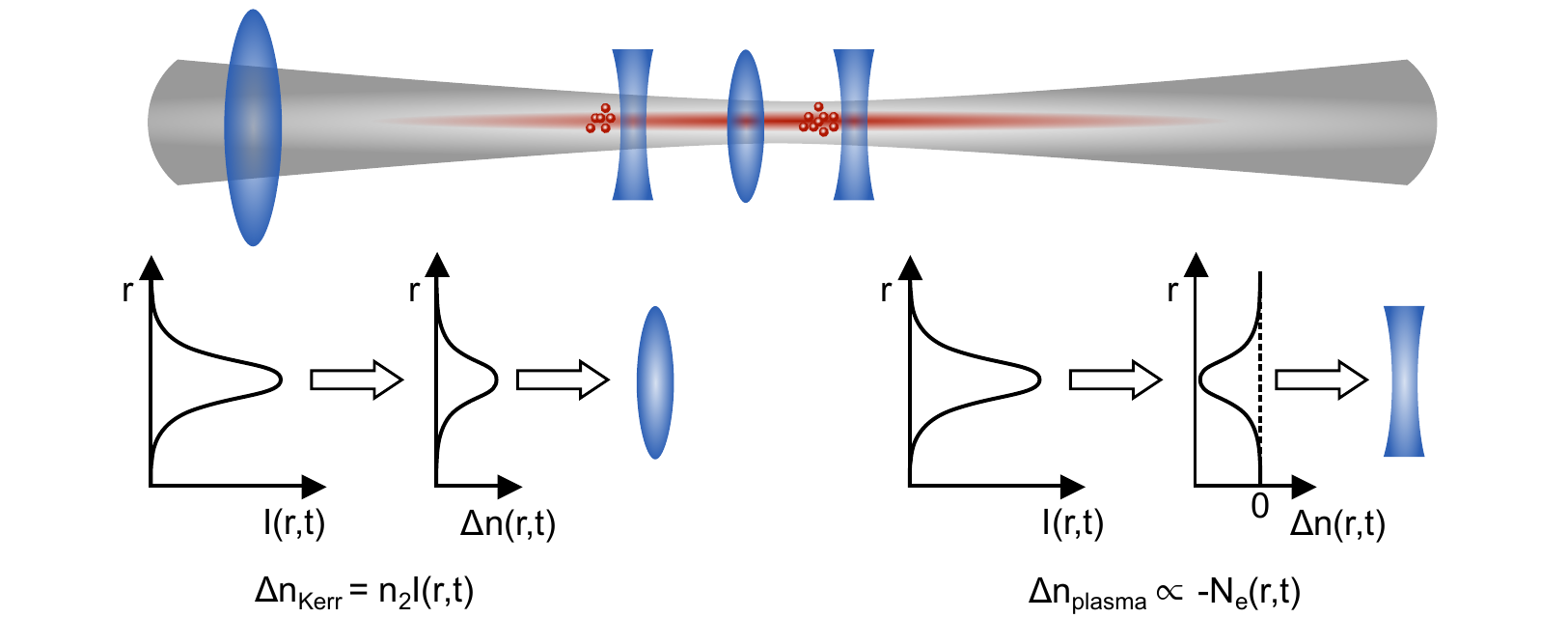}
    \caption[Schematic of a plasma filament]{Schematic of a plasma filament, illustrating a dynamic interplay between \ac{nl} focusing and plasma defocusing. The Kerr lens effect adds a positive refractive index  scaling with the \ac{nl} index of refraction $n_2$ and pulse intensity $I(r,t)$, wherein $t$ corresponds to the \textit{time slice} of the propagating pulse. Self-focusing is counter balanced by the negative additive of the refractive index originating from photoionized electrons, with $\Delta n_{\mathrm{plasma}}\cong - 4\pi e^2 N_e(t)/(2 m_e \omega_0^2)$ referring to a change in the index of refraction of a pulse slice propagating in a plasma \cite{chin_femtosecond:2010}. Figure is adapted from \cite{Valentina:2018}. }
\label{fig:filament}
\end{figure}
Note that, the formation of such a directional current surge will not happen only once, but several times during propagation when the plasma filament is formed. 
A filament is defined as the propagation distance for which a high energy beam experiences intensity clamping.
Within this region, the beam undergoes an alternation of self-focusing and successive  de-focusing.
Self-focusing is due to the Kerr lens effect and can only be observed if the peak power of a pulse overcomes the so called critical power $P_\mathrm{crit}$.
In this limit, the natural linear diffraction of the pulse is just balanced by self-focusing.
If the peak power is only very slightly higher than $P_\mathrm{crit}$, the \ac{gvd} will lengthen the pulse after a short distance of propagation, lowering the peak power to $P_{\mathrm{peak}}<P_\mathrm{crit}$ and the pulse will again diverge slowly through diffraction.
When both linear diffraction and \ac{gvd} are overcome by self-focusing, the beam will keep on focusing to a small spot until ionization occurs. 
Thus, in the case of intense fs-pulses, the main de-focusing mechanism is caused by plasma diffraction, which adds a negative contribution to the refractive index \cite{chin_femtosecond:2010}, as sketched in Fig.\ref{fig:filament}.\\
The relative phase delay $\theta$ and its optimized value of $\pi/2$ refer to the phase difference between fundamental and SH inside the filament. 
However, this relative phase changes during propagation $d$ inside the filament as
\begin{equation}
\theta=\omega \left(n_{\omega}-n_{2\omega}\right)d/c+\theta_0,
\label{equ:RelPhase}
\end{equation} 
where $n_{\omega}$  and $n_{2\omega}$ are the refractive indices of the air-plasma filament, $c$ is the speed of light and $\theta_0$ is the initial relative phase at the starting point of the filament. 
Moreover, the refractive index inside the filament changes with respect to the generated plasma density and optical Kerr effect \cite{PlasmaWidth:2013}. Though, the Kerr effect is often neglected and only the contribution of air and plasma dispersion  is taken into account with \cite{KimConical:2012,Gorodetsky:2014}
\begin{equation}
n_{\mathrm{filament},\omega}=n_{\mathrm{air},\omega}+ n_{\mathrm{plasma},\omega} \quad \textrm{and} \quad n_{\mathrm{plasma},\omega}\approx \sqrt{1-\frac{\omega_p^2}{\omega^2}},
\label{equ:plasmaRef}
\end{equation}
where $\omega_p=\sqrt{4\pi e^2 N_e(t)/m_e}$ is the plasma frequency \cite{chin_femtosecond:2010}.
When electrons from a neutral plasma consisting of negatively charged electrons and positively charged ions are displaced from their equilibrium position, the Coulomb force pulls the electrons back, acting as a restoring force. As the electrons will overshoot with respect to their equilibrium position,  the electrons start to oscillate with $\omega_p$.
Thus, because of this filament dispersion, the relative phase $\theta$ changes from 0 to $\pi$ over a de-phasing length $l_d=(\lambda/2)/(n_{\omega}-n_{2\omega})$, with $\lambda$ as the optical wavelength of the fundamental pulse.
Because the refractive index inside the filament depends on the plasma density, the de-phasing length elongates for less plasma content.
As a consequence, because of this de-phasing, THz emission generated at different points of the filament experience a phase shift and change in polarity, which ultimately leads to the typical conical emission in the far field, wherein the lower THz frequencies (longer wavelength) are found in the outer part.
\begin{figure}[htb]
    \centering
    \includegraphics[width=\linewidth]{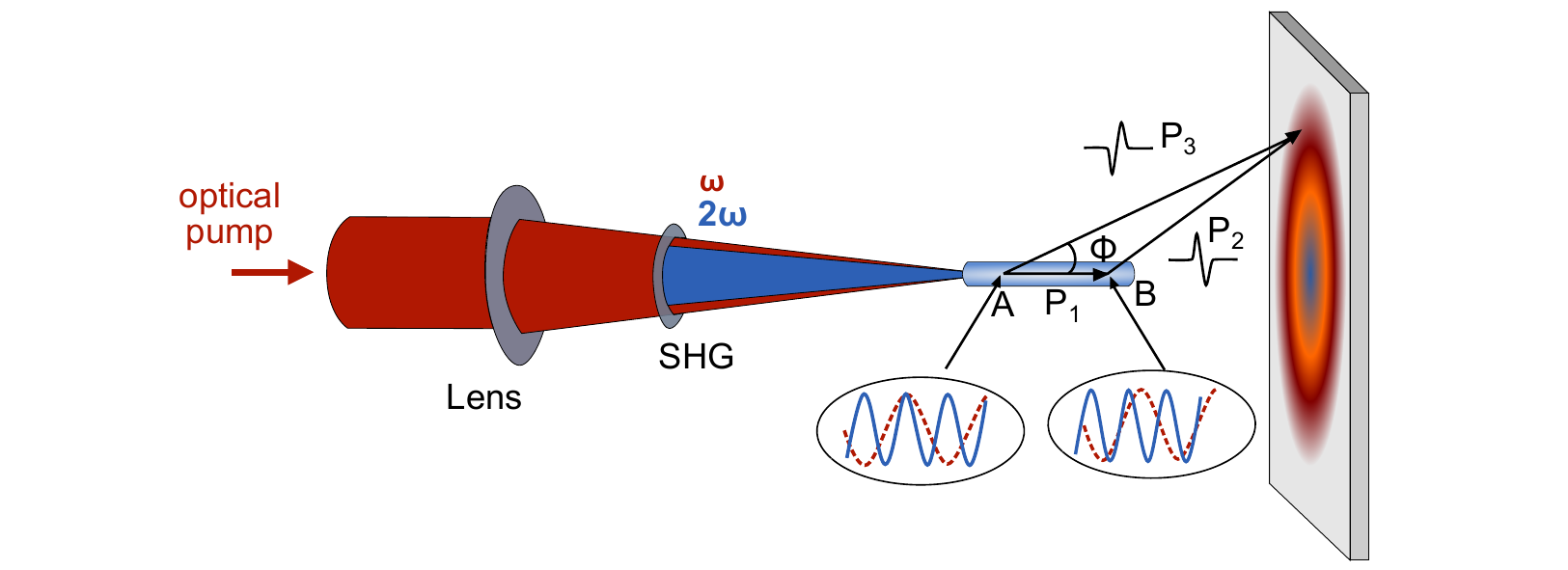}
    \caption[Schematic of conical THz emission from a plasma filament]{Schematic of conical THz emission from a long two-color plasma filament, adapted from \cite{KimConical:2012}. The THz waves produced at points A and B interfere constructively if the path difference $\Delta l=P_3-(P_1+P_2)=(m+1/2)\Gamma$, where $m=0, 1, 2, ...$ and  $\Gamma$ is the THz wavelength. For $m=0$, the angle for constructive interference is given by $\cos\Phi\approx 1-\Gamma/(2l_d)$. }
\label{fig:plasma_dephasing}
\end{figure}
The conical emission pattern can be intuitively understood (see Fig.\ref{fig:plasma_dephasing}) by naturally occurring off-axis constructive interference between locally generated THz waves. 
The de-phasing length for $\omega$ and $2\omega$ pulses (fundamental wavelength of $\lambda = 800 \mathrm{nm}$) in a plasma string with typical average plasma densities is in the order of 10--\SI{20}{mm} \cite{Gorodetsky:2014}.
Depending on the filament length (several times the de-phasing length) and emitted THz spectrum, the far field profile result in conical emission angles of $\leq$ \SI{10}{^{\circ}} \cite{KimConical:2012, Gorodetsky:2014}.
In the limit where the plasma filament length is significantly shorter than the de-phasing length, Kim \textit{et al.} \cite{KimConical:2012} observed THz waves with the same polarity and constructive interference in the forward direction, resulting in on-axis THz radiation, but with a very narrow THz spectral bandwidth and low THz yield.
They further observe a continuous enhancement of the THz output (in a conical emission pattern) by simply increasing the filament length, until $\omega$ and $2\omega$ pulses do not temporally overlap.
In the case of THz generation of an ultra broadband THz pulse, Gorodetsky and co-workes \cite{Gorodetsky:2014} demonstrated an absence of on-axis components, even for a filament size shorter than the de-phasing length.
They further elucidate that the angle of the conical emission drops with decreasing plasma density or increasing filament length. 
As it is the case in ref \cite{KimConical:2012}, the total THz yield increases almost linearly with the filament length, which indicates the feasibility to effectively increase the THz
output by extending the filament length with a constant plasma density.

Another crucial parameter for efficient THz generation is the polarization of the fundamental and SH harmonic field, which does not only affect the THz yield, but further influences the polarization state of the THz output.
In general, it is widely accepted that most efficient THz generation is achieved when the polarization of fundamental and SH fields are parallel \cite{Kim:2009, Thiele:2017}.
Although THz radiation can still be generated with cross polarized fundamental and SH pulses, the contribution is minor compared to the parallel case.
Nonetheless, as mentioned before, the THz yield from linear polarized pump pulses strongly depends on the relative phase between fundamental and SH.
In a microscopic (single-atom) model, THz radiation would be solely polarized along the direction of the current vector of the photo-ionized electrons.
Nonetheless, in the case of long filaments, the polarisation state of the radiated THz pulse evolves from linear to elliptical with respect to the filament length. 
This behaviour is attributed to (i) air-plasma dispersion, causing a relative phase shift between the $\omega$ and $2\omega$ pulses along the plasma filament, and (ii) cross-phase modulation (XPM) between the $2\omega$ and stronger $\omega$ pulse, which adds polarization changes. 
In turn, the modulation of the relative phase and polarization state change the direction of the local plasma current and thus, THz polarization.
In addition, the optical and THz pulses propagate with different velocities in the plasma filament, causing the local THz waves emitted from different parts of the plasma to arrive at different times in the far field. 
These combined effects generate a THz pulse in the far field whose polarization direction rotates with time \cite{You:2013}.\\
However, in the case of high plasma densities Meng \textit{et al.} \cite{Meng:2016}, and later Tailliez \textit{et al.} \cite{Tailliez}, experimentally demonstrated a 5 times higher efficiency of THz emission for circularly polarized laser fields with same helicity compared to linearly (parallel) polarized fields.
This enhancement is explained based on the analysis of electron tunnelling ionization and subsequent dynamics. 
It is well understood that atoms or molecules can be more easily ionized in laser fields with linear polarisation than with circular polarisation \cite{Ionization:2006}.
Thus, in the case of low laser intensities ($\sim 1\times 10^{14}$ \SI{}{W/cm^2} for a \SI{800}{nm} fundamental pulse \cite{Meng:2016}) much higher energy densities are created for linearly polarized light than for circularly polarized light, resulting in larger THz energies.
In turn, for high laser intensities ($\sim 5\times 10^{14}$ \SI{}{W/cm^2}), the gaseous atoms are fully ionized and the electron density saturates for both polarizations with the same peak intensity.
Nonetheless, for linearly polarized light, complete ionization occurs at the leading edge of the pulse and the residual pulse does not contribute to the plasma current but experiences plasma scattering instead. 
In turn, in the case of circularly polarized pulses, the saturation ionization instance is closer to the peak of the laser field. 
This delayed saturation of ionization leads to a higher vector potential at the instance of newly born electrons and thus, higher drift velocity. 
Consequently, because the plasma density for linearly and circularly polarized pulses are the same, but  the electron drift velocity is larger for the latter case, circular polarized light results in a higher plasma current and THz energies.\\
In addition, they observed that the dependence of THz yield with respect to the relative phase between fundamental and \ac{sh} almost vanishes in the case of circular polarization. 
Such an insensitivity can be attributed to the ionization probability, which is independent of the relative phase for circularly polarized light, according to theoretical studies of Song \textit{et al.} \cite{Song:2021}. 

In spite of that, the THz yield does not only depend on the polarisation, relative phase shift and pulse energy of the fundamental and its \ac{sh} pulse, but on the gas pressure and focusing geometry, which further effects the length of the filament, plasma volume and therefore THz generation efficiency.
Most importantly, it has been demonstrated that the THz energy strongly saturates with laser intensity and gas pressure \cite{PlasmaWidthKim:2008,PlasmaOhKim:2013}. 
In the case of a tight focusing geometry, resulting in  short intensity clamping, a dense plasma is formed and strongly defocuses the high-energy laser pulse. 
Thus, the laser intensity is not effectively coupled into the plasma channel. 
Only the front part of the pulse is used for photoionization, while the tailing edge does not contribute to plasma generation but gets scattered instead. 
As a consequence, the THz emission becomes less and strongly saturates with  increasing laser input intensity. 
In turn, such ionization induced defocusing can be reduced with weaker focusing conditions in order to elongate the filament and hence increase the plasma volume. 
Though, this technique is limited by (i) a critical plasma density, \textit{i.e.,} with extremely weak focusing the plasma density drops significantly, resulting in reduced THz energy, and (ii) macroscopic phase-matching becomes crucial for efficient THz generation since the fundamental, SH and THz pulse do not exhibit the same propagation velocity.

Ultimately, optimization of the optical- to- THz conversion efficiency in the case of near-IR driving pulses centred at around \SI{800}{nm} is fundamentally limited by dispersion of the gas medium, plasma volume and symmetry of the combined laser field.
In order to overcome these constrains, long wavelength pump sources are necessary.

\section{Long-Wavelength Drivers}
\label{sec:LWD}

\subsection{Larger Plasma Volume}

THz generation from long wavelength driving pulses is supported by the physical mechanism of laser filamentation.
A commonly accepted definition of filamentation is a dynamic interplay between self-focusing and diffraction, resulting in self-channelling that can persist over many meters.
Because the critical power for Kerr self-focusing scales as $P_{\mathrm{crit}} \propto \lambda^2$ and is hence higher for long wavelength driving pulses, it is possible to deliver more energy to the filament. 
In particular, the critical power for self-focusing for \SI{800}{nm} driving pulses in ambient air, in dependence on the pulse duration, was evaluated to be  in the order of $\sim 5-10\ \mathrm{GW}$ \cite{Liu:2005}, whereas in the case of \SI{3.9}{\mu m} pulses, the critical power was estimated to $\sim 125-250\ \mathrm{GW}$ \cite{Shumakova:2018, Kartashov:2013}.
Because the photoinoization rate is lower for long wavelength pulses, the onset of saturation (wherein the gaseous atoms are fully ionized) is at significantly higher intensities.
Thus, the released electrons -with a birth time close to the intensity peak- experience a larger drift velocity resulting in a positive effect on the generated photo current.
Moreover, the filament volume increases with respect to to the central wavelength of the driving pulse.
An intense femtosecond pulse will start to self-focus at a distance $z_f$ from the beginning of propagation in the medium according to the Marburger formula \cite{Marburger:1975}
\begin{equation}
z_f=\frac{0.367 k a_0^2}{\sqrt{\left[\sqrt{\frac{P}{P_{\mathrm{crit}}}}-0.852\right]^2-0.0219}},
\label{equ:Marburger}
\end{equation}  
where $k=2\pi/\lambda$ is the wave vector, $a_0$ is the radius of the beam profile at 1/e level of the intensity and $P$ the peak power of the pulse.
In laboratory conditions, filamentation is usually assisted by external focusing to force self-focusing within the limited length of the medium.
The new starting point $z_f'$ of the filament appears before the position of the geometrical focus and satisfies the lens transformation equation:$ 1/z_f+1/f=1/z_f'$, with $f$ as the geometrical focal length.
For a long focal length, the filament occurs before the geometrical focus and the filament length $L_f=f-z_f'$ is roughly the distance between the self-focusing position $z_f'$ and geometrical focus \cite{chin_femtosecond:2010}.
\begin{figure}[htb]
    \centering
    \includegraphics[width=\linewidth]{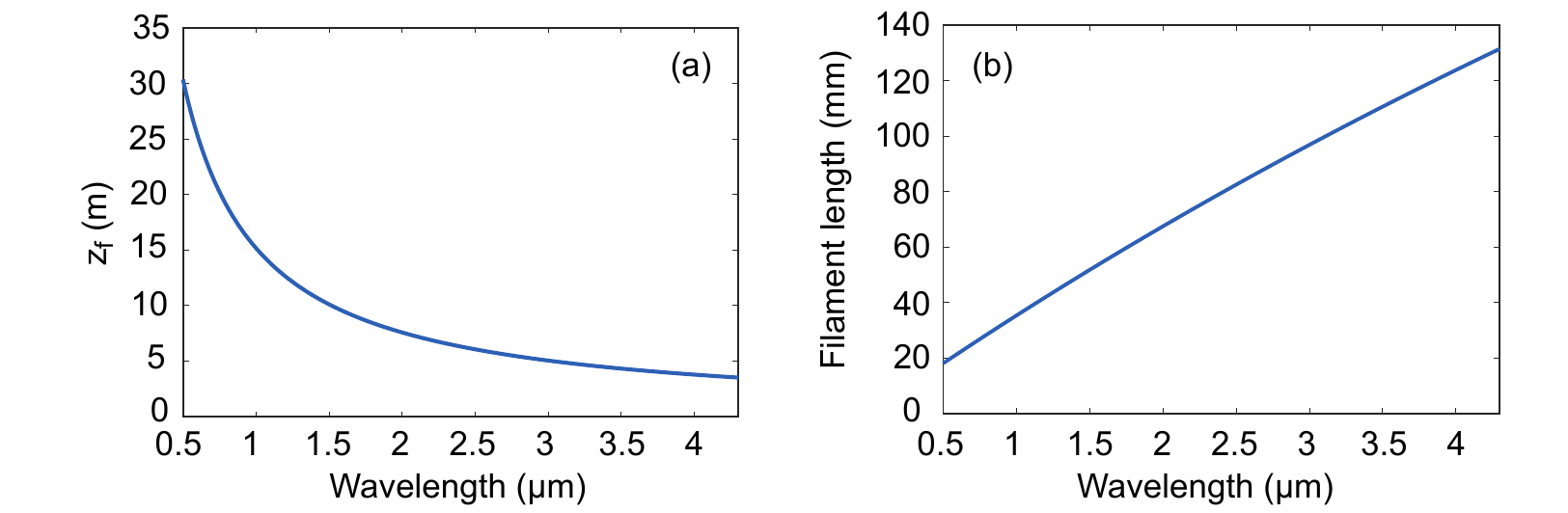}
    \caption[Filament length]{(a) Self focusing distance without external focusing of a collimated beam with radius $a_0$=\SI{5}{cm} and peak power $P=1.5P_{\mathrm{crit}}$. (b) Resulting filament length for a loose focusing geometry of $f=$ 75 cm.}
\label{fig:filamentLength}
\end{figure}
Figure \ref{fig:filamentLength}(a) shows the self focusing distance without external focusing ($f\rightarrow\infty$) with respect to the wavelength for a beam radius of $a_0= 1.5\ \mathrm{mm}$ and peak power of $P=1.5P_{\mathrm{crit}}$.
The filament length for a focal distance of $f=75\ \mathrm{cm}$ is shown in Fig.\ref{fig:filamentLength}(b).
When the geometrical focal length becomes shorter, the filament extends into the geometrical focus and the formula $L_f=f-z_f'$ no longer holds.
However, for the case of tight focusing with $f\ll z_f$, Geints \textit{et al.}\cite{Geints:2014} found that the filament length scales linearly with the wavelength, is proportional to the ratio between the peak power and critical power and is inversely dependent on the beam radius
\begin{equation}
L_f \propto \lambda a_0^{-2}\sqrt{P/P_{\mathrm{crit}}}.
\end{equation}
An analytical expression for the filament diameter $D_f$ can be deduced from equating diffraction, as well as Kerr and ionization response of the medium, and can be described by \cite{Skupin:2007}
\begin{equation}
D_f \approx \frac{\pi}{\sqrt{2k^2 \tilde{n_2}I_{\mathrm{max}}/n_0}} \propto \lambda,
\label{filamentD}
\end{equation} 
with $\tilde{n_2}$ as the maximal effective Kerr index, $n_0$ the linear refractive index and $I_{\mathrm{max}}$ as the clamped intensity which depends on the ionization rate.
Equation \eqref{filamentD} is related to a collimated beam without external focusing elements.
Theoretical studies \cite{fedorov2018generation} on  THz radiation by two-color filamentation  with different wavelengths confirm a linear dependence of the filament diameter with respect to the fundamental wavelength, when the laser pulses are focused by a lens with focal distance $f=200\ \mathrm{mm}$. 
In the simulations, the peak power of the fundamental pulses with a central wavelength ranging from $0.6 - 1.6\ \mathrm{\mu m}$  (100 fs FWHM pulse duration) was set to $P=1.2P_{\mathrm{crit}}$.
Consequently, due to the elongated filament  and larger diameter, a higher plasma content can be generated with long wavelength driving pulses.

Clerici \textit{et al.} \cite{LongWLplasma:2013} have been one of the first who investigated, theoretically as well as experimentally, THz emission generated by two-color plasma filaments of long wavelength driving pulses ranging from  \SI{0.8}{\mu m} to \SI{2.02}{\mu m}.
While sweeping the central wavelength of a commercial \ac{opa} from \SI{1.225}{\mu m} to \SI{2.02}{\mu m} (the \SI{800}{nm} pulse came directly from the laser amplifier), all other pulse parameters such as energy (\SI{400}{\micro J} of fundamental and \SI{20}{\micro J} of \ac{sh}), pulse duration ($60\pm5\ \mathrm{fs}$), as well as focusing condition (off- axis parabolic mirror with a focal length of \SI{10}{cm}) were fixed during the experiments.
They observe an increase in THz energy which scales with the central wavelength as $\lambda^{4.6}$  from \SI{0.8}{\mu m} to \SI{1.8}{\mu m}. 
For longer wavelength, the THz energy suddenly dropped.
The discrepancy from the $\lambda^2$- scaling law, as predicted by the transient photo current model, is described as follows:
in the case of tunnel ionization, which is independent of the wavelength, the photocurrent depends linearly on the wavelength (see Eq.\eqref{equ:dirftVel} and \eqref{equ:currentDens}), which in turn resluts in a quadratic scaling of the radiated THz energy with the pump wavelength.
This simple model doesn't take the wavelength dependent plasma volume into account, wherein both the filament length as well as the radius scale with $\lambda$.
By including these parameters in the numerical simulation, Clerici \textit{et al.} demonstrate a remarkable fit with the experimental data.
Thus, the outcome can be interpreted as a combined result of both the fundamental $\lambda^2$-law resulting from the plasma currents and the effect of the wavelength dependent plasma volume.
Moreover, by including the $\lambda^{-2}$ dependence on the intensity ($I_{\mathrm{peak}}=2P_{\mathrm{peak}}/(\pi w_2^2)$, where $w_2=\lambda f/(\pi w_1)$ is the beam waist after focusing the beam with radius $w_1$ by a lens with focal distance $f$, and $P_{\mathrm{peak}}$ is the peak power), they can successfully reproduce the sudden drop of THz emission for $ > 1.8\ \mathrm{\mu m}$ driving pulses.
Simply, at $\sim 1.8\ \mathrm{\mu m}$, the threshold is reached for which the ionization probability drops below 100\%.
Thus, they find that the wavelength-scaling of $\lambda^{4.6}$ is not universal but is related to the photocurrent amplitudes combined with the filament volume and linear focusing conditions used in their work.
Nonetheless, the THz field amplitude generated by a \SI{1.85}{\mu m} pump pulse reached \SI{4.4}{MV/cm} with a conversion efficiency from mid-IR pump to a single-cycle sub-10 THz pulse of $>10^{-3}$, which is more than one order of magnitude higher than in the case of  THz generation with \SI{800}{nm} pump pulses. The achieved THz field strength was the highest value reported for a table-top source at this time.

Later, Nguyen \textit{et al.} \cite{Nguyen:2019} repeated the experiment in a similar wavelength range with two different laser sources and varying laser parameters. They experimentally observe a THz energy scaling of  $\lambda^{\alpha}$ with a large spread of the exponential factor of $5.6 \leq \alpha \leq 14.3$.
As it follows from  their simulations, the distinct exponents can be attributed to the envelope laser parameters such as the beam width, pulse duration, energy ratio \ac{sh}/F between \ac{sh} and fundamental F pulse, and the relative phase between the two color fields when tuning the carrier wavelength of the \ac{opa}.
Moreover, they theoretically reproduced experimental results of Clerici \textit{et al.} \cite{LongWLplasma:2013},  where the pulse duration and \ac{sh}/F energy ratio was reported to be constant during the  experiment.
A phase slippage was identified as one of the main reasons for the deviation from the expected $\lambda^2$ scaling law.
During the experiment, the  \ac{sh} crystal is placed at a certain distance away from the focal plane in order to provide an optimal  relative phase of $\pi/2$  between the fundamental and SH pulse.
As the central wavelength is changed, the \ac{sh} crystal is rotated, leading to a strong phase shift at the focus, which can ultimately cause a sudden drop in the THz signal for specific wavelengths.\\
Thus, in order to accurately describe the scaling of THz energy with respect to the central wavelength of the driving pulse, several parameters have to be taken into account, such as the elongated plasma volume, laser pulse dimensions and dispersion of the medium being responsible for the change in relative phase between the fundamental and SH pulse.
From the practical point of view during the experiments, it is basically impossible to keep all of these parameters constant at the same time.

\subsection{Increased Asymmetry}

Although wavelength scaling laws are not universal with respect to the laser parameters \cite{Nguyen:2017}, similar theoretical results were reported by Zhang \textit{et al.}\cite{Zhang:2017_incommensurate},
wherein they control strong tunable THz emission with optimal incommensurate multi-color
laser fields. They use the interference (or local current) model proposed by 
Babushkin \textit{et al.}\cite{Babushkin:2011} to describe the mechanism of the THz frequency tunability. 
Here, the  THz wave results from a linear superposition of many individual ionization events, with ionization amplitudes $C_{\mathrm{n}}=-e\delta n_{en}v_d(t_\mathrm{n})$, with $\delta n_{\mathrm{en}}$ and $v_d(t_\mathrm{n})$ as the increment of electron density and drift velocity for the \textit{n}-th ionisation event, respectively.
They find a wavelength scaling of $\lambda^{2.62}$ for the THz yield of THz pulses with a power spectrum centred at 15 THz, when the pulse duration and peak intensity of the driving pulses are kept constant.
An even higher scaling law of $\lambda^{3.11}$ is found for 30 THz pulses.
An increase in THz yield with respect to the central wavelength of the driving pulse can be understood as follows: although the total ionization events are reduced in the field of long wavelength drivers due to the reduced number of oscillations for a  constant pulse duration, the number of ionization events with dominant contribution to the total THz radiation is maintained.
 The amplitude for each ionization event is enlarged due to the enhanced increment of electron density $\delta n_{en}$ and velocity $v_d(t_n)$ for consecutive ionization events.
In other words, the difference in the field amplitude of the two-color laser field is larger in between every optical cycle in the case of longer wavelength drivers which leads to a higher field asymmetry and a steeper slope of the current density with respect to infinitesimally small steps in time. 
\begin{figure}[htb]
    \centering
    \includegraphics[width=\linewidth]{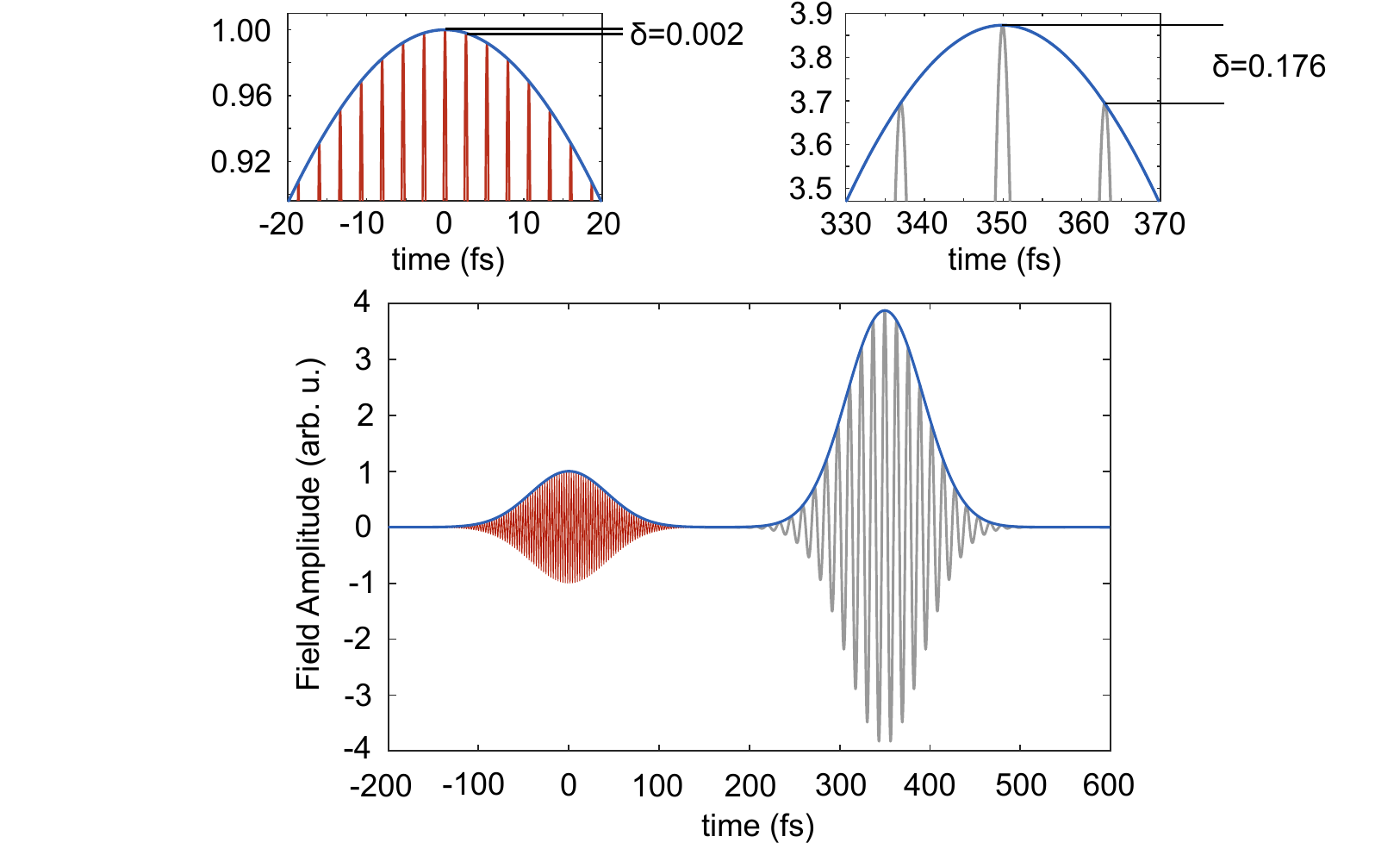}
    \caption[Change in field amplitudes for near- and mid-IR pulses]{Field amplitudes for 100-fs pulses with a central wavelength of \SI{800}{nm} (red) and \SI{3.9}{\mu m} (grey). The magnified view on top reveals a change in field amplitude of subsequent global maxima of 0.002 in the case of the near-IR driving pulse, and 0.176 for the long wavelength driver, indicating a higher field asymmetry for the latter case.}
\label{fig:FieldAmplitudes}
\end{figure}
A change in field amplitude is illustrated inf Fig.\ref{fig:FieldAmplitudes} for two pulses with a central wavelength of \SI{800}{nm} and \SI{3.9}{\mu m}. In both cases, the pulse duration at \ac{fwhm} is \SI{100}{fs}, the maximum of the \SI{3.9}{\mu m} pulse envelope is chosen in order to account for the $\lambda^2$ scaling of the critical power of self-focusing. 
Close to the peak of the pulse, the change in field amplitude between subsequent global maxima is 88 times larger in the case of the mid-IR pulse as compared to the \SI{800}{nm} driver.
Even if the peak field amplitudes would be equal, the change in field amplitude would be more than 22 times larger for the mid-IR pulse.
Accordingly, due to the enhanced asymmetry between consecutive ionization events,  a more intense THz emission is generated. \\
The same argument of an increased laser field asymmetry being responsible for an enhanced THz yield is found in the case of THz generation by a single-color filament of long wavelength drivers containing only a few optical cycles. 
Wang \textit{et al.}\cite{WangMIR:2011} theoretically predicted an enhancement of the THz amplitude by 35 times as the laser wavelength increases from \SI{1}{\mu m} to \SI{4}{\mu m}. 
In addition, it was shown that the laser intensity threshold for THz emission is lower for longer laser wavelengths, and it was demonstrated that a \SI{5}{MV/cm} THz field can be produced by a \SI{4}{\mu m} laser with an intensity of $10^{15}$ \SI{}{W/cm^2}.
This is because with an increase in laser wavelength, the number of laser cycles  needed to ionize the gas completely is decreased, the electron number distribution versus the electron birth position tends to be more highly asymmetric, and the average drift velocity becomes larger, resulting in strong net current and subsequent powerful THz radiation. 

Moreover, Mitrofanov \textit{et al.}\cite{MitrofanovHH:2015} reported on higher order harmonic generation in a single channel filament of a linearly polarized intense mid-IR pulse centred at \SI{3.9}{\mu m}. 
Harmonics were detected up to the 9th order in ambient air. 
In the case of elliptically polarized light, harmonic generation in ambient air could be measured effortless up to the 7th order \cite{Shumakova:2019}.
Such higher order harmonics, propagating with the fundamental and SH pulse, are identified theoretically \cite{Fedorov:2018} as additional field symmetry breaking factor which further boosts THz generation.

\subsection{Lower Dispersion}

\begin{figure}[htb]
    \centering
    \includegraphics[width=\linewidth]{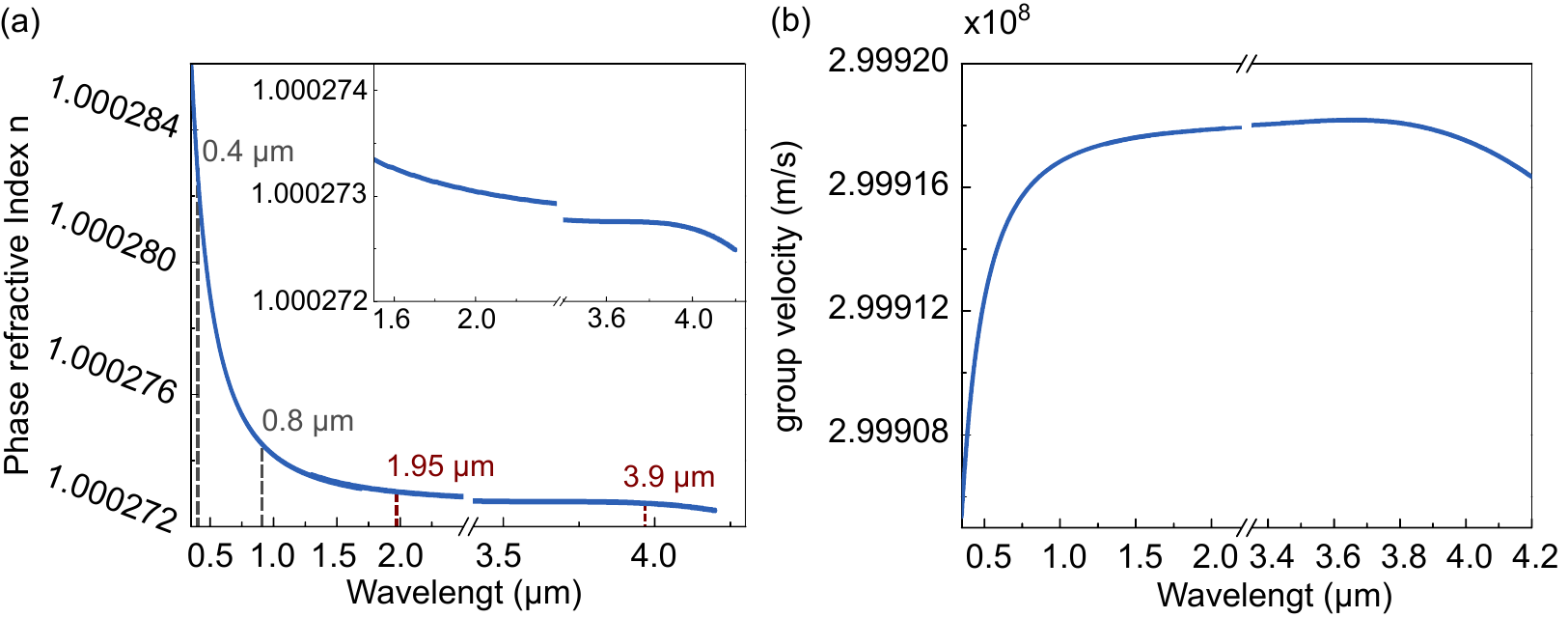}
    \caption[Phase refractive index of dry air]{(a) Phase refractive index and (b) group velocity of dry air (at $15\ \mathrm{^{\circ}C}$, 101.325 kPa). Coefficients for the Sellmeier equation in the visible and near-IR spectral range ($0.23-1.69\ \mathrm{\mu m}$) are taken from ref \cite{AirCiddor:1996}. Data for the mid-IR range is provided by ref \cite{AirMathar:2007}. The inset of (a) shows a magnified view on the mid-IR spectral range.}
\label{fig:airDispersion}
\end{figure}

Dispersion of ambient air is significantly smaller in the mid-IR spectral range, as shown in Fig.\ref{fig:airDispersion}, which leads to a smaller walk-off between the fundamental and SH.
After one meter of propagation in air of a fundamental pulse with central wavelengths of \SI{800}{nm}, the second harmonic will lag behind by \SI{82}{fs}. 
In comparison, the temporal walk-off between \SI{3.9}{\mu m} and \SI{1.95}{\mu m} only accounts for \SI{2.2}{fs} \cite{FedorovDisp:2020}.
Moreover, the small walk off between the THz and fundamental pulse further explains the better directionality of THz radiation in the case of mid-IR pulses.
In combination with the elongated filament length, this results in a reduced angle of THz emission cone of $\sim 2^{\circ}$ for \SI{3.9}{\mu m} compared to $\sim 7^{\circ}$ for \SI{800}{nm}, according to the theoretical work.
Thus, focusing of the THz beam will be easier and ensures a smaller focal spot size.

Overall, Fedorov \textit{et al.} \cite{Fedorov:2018} theoretically compared the performance of THz generation and filament characteristics between \SI{800}{nm} and \SI{3.9}{\mu m} driving pulses with the same pulse duration.
The pulse energy was chosen in such a way that the power of the corresponding single-color pulse is comparable in both cases and set to 1.2$P_{\mathrm{crit}}$.
It was found that a $\sim$ 3 times longer plasma channel, containing $\sim$ 45 times more free electrons in the case of mid-IR drivers is generated.
In the simulations, they further included higher harmonics of up to the 15th order, generated during propagation in air, which contributes to the field symmetry breaking, and a smaller walk off between fundamental and \ac{sh}.
The combination of all these beneficial factors resulted in an astonishing predicted conversion efficiency of 6.7\% for \SI{3.9}{\mu m}, compared to 0.06\% for \SI{800}{nm} pulses.

\section{Experimental setup}
\label{sec:plasmaSetup}
\begin{figure}[hbt]
    \centering
    \includegraphics[width=\linewidth]{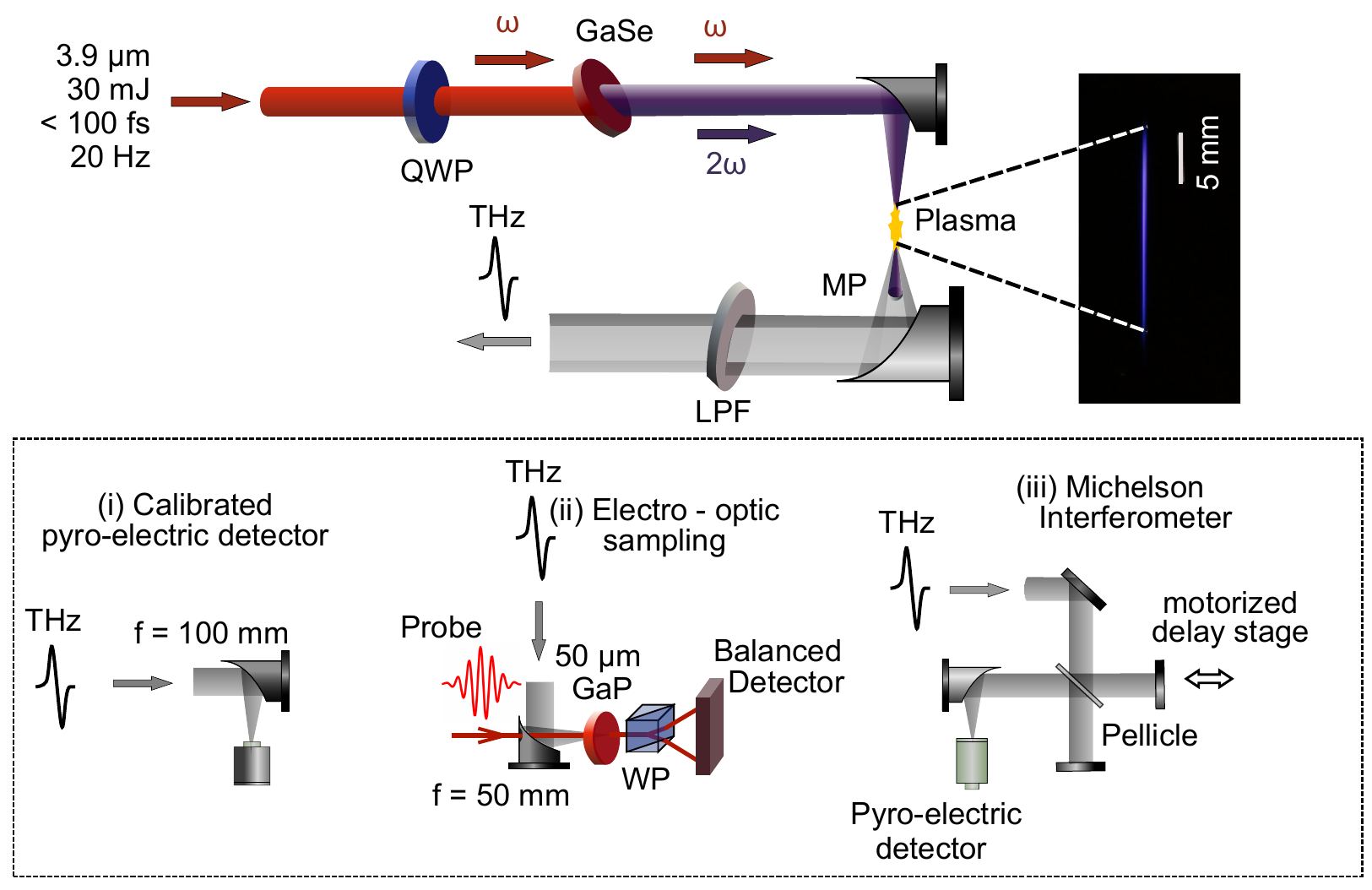}
    \caption[Experimental setup for THz generation by two-color plasma filaments and THz diagnostics]{Experimental setup for THz generation by two-color plasma filaments (top) and THz diagnostics (bottom), consisting of a Michelson interferometer, EOS and a calibrated pyro-electric detector. QWP - quarter waveplate; MP - metal plate; LPF - long pass filter; WP - Wollaston prism }
\label{fig:PlasmaSetup}
\end{figure}

As described in section \ref{sec:OPCPA}, the \ac{opcpa} system generates more than \SI{30}{mJ} of pulse energy at a central wavelength of \SI{3.9}{\mu m}. The pulse duration and temporal chirp can be controlled with a grating compression and enables sub \SI{100}{fs} pulses.
The system operates at a repetition rate of \SI{20}{Hz}.
A simple in-line schematic of the experimental setup is shown in Fig.\ref{fig:PlasmaSetup}.
The fundamental pulse is steered through a quarter wave plate (QWP), in order to control the ellipticity of the field polarization. 
\ac{shg} is conducted in a \SI{100}{\mu m} thick z-cut GaSe crystal with a clear aperture of \SI{7}{mm}. The negative uniaxial SH  crystal is tilted to obtain type-II phase matching conditions, wherein two of the three waves involved are polarized extraordinarily ($\omega(\mathrm{e})+\omega(\mathrm{o})\rightarrow 2\omega(\mathrm{e})$). 
The configuration is chosen to achieve a maximal mutual projection of the polarizations of the fundamental and the second harmonic beam on the same axis. 
Parallel polarizations of $\omega$ and $2\omega$ fields are known to be the optimum composition for efficient THz generation.
The combined laser field is then focused with a parabolic gold mirror with a focal distance of 150 mm in ambient air to produce a plasma channel.
Because the mid-IR (\SI{3.9}{\mu m} $\overset{\scriptscriptstyle\wedge}{=}$ \SI{76.87}{THz} and \SI{1.95}{\mu m} $\overset{\scriptscriptstyle\wedge}{=}$ \SI{153.75}{THz}) and THz pulse are spectrally close, separation of the generated THz radiation from the pump spectrum is not trivial. 
We therefore use a combination of spatial and frequency filters.
A metal disc (MD) is inserted before the collimating mirror to block the central part of the beam, which mainly consists of the fundamental pulse, SH and generated supercontinuum.
The conically emitted THz radiation passes around it.
Residual driving pulses are blocked after the collimating mirror with a 5 mm thick \ac{hdpe} plate.
The THz pulse is then steered to the diagnostic setup, which mainly consists of \ac{eos} and a Michelson interferometer to measure the THz transients and power spectrum, as described in section \ref{sec:Spectrum}.
The THz energy is measured with a calibrated pyroelectric detector (PED, SPI-A-62-THz Gentec-EO).
To prevent the detector from saturation, additional filters are used in order to attenuate the intense THz radiation.
The set of long pass filters comprise of another 5 mm HDPE plate, one 2 mm thick high resistivity float zone Silicon wafer, and one 0.5 mm thick low resistivity black Silicon wafer.
Transmission spectral of the applied long pass filters are shown in Fig.\ref{fig:Filter}.

\section{Results and Discussion}
\subsection{Polarization States}
\begin{figure}[htb]
    \centering
    \includegraphics[width=\linewidth]{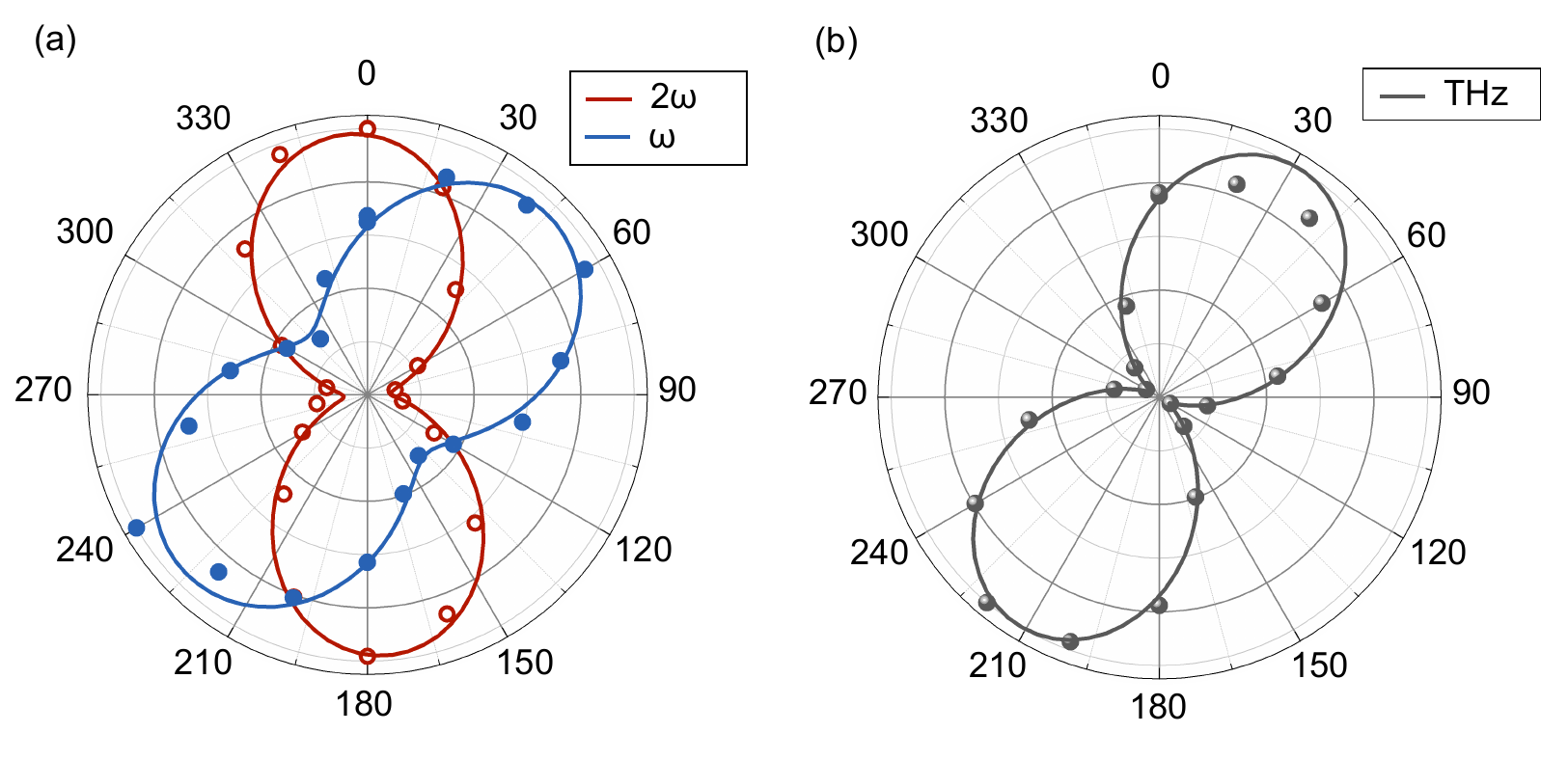}
    \caption[Polarization state of the SH, fundamental and generated THz pulse]{Polarization state of (a) the SH (red) and fundamental (blue) pulse with the main axes of the polarization ellipse rotated by \SI{356}{^{\circ}} and \SI{43}{^{\circ}}, respectively. (b) Resulting linearly polarized THz pulse with a polarization plane rotated by  \SI{33}{^{\circ}}. Experimental data is presented by dots, while solid lines refer to a fit function to guide the eye.}
\label{fig:PolarizationStates}
\end{figure}

For conventional pump schematics with a Ti:Sapphire system centred around 800 nm it is common practise to use a BBO crystal with type-I phase matching conditions to generate the SH.
The resulting orthogonal polarization of fundamental and SH can be easily corrected for parallel alignment with a $\lambda$/2-waveplate for the fundamental pulse which may simultaneously  act as $\lambda$ plate for the $2\omega$ pulse.
Although the optical coating and plate thickness needs to be carefully designed, such waveplates do exist in the near-IR and visible range.
In contrast, in the case of mid-IR pulses, such phase retarders are  not (yet) available.
Alternatively, in order to maximize THz emission, the GaSe crystal is rotated and tilted (close to the phase matching angle of \SI{17.3}{^{\circ}}) for type-II interaction. 
The angle between the polarization states of the fundamental and the SH pulse are then close to \SI{45}{^{\circ}} (see Fig.\ref{fig:PolarizationStates}(a)), which gives a maximal projection of the polarizations.
In addition, a \ac{qwp} is used to control the ellipticity of the polarization states.

Although the emerging ellipticity of the driving pulses might seem to be detrimental for efficient THz generation on a first sight, it is indeed advantageous.
As elucidated in section \ref{sec:TPCM}, a relative phase of $\pi/2$ between the fundamental and SH is needed to gain the highest asymmetry of the combined electric field.
The relative phase is basically determined by the difference in refractive index and the common propagation path of fundamental and \ac{sh}, as shown in Eq.\eqref{equ:RelPhase}, wherein the common path length can be controlled by changing the distance between the \ac{shg} crystal and plasma source.
In the case of a Ti:Sapphire system, the relative phase between the fundamental and SH pulse can be shifted from 0 to $\pi/2$  by moving the crystal by only 2 cm \cite{Kim:2012}.
However, as mentioned before, dispersion in the mid-IR spectral range is very small.
Thus, several meters of a common propagation path distance would be necessary to produce the desired relative phase.
Instead, it is possible to use elliptically polarized light in order to overcome this obstacle. 
Elliptical polarized light is the superposition of two plane waves with mutually orthogonal polarization, shifted in phase by $\pi/2$. 
Hence, at any given time, there is an electric field component of the fundamental pulse which is phase shifted by $\pi/2$ with respect to the SH pulse. 

Moreover, in the case of intense pump pulses, circularly polarized light can enhance the THz conversion efficiency because ionization events will be only initiated by the central part of the driving pulse, which can decrease plasma scattering of the trailing edge and increase the coupling efficiency into the plasma channel \cite{Tailliez, Meng:2016}.
Thus, the ellipticity of the mid-IR driving pulse can be another indicator for the observed extraordinary conversion efficiency, as presented in the following section.

Consequently, the experimental setup can be empirically optimized for a maximum mutual projection of the field components and suitable relative phase shift, resulting in a linearly polarized THz pulse with a polarization plane in between the major polarization axis of $\omega$ and $2\omega$ fields (Fig.\ref{fig:PolarizationStates}(b)).
The polarization of the generated THz pulse is thereby characterized by placing a rotating wire grid polarizer (Tydex) before the focusing mirror in front of the pyro-electric detector. 

\subsection{THz Conversion Efficiency}

\begin{figure}[hbt]
    \centering
    \includegraphics[width=\linewidth]{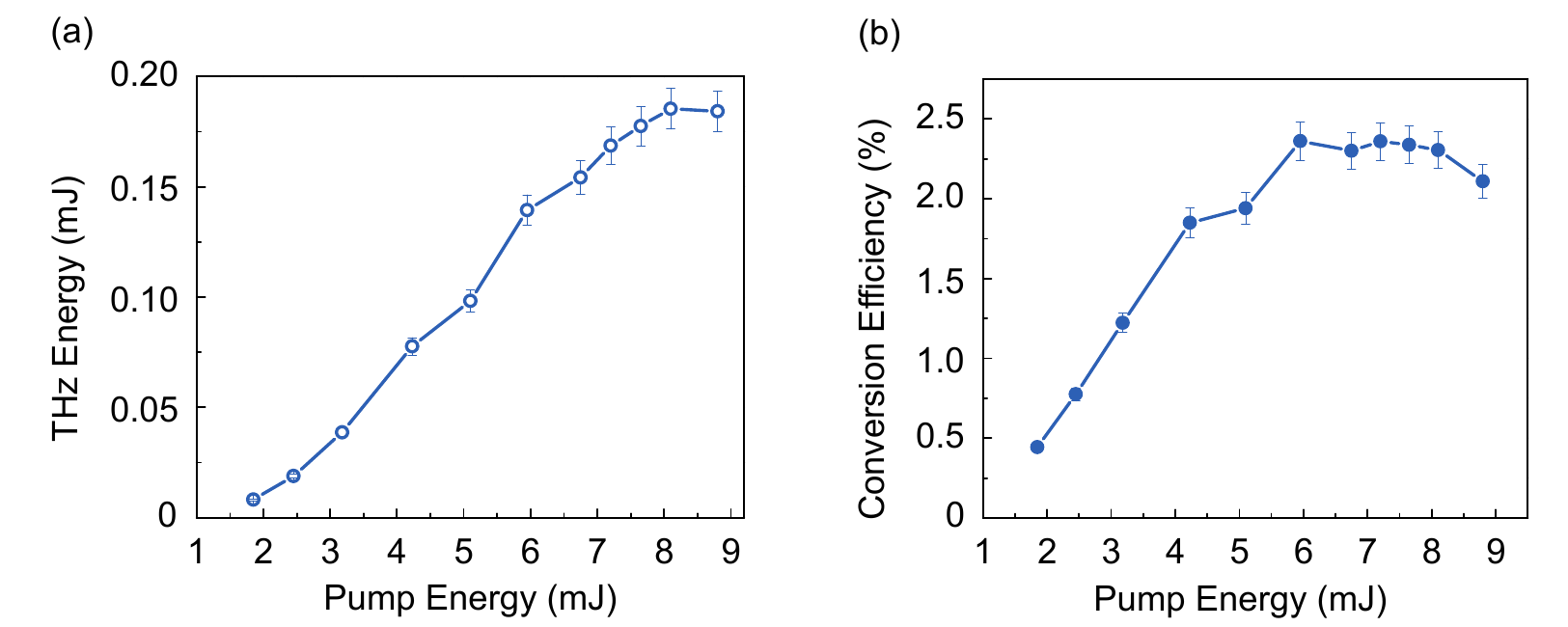}
    \caption[THz energy and conversion efficiency in two-color plasma filaments]{THz energy (a) and conversion efficiency (b) in dependence of the pump energy, corresponding to the pulse energy of the combined $\omega-2\omega$ laser field after GaSe.}
\label{fig:EnergyTHzPlasma}
\end{figure}

To measure the  THz energy and conversion efficiency, the filtered THz pulses are focused onto the pyro-electric detector with a parabolic mirror with a focal distance of 100 mm. The corrected THz energy and conversion efficiency with respect to the pump energy, consisting of fundamental and SH pulses after the GaSe crystal, are shown in Fig.\ref{fig:EnergyTHzPlasma}. 
The THz energy is thereby restored by taking the frequency-dependent transmission coefficients and detector calibration (as described in section \ref{sec:Pyro}) into account.
As the pump energy increases, an extremely high THz pulse energy of 0.185 mJ and record conversion efficiency of 2.36\% can be reached, which is more than two orders of magnitude higher as compared to typical values reported for \SI{0.8}{\mu m} driving pulses. 

In order to verify that the detected energies solely originate from THz radiation generated in the plasma filament, several experiments are conducted.
First of all, using only the metal plate and a 5 mm HDPE sheet, the THz source is actually strong enough to be measured with a standard mid-IR laser power meter. 
The measured THz energies attained from the power meter and pyro-electric detector are thereby in the same order of magnitude.
However, because the pyro-electric detector is calibrated for the THz spectrum, the acquired signal is more reliable and used for the presented data.
To exclude the possibility that the signal on the power meter emerges from residual pump pulses which are transmitted through the filters, THz generation is suppressed by either detuning the compressor of the \ac{opcpa} or by misaligning the \ac{sh} crystal. 
In both cases, the signal on the power meter disappears.
Second, GaSe is known for efficient THz generation by optical rectification.
To rule out THz radiation arising from the NL crystal under the phase matching conditions used during the experiments, a HDPE filter is placed immediately after the GaSe crystal allowing any THz radiation produced by the crystal to pass, while the blocked laser beam does not propagate further to create a filament.
The resulting absence of any measurable signal on the pyro-electric detector beyond the noise limit confirms that all the recorded THz radiation is produced by the filament.
\begin{figure}[htb]
    \centering
    \includegraphics[width=\linewidth]{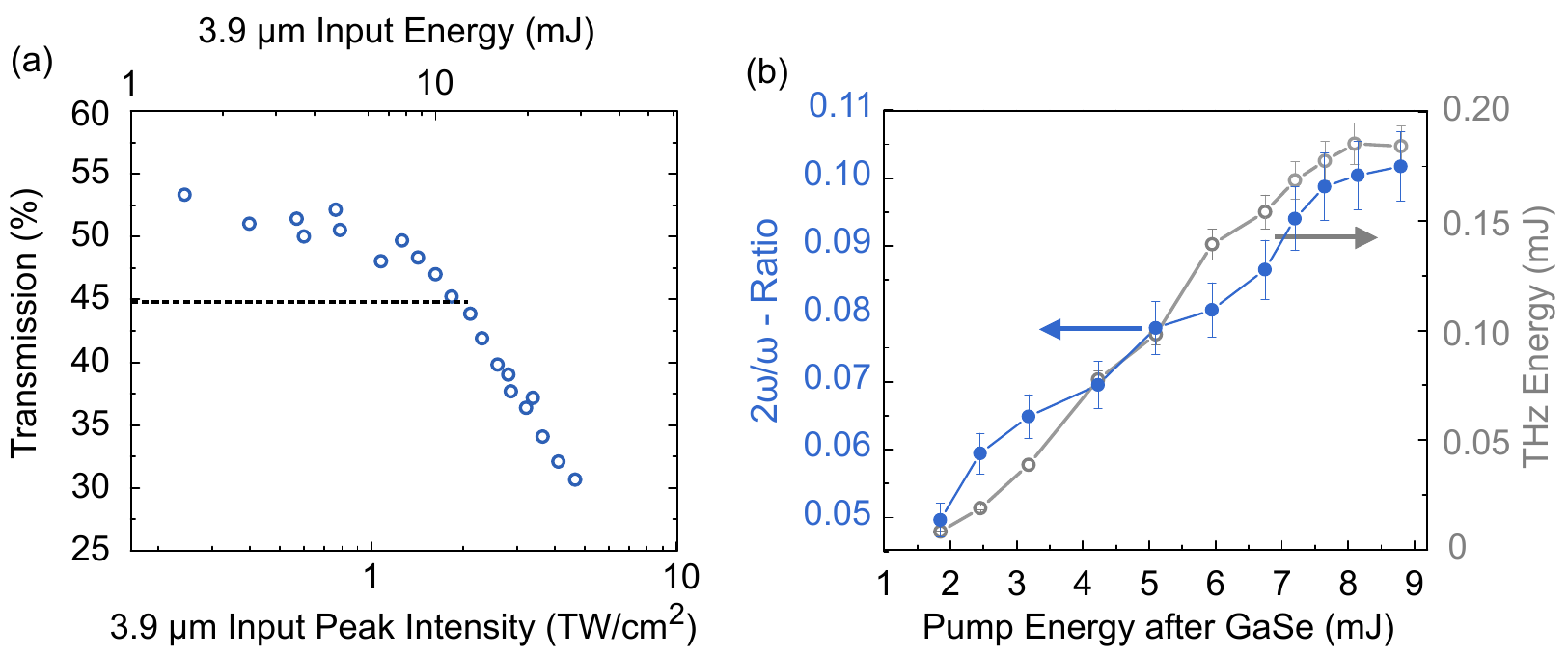}
    \caption[Measured transmission of a \SI{100}{\mu m} thick GaSe crystal with respect to the input energy of the fundamental driving pulse at the crystal surface]{(a) Measured transmission of a \SI{100}{\mu m} thick GaSe crystal with respect to the input energy of the fundamental driving pulse at the crystal surface (top x-axis) and peak intensity (bottom x-axis) for a pulse duration of 100 fs and 1.4 mm beam radius at FWHM level. Both are shown in logarithmic scale. The horizontal line indicates a transmission level of 10\% of its initial value (b) Ratio of fundamental and SH pulse (blue dots) and THz energy (grey circles) in dependence of the pump pulse energy measured after GaSe .} 
\label{fig:GaSeTransmission}
\end{figure}
 
Note that, despite the achieved THz conversion efficiency, we believe that we are not yet fundamentally restricted to reach higher values.
Particularly, we are limited by the clear aperture of the \ac{sh} crystal and surface reflection from the crystal, which are technical problems that can be overcome. 
Fig.\ref{fig:GaSeTransmission}(a) depicts the transmission of the \SI{100}{\mu m} thick GaSe crystal with respect to the pump intensity and pump energy of the fundamental driving pulse centred at \SI{3.9}{\mu m}.
As the crystal is uncoated, Fresnel reflection losses from the crystal surfaces cause a reduced transmission.
Moreover, for high pump energies, the GaSe crystal exhibits noticeable multi-photon absorption or tunnel ionization. 
However, as long as the transparency level is kept above 10\% of its initial value (as visualized by the horizontal line in Fig.\ref{fig:GaSeTransmission}(a)), the reduced transparency is reversible \cite{FengGaSe:2014}.
Significantly larger pump intensities lead to irreversible damage of the crystal wherein Selenide starts to evaporate and a solidified Gallium structure appears on the crystal surface.
Nevertheless, although we are operating in a pump regime where the transmission has dropped to less than 58\% of its initial value, and dark spots become visible on the crystal surface, the energy ratio between fundamental and \ac{sh} pulse contentiously increases with respect to the pulse energy after the GaSe crystal, as shown by the blue dots in Fig.\ref{fig:GaSeTransmission}(b).
For energies $>7\ \mathrm{mJ}$ the curve flattens, similar to the measured THz energy (grey circles), indicating a possible reason for the onset of saturation of the THz conversion efficiency depicted in Fig.\ref{fig:EnergyTHzPlasma}(b).
Thus, for a GaSe crystal with a clear aperture of 7 mm, the maximum achievable energy of the two-color laser pulses is limited to 8.8 mJ in the current experiment.
However, nowadays, GaSe crystals with a diameter of 2 cm are commercially available, which would lead to an increased pump energy by 8 times before permanently damaging the crystal.

\subsection{THz Transient}
\label{sec:PlasmaTHzTransient}
\begin{figure}[htb]
    \centering
    \includegraphics[width=\linewidth]{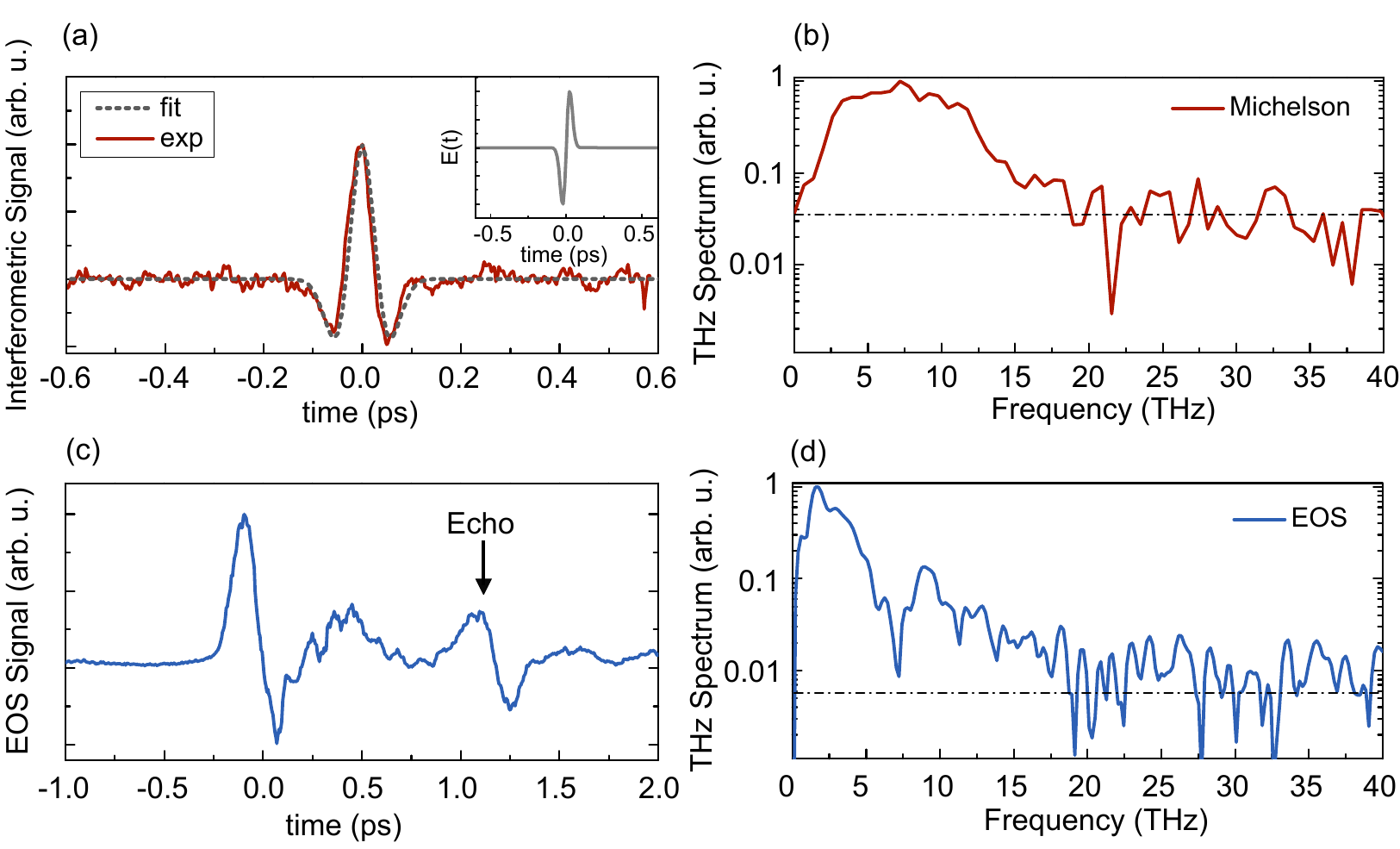}
    \caption[THz spectrum measured with a Michelson interferometer and electro-optical sampling]{(a) Autocorrelation signal measured with a Michelson interferometer (red solid line) and fitted interferometric signal (grey dashed line), calculated from the temporal pulse profile shown in the inset. (b) Related Fourier transformation. (c) EOS signal and (c) corresponding THz spectrum.} 
\label{fig:PlasmaTHzSpectra}
\end{figure}

The spectral characterization of the generated THz pulse is performed with two different methods, namely with EOS and with a field autocorrelation technique based on a Michelson interferometer.
Although the latter is an incoherent Fourier transform spectrometer which lacks phase information of  the THz transient, it allows to detect all spectral components, as discussed in section \ref{sec:Michelson}.
A schematic of the experimental setup is shown in Fig.\ref{fig:PlasmaSetup}.
After the two replica of the THz pulse are recombined on the pellicle beam splitter, the interfered THz radiation is focused on the pyro-electric detector by a parabolic mirror with a focal distance of 100 mm.
The same set of long pass filters, which were already used for energy and polarization measurements, are applied to protect the detector from saturation.
Figure \ref{fig:PlasmaTHzSpectra}(a) shows the recorded THz interferometric signal (red line) averaged over 5 consecutive scans.
The grey dashed line depicts a nonlinear curve fitting of the THz signal with the autocorrelation result (see Eq.\eqref{Equ:Michelson}) from the temporal profile of a THz pulse of \cite{TasosEfield:2016}
\begin{equation}
E_{\mathrm{t}}(t)=E_0\frac{t}{t_0}\exp\left(\frac{-t^2}{t_0^2}\right),
\end{equation}
where $E_0$ is the amplitude normalizing $E_{\mathrm{t}}(t)$ to unity and $t_0$ is the fitting parameter, accounting for \SI{32.91}{fs}.
The model THz electric field used to fit the autocorrelation signal is shown in the inset of Figure \ref{fig:PlasmaTHzSpectra}(a). 
The corresponding THz spectrum, obtained by Fourier transforming the recorded signal, is depicted in Fig.\ref{fig:PlasmaTHzSpectra}(b). 
The noise level is indicated by the dashed line.
The autocorrelation measurement reveals a spectral bandwidth of more than 15 THz, centred at around 7.5 THz.
Note that, detection of higher frequencies is prevented by the low resistivity black Silicon wafer installed in front of the detector (see the transmission spectrum in Fig.\ref{fig:Filter}).

In turn, EOS allows to measure the THz transient coherently, including phase information.
Nonetheless, in the case of a broadband THz pulse, detection of all spectral components is not trivial due to strong THz absorption in the EO crystal and phase mismatch between the THz and probe pulse, as reviewed in section \ref{sec:EOS}.
Here, the generated THz pulse is focused on a \SI{50}{\mu m} thick GaP crystal by a parabolic mirror with the focal distance of \SI{50}{mm}. 
The synchronised probe pulse (central wavelength of 680 nm and pulse duration of 40 fs), is spatially overlapped with the THz pulse by a focusing lens (f=150 mm) through a hole in the center of the parabolic mirror.
To ensure linear response of the GaP crystal, the intensity of the probe pulse is controlled with a neutral density filter and the THz intensity is reduced by two 5 mm thick HDPE filters and a pair of wire-grid polarizers.
To avoid water absorption from ambient air, the \ac{eos} setup was purged with Nitrogen resulting in a reduced humidity of 2.2\%.
The sampled THz transient is depicted in Fig.\ref{fig:PlasmaTHzSpectra}(c), averaged over 3 consecutive scans.  
The well pronounced post-pulse at > 1 ps is attributed to an echo caused by a surface reflection of the THz pulse at the GaP crystal.
The corresponding THz spectrum is shown in Fig.\ref{fig:PlasmaTHzSpectra}(c) (blue line), with a dashed line indicating the noise level  measured under the same experimental conditions with the THz beam blocked.
Due to the excitation of transverse-optical lattice vibration of the detection crystal, the detection bandwidth is limited mainly to the lower part of the THz spectrum (<8 THz). 
However, due to the small crystal thickness and the high dynamic range of the detection technique, higher frequencies (up to 20 THz) are resolved, indicating the ultra-broadband nature of our source.

In order to estimate the amplitude of the THz electric field, it is necessary to obtain data on the spatial beam shape and temporal pulse profile, as well as on the THz energy, as elucidated in section \ref{sec:Retrieval}.
The beam radius is measured with the knife edge method at the focal plane of a 50 mm parabolic mirror and accounts for \SI{59.3}{\mu m} at 1/$e$ level.
If the THz waveform recorded by EOS is used, the electric field strength can be approximated to $\sim$\SI{100}{MV/cm} for a THz energy of \SI{0.185}{mJ}.
However, since the detection bandwidth of this technique is limited, the retrieved THz pulse duration will be overestimated leading to the underestimation of the THz field amplitude.
Alternatively, the temporal profile $E_{\mathrm{t}}(t)$, retrieved from the Michelson interferometer, can be used for the time integral $G_{\mathrm{t}}$ in Eq.\eqref{equ:FielAmplidtude}, providing an upper limit of the THz field strength.  
Thus, for the two temporal shapes, it can be concluded that the THz electric field amplitude lies in between 100 and \SI{150}{MV/cm} with corresponding magnetic field strengths between 33 and \SI{50}{T}, which are, to the best of our knowledge, the highest values ever reported for a table-top THz source.

\section{Conclusion}
In summary, we have demonstrated for the first time, an extremely efficient generation of high power THz pulses from mid-IR two-color laser filaments, confirming theoretical predictions on the virtue of long wavelength driving sources.
An extremely high  THz conversion efficiency of 2.36\% could be achieved, which is  more than two orders of magnitude higher compared to typical values obtained with two-color pulses from Ti:sapphire lasers.
The single cycle THz pulse promotes energies of up to \SI{0.185}{mJ} and an electric field strength exceeding \SI{100}{MV/cm}.
We attribute the beneficial outcome to stronger photo-currents from larger ponderomotive forces, an increased electric field asymmetry of the two-color field, additional symmetry breaking due to generated low order harmonics, a larger plasma volume and smaller temporal walk-off between the fundamental, SH and THz pulse.

Nevertheless, despite the remarkable performance of the current THz source, the limits are not yet reached and further improvements can be expected.
For example, technical issues concern the use of spectral filters that absorb a part of the THz radiation, the control of the polarization of the driving laser and its second harmonic, and the energy losses in the SH NL crystal.
Theoretical simulations \cite{Fedorov:2018} predict multi-mJ THz pulses with peak electric and magnetic fields to the gigavolt per centimeter and kilotesla level if to merely increase the input laser pump energies in the range of a few tens of millijoules.
THz pulses generated in this work, and envisioned future THz sources driven by mid-IR pulses, pave the way for NL THz optics with extreme field strengths.

%% file: Chapters/XPM.tex
\chapter{Cross-phase Modulation and THz induced PL in Semiconductors}
\label{ch:XPM}
The extraordinary THz source presented in the previous chapter allows to enter the realm of NL THz optics, wherein THz radiation is used as a driver to trigger transient phase transitions or manipulate the four states of matter.

The following chapter represents  proof of principle experiments confirming the extreme THz fields sufficient to modulate the optical properties of bulk semiconductors.
Namely, we demonstrate modulation of a visible probe pulse through \acf{xpm} driven by strong THz fields in  a \SI{1}{mm} thick ZnTe crysta, and report on THz induced \acf{pl} in ZnTe as well as ZnSe.
The main results can be found in ref \citeA{KoulouklidisA:20} and  \citeB{THzInduced:19}.

\section{THz induced Cross-Phase Modulation}
\ac{xpm} refers to a change in optical phase of an electromagnetic wave due to the interaction with another wave in a NL medium.
This phenomena is typically observed in a NL Kerr medium exhibiting a large $\chi^{(3)}$ nonlinearity, in which the phase of a weak probe changes
proportionally to the intensity of a strong excitation pulse,
usually containing many oscillation cycles of optical fields.
The intensity $I^{(1)}$ of beam 1 thereby alters the refractive index for beam 2 by $\Delta n^{(2)}=2n_2I^{(1)}$, with $n_2$ as the NL index.
However, Shen and co-workers  \cite{ShenXOM:2007, ShenXOM:2010} first demonstrated a second-order XPM of a visible probe induced by a single cycle THz pulse in a \SI{0.5}{mm} thick ZnTe crystal. 
They attribute the observed spectral broadening and shifting of the probe pulse to the Pockels effect, which depends on the electric field strength rather than the intensity, and confirm their observation with a time-dependent phase shift-model.
The change in refractive index can be expressed as \cite{Boyd:2003} 
\begin{equation}
n=n_0+\Delta n_1+\Delta n_2, \quad \mathrm{with} \quad \Delta n_1=\frac{\chi^{(2)}E_{\mathrm{THz}(t)}}{n_0}  \quad \mathrm{and} \quad \Delta n_2=\frac{3\chi^{(3)}|E_{\mathrm{THz}}(t)|^2}{2n_0},
\label{equ:n0}
\end{equation}
where $n_0$ is the linear refractive index of the EO crystal, $\Delta n_1$ and $\Delta n_2$ are the change in refractive index caused by the Pockels and Kerr effect, respectively.
If temporal walk-off between the THz and probe pulse is neglected, the change in refractive index affects the phase of the probe as
\begin{equation}
\Delta \phi(t)=\frac{2\pi}{\lambda_0}\int_0^L\Delta n \left(t\right) dz,
\end{equation}
with $\lambda_0$ as the central wavelength  of the probe pulse, and $L$ the crystal thickness.
When the probe pulse is short compared to the time variation of the THz waveform, the phase retardation accumulated in the EO crystal can be decomposed in a Taylor expansion
\begin{equation}
\Delta \phi(t) =\Delta\phi(t_0)+\frac{d\Delta\phi}{dt}\bigg|_{t_0}(t-t_0)+\frac{1}{2}\frac{d^2\Delta\phi}{dt^2}\bigg|_{t_0}\left(t-t_0\right)^2+\cdots, 
\label{equ:TaylorPhase}
\end{equation}
where $t-t_0$ is limited to the temporal extent of the probe pulse, relative to the electric field at time $t_0$.
For an applied THz electric field of $E_{\mathrm{THz}}=5 \times 10^7$ \SI{}{V/m} Shen \textit{et al.}\cite{ShenXOM:2007} calculated a ratio $\Delta n_2/\Delta n_1 \approx 0.2$ for material parameters of ZnTe \cite{CaumesZnTe:2002} and hence neglected the  contribution from the Kerr effect.
As a consequence, for the Pockels effect only, Eq.\eqref{equ:TaylorPhase} in combination with Eq.\eqref{equ:n0} can be approximated to 
\begin{equation}
\Delta \phi(t) \propto E_{\mathrm{THz}}(t_0) + \frac{dE_{\mathrm{THz}}}{dt}\bigg|_{t_0}(t-t_0)+\frac{1}{2}\frac{d^2E_{\mathrm{THz}}}{dt^2}\bigg|_{t_0}\left(t-t_0\right)^2+\cdots,
\label{equ:TaylorPhase2}
\end{equation}
where $\Delta\phi(t_0)$ represents the phase retardation proportional to $E_{\mathrm{THz}}(t_0)$ measured with conventional EOS.
Because the instantaneous probe frequency is given by $\omega(t)=\omega_0-d\Delta\phi(t)/dt$, with $\omega_0$ as the carrier frequency, the second term in Eq.\eqref{equ:TaylorPhase} describes a frequency shift which is proportional to the first derivative of the THz field transient.
The third term corresponds to a linear chirp resulting in spectral broadening of the probe pulse. 
Accordingly, the time-shift model predicts a spectral broadening caused by the quadratic phase modulation occurring at the crest and troughs of the THz transient. 

Vicario and co-workers \cite{VicarioXPM:2017} explored for the first time THz induced XPM in a \SI{50}{\mu m} thick GaP crystal with significantly stronger THz field strengths in the order of \SI{25}{MV/cm} ($2.5 \times 10^9$ \SI{}{V/m}).
Due to the large THz field amplitude, in addition to the Pockels effect, phase retardation of the probe pulse is strongly assisted by the NL Kerr effect, resulting in spectral broadening of the probe pulse (centred at \SI{800}{nm}) by more than a factor of 5 at the maximum of the THz field.
Despite the large THz field, the time-shift model developed by Shen \textit{et al.}\cite{ShenXOM:2007} agrees surprisingly well with their experimental findings, wherein they observe a clear correlation between the shift of  the central wavelength and the first derivative of the THz field, while the maxima of the spectral width follow the THz field.
\begin{figure}[h]
    \centering
    \includegraphics[width=\linewidth]{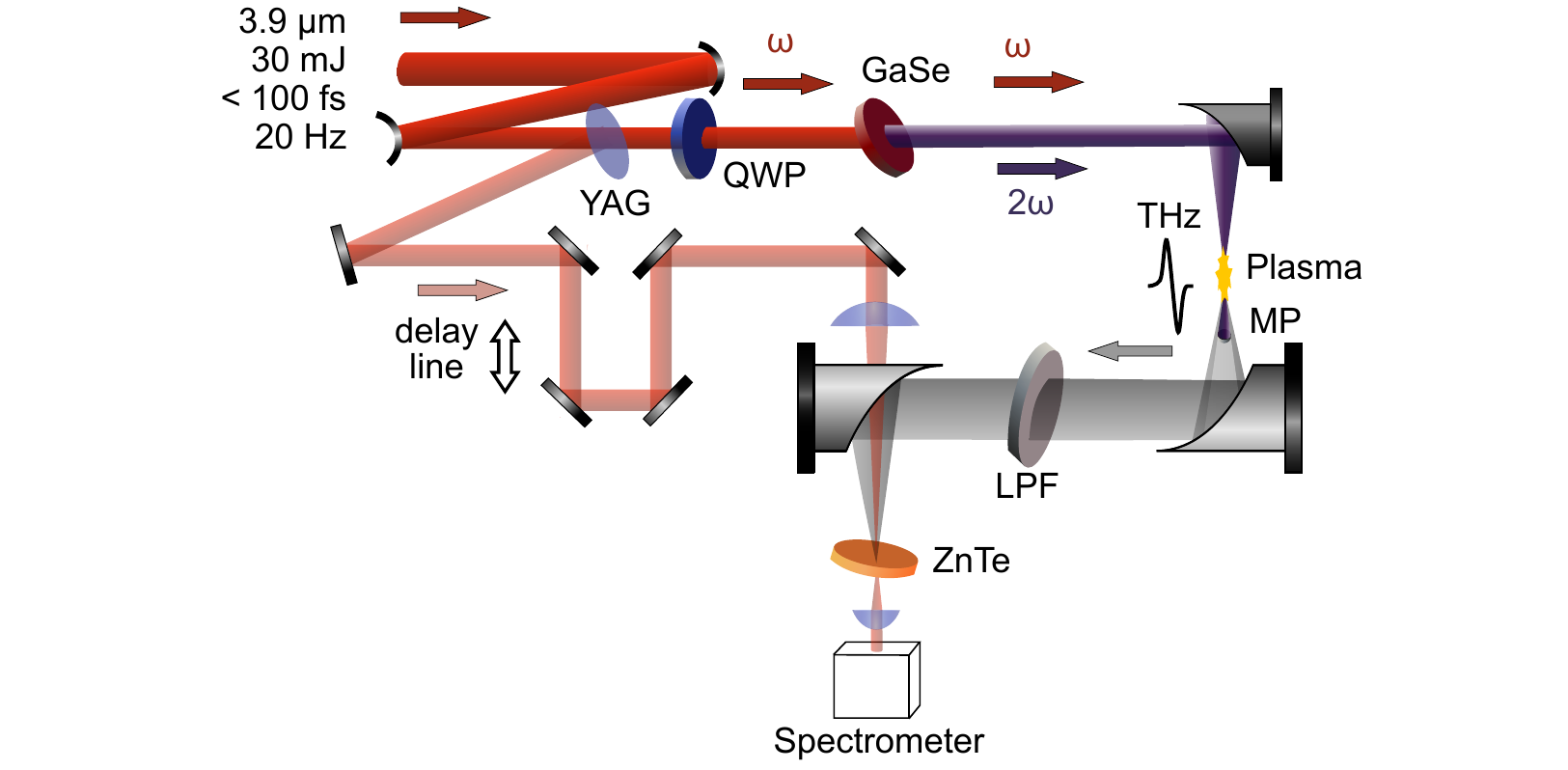}
    \caption[Experimental setup to observe THz induced XPM]{Experimental setup to observe THz induced XPM in a \SI{1}{mm} thick ZnTe crystal. A small fraction of the mid-IR pump, which accumulated low order harmonic generation during propagation in air, is split with a thin YAG plate in order to use it as a visible probe.} 
\label{fig:XPM_Setup}
\end{figure}

In this work, in order to illustrate the strength of the THz field generated by the mid-IR filament, we focus the THz pulse into a \SI{1}{mm} thick ZnTe crystal together with a synchronised probe pulse. 
The experimental setup is depicted in Fig.\ref{fig:XPM_Setup}.
After shaping the mid-IR beam profile with a set of spherical mirrors,
a fraction of the pump is reflected with a thin YAG plate, of which a spectral portion is used as a probe.
The probe pulse is generated in the vicinity of the fifth harmonic of the mid-IR pump during propagation in air, with a central wavelength of \SI{761}{nm}.
In order to obtain the largest possible THz field strength at the ZnTe crystal, we reduce the number of long pass filters and use only  a single 5 mm thick HDPE plate, resulting in an estimated peak amplitude of \SI{57}{MV/cm}.
The large elctric field leads to a ratio between the refractive index change caused by the Pockels and Kerr effect of $\Delta n_2/\Delta n_1 \approx 28.5$, indicating a large contribution from a third order NL phase retardation.
The time delay between THz and probe pulse is controlled with a high precision delay stage.
The probe pulse is collected with a \SI{15}{mm} lens after the ZnTe crystal and send to a spectrometer.
Prior to the experiments on THz induced XPM, the THz transient is recorded with EOS in the same \SI{1}{mm} thick ZnTe crystal, but with an attenuated THz radiation and probe pulse centred at \SI{680}{nm}, as described in section \ref{sec:PlasmaTHzTransient}.
\begin{figure}[h]
    \centering
    \includegraphics[width=\linewidth]{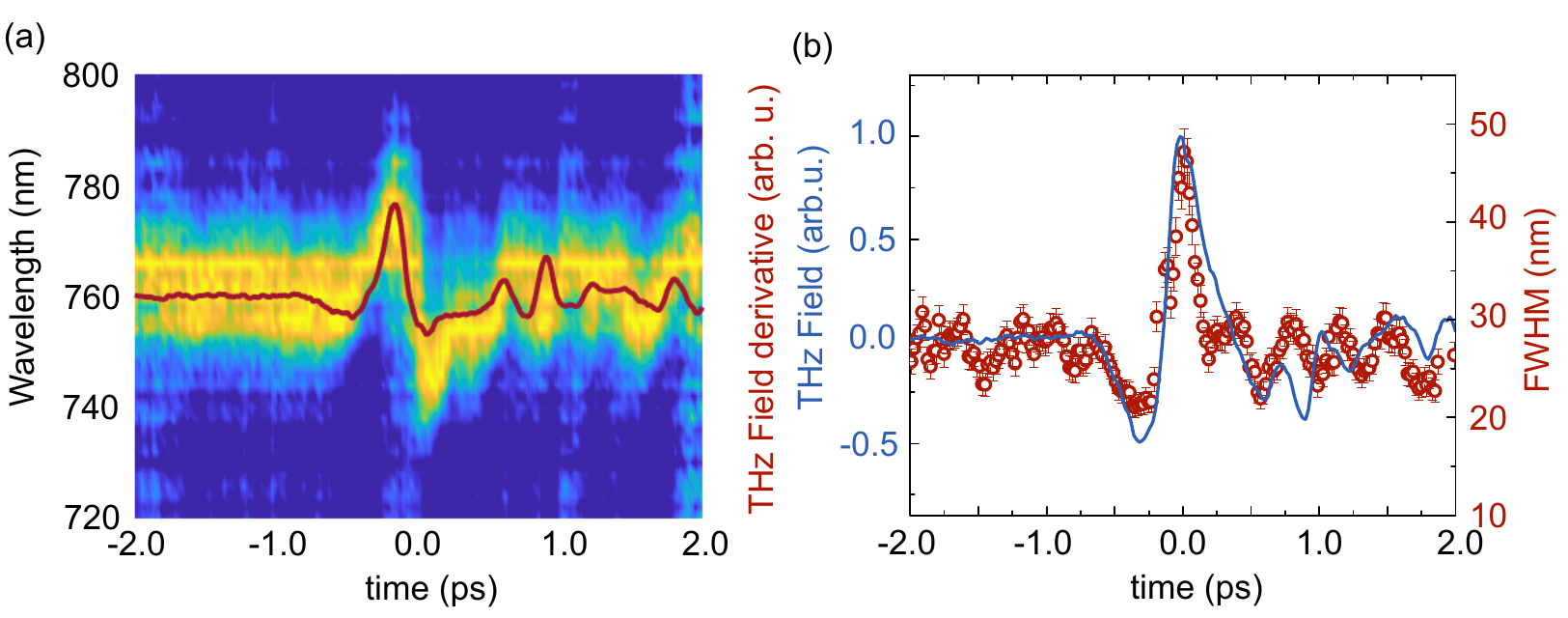}
    \caption[Probe spectra with respect to the time delay between THz and probe pulse, experiencing XPM]{(a) Probe spectra with respect to the time delay between THz and probe pulse. The spectral shift of the central wavelength follows the first derivative of the THz field (red line). (b) THz transient (blue line) measured with the same 1 mm thick ZnTe crystal and spectral width (red dots) of the probe pulse as a function of the time delay between the THz and probe pulses.} 
\label{fig:XPM_Trace}
\end{figure}

Fig.\ref{fig:XPM_Trace}(a) shows the spectrum of the probe pulse with respect the relative time delay between the probe and the THz pulse, exhibiting obvious spectral modulations when the two pulses overlap.
As predicted by the time-shift model, the spectral shift clearly follows the first derivative of the THz transient, shown as red overlay in Fig.\ref{fig:XPM_Trace}(a), wherein the central frequency of the probe pulse initially undergoes a red shift which then gives way to a blue shift, both about \SI{12}{nm}.
In agreement with observations reported in ref \cite{VicarioXPM:2017}, spectral broadening (as it follows from the FWHM of the spectrum) of the probe pulse occurs at the crest and troughs of the THz transient, as shown in Fig.\ref{fig:XPM_Trace}(b).
Despite the large THz field applied to the crystal, the spectral width is broadened by less than a factor of 2 (as compared to a factor of 5 reported in ref \cite{VicarioXPM:2017}).
A possible reason for the reduced broadening and spectral shift can be attributed to a significant walk-off between the THz and probe pulse in the thick NL crystal.
With imperfect group-velocity matching, the probe pulse slips across different portions of THz field such that phase modulation becomes degraded. 
For ZnTe, the calculated walk-off parameter is \SI{1.05}{ps/mm} \cite{ShenXOM:2007} and the measured value is \SI{0.4}{ps/mm}\cite{Wu:1996}, in the case of a \SI{790}{nm} probe and \SI{1}{THz} pump pulse.
Thus, for a \SI{1}{mm} thick crystal and single cycle THz pulse with a duration of less than \SI{1}{ps}, the THz and probe pulse will be almost entirely separated in time.
Nonetheless, regardless of the detrimental material parameters, the presented THz induced spectral modulation of the visible probe pulse demonstrates the strength of the generated THz trasnient and feasibility for NL THz spectroscopy. 

\section{THz induced Photoluminescence in Semiconductors}

THz induced PL has already been reported in colloidal \acp{qd}  \cite{Pein:2017} deposited on a micro slit array used to enhance the incident THz electric field, and in a cryogenically cooled GaAs quantum well (QW) structure \cite{HiroriPL:2011}.
In the first case, the THz electric field causes a voltage drop between adjacent QDs. 
If this potential difference between the  neighbouring \acp{qd} is equal to or greater than the QD band gap, valence band electrons can tunnel into the conduction band of neighbouring QDs leading to a population of electrons and holes which, when paired in a QD, lead to exciton formation and subsequent luminescence.
\begin{figure}[htb]
    \centering
    \includegraphics[width=\linewidth]{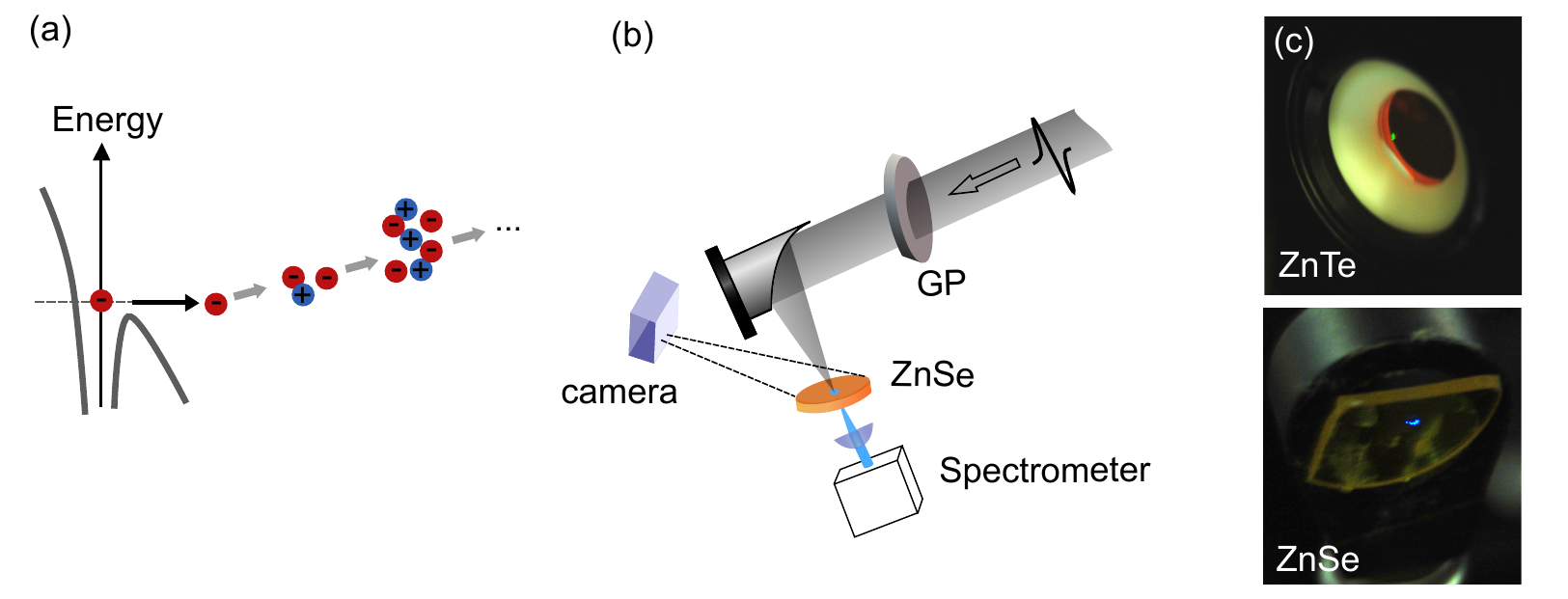}
    \caption[Experimental setup for THz induced PL in ZnSe]{(a) Schematic of field induced impact ionization. (b) Experimental setup for THz induced PL in ZnSe. The THz field amplitude is controlled with a pair of grid polarizer (GP). Luminescence spectra are recorded behind the crystal, pictures are taken from the front side. (c) Images of THz induced PL emission from ZnTe (top) and ZnSe (bottom) crystals taken by a digital camera.} 
\label{fig:PLinduced}
\end{figure}

In the latter case of GaAs QWs, the detected near-IR luminescence is attributed to impact ionization initiated by an intense near-half cycle THz transient.
A schematic of the process is shown in Fig.\ref{fig:PLinduced}(a).
In the vicinity of an impurity donor, the large ponderomotive energies (i.e. mean kinetic energy of a charged particle in an oscillating electric field ) associated with high-field THz pulses (proportional to $\lambda ^2$) can modify the interatomic potential such that electrons tunnel into the continuum \cite{HiroriPL:2011, WenPL:2008, Nordstrom:1998}.
Such a freed electron in the conduction band, with kinetic energy gained from the electric field, can create an \textit{e-h} pair by losing energy when it knocks out a bound electron from its bound state ($e_1\rightarrow e_1'+e_2+h_2$).
Subsequently, both the electron that lost energy to create the \textit{e-h} pair and the created \textit{e-h} pair can, in turn, gain energy from the field and contribute to another ionization event.
Such impact ionization and thus carrier multiplication will continue as long as the energy of the freed electrons is larger than the band gab energy.
Finally, the impulsive THz electric field allows the unbound \textit{e-h} pairs to form excitons which decay radiatively because the electric field that would otherwise ionize the exciton is absent just after the pairs are generated \cite{HiroriExciton:2010, Watanabe:2011,Miller:1985}.

In this work, we demonstrate for the first time, THz induced PL in two direct band gap semiconductors, namely in bulk ZnTe and ZnSe with a thickness of 1 mm, without any field enhancing structure at room temperature. 
The experimental setup is shown in Fig.\ref{fig:PLinduced}(b), wherein THz radiation is simply focused with a parabolic mirror (f = 50 mm), resulting in a maximum THz field strength of \SI{57}{MV/cm} at the crystal position.
Due to the surprisingly bright PL radiation, the emission can be recorded with a conventional digital camera (see Fig.\ref{fig:PLinduced}(c)). 
PL spectra are collected by a short lens (f = 15 mm) and guided to the spectrometer.
Because of the large THz absorption, only emission from the crystal surface can be collected.
A pair of wire-grid polarizers is used to control the THz electric field strength.
Similar to the previous chapter, in order to verify that the PL emission solely originates from THz radiation, we suppress THz generation either by detuning the grating compressor of the OPCPA system, or the second harmonic crystal is misaligned from the optimal conditions for efficient THz generation. PL emission disappears in both cases.
\begin{figure}[htb]
    \centering
    \includegraphics[width=\linewidth]{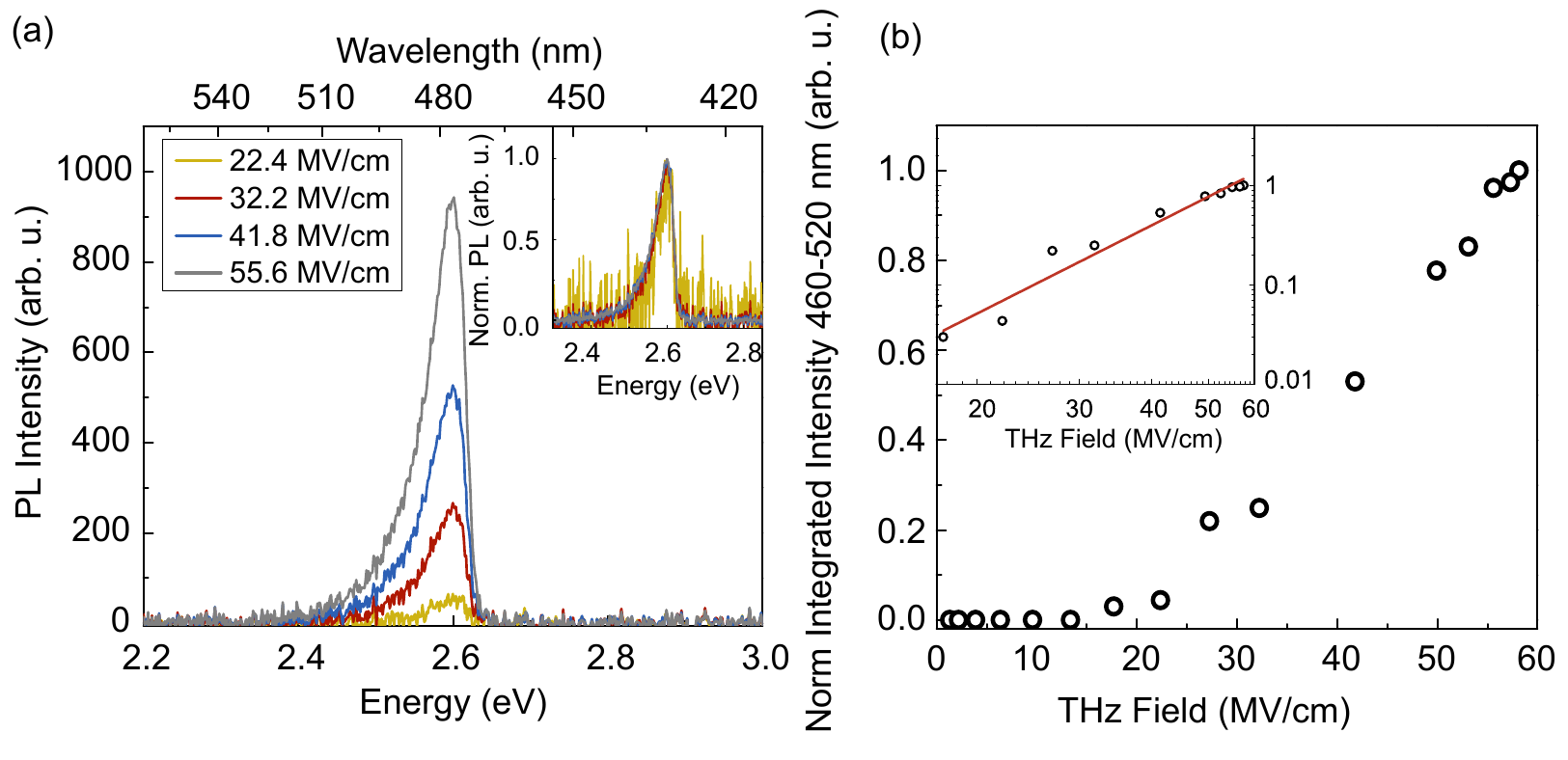}
    \caption[THz induced PL in a \SI{1}{mm}  ZnSe crystal for different electric field strengths]{(a) THz induced PL in a \SI{1}{mm}  ZnSe crystal for different electric field strengths (see legend) and corresponding normalized spectra in the inset.  (b) PL spectra as shown in (a) integrated from \SI{460}{nm} to \SI{520}{nm} and normalized to 1. The inset depicts a log-log scale which reveals a field dependence of $\propto E_{\mathrm{THz}}^3$.
    } 
\label{fig:PLinducedSpectra}
\end{figure}

Although the central photon energy of  the THz pulse ($\sim$ \SI{31}{meV}) is about 87 times lower than the band gap energy of ZnSe (\SI{2.7}{eV}),
Fig.\ref{fig:PLinducedSpectra}(a) depicts PL spectra for different THz field amplitudes (see legend) and corresponding normalized spectra in the inset. 
The spectral shape does not change with respect to the field strength, suggesting that excitation phenomena arising from dense electron-hole plasma inside the crystal do not occur at these excitation intensities \cite{PLshift:1987}.
The maximum PL emission at \SI{2.6}{eV} (\SI{477}{nm}), which is Stoke shifted from the band gap energy by \SI{100}{meV}, increases as a function of the applied THz field. 
As the luminescence is proportional to the created \textit{e-h} pair density \cite{Kaindl:2009}, the peak PL intensity is plotted in Fig.\ref{fig:PLinducedSpectra}(b) with respect to the THz field strength.
In the case of a GaAs quantum well structure measured at a temperature of 10 K, Hirori \textit{et al.} \cite{HiroriPL:2011} reported on a carrier density $N$  with an extreme NL dependence on the THz field of $N \propto E_{\mathrm{THz}}^8$.
Such a strong nonlinearity is not observed in this work.
Despite the different dimensionality of the samples (\textit{i.e.} 2D quantum well compared to 3D bulk material), we attribute the discrepancy to a dominant free-carrier THz absorption from thermally-excited carriers at room temperature.
Moreover, Hirori and co-workers further showed a drastic decrease of luminescence signal for an increasing temperature, measurable up to 140 K.
Thus, the observed luminescence spectra in this work at room temperature in bulk ZnSe are an unambiguous evidence of the extreme THz field strength generated with mid-IR two-color plasma filaments and paves the way for further investigations on NL strong field induced carrier dynamics in semiconductor materials.

%% file: Chapters/Summary.tex
\chapter{Summary and Conclusion}
\label{ch:Sum}

In summary,  this work illustrates the great potential of  mid-IR drivers to generate intense THz radiation by \ac{or} in organic crystals or by inducing currents in two-color plasma filaments,  and demonstrates their capability to prompt THz induced NL phenomena in confined materials such as quantum dots as well as in bulk semiconductors.

When the organic crystal DAST is pumped with fs-pulses centred at either \SI{3.9}{\mu m} or its \ac{sh} at \SI{1.95}{\mu m},  \ac{mpa} is suppressed for pump energy densities of more than \SI{100}{mJ/cm^2}  and \SI{80}{mJ/cm^2},  respectively,  which is almost five times higher than the crystal damage threshold for pump pulses generated by conventional drivers operating  at the telecommunication wavelength of  \SI{1.5}{\mu m}.
Despite the high linear absorption at \SI{3.9}{\mu m},   a surprisingly high conversion efficiencie of 1.5 \% without a distinct onset of saturation is observed.
The remarkable conversion efficiency is attributed to a possible resonantly enhanced  electro-optic coefficient in the vicinity of \SI{3.9}{\mu m},  which is evident by the strong linear absorption. 
The combination of large THz energies of more than \SI{100}{\mu J},  a broad spectrum of \SI{4.2}{THz}  at FWHM centred at \SI{2.1}{THz} and small focused beam radius of  \SI{107}{\mu m}  ensues electric field strengths exceeding  \SI{40}{MV/cm}.

In  the case of a \SI{1.95}{\mu m} driving pulses, optical parameters such as linear transmission,  electro-optic coefficient and phase matching conditions are comparable to \SI{1.5}{\mu m} pump sources.
However, the suppression of \ac{mpa}  allows to pump the crystal with significantly higher energy densities. 
Together with cascaded \ac{or} , it provokes extraordinarily high THz conversion efficiencies approaching 6\%.
Cascaded effects are thereby identified as both, a mechanism through which a high conversion efficiency is achieved and a limiting factor for farther THz generation causing an onset of saturation, when the newly generated spectral components of the driving pulse promote phase mismatch between the group velocity of the pump and THz phase velocity.
Nonetheless,  the observed conversion efficiency  is to the best of our knowledge, the highest value ever reported  for \ac{or}  at room temperature. 

Applying the generated THz radiation to  heterostructure semiconductor materials, we report on THz induced ultrafast \ac{ea} switching.
\ac{ea} modulators based on the \ac{qcse} are critical for increasing the speed of optical networks. 
Tbit/s interconnect rates per single channel should become feasible with THz waveforms applied as a driving field. 
In a nano scale device, the required absorption modulation depth can only be induced by THz field strengths  in the order of MV/cm. 
Due to the lack of intense THz emitters, up until recently, a THz-induced Stark-effect could only be demonstrated in local field enhancement geometries.
Using our \SI{3.9}{\mu m}  driven THz source, we succeed, for the first time, in a direct all-optical encoding of a free-space THz-signal onto an optical signal that probes the absorption of CdSe/CdS core/shell \acp{qd}. 
The simplified geometry, that strips the system of numerous artefacts found in enhancement geometries, such as electrode-driven charge injection, enables us to map the undistorted ultrafast carrier dynamics.
A beneficial energy band alignment in heterostructure nano-crystals enables a 15\% transmission modulation, the highest value ever reported for solution-processed materials at room temperature. 
By changing the parameters of the applied THz waveform, we demonstrate the feasibility to control the amplitude of the optical probe with a contrast of more than 6 dB, which corresponds to the performance of state-of-the art quantum-well \ac{ea} modulators with electric current injection.
Furthermore, after eliminating the complications introduced by the previously indispensable field enhancement structures, we were able to adopt a simple and intuitive theoretical model which matches the experimental data remarkable well.
This allows us to explain the experimental findings as well as to propose a route to further  improve the signal contrast and transmission modulation, namely by utilizing \acp{qd} with thicker shells in order to exploit a type-II energy band alignment for next generation ultrafast optical switches.

We further report on THz generation in two-color mid-IR laser filaments.
Although theoretical predictions  anticipate large optical-to THz conversion efficiencies for long wavelength drivers since more than a decade, experimental proof was still missing due to the lack of intense mid-IR pump sources.
In this work, we use the output of a home-built high power \ac{opcpa} system operating at a wavelength of  \SI{3.9}{\mu m}, combined with its \ac{sh} to produce a plasma filament in ambient air.
In our experiments with mid-IR two-color laser filaments we observe the generation of \SI{0.185}{mJ} single cycle THz pulses with unprecedented THz conversion efficiency of 2.36\%, which is more than two orders of magnitude higher as compared to typical values obtained with Ti:Sapphire lasers operating at a fundamental wavelength of \SI{0.8}{\mu m}.
The generated electric field strength exceeds \SI{100}{MV/cm}, which is the highest value ever reported for a table-top THz source.
The experimental findings are supported by theoretical simulations, which attribute the beneficial outcome  to stronger photo-currents from larger ponderomotive forces and increased electric field asymmetry of the two-color field generated by mid-IR pulses. 
Additional field symmetry breaking is caused by the generated odd harmonics when the fundamental pulse propagates in air. 
Furthermore, because the critical power of self-focusing scale as $\lambda^2$ and is thus  larger for long-wavelength driving pulses, it is possible to deliver more energy to a single filament, exhibiting a wider and longer plasma string volume. 
In combination with a smaller temporal walk-off between the fundamental, SH and THz wave, this leads to an additional increase of the THz signal.
Despite the large THz energies already attained in the current setting, even higher values can be expected if to overcome  technical limitations, such as spectral filters that absorb a part of the THz radiation, difficulties in controlling  the polarization of the driving laser and its \ac{sh}, as well as limited aperture size  and energy losses in the \ac{sh} nonlinear crystal.

Finally, in order to further confirm the exceptional THz field strength generated in mid-IR two-color plasma filaments, we conduct two proof of principle experiments which illustrate the capability to modulate the optical properties of bulk semiconductors.
For the first demonstration, we focus the THz pulse with a field amplitude of \SI{57}{MV/cm} into a \SI{1}{mm} thick ZnTe crystal, together with a visible probe centred at  \SI{761}{nm}. 
When the THz and probe pulses overlap in time, the spectrum of the probe pulse becomes strongly modulated.
We observe a shift of the central wavelength to the blue and red side side of about \SI{12}{nm}, following the first derivative of the THz transient,  as well as a change of the spectral width by almost a factor of two with respect to the THz field strength.
The observed spectral transformation is associated with THz induced XPM originating from the Pockels as well es NL Kerr effect.
In the second example, we present for the first time, excitonic PL from two bulk semiconductors (ZnSe and ZnTe) induced by intense free space THz fields.
The bright PL emission from both semiconductors is strong enough to be detected by a simple CCD camera and scales nonlinearly with the incident THz field, which is attributed to impact ionization and subsequent radiative decay of electron-hole recombination.

Thus, the achieved THz radiation generated by intense mid-IR pulses and consecutive applications in this work set new milestones in the next frontier of NL THz spectroscopy.
The presented results touch upon several fast-developing areas of the current technological revolution in optical science, such as non-linear THz optics, high-speed optical communication systems, and nano device technology.
Moreover,  the attained quasi-static ultrashort electric and magnetic bursts with extreme field strengths can enable free space extreme nonlinear and relativistic science.

%% file: Struktur/acknow.tex
\addchap{Acknowledgement}
\vspace{0.5cm}

First and foremost I would like to thank my advisor Dr. Audrius Pug\v{z}lys who taught me how to fight with and against our home-built high-power mid-IR laser system. 
I appreciate fruitful discussions and invaluable advice, his helping hand  in the lab, his tricks and workarounds,  as well as his continuous support for conference proceedings and journal publications.
 I would like to express my gratitude to the head of our research group Univ. Prof. Dr. Andrius Baltu\v{s}ka for giving me the opportunity to work in such an exciting research area, for his help on major publications and for the possibility to attend and participate in a large number of international conferences.
Although seemingly in the background, another highly appreciated supportive hand was given by the former head of the Photonics Institute Ao.Univ.Prof.i.R. Dr. Georg Reider, who always encouraged me to trust in my scientific abilities.   
Furthermore, I have to thank my former advisor of my Master thesis Prof. Dr. Wolfgang Heiss, who ignited my enthusiasm for fundamental science.
   
During my PhD studies, I was fortunate to not only learn and gather experience from my advisor and colleagues,  but to further be part of many fascinating, international and interdisciplinary collaborations, which resulted in interesting findings presented in this thesis.\\
I would like to express my appreciation to the research group of Prof. Dr. Stelios Tzortzakis from the University of Crete and Texas A\&M University at Qatar, who initiated the project on THz generation in mid-IR  two-color plasma filamentation and took the key role in the publishing process of the experimental outcome.
Moreover, I am particularly grateful to Dr. Anastasios D. Koulouklidis for giving me first insights on THz generation and characterization, for teaching me handy tricks at the setup and for his endless effort during the experimental campaign, including collective night shifts in the lab.\\
I would also like to thank Dr. Mostafa Shalaby, CEO of Swiss Terahertz and PI at Beijing Key THz Lab, for his collaboration on THz generation by optical rectification of mid-IR pulses in DAST, for providing the organic crystals, THz filters and additional crucial experimental components, as well as for conducive discussions and advise.\\
I further want to acknowledge the research group of Prof. Dr. Maksym Kovalenko from ETH Zürich for the collaboration on THz induced electro-absorption modulation in colloidal quantum dots.
Especially Dr. Dmitry Dirin for the synthesis of CdSe/CdS quantum dots, and Dr. Simon Christian Böhme for elucidative discussions and support during the writing process of the manuscript.
In addition, I'd like to thank Dominik Kreil from JKU Linz for his theoretical investigations and simulations on the quantum confined Stark effect in heterostructure quantum dots with respect to the energy band alignment. \\
Several experiments and collaborations were outside of the scope of this thesis, but nevertheless played an important role in my education and understanding in applications of high-power laser systems. 
Among those cooperation, I'm especially thankful to Dr. Linus Feder, Dr. Daniel Woodbury and Prof. Pavel Polynkin.

Throughout the time at the Photonics Institute, I furthermore received a lot of support from all of my colleagues. 
Along with others, I would like to address special recognition to Rokas Jutas for his contribution on electro-absorption modulation in colloidal quantum dots and for his companionship in the lab. Ignas Astrauskas,  Edgar Kaksis and Tobias Flöry for their  technical support and shared equipment, 
Andreas Klein as our former in-house electrical engineer, as well as Melanie Molnar and Christina Huber - without their administrative work the Photonics Institute would  fall apart. 

Last but not least, I would like to express my heartfelt gratitude to my family and friends, with special thanks to my father for his unconditional support.
In addition, I can't help but have to thank my  friends, who backed me up during so many years.
Including, but not limited to, Arevik Hakobian for always being smarter, Gayaneh Issayan who taught me to toughen up, Mageritha Matzer for being a feisty companion and always ready for adventures, Joachim Schütz for always being there for me no matter what, Dominik Kreil for continuously challenging me and never getting tired of explaining a theorist's world to an experimental physicist, Tina Mairhuber for sharing the first beer at the university with me and for those which followed, and Laura Stark for her endless love, for making Vienna a home and for watching my back when ever things got rough.